\RequirePackage{ifpdf}
\ifpdf 
\documentclass[pdftex]{sigma}
\else
\documentclass{sigma}
\fi

\usepackage{mathrsfs}
\usepackage{amsfonts,upgreek}

\def\bm{\fam9}

\def\beq{\begin{equation}}
\def\eeq{\end{equation}}
\def\beqa{\begin{eqnarray}}
\def\eeqa{\end{eqnarray}}
\newcommand\BA{\begin{array}}
\newcommand\EA{\end{array}}

\def\D{{\bm D}}
\def\F{{\bm F}}

\def\ve{{\bm  e}}

\def\vr{{\bm  r}}

\def\a{{\alpha }}
\def\b{{\beta }}
\def\g{{\gamma }}
\def\vr{{\bm  r}}

\newcommand{\e}{{\sf \epsilon
\hspace*{-0.4ex}\rule{0.08ex}%
{0.8ex}\hspace*{0.5ex}}}
\newcommand{\se}{{\sf \epsilon
\hspace*{-0.4ex}\rule{0.08ex}%
{0.6ex}\hspace*{0.4ex}}}

\numberwithin{equation}{section}

\begin{document}

\allowdisplaybreaks

\renewcommand{\thefootnote}{$\star$}

\renewcommand{\PaperNumber}{009}

\FirstPageHeading

\ShortArticleName{Self-Consistent-Field Method and $\tau$-Functional Method}

\ArticleName{Self-Consistent-Field Method\\ and $\boldsymbol{\tau}$-Functional Method
on Group Manifold\\ in Soliton Theory:
a Review and New Results\footnote{This paper is a
contribution to the Special Issue on Kac--Moody Algebras and Applications. The
full collection is available at
\href{http://www.emis.de/journals/SIGMA/Kac-Moody_algebras.html}{http://www.emis.de/journals/SIGMA/Kac-Moody{\_}algebras.html}}}

\Author{Seiya NISHIYAMA~$^\dag$,
Jo\~ao da PROVID\^{E}NCIA~$^\dag$,
Constan\c{c}a PROVID\^{E}NCIA~$^\dag$,\\
Fl\'avio CORDEIRO~$^\ddag$ and Takao KOMATSU~$^\S$}

\AuthorNameForHeading{S. Nishiyama, J. da Provid\^{e}ncia,
C. Provid\^{e}ncia, F. Cordeiro and T. Komatsu}

\Address{$^\dag$~Centro de F\'\i sica Te\' orica,
Departamento de F\'\i sica, Universidade de Coimbra,\\
\hphantom{$^\dag$}~P-3004-516 Coimbra, Portugal}
\EmailD{\href{mailto:seikoaquarius@ybb.ne.jp}{seikoaquarius@ybb.ne.jp}, \href{mailto:providencia@teor.fis.uc.pt}{providencia@teor.fis.uc.pt},
\href{mailto:cp@teor.fis.uc.pt}{cp@teor.fis.uc.pt}}

\Address{$^\ddag$~Mathematical Institute, Oxford OX1 3LB, UK}
\EmailD{\href{mailto:cordeiro@maths.ox.ac.uk}{cordeiro@maths.ox.ac.uk}}
\Address{$^\S$~3-29-12 Shioya-cho, Tarumi-ku, Kobe 655-0872, Japan}
\EmailD{\href{mailto:tkomatu@imail.plala.or.jp}{tkomatu@imail.plala.or.jp}}

\ArticleDates{Received September 05, 2008, in f\/inal form January 10,
2009; Published online January 22, 2009}

\Abstract{The maximally-decoupled method
has been considered as a theory to apply an basic idea of
an integrability condition
to certain multiple parametrized symmetries.
The method is regarded as a mathematical tool
to describe a symmetry of a collective subma\-ni\-fold
in which a canonicity condition
makes the collective variables
to be an orthogonal coordinate-system.
For this aim
we adopt a concept of curvature unfamiliar
in the conventional time-dependent (TD) self-consistent f\/ield (SCF) theory.
Our basic idea lies in the introduction of a sort of
Lagrange manner familiar to f\/luid dynamics to
describe a collective coordinate-system.
This manner enables us to take a one-form
which is linearly composed of a TD SCF Hamiltonian and
inf\/initesimal generators induced by
collective variable dif\/ferentials of a canonical transformation
on a group.
The integrability condition of the system read
the curvature $C= 0$.
Our method is constructed manifesting itself
the structure of the group under consideration.
To go beyond the maximaly-decoupled method,
we have aimed to construct an SCF theory, i.e.,
$\upsilon$ (external parameter)-dependent Hartree--Fock (HF) theory.
Toward such an ultimate goal,
the $\upsilon$-HF theory has been reconstructed on
an af\/f\/ine Kac--Moody algebra along the soliton theory,
using inf\/inite-dimensional fermion.
An inf\/inite-dimensional fermion operator is introduced through
a Laurent expansion of f\/inite-dimensional fermion operators
with respect to degrees of freedom of the fermions
related to a~$\upsilon$-dependent potential with a $\Upsilon$-periodicity.
A bilinear equation for the $\upsilon$-HF theory
has been transcribed
onto the corresponding $\tau$-function
using the regular representation for the group and the Schur-polynomials.
The $\upsilon$-HF SCF theory on an inf\/inite-dimensional Fock space
$F_\infty$
leads to a dynamics on an inf\/inite-dimensional Grassmannian~${\rm Gr}_\infty$
and may describe more precisely such a dynamics on the group manifold.
A f\/inite-dimensional Grassmannian
is identif\/ied with a~${\rm Gr}_\infty$
which is af\/f\/iliated with the group manifold obtained
by reducting $gl(\infty)$ to~$sl(N)$ and $su(N)$.
As an illustration
we will study an inf\/inite-dimensional matrix model
extended from the f\/inite-dimensional $su(2)$
Lipkin--Meshkov--Glick model
which is a famous exactly-solvable model.
}


\Classification{37K10; 37K30; 37K40; 37K65}

{\renewcommand{\baselinestretch}{0.9}\small
\tableofcontents

}

\newpage

\renewcommand{\baselinestretch}{1.0}


\renewcommand{\thefootnote}{\arabic{footnote}}
\setcounter{footnote}{0}


\begin{flushright}
\begin{minipage}{14cm}
\small
An original version of this work
was f\/irst presented by
S.~Nishiyama at the {\it Sixth International Wigner Symposium}
held in Bogazici University,
Istanbul, Turkey,  August 16--22, 1999
\cite{WigSym99}
and has been presented by
S.~Nishiyama at the {\it YITP Workshop on
Fundamental Problems and Applications of
Quantum Field Theory ``Topological aspects of quantum field theory''}
Yukawa Institute for Theoretical Physics, Kyoto University,
December 14--16, 2006.
A preliminary version of
the recent work has been presented by S.~Nishiyama
at the {\it KEK String Workshop} (poster session) 2008
held at
High Energy Accelerator Research Organization, KEK, March 4--6, 2008.
\end{minipage}
\end{flushright}

\section{Introduction}\label{section1}

\subsection{Historical background on microscopic study of nuclear collective motions}\label{section1.1}

\looseness=1
A standard description of
fermion many-body systems
starts with the most basic approxi\-mation
that is based on an independent-particle (IP) picture, i.e.,
self-consistent-f\/ield (SCF) approxi\-mation for the fermions.
The Hartree--Fock theory (HFT) is
typical one of such an approximation
for ground states of the systems.
Excited states are treated with the random phase approximation (RPA).
The time dependent Hartree--Fock (TDHF) equation
and
time dependent Hartree--Bogoliubov (TDHB) equation
are nonlinear equations owing to their SCF characters
and may have no unique solution.
The HFT and HBT are given
by variational method to optimize energy expectation value
by a Slater determinant (S-det) and an HB wave function,
respectively~\cite{RS.80}.
Particle-hole (p-h) operators
of the fermions with $N$ single-particle states
form a~Lie algebra~$u(N)$~\cite{Fu.Int.J.Quantum.Chem.81}
and generate a Thouless transformation~\cite{Th.60}
which induces a representation of the corresponding Lie group $U(N)$.
The $U(N)$ canonical transformation
transforms an S-det with $M$ particles to another S-det.
Any S-det is obtained
by such a transformation of a~given reference S-det, i.e.,
Thouless theorem
provides an exact wave function of fermion state vector which
is the generalized coherent state representation (CS rep)
of $U(N)$ Lie group
\cite{Pere.72}.
Following Yamamura and Kuriyama
\cite{YK.87},
we give a brief history of
methods extracting collective motions out
of fully parametrized TDHF/TDHB manifolds in SCF.
Arvieu and Veneroni, and Baranger and independently Marumori
have proposed a theory for spherical even nuclei
\cite{AM.60}
called quasi-particle RPA (QRPA)
and it has been a standard approximation
for the excited states of the systems.
In nuclei,
there exist a short-range correlation and a long-range one
\cite{Mo.59}.
The former is induced by a pairing interaction
and generates a superconducting state.
The excited state is classif\/ied by a seniority-scheme
and described in terms of quasi-particles given by
the BCS-Bogoliubov theory~\cite{Bogo.59}.
The latter is occured by p-h interactions
and gives rise to collective motions
related to a~density f\/luctuation around equilibrium states.
The p-h RPA (RPA) describes such collective motions
like vibrational and rotational motions.
It, however, stands on a~harmonic approximation
and should be extended to take some nonlinear ef\/fects into account.
To solve such a problem,
the boson expansion HB theory (BEHBT)
has been developed by Belyaev and Zelevinsky
\cite{BZ.62},
and Marumori, Yamamura and Tokunaga
\cite{MYT.64}.
The essence of the BEHBT is to express the fermion-pairs
in terms of boson operators keeping a pure boson-character.
The boson representation
is constructed to reproduce the Lie algebra of the fermion-pairs.
The state vector in the fermion Fock space corresponds to
the one in the boson Fock space by one-to-one mapping.
Such a boson representation
makes any transition-matrix-valued quantity
for the boson-state vectors coincide with
that for the fermion-state ones.
The algebra of fermion-pairs
and the boson representation
have been extensively investigated.
The fermion-pairs form an algebra $so(2N)$.
As for the boson representation,
e.g.,
da Provid\^encia and Weneser and Marshalek
\cite{PWM.68}
have proposed boson operators
basing on
p-h pairs forming an algebra $su(N)$.
By Fukutome, Yamamura and Nishiyama
\cite{YN.76,FYN.77},
the fermions were found
to span the algebras $so(2N \!+\! 1)$ and $so(2N \!+\! 2)$
accompanying with $u(N \!+\! 1)$.
The BET expressed by Schwinger-type and Dyson-type bosons
has been intensively studied by Fukutome and Nishiyama
\cite{Fu.77, Fu.81,NK.84.87,Nishi.98}.
However, the above BET's themselves do not contain any scheme
under which
collective degree of freedom can be selected from the
whole degrees of freedom.

On the contrary,
we have a traditional approach
to the microscopic theory of collective motion,
the TDHF theory (TDHFT) and TDHB theory (TDHBT), e.g.,
\cite{Nishi.81,Nishi.83}.
The pioneering idea of the TDHFT was
suggested by Marumori
\cite{AM.60}
for the case of small amplitude vibrational motions.
Using this idea, one can determine
the time dependence of any physical quantity, e.g.,
frequency of the small f\/luctuation around a static HF/HB f\/ield.
The equation for the frequency has the same form as that given by RPA.
A quantum energy given by this method
means an excitation energy of the f\/irst excited state.
Then, the RPA is a possible quantization
of the TDHFT/TDHBT in the small amplitude limit.
In fact, as was proved by Marshalek and Horzwarth
\cite{MH.72},
the BEHBT is reduced to the TDHBT
under the replacement of boson operators
with classical canonical variables.
Using a canonical transformation in a classical mechanics,
it is expected to obtain a scheme for choosing
the collective degree of freedom in the SCF.
Historically, there was another stream, i.e.,
an adiabatic perturbation approach.
This approach starts from an assumption
that the speed of collective motion is much slower than that of
any other non-collective motion.
At an early stage of the study of this stream,
the adiabatic treatment of the TDHFT (ATDHFT)
was presented by
Thouless and Valatin~\cite{TV.59}.
Such a~theory has a feature common
to the one of the theory for large-amplitude collective motion.
Later the ATDHFT
was developed mainly
by Baranger and Veneroni, Brink, Villars, Goeke and Reinhard,
and Mukherjee and Pal~\cite{BV.78}.
The most important point
of the ATDHFT by Villars is in
introducing a ``collective path'' into a phase space.
A collective motion corresponds to a trajectory
in the phase space which moves along the collective path.
Standing on the same spirit, Holtzwarth and Yukawa,
Rowe and Bassermann~\cite{HY.74},
gave the TDHFT
and
Marumori, Maskawa, Sakata and Kuriyama
so-called ``maximally decoupled'' method in a canonical form~\cite{Ma.80}.
So, various techniques of classical mechanics are useful
and then canonical quantization is expected.
By solving the equation of collective path,
one can obtain some corrections to the TDHF result.
The TDHFT has a possibility to illustrate
not only collective modes but also intrinsic modes.
However, the following three  points remain to be solved yet:
(i) to determine a microscopic structure of collective motion,
which may be a superposition of each particle motion,
in relation to dynamics under consideration
(ii) to determine IP motion which should be
orthogonal to collective motion
and
(iii) to give a coupling between both the motions.
The canonical-formed TDHFT enables us to select
the collective motion in relation to the dynamics,
though it makes no role to take IP motion into account,
because the TD S-det
contains only canonical variables to represent the collective motion.
Along the same way as the TDHFT,
Yamamura and Kuriyama have extended the TDHFT to that on
a fermion CS constructed on the TD S-det.
The CS rep contains not only
the usual canonical variables
but also the Grassmann variables.
A classical image of fermions can be obtained by regarding
the Grassmann variables as canonical ones~\cite{Grass variable}.
The constraints governing the variables to remove
the overcounted degrees of freedom were decided
under the physical consideration.
Owing to the Dirac's canonical theory for a constrained system,
the TDHFT was successfully developed
for a unif\/ied description of collective and IP
motions  in the classical mechanics~\cite{YK.81}.


\vspace{-2mm}

\subsection{Viewpoint of symmetry of evolution equations}\label{section1.2}

\looseness=-1
The TDHF/TDHB can be summarized to f\/ind
optimal coordinate-systems on a group manifold
basing on
Lie algebras of the f\/inite-dimensional fermion-pairs
and to describe dynamics on the manifold.
The boson operators in BET are generators occurring in
the coordinate system of tangent space on the manifold
in the fermion Fock space.
But the BET's themselves do not contain any scheme
under which
collective degrees of freedom can be selected from the
whole degrees of freedom.
Approaches to collective motions by the TDHFT suggest
that
the coordinate system on which collective motions is
describable deeply relates not only to the global symmetry
of the f\/inite-dimensional group manifold itself
but also to {\em hidden local symmetries},
besides the Hamiltonian.
Various collective motions may be well understood
by taking the local symmetries into account.
The local symmetries may be closely connected with
inf\/inite-dimensional Lie algebras.
However,
there has been little attempts to {\em manifestly}
understand collective motions
in relation to the local symmetries.
From the viewpoint of symmetry of evolution equations,
we will study the algebro-geometric structures
toward a unif\/ied understanding
of both the collective and IP motions.

The f\/irst issue is to investigate
fundamental ``curvature equations''
to extract collective submanifolds
out of the full TDHF/TDHB manifold.
We show that the expression in a quasi-particle frame (QPF)
of the zero-curvature equations described later
becomes the nonlinear RPA
which is the natural extension of the usual RPA.
We abbreviate RPA and QRPA to only RPA.
We had at f\/irst started from a question
whether soliton equations exist in the TDHF/TDHB manifold or do not,
in spite of the dif\/ference that
the solitons are described
in terms of inf\/inite degrees of freedom
and the RPA in terms of f\/inite ones.
We had met with
the inverse-scattering-transform method by AKNS
\cite{AKH.74}
and
the dif\/ferential geometrical approaches
on group manifolds
\cite{Satt.82}.
An integrable system is explained by the zero-curvature, i.e.,
integrability condition of connection
on the corresponding Lie group.
Approaches to collective motions had been little
from the viewpoint of the curvature.
If a collective submanifold is a collection of collective paths,
an inf\/initesimal condition to transfer a path to another
may be nothing but the integrability condition
for the submanifold with respect
to a parameter time $t$ describing a trajectory
of an SCF Hamiltonian
and to other parameters specifying any point
on the submanifold.
However the trajectory of the SCF Hamiltonian
is unable to remain on the manifold.
Then the curvature may be
able to work as a criterion of ef\/fectiveness
of the collective submanifold.
From a wide viewpoint of symmetry
the RPA is extended to any point on the manifold
because
an equilibrium state which we select as a starting point
must be equipotent with any other point on the manifold.
The well-known RPA had been introduced
as a linear approximation to treat excited states
around a ground state (the equilibrium state),
which is essentially a harmonic approximation.
When an amplitude of oscillation becomes larger
and then an anharmonicity appears,
then we have to treat the anharmonicity
by taking nonlinear ef\/fects in the equation of motion into account.
It is shown that
{\em equations defining the curvature}
of the collective submanifold
becomes
{\em fundamental equations} to treat the anharmonicity.
We call them ``the formal RPA equation''.
It will be useful to understand
algebro-geometric meanings of large-amplitude collective motions.

The second issue is
to go beyond the perturbative method with respect to the
collective variables
\cite{Ma.80}.
For this aim,
we investigate an interrelation between
the SCF method (SCFM) extracting {\em collective motions}
and $\tau$-functional method ($\tau$-FM)
\cite{JM.83}
constructing {\em integrable equations}
in solitons.
In a soliton theory on a group manifold,
transformation groups governing solutions for soliton equations
become inf\/inite-dimensional Lie groups
whose generators of the corresponding Lie algebras
are expressed as
inf\/inite-order dif\/ferential operators of
af\/f\/ine Kac--Moody algebras.
An inf\/inite-dimensional fermion Fock space $F_{\infty }$
is realized in terms of
a space of complex polynomial algebra.
The inf\/inite-dimensional fermions
are given in terms of the inf\/inite-order dif\/ferential operators
and the soliton equation is nothing but
the dif\/ferential equation to determine
the group orbit of the highest weight vector
in the $F_{\infty }$
\cite{JM.83}.
The generalized CS rep gives a key to elucidate relationship of
a HF wave function to a $\tau$-function in the soliton theory.
This has been pointed out f\/irst by D'Ariano and Rasetti
\cite{AR.85}
for an inf\/inite-dimensional harmonic electron gas.
Standing on their observation,
for the SCFM
one can give a theoretical frame
for an integrable sub-dynamics
on an abstract $F_{ \infty }$.
The relation between
SCFM in f\/inite-dimensional fermions
and $\tau$-FM in inf\/inite ones,
however, has not been investigated
because dynamical descriptions of fermion systems
by them have looked very dif\/ferent manners.
In the papers
\cite{KN.00,KN.01,NK.02,NPK.05,WigSym99},
we have f\/irst tried to clarify it
using SCFM on $U(N)$ group and $\tau$-FM on that group.
To attain this object
we will have to solve the following main problems:
f\/irst, how we embed the f\/inite-dimensional fermion system
into a certain inf\/inite one
and how we rebuilt the TDHFT on it;
second, how any algebraic mechanism works behind
particle and collective motions
and how any relation between collective variables
and a spectral parameter in soliton theory is there;
last, how the SCF Hamiltonian selects various subgroup-orbits
and
how a collective submanifold is made from them
and further
how the submanifold relates to the formal RPA.
To understand microscopically cooperative phenomena,
the concept of collective motion is introduced
in relation to a TD variation of SC mean-f\/ield.
IP motion is described
in terms of particles referring to a stationary mean-f\/ield.
The variation of a TD mean-f\/ield gives rise to couplings
between collective and IP motions
and couplings among quantum f\/luctuations of the TD mean-f\/ield itself
\cite{YK.87},
while
in $\tau$-FM
a soliton equation is derived as follows:
Consider an inf\/inite-dimensional Lie algebra
and its representation on a functional space.
The group-orbit of the highest weight vector becomes
an {\em infinite-dimensional Grassmannian} $G_{\!\infty }$.
The bilinear equation (Pl\"{u}cker relation)
is nothing else than the soliton equation.
This means that a solution space of the soliton equation
corresponds to a group-orbit of the vacuum state.
The SCFM does not use the Pl\"{u}cker relation
in the context of a bilinear dif\/ferential equation
def\/ining {\em finite-dimensional Grassmannian} $G_M$
but seems to use implicitly such a relation.
In the SCFM
a physical concept of quasi-particle and vacuum and
a coset space is used instead.
If we develop a perturbative theory
for large-amplitude collective motion
\cite{Ma.80},
an inf\/inite-dimensional Lie algebra
might been necessarily used.
The sub-group orbits consisting of several
{\em loop-group} paths
\cite{PS.86}
classif\/ied by the Pl\"{u}cker relation
exist innumerably in $G_M$
so that the SCFM is related
to the soliton theo\-ry in $G_{ \infty }$.
The Pl\"{u}cker relation in a coset space
$\frac{U(N)}{U(M) \times U(N  -  M)}$
\cite{NMO.04}
becomes analogous with
the {\em Hirota's bilinear form}
\cite{Sa.81,Hi.76}.
Toward an ultimate goal
we aim to reconstruct a theoretical frame for
a $\upsilon$ (external parameter)-dependent SCFM
to describe more precisely
the dynamics on the~$F_{\infty }$.
In the abstract fermion Fock space,
we f\/ind common features in both SCFM and $\tau$-FM.
(i) {\em Each solution space} is described as
{\em Grassmannian} that is
group orbit of the corresponding vacuum state.
(ii) The former may implicitly explain the Pl\"{u}cker relation
not in terms of
bilinear dif\/ferential equations def\/ining $G_M$
but in terms of
the physical concept of quasi-particle and vacuum and
mathematical language of coset space and coset variable.
The various BETs
are built on the Pl\"{u}cker relation to hold the Grassmannian.
The latter asserts that
the soliton equations are nothing but
the bilinear dif\/ferential equations
giving {\em a boson representation of the Pl\"{u}cker relation}.
The relation, however,
has been unsatisfactorily investigated yet within the framework
of the usual SCFM.
We study it and show that
both the methods stand on the common features,
Pl\"{u}cker relation or bilinear dif\/ferential equation
def\/ining the Grassmannian.
On the contrary, we observe dif\/ferent points:
(i) The former is built
on a {\em finite}-dimensional Lie algebra but the latter
on an {\em infinite}-dimensional one.
(ii) The former has an SCF Hamiltonian
consisting of a fermion one-body operator,
which is derived from
a functional derivative of an expectation value of
a fermion Hamiltonian by a ground-state wave function.
The latter introduces artif\/icially
{\em a fermion Hamiltonian} of
one-body type operator as {\em a boson mapping operator}
from states
on fermion Fock space to corresponding ones
on $\tau$-functional space ($\tau$-FS).

The last issue is,
despite a dif\/ference due to the dimension of fermions,
to aim at obtaining a {\em close connection} between
{\em concept of mean-field potential} and {\em gauge of fermions}
inherent in the SCFM and at making a role of a loop group~\cite{PS.86}
to be clear.
Through the observation,
we construct inf\/inite-dimensional fermion operators
from the f\/inite-dimensional ones
by Laurent expansion with respect to a circle $S^1$.
Then with the use of
an af\/f\/ine Kac--Moody (KM) algebra according to
the idea of Dirac's positron theory~\cite{Dirac.58},
we rebuilt a TDHFT in $F_\infty$.
The TDHFT results in a gauge theory of fermions
and the collective motion,
f\/luctuation of the mean-f\/ield potential,
appears as
the motion of fermion gauges with a common factor.
The physical concept of the quasi-particle and vacuum in
the SCFM on the $S^1$
connects to the ``Pl\"{u}cker relations'' due to
the Dirac theory, in other words,
the algebraic mechanism extracting
various sub-group orbits consisting of {\em loop}
path out of the full TDHF manifold
is just the ``Hirota's bilinear form''~\cite{Hi.76}
which is an $su(N) (\in sl(N))$ reduction of $gl(N)$
in the $\tau$-FM.
As a result,
it is shown that
an inf\/inite-dimensional fermion many-body system
is also realizable
in a f\/inite-dimensional one
and that
roles of the soliton equation
(Pl\"{u}cker relation) and the TDHF equation are made clear.
We also understand
an SCF dynamics through gauge of interacting
inf\/inite-dimensional fermions.
A bilinear equation for the $\upsilon$-HFT has been transcribed
onto the corresponding $\tau$-function
using the regular representation for the group and the Schur polynomials.
The $\upsilon$-HF SCFM on an inf\/inite-dimensional Fock space $F_\infty$
leads to a dynamics on an inf\/inite-dimensional Grassmannian ${\rm Gr}_\infty$
and may describe more precisely such a dynamics on the group manifold.
A f\/inite-dimensional Grassmannian
is identif\/ied with a ${\rm Gr}_\infty$
which is af\/f\/iliated with the group manifold obtained
by reducting $gl(\infty)$ to $sl(N)$ and $su(N)$.
We have given explicit expressions for
Laurent coef\/f\/icients of
soliton solutions for $\widehat{sl}(N)$ and $\widehat{su}(N)$
on the ${\rm Gr}_\infty$
using Chevalley bases for $sl(N)$ and $su(N)$~\cite{NishiProviKoma2.07}.
As an illustration
we will attempt to make a $\upsilon$-HFT approach to
an inf\/inite-dimensional matrix model
extended from the f\/inite-dimensional $su(2)$
Lipkin--Meshkov--Glick (LMG) model~\cite{LMG.65}.
For this aim, we give an af\/f\/ine KM algebra $\widehat{sl}(2,C)$
(complexif\/ication of~$\widehat{su}(2))$
to which the LMG generators subject,
and their $\uptau$ representations and the $\sigma_K$ mappings for them.
We can represent an inf\/inite-dimensional matrix of the LMG Hamiltonian
and its HF Hamiltonian
in terms of the Schur polynomials.
Its inf\/inite-dimensional HF operator is also given
through the mapping $\sigma_M$ for $\psi_i\psi_j^*$ of
inf\/inite-dimensional fermions
$\psi_i$ and $\psi_i^*$,
which is expressed
by the Schur polynomials $S_k(x)$ and $S_k(\partial_x)$.
Further its $\tau$-function for a simple case
is provided by the Pl\"{u}cker coordinates and Schur polynomials.

In Section~\ref{section2},
we propose curvature equations as
fundamental equations
to extract a collective submanifold
out of the full TDHB manifold.
Basing on these ideas, we construct
the curvature equations and study
the relation between the maximal decoupled method
and the curvature equations.
We further investigate the role of
the non-zero curvature arising from the residual Hamiltonian.
Making use of the expression of
the zero-curvature equations in the QPF,
we f\/ind the formal RPA equation.
In Section~\ref{section3},
we present a simply unif\/ied aspect for the SCFM and the $\tau$-FM
and show a simple idea connecting both the methods.
We study the algebraic relation between coset coordinate
and Pl\"{u}cker coordinate.
Basing on the above idea,
we attempt to rebuilt the TDHFT in $\tau$-FS.
We introduce $\upsilon$-dependent
inf\/inite-dimensional fermion operators and a $F_\infty$
through Laurent expansion with respect to
the degrees of freedom of the original fermions.
The algebraic relation between both the methods
is manifestly described.
We embed a HF $u(N)$ Lie algebra into a $gl(\infty)$ by means of
inf\/inite-dimensional fermions.
The $\upsilon$-SCFM in $\tau$-FS is developed.
The role of the shift operators in the $\tau$-FM is studied.
As an illustration,
explicit expressions for Laurent coef\/f\/icients of soliton solutions
for~$\widehat{sl}(N)$ and~$\widehat{su}(N)$
are presented.
A problem related to a nonlinear Schr\"{o}dinger equation
is also discussed.
In Section~\ref{section4},
we construct a formal RPA equation on $F_\infty$
and also argue about the relation between a $loop$ collective path
and a formal RPAEQ.
Consequently, it can be proved that the usual perturbative method
with respect to periodic collective variables
in the TDHFT is involved in the present method which
aims for constructing the TDHFT on the af\/f\/ine KM algebra.
In Section~\ref{section5},
we introduce inf\/inite-dimensional
``particle'' and ``hole'' operators
and operators
${\widehat{K}}_{0}$ and ${\widehat{K}}_{\pm }$
def\/ined by inf\/inite-dimensional ``particle-hole'' pair operators.
Using these operators,
we construct an inf\/inite-dimensional Heisenberg subalgebra of
the af\/f\/ine KM algebra $\widehat{sl}(2,C)$.
The LMG Hamiltonian and its HF Hamiltonian are expressed in terms of
the Heisenberg basic-elements whose representations
are isomorphic to those in the corresponding boson space.
They are given in terms of inf\/inite numbers of
variables~$x_k$ and derivatives~$\partial_{x_k}$ through
the Schur polynomials~$S_k(x)$.
We give also an inf\/inite-dimensional representation of
$SU(2N)_{\infty }$ transformation of the particle and hole operators.
Finally,
in Section~\ref{section6},
we summarize and discuss the results and future problems.
We give Appendices~\ref{appendixA}--\ref{appendixJ}.
Especially in Appendix~\ref{appendixJ},
we show an explicit expression for Pl\"{u}cker coordinate for the LMG model
and calculate a quantity,
$\det (1_N  +  p^\dag p)~(p:\mbox{coset variable})$ for the LMG model,
in terms of the Schur polynomials.

\newpage


\section{Integrability conditions and collective submanifolds}\label{section2}




\subsection{Introduction}\label{section2.1}

Let us consider an abstract evolution equation
$\partial_t u(t) = K (u(t))$ for $u$,
which is dependent only on a parameter, time $t$.
If there exists a symmetry operation
to transfer a solution for $u$ to another one,
then
introducing another parameter $s$ specifying various solutions,
we can derive another form of evolution equation
with respect to $s$,
$\partial_s u(t,s) = \overline{K}(u(t,s))$
for which we should want to search.
The inf\/initesimal condition for the existence of such a symmetry
appears as the well-known integrability condition
$
\partial_s K(u(t,s)) = \partial_t \overline{K}(u(t,s))
$.
The ``maximally decoupled" method proposed by Marumori et al.~\cite{Ma.80},
invariance principle of Schr\"{o}dinger equation and
canonicity condition,
can be considered as a theory to apply the above basic idea to
certain multiple parametrized symmetries.
The method is regarded as a mathematical tool
to describe the symmetries of the collective submanifold
in terms of $t$ and collective variables,
in which the canonicity conditions
make the collective variables
to be orthogonal coordinate-systems.

Therefore we adopt a concept of curvature unfamiliar
in the conventional TDHBT.
The reason why we take such a thing is the following:
let us consider a description of motions of systems
on a group manifold.
An arbitrary state of the system induced by a transitive
group action corresponds to any point of the full
group parameter space and therefore its time evolution
is represented by an integral curve in this space.
In the whole representation space adopted,
we assume the existence of $2m$ parameters specifying
the proper subspace in which the original motion
of the system can be approximated well,
the existence of the well-def\/ined symmetries.
Suppose we start from a given point on a space,
which consists of $t$ and the $2m$ parameters, and end
at the same point again along the closed curve.
Then we have the value
of the group parameter dif\/ferent from the one at an initial point
on the proper subspace.
We search for some quantities characterizing
the dif\/ference of the value. For our aim, we introduce
a dif\/ferential geometrical viewpoint.
The our basic idea lies in the introduction of a~sort of
Lagrange manner familiar to f\/luid dynamics to
describe collective coordinate systems.
This manner enables us to take a one-form
which is linearly composed of TDHB Hamiltonian and
inf\/initesimal generators induced by
collective variable dif\/ferentials of
an $SO(2N)$ canonical transformation.
The integrability conditions of the system read
the curvature $C =  0$.
Our methods are constructed manifesting themselves
the structure of the group under consideration
to make easy to understand physical
characters at any point on the group manifold.


\subsection{Integrability conditions}\label{section2.2}

We consider many fermion systems with pair correlations.
Let $c_\alpha$ and $c_\alpha^\dagger$ ($\alpha = 1,\dots, N$)
be the annihilation-creation operators of the fermion.
Owing to the anti-commutation relations among them,
some sets of fermion operators with simple construction
become the basis of a Lie algebra.
The operators in the fermion $so(2N)$ Lie algebra,
$
\bar{E}^\alpha_{~\beta }
=
c_\alpha^\dagger c_\beta -\frac{1}{2}\delta_{\alpha \beta }$,
$
\bar{E}_{\alpha \beta }
=
c_\alpha c_\beta$,
$
\bar{E}^{\alpha \beta }
=
c_\alpha^\dagger c_\beta^\dagger
$
generate a canonical transformation $U(g)$
(the Bogoliubov transformation~\cite{Bogo.59})
which is specif\/ied by an $SO(2N)$ matrix $g$:
\begin{gather}
[d, d^\dagger]
=
U(g)[c, c^\dagger]U^\dagger (g)
=
[c, c^\dagger]g, \qquad
g
=
\left[
\begin{array}{cc}
a & b^\star \\
b & a^\star
\end{array}
\right],
\qquad gg^\dagger = g^\dagger g = 1_{2N},\label{SO2NCanonicalTrans} \\
U^{-1}(g) = U^\dagger (g),\qquad
U(g)U(g^\prime) = U(g g^\prime),
\nonumber
\end{gather}
where
$(c,  c^\dagger)  =  ((c_\alpha),  (c_\alpha^\dagger))$
and
$(d,   d^\dagger) = ((d_i), (d_i^\dagger))$
are $2N$-dimensional row vectors and
$a = (a^\alpha_{~i})$
and
$b = (b^\alpha_{~i})$
$(i = 1, \dots ,N)$
are $N \times N$ matrices.
$1_{2N}$ is a $2N$-dimensional unit matrix.
The symbols
$\dagger$, $\star$ and $\mbox{\scriptsize{T}}$
mean
the hermitian conjugate, the complex conjugation and the transposition,
respectively.
The explicit expressions of the canonical
transformations are given by Fukutome for the various types
of the fermion Lie algebra
\cite{Fu.81}.
The fermion Lie operators of the quasi-particles
$(E^i_{~j},E_{ij}, E^{ij})$ are constructed from the operators
$d$ and $d^\dagger$
in~(\ref{SO2NCanonicalTrans})
by the same way as
the one to def\/ine the set
$(\bar{E}^\alpha_{~\beta },\bar{E}_{\alpha \beta }
,\bar{E}^{\alpha \beta })$.
The set $E$ in the quasi-particle frame is transformed
into the set $\bar{E}$ in the particle frame as follows:
\begin{equation}
\left[
\begin{array}{cc}
E^\bullet_{~\bullet } & E^{\bullet\bullet } \\
E_{\bullet\bullet } & -E^\bullet_{~\bullet }{}^\dagger
\end{array}
\right]
=
g^\dagger
\left[
\begin{array}{cc}
\bar{E}^\bullet_{~\bullet } & \bar{E}^{\bullet \bullet } \\
\bar{E}_{\bullet \bullet }
& -\bar{E}^\bullet_{~\bullet }{}^\dagger
\end{array}
\right] g .
\label{E-E}
\end{equation}
The $E^\bullet_{~\bullet } = (E^i_{~j})$
and
$\bar{E}^\bullet_{~\bullet } = (\bar{E}^\alpha_{~\beta })$ etc.
are $N  \times  N$ matrices.
A TDHB Hamiltonian of the system is given by
\begin{eqnarray}
H_{\rm HB}
=
\tfrac{1}{2} [c, c^\dagger] {\cal F}
\left[
\begin{array}{l}
c^\dagger \\
c
\end{array}
\right],
\qquad
{\cal F}
=
\left[
\begin{array}{cc}
-F^\star & -D^\star \\
 D       &  F
\end{array}
\right]
=
{\cal F}^\dagger ,
\label{HB-matrix}
\end{eqnarray}
where the HB matrices
$F = (F_{\alpha \beta })$ and $D = (D_{\alpha \beta })$
are related to the quasi-particle
vacuum expectation values of the Lie operators
$\langle \bar{E} \rangle$
as
\begin{gather*}
F_{\alpha\beta }
=
h_{\alpha \beta }+[\alpha\beta|\gamma\delta]
(\langle \bar{E}^\gamma_{~\delta } \rangle
+\tfrac{1}{2}\delta_{\gamma\delta })  \qquad
(F^\dagger = F),\nonumber\\
D_{\alpha \beta }
=
\tfrac{1}{2}
[\alpha\gamma|\beta\delta]
\langle \bar{E}_{\delta \gamma }\rangle  \qquad
(D^{\mbox{\scriptsize{T}}} = -D), \\ 
[\alpha\beta|\gamma\delta]=-[\alpha\delta|\gamma\beta]
=[\gamma\delta|\alpha\beta]=[\beta\alpha|\delta\gamma]^\star.
\nonumber
\end{gather*}
The quantities $h_{\alpha\beta}$ and $[\alpha\beta|\gamma\delta]$
are the matrix element of the single-particle Hamiltonian
and the antisymmetrized one of the interaction potential, respectively.
Here and hereafter we use the dummy index
convention to take summation over the repeated index.

Let $|0 \rangle$ be the free-particle vacuum
satisfying $c_\alpha|0 \rangle = 0$.
The $SO(2N)$(HB) wave function $|\phi(\check{g})\rangle$
is constructed by a transitive action of the $SO(2N)$
canonical transformation $U(\check{g})$ on $|0 \rangle$:
$|\phi(\check{g})\rangle = U^{-1}(\check{g})|0 \rangle$,
$\check{g} \in SO(2N)$.
In the conventional TDHBT,
the TD wave function $|\phi(\check{g})\rangle$
is given through that of the TD group parameters
$a~(a^\star)$ and $b~(b^\star)$.
They characterize the TD self-consistent mean HB f\/ields $F$ and $D$
whose dynamical changes induce
the collective motions of the many fermion systems.
As was made in the TDHF case~\cite{Ma.80}
and~\cite{YK.81},
we introduce a~TD $SO(2N)$ canonical transformation
$
U(\check{g})
=
U[\check{g}(\check{\Lambda }(t),  \check{\Lambda }^\star (t))]
$.
A set of TD complex variables
$(\check{\Lambda }(t), \check{\Lambda }^\star(t))
=
(\check{\Lambda }_n(t),\check{\Lambda }_n^\star(t);\;
n = 1,\dots,m)$
associated with the collective motions specif\/ies the group parameters.
The number $m$ is assumed to be much smaller
than the order of the $SO(2N)$ Lie algebra, which
means there exist only a few ``collective degrees of freedom''.
The above is the natural extension of the method
in TDHF case to the TDHB case.

However, dif\/fering from the above usual manner,
we have another way, may be called
a~{\em Lag\-range-like manner}, to introduce
a set of complex variables.
This is realized if we regard
the above-mentioned variables
$(\check{\Lambda }(t), \check{\Lambda }^\star(t))$
as functions of independent variables
$(\Lambda, \Lambda^\star)
=
(\Lambda_n, \Lambda_n^\star)$ and~$t$,
where {\em time-independent variables} $(\Lambda, \Lambda^\star)$
are introduced as local coordinates to specify any point
of a $2m$-dimensional collective submanifold.
A collective motion in the $2m$-dimensional manifold is
possibly determined in the usual manner if we could know
the explicit forms of~$\check{\Lambda }$ and~$\check{\Lambda }^\star$
in terms of~$(\Lambda, \Lambda^\star)$ and~$t$.
The above manner seems to be very analogous to
the Lagrange manner in the f\/luid dynamics.
The pair of variables $(\Lambda, \Lambda^\star)$
specif\/ies variations of the SCF associated with
the collective motion described by a pair of
collective coordinates $\alpha$ and their conjugate~$\pi$
in the {\em Lagrange-like manner},
$
\widehat{\alpha }
=
\frac{1}{\sqrt{2}}
(\Lambda^\star + \Lambda)$
and
$\widehat{\pi }
=
i\frac{1}{\sqrt{2}}
(\Lambda^\star - \Lambda)
$\cite{Ma.80}.
Thus, the $SO(2N)$ canonical transformation is rewritten as
$
U(\check{g})
=
U[g(\Lambda, \Lambda^\star, t)]  \in  SO(2N)
$.
Notice that a functional form
$\check{g}(\check{\Lambda }(t), \check{\Lambda }^\star (t))$
changes into another form
$g(\Lambda, \Lambda^\star,t)$
due to an adoption of the Lagrange-like manner.
This manner enables us to take a one-form
$\Omega$ which is linearly composed of the inf\/initesimal
generators induced by the time dif\/ferential
and the collective variable ones
$(\partial_t, \partial_\Lambda, \partial_{\Lambda^\star })$
of the $SO(2N)$ canonical transformation
$U[g(\Lambda, \Lambda^\star,t)]$.
By introducing the one-form~$\Omega$,
it is possible to search for the collective path
and the collective hamiltonian almost
separated from other remaining degrees
of freedom of the systems.
It may be achieved to study the integrability
conditions of our systems
which are expressed as the set of
the Lie-algebra-valued equations.

We def\/ine the Lie-algebra-valued
inf\/initesimal generators of collective submanifolds
as follows:
\begin{gather}
H_{\rm c}/\hbar
\stackrel{d}{=}
(i\partial_t U^{-1}(g))U(g), \nonumber\\
O_n^\dagger
\stackrel{d}{=}
(i\partial_{\Lambda_n} U^{-1}(g))U(g) ,\qquad
O_n
\stackrel{d}{=}
(i\partial_{\Lambda_{n^\star }} U^{-1}(g))U(g) .
\label{HcdefOcdef}
\end{gather}
Here and hereafter,
for simplicity we abbreviate
$g(\Lambda, \Lambda^\star, t)$ as $g$.
The explicit form of the inf\/initesimal generators for the TDHF
was f\/irst given by Yamamura and Kuriyama~\cite{YK.81}.
In our TDHB, the Lie-algebra-valued inf\/initesimal
generators are expressed by the trace form as
\begin{gather}
H_{\rm c}/\hbar
=
-\tfrac{1}{2}\mbox{Tr}
\left\{
(i\partial_tg\cdot g^\dagger)
\left[
\begin{array}{cc}
\bar{E}^\bullet_{~\bullet } & \bar{E}^{\bullet \bullet } \\
\bar{E}_{\bullet \bullet } & -\bar{E}^\bullet_{~\bullet }{}
^\dagger
\end{array}
\right] \right\}
=
\tfrac{1}{2}[c, c^\dagger](i\partial_tg \cdot g^\dagger)
\left[
\begin{array}{l}
c^\dagger \\
c
\end{array}
\right] ,
\label{HcTr-express}
\\
O_n^\dagger
=
- \tfrac{1}{2}\mbox{Tr}
\left\{(i\partial_{\Lambda_n}g\cdot g^\dagger)
\left[
\begin{array}{cc}
\bar{E}^\bullet_{~\bullet } & \bar{E}^{\bullet \bullet } \\
\bar{E}_{\bullet \bullet } & -\bar{E}^\bullet_{~\bullet }{}
^\dagger
\end{array}
\right] \right\}
=
 \tfrac{1}{2}
[c,c^\dagger](i\partial_{\Lambda_n}
g\cdot g^\dagger)
\left[
\begin{array}{l}
c^\dagger \\
c
\end{array}
\right] , \nonumber\\
O_n
=
- \tfrac{1}{2}\mbox{Tr}
\left\{(i\partial_{\Lambda_n^\star }
g\cdot g^\dagger)
\left[
\begin{array}{cc}
\bar{E}^\bullet_{~\bullet } & \bar{E}^{\bullet \bullet } \\
\bar{E}_{\bullet \bullet } & -\bar{E}^\bullet_{~\bullet }{}
^\dagger
\end{array}
\right] \right\}
=
\tfrac{1}{2}
[c,c^\dagger](i\partial_{\Lambda_n^\star }g\cdot g^\dagger)
\left[
\begin{array}{l}
c^\dagger \\
c
\end{array}
\right] .
\label{OcTr-express}
\end{gather}

Multiplying the $SO(2N)$ wave function $|\phi(g)\rangle$
on the both sides of
(\ref{HcdefOcdef}),
we get a set of equations on the $so(2N)$ Lie algebra:
\begin{gather}
D_t|\phi(g)\rangle
\stackrel{d}{=}
(\partial_t+iH_{\rm c}/\hbar)
|\phi(g)\rangle
= 0 ,\nonumber\\
D_{\Lambda_n}|\phi(g)\rangle
\stackrel{d}{=}
(\partial_{\Lambda_n}+iO_n^\dagger)
|\phi(g)\rangle=0 , \qquad
D_{\Lambda_n^\star }|\phi(g)\rangle
\stackrel{d}{=}
(\partial_{\Lambda_n^\star }+iO_n)
|\phi(g)\rangle
= 0 .
\label{D-tD-Lambda}
\end{gather}
We regard these equations (\ref{D-tD-Lambda}) as
partial dif\/ferential equations for $|\phi(g)\rangle$.
In order to discuss the conditions under which the
dif\/ferential equations (\ref{D-tD-Lambda}) can be solved,
the mathematical method well known as integrability
conditions is useful.
For this aim, we take a one-form $\Omega$
linearly composed of the inf\/initesimal generators
(\ref{HcdefOcdef}):
$
\Omega
=
-
i(H_{\rm c}/\hbar \cdot dt
+
O_n^\dagger \cdot d\Lambda_n
+
O_n \cdot d\Lambda_n^\star)
$.
With the aid of the $\Omega$, the integrability
conditions of the system read
$
C
 \stackrel{d}{=}
d\Omega - \Omega\wedge \Omega = 0
$,
where $d$ and $\wedge$ denote the exterior dif\/ferentiation
and the exterior product, respectively.
From the dif\/ferential geometrical viewpoint,
the quantity $C$ means
the curvature of a connection.
Then the integrability conditions may be interpreted as
the vanishing of the curvature of the connection
$(D_t, D_{\Lambda_n}, D_{\Lambda_n}^\star)$.
The detailed structure of the curvature is calculated to be
\begin{gather*}
C
=
 C_{t,\Lambda_n}d\Lambda_n  \wedge  dt
+
C_{t,\Lambda_n^\star }d\Lambda_n^\star  \wedge  dt
+
C_{\Lambda_{n^\prime },\Lambda_n^\star }
d\Lambda_n^\star  \wedge  d\Lambda_{n^\prime } \nonumber\\
\phantom{C=}{}
+
 \tfrac{1}{2} C_{\Lambda_{n^\prime },\Lambda_n}
d\Lambda_n  \wedge  d\Lambda_{n^\prime }
+
 \tfrac{1}{2} C_{\Lambda_{n^\prime }^\star,\Lambda_n^\star }
d\Lambda_n^\star  \wedge  d\Lambda_{n^\prime }^\star ,
\end{gather*}
where
\begin{gather}
C_{t,\Lambda_n}
\stackrel{d}{=}
[D_t, D_{\Lambda_n}]
=
 i\partial_tO_n^\dagger -i\partial_{\Lambda_n}
H_{\rm c}/\hbar
+[O_n^\dagger,H_{\rm c}/\hbar] , \nonumber\\
C_{t,\Lambda_n^\star }
\stackrel{d}{=}
[D_t, D_{\Lambda_n^\star }]
=
 i\partial_tO_n-i\partial_{\Lambda_n^\star }
 H_{\rm c}/\hbar
+[O_n,H_{\rm c}/\hbar] , \nonumber\\
C_{\Lambda_{n^\prime },\Lambda_n^\star }
\stackrel{d}{=}
[D_{\Lambda_{n^\prime }}, D_{\Lambda_n^\star }]
=
 i\partial_{\Lambda_{n^\prime }}O_n
-i\partial_{\Lambda_n^\star }O_{n^\prime }^\dagger
+[O_n,O_{n^\prime }^\dagger] , \label{Cexplicit}\\
C_{\Lambda_{n^\prime },\Lambda_n}
\stackrel{d}{=}
[D_{\Lambda_{n^\prime }}, D_{\Lambda_n}]
=
 i\partial_{\Lambda_{n^\prime }}O_n^\dagger
-i\partial_{\Lambda_n} O_{n^\prime }^\dagger
+[O_n^\dagger,O_{n^\prime }^\dagger] , \nonumber\\
C_{\Lambda_{n^\prime }^\star,\Lambda_n^\star }
\stackrel{d}{=}
[D_{\Lambda_{n^\prime }^\star }, D_{\Lambda_n^\star }]
=
 i\partial_{\Lambda_{n^\prime }^\star } O_n
-i\partial_{\Lambda_n^\star } O_{n^\prime }
+[O_n,O_{n^\prime }] .
\nonumber
\end{gather}
The vanishing of the curvature $C$ means
$C_{\bullet , \bullet } = 0$.

Finally with the use of the explicit forms
of
(\ref{HcTr-express}) and (\ref{OcTr-express}),
we can get the set of Lie-algebra-valued equations
as the integrability conditions
of partial dif\/ferential equations
(\ref{D-tD-Lambda})
\begin{alignat}{3}
& C_{t,\Lambda_n}
=
\tfrac{1}{2}[c,c^\dagger]
{\cal C}_{t,\Lambda_n}
\left[
\begin{array}{l}
c^\dagger \\
c
\end{array}
\right] , \qquad &&
{\cal C}_{t,\Lambda_n}
=
i\partial_t\theta_n^\dagger -i\partial_{\Lambda_n}
{\cal F}_{\rm c}/\hbar
+[\theta_n^\dagger,{\cal F}_{\rm c}/\hbar] ,& \nonumber\\
& C_{t,\Lambda_n^\star }
=
 \tfrac{1}{2}[c,c^\dagger]
{\cal C}_{t,\Lambda_n^\star }
\left[
\begin{array}{l}
c^\dagger \\
c
\end{array}
\right] , \qquad &&
{\cal C}_{t,\Lambda_n^\star }
=
i\partial_t\theta_n-i\partial_{\Lambda_n^\star }
{\cal F}_{\rm c}/\hbar
+[\theta_n,{\cal F}_{\rm c}/\hbar] , & \nonumber\\
& C_{\Lambda_{n^\prime },\Lambda_n^\star }
=
\tfrac{1}{2}[c,c^\dagger]
{\cal C}_{\Lambda_{n^\prime },\Lambda_n^\star }
\left[
\begin{array}{l}
c^\dagger\\
c
\end{array}
\right] ,\qquad &&
{\cal C}_{\Lambda_{n^\prime },\Lambda_n^\star }
=
 i\partial_{\Lambda_{n^\prime }}\theta_n
-i\partial_{\Lambda_n^\star }\theta_{n^\prime }^\dagger
+[\theta_n,\theta_{n^\prime }^\dagger] , & \label{Cc-val}\\
&
C_{\Lambda_{n^\prime },\Lambda_n}
=
\tfrac{1}{2}[c,c^\dagger]
{\cal C}_{\Lambda_{n^\prime },\Lambda_n}
\left[
\begin{array}{l}
c^\dagger \\
c
\end{array}
\right] ,\qquad &&
{\cal C}_{\Lambda_{n^\prime },\Lambda_n}
=
 i\partial_{\Lambda_{n^\prime }}\theta_n^\dagger
-i\partial_{\Lambda_n}\theta_{n^\prime }^\dagger
+[\theta_n^\dagger,\theta_{n^\prime }^\dagger] , & \nonumber\\
&
C_{\Lambda_{n^\prime }^\star,\Lambda_n^\star }
=
\tfrac{1}{2}[c,c^\dagger]
{\cal C}_{\Lambda_{n^\prime }^\star,\Lambda_n^\star }
\left[
\begin{array}{l}
c^{\dagger }\\
c
\end{array}
\right] ,\qquad &&
{\cal C}_{\Lambda_{n^\prime }^\star,\Lambda_n^\star }
=
 i\partial_{\Lambda_{n^\prime }^\star } \theta_n
-i\partial_{\Lambda_n^\star }\theta_{n^\prime }
+[\theta_n,\theta_{n^\prime }] . &\nonumber
\end{alignat}
Here the quantities ${\cal F}_{\rm c}$,
$\theta_n^\dagger$, $\theta_n$
are def\/ined through partial dif\/ferential equations,
\begin{gather}
i\hbar \partial_t g = {\cal F}_{\rm c} g
\qquad \mbox{and} \qquad
i\partial_{\Lambda_n} g = \theta_n^\dagger g , \qquad
i\partial_{\Lambda_n^\star }g = \theta_n g .
\label{TDFcSDtheta}
\end{gather}
The quantity ${\cal C}_{\bullet , \bullet }$
may be naturally regarded as the curvature
of the connection on the group manifold.
The reason becomes clear
if we take the following procedure quite parallel
with the above:
Starting from~(\ref{TDFcSDtheta}),
we are led to a set of partial dif\/ferential equations
on the $SO(2N)$ Lie group,
\begin{gather}
{\cal D}_t g
\stackrel{d}{=}
(\partial_t+i{\cal F}_{\rm c}/\hbar) g = 0 ,\nonumber\\
{\cal D}_{\Lambda_n} g
\stackrel{d}{=}
(\partial_{\Lambda_n}+i\theta_n^\dagger) g = 0 ,\qquad
{\cal D}_{\Lambda_n^\star }g
\stackrel{d}{=}
(\partial_{\Lambda_n^\star }+i\theta_n) g = 0 .
\label{D-tD-Lambdag}
\end{gather}

The curvature
$
{\cal C}_{\bullet , \bullet }$ $(\stackrel{d}{=}
[{\cal D}_\bullet, {\cal D}_\bullet])
$
of the connection
$({\cal D}_t, {\cal D}_{\Lambda_n}, {\cal D}_{\Lambda_n^\star })$
is easily shown to be equivalent to
the quantity ${\cal C}_{\bullet , \bullet }$ in~(\ref{Cc-val}).
The above set of the Lie-algebra-valued equations
(\ref{Cc-val})
evidently leads us to putting all the curvatures
${\cal C}_{\bullet , \bullet }$ in
(\ref{Cc-val})
equal to zero.
On the other hand, the TDHB Hamiltonian
(\ref{HB-matrix}),
being the full Hamiltonian on the full~$SO(2N)$
wave function space, can be represented
in the same form as (\ref{HcdefOcdef}),
$
H_{\rm HB}/\hbar
=
(i\partial_t U^{-1}(g^\prime))U(g^\prime)
$,
where $g^\prime$ is any point on the $SO(2N)$
group manifold.
This Hamiltonian is also
transformed into the same form as
(\ref{HcTr-express}).
It is self-evident that the above fact leads us to
the well-known TDHBEQ,
$
i\hbar\partial_t g^\prime = {\cal F}g^\prime
$.
The full TDHB Hamiltonian can be decomposed
into two components at the reference point
$g^\prime = g$:
\begin{gather*}
H_{\rm  HB}|_{U^{-1}(g\prime) = U^{-1}(g)}
= H_c + H_{\rm  res} ,\qquad
{\cal F}|_{{g^\prime } = g}
=
{\cal F}_c+{\cal F}_{\rm  res} ,
\end{gather*}
where the second part
$H_{\rm  res}
({\cal F}_{\rm  res})$
means a residual component out of a {\em well-defined}
collective submanifold for which we should search now.

For our purpose, let us introduce another curvature
${\cal C}^\prime_{t,\Lambda_n}$ and
${\cal C}^\prime_{t,\Lambda_n^\star }$
with the same forms as those
in (\ref{Cc-val}),
except that the Hamiltonian ${\cal F}_{\rm c}$ is replaced
by
${\cal F}|_{{g^\prime } = g}$
$(= {\cal F}_{\rm  c}
+
{\cal F}_{\rm res})$.
The quasi-particle vacuum expectation values
of the Lie-algebra-valued curvatures are easily calculated~as
\begin{gather}
{\langle C^\prime_{t,\Lambda_n} \rangle }_g
=
-i\partial_{\Lambda_n}
{\langle H_{\rm res}/\hbar \rangle }_g ,\qquad
{\langle C^\prime_{t,\Lambda_n^\star } \rangle }_g
=
-i\partial_{\Lambda_n^\star }
{\langle H_{\rm res}/\hbar \rangle }_g ,
\label{C-res}
\\
{\langle H_{\rm res} \rangle }_g
=
-\tfrac{1}{4}\mbox{Tr}
\left\{
g
\left[
\begin{array}{rr}
-1_N & 0 \\
0  & 1_N
\end{array}
\right]
g^\dagger({\cal F}_g - i\hbar\partial_tg \cdot g^\dagger)
\right\} ,
\label{Hres-expectation}
\end{gather}
where we have used $\langle C_{t,\Lambda_n} \rangle = 0$ and
$\langle C_{t,\Lambda_n^\star } \rangle = 0$.
The above equations
(\ref{C-res}) and (\ref{Hres-expectation})
are interpreted that the values of
${\langle {C^\prime }_{t,\Lambda_n} \rangle }_g$ and
${\langle {C^\prime }_{t,\Lambda_n^\star } \rangle }_g$
represent the {\em gradient} of energy of the residual Hamiltonian
in the $2m$-dimensional manifold.
Suppose there exists the {\em well-defined}
collective submanifold.
Then it will be not so wrong to deduce the following remarks:
the energy value of the residual Hamiltonian becomes
almost constant on the collective submanifold, i.e.,
\begin{gather*}
\delta_g\langle H_{\rm res} \rangle_g
\cong 0
\qquad \mbox{and} \qquad
\partial_{\Lambda_n}
\langle H_{{\rm res}} \rangle_g \cong 0 ,\qquad
\partial_{\Lambda_n^\star }
\langle H_{{\rm res}} \rangle_g \cong 0 ,
\end{gather*}
where $\delta_g$ means $g$-variation, regarding $g$
as function of $(\Lambda, \Lambda^\star)$ and $t$.
It may be achieved if we should determine $g$
(collective path) and
${\cal F}_{\rm c}$
(collective Hamiltonian)
through auxiliary quantity $(\theta, \theta^\dagger)$
so as to satisfy
$H_{\rm c} +{\rm const}
=
H_{\rm HB}$ as far as possible.
Putting
${\cal F}_{\rm c} = {\cal F}$
in
(\ref{Cc-val}),
we seek for $g$ and ${\cal F}_{\rm c}$ satisfying
\begin{gather}
{\cal C}_{t,\Lambda_n} \cong 0 ,
 \qquad {\cal C}_{t,\Lambda_n^\star } \cong 0 ,\qquad
{\cal C}_{\Lambda_{n^\prime },\Lambda_n^\star } = 0 ,
\qquad {\cal C}_{\Lambda_{n^\prime },\Lambda_n} = 0 ,
\qquad {\cal C}_{\Lambda_{n^\prime }^\star,\Lambda_n^\star } = 0 .
\label{Cnearly0}
\end{gather}
The set of the equations ${\cal C}_{\bullet,\bullet }=0$
makes an essential role to determine the collective submanifold
in the TDHBT.
The set of the equations~(\ref{Cnearly0}) and (\ref{D-tD-Lambdag})
becomes our fundamental equation
for describing the collective motions,
under the restrictions~(\ref{restriction}).

If we want to describe the collective motions
through the TD complex variables
$(\check{\Lambda }(t), \check{\Lambda }^\star(t))$
in the usual manner, we must inevitably know
$\check{\Lambda }$ and $\check{\Lambda }^\star$
as functions of
$(\Lambda, \Lambda^\star)$ and $t$.
For this aim,
it is necessary to discuss the correspondence of
the Lagrange-like manner to the usual one.

First let us def\/ine the Lie-algebra-valued inf\/initesimal
generator of collective submanifolds~as
\begin{gather*}
\check{O}_n^\dagger
\stackrel{d}{=}
(i\partial_{\check{\Lambda }_n}U^{-1}(\check{g}))U(\check{g}) ,\qquad
\check{O}_n
\stackrel{d}{=}
(i\partial_{\check{\Lambda }_n^\star }U^{-1}(\check{g}))U(\check{g}) , \qquad
(\check{g} \in g) ,
\end{gather*}
whose form is the same as the one in (\ref{HcdefOcdef}).
To guarantee
$\check{\Lambda }_n(t)$
and
$\check{\Lambda }_n^\star(t)$
to be canonical, according to
\cite{Ma.80,YK.81},
we set up the following expectation values with use of
the $SO(2N)$ (HB) wave function $|\phi(\check{g})\rangle$:
\begin{gather}
\langle\phi(\check g)|i\partial_{{\check \Lambda }_n}
|\phi(\check g)\rangle
=
\langle\phi(\check g)|{\check O}_n^\dagger
|\phi(\check g)\rangle
=
i  \tfrac{1}{2}{\check \Lambda }_n^\star ,\nonumber\\
\langle\phi(\check g)|i\partial_{{\check \Lambda }_n^\star }
|\phi(\check g)\rangle
=
\langle\phi(\check g)|{\check O}_n
|\phi(\check g)\rangle
=
-i  \tfrac{1}{2}{\check \Lambda }_n.
\label{canonikalcondition}
\end{gather}
The above relation leads us to the {\em weak}
canonical commutation relation
\begin{gather}
\langle\phi(\check{g})|
[\check{O}_n , \check{O}_{n^\prime }^\dagger]
|\phi(\check{g})\rangle
=\delta_{nn^\prime } ,\nonumber\\
\langle\phi(\check{g})|
[\check{O}_n^\dagger , \check{O}_{n^\prime }^\dagger]
|\phi(\check{g})\rangle
= 0 , \qquad
\langle\phi(\check{g})|[\check{O}_n, \check{O}_{n^\prime }]
|\phi(\check{g})\rangle
= 0
\qquad (n, n^\prime = 1,\dots,m)
\label{weakorthgonality}
\end{gather}
the proof of which was shown in
\cite{Ma.80}
and
\cite{YK.81}.

Using (\ref{D-tD-Lambda}),
the collective Hamiltonian
$H_{\rm c}/\hbar$
and
the inf\/initesimal generators $O_n^\dagger$ and  $O_n$
in the Lagrange-like manner are expressed in terms of
inf\/initesimal ones $\check{O}_n^\dagger$ and $\check{O}_n$
in the usual way as follows:
\begin{gather}
H_{\rm c}/\hbar
=
\partial_t\check{\Lambda }_n\check{O}_n^\dagger
+\partial_t{\check{\Lambda }_n^\star }\check{O}_n,\nonumber\\
O_n^\dagger
=
\partial_{\Lambda_n}\check{\Lambda }_{n^\prime }
\check{O}_{n^\prime }^\dagger
+\partial_{\Lambda_n}{\check{\Lambda }_{n^\prime }^\star }
\check{O}_{n^\prime } ,\qquad
O_n
=
\partial_{\Lambda_n^\star }\check{\Lambda }_{n^\prime }
\check{O}_{n^\prime }^\dagger
+\partial_{\Lambda_n^\star }{\check{\Lambda }_{n^\prime }^\star }
\check{O}_{n^\prime } .
\label{LagHetstotheusualone}
\end{gather}
Substituting (\ref{LagHetstotheusualone})
into (\ref{Cexplicit}),
it is easy to evaluate the expectation values
of the Lie-algebra-valued curvatures
$C_{\bullet,\bullet }$ by the $SO(2N)$ (HB) wave function
$|\phi[\check{g}(\check{\Lambda }(t),\check{\Lambda }^\star(t))]
\rangle $ ($=|\phi[g(\Lambda,\Lambda^\star,t)]\rangle$).
A {\em weak\/} integrability condition requiring
the expectation values
$\langle\phi(\check{g})|C_{\bullet,\bullet }|\phi(\check{g})\rangle
=0$ yields the following set of partial dif\/ferential equations with aid of
the quasi-particle vacuum property, $d|\phi(g)\rangle=0$:
\begin{gather}
\partial_{\Lambda_n}\check{\Lambda }_{n^\prime }
\partial_t\check{\Lambda }_{n^\prime }^\star
-
\partial_{\Lambda_n}\check{\Lambda }_{n^\prime }^\star
\partial_t\check{\Lambda }_{n^\prime }
=
\partial_{\Lambda_n^\star }\check{\Lambda }_{n^\prime }
\partial_t\check{\Lambda }_{n^\prime }^\star
-
\partial_{\Lambda_n^\star}\check{\Lambda }_{n^\prime }^\star
\partial_t\check{\Lambda }_{n^\prime }
=
 \tfrac{1}{4}\mbox{Tr}
\{{\cal R}(g)[\theta_n^\dagger,
{\cal F}_{\rm c} /\hbar]\} ,\label{Lagrangebracketandweakcondition1}
\\
\partial_{\Lambda_n^\star }\check{\Lambda }_{n^{\prime\prime }}
\partial_{\Lambda_{n^\prime }}\check{\Lambda }_{n^{\prime\prime }}^\star
-
\partial_{\Lambda_n^\star }\check{\Lambda }_{n^{\prime\prime }}^\star
\partial_{\Lambda_{n^\prime }}\check{\Lambda }_{n^{\prime\prime }}
=
\tfrac{1}{4}\mbox{Tr}
\{{\cal R}(g)[\theta_n,\theta_{n^\prime }^\dagger]\} ,\nonumber\\
\partial_{\Lambda_n}\check{\Lambda }_{n^{\prime\prime }}
\partial_{\Lambda_{n^\prime }}\check{\Lambda }_{n^{\prime\prime }}^\star
-
\partial_{\Lambda_n}\check{\Lambda }_{n^{\prime\prime }}^\star
\partial_{\Lambda_{n^\prime }}\check{\Lambda }_{n^{\prime\prime }}
=
\tfrac{1}{4}\mbox{Tr}
\{{\cal R}(g)[\theta_n^\dagger,\theta_{n^\prime }^\dagger]\} , \label{Lagrangebracketandweakcondition2}\\
\partial_{\Lambda_n^\star }\check{\Lambda }_{n^{\prime\prime }}
\partial_{\Lambda_{n^\prime }^\star }\check{\Lambda }_{n^{\prime\prime }}^\star
-
\partial_{\Lambda_n^\star }\check{\Lambda }_{n^{\prime\prime }}^\star
\partial_{\Lambda_{n^\prime }^\star }\check{\Lambda }_{n^{\prime\prime }}
=
 \tfrac{1}{4}\mbox{Tr}
\{{\cal R}(g)[\theta_n,\theta_{n^\prime }]\} ,
\nonumber
\end{gather}
where an $SO(2N)$ (HB) density matrix ${\cal R}(g)$
is def\/ined as
\begin{gather}
{\cal R}(g)
\stackrel{d}{=}
g
\left[
\begin{array}{rr}
-1_N & 0 \\
0 & 1_N
\end{array}
\right]
g^\dagger ,
\qquad {\cal R}^\dagger(g) = {\cal R}(g) ,
\qquad {\cal R}^{2}(g) = 1_{2N} ,
\label{densityRg}
\end{gather}
in which
$g$ becomes function of
the complex variables $(\Lambda,\Lambda^\star)$ and $t$.
We here have used the transformation property (\ref{E-E}),
the trace formulae
equations (\ref{HcTr-express}) and (\ref{OcTr-express})
and the dif\/ferential formulae, i.e.,
\begin{alignat*}{3}
& \langle\phi(\check{g})|i\partial_{\check{\Lambda }_{n^\prime }}
\check{O}_n
|\phi(\check{g})\rangle
=
-  \tfrac{1}{2} \delta_{nn^\prime } ,\qquad &&
\langle\phi(\check{g})|i\partial_{\check{\Lambda }_{n^\prime }^\star }
\check{O}_n
^\dagger|\phi(\check{g})\rangle
=
\tfrac{1}{2}\delta_{nn^\prime } ,& \nonumber\\
& \langle\phi(\check{g})|i\partial_{\check{\Lambda }_{n^\prime }}
\check{O}_n
^\dagger|\phi(\check{g})\rangle = 0 ,\qquad &&
\langle\phi(\check{g})|i\partial_{\check{\Lambda }_{n^\prime }^\star }
\check{O}_n
|\phi(\check{g})\rangle = 0 ,&
\end{alignat*}
which owe to the canonicity condition
(\ref{canonikalcondition}) and
{\em weak} canonical commutation relation
(\ref{weakorthgonality}).

Through the above procedure,
as a f\/inal goal,
we get the correspondence of the Lagrange-like manner
to the usual one.
We have no unknown quantities in the r.h.s.\ of
equations~(\ref{Lagrangebracketandweakcondition1})
and~(\ref{Lagrangebracketandweakcondition2}),
if we could completely solve our fundamental equations
to describe the collective motion.
Then we come up to be able to know in principle
the explicit forms of
$(\check{\Lambda }, \check{\Lambda }^\star)$
in terms of
$(\Lambda, \Lambda^\star)$ and $t$
by solving the partial dif\/ferential equations
(\ref{Lagrangebracketandweakcondition1})
and~(\ref{Lagrangebracketandweakcondition2}).
However we should take enough notice of roles
dif\/ferent from each other made by
equations~(\ref{Lagrangebracketandweakcondition1})
and~(\ref{Lagrangebracketandweakcondition2}), respectively,
to construct the solutions.
Especially, it turns out that the l.h.s.\ in~(\ref{Lagrangebracketandweakcondition2})
has a close connection with Lagrange bracket.
From the outset we have set up the canonicity condition
to guarantee the complex variables
$(\check{\Lambda }, \check{\Lambda }^\star)$
in the usual manner to be canonical.
Thus the variables
$(\check{\Lambda }, \check{\Lambda }^\star)$
are interpreted as functions giving
a canonical transformation from
$(\check{\Lambda }, \check{\Lambda }^\star)$ to
another complex variables $(\Lambda, \Lambda^\star)$
in the Lagrange-like manner.
From this interpretation, we see that
the canonical invariance requirements impose the following
restrictions on the r.h.s.\ of~(\ref{Lagrangebracketandweakcondition2}):
\begin{gather}
- \tfrac{1}{4}\mbox{Tr}
\{{\cal R}(g)
[\theta_n,\theta_{n^\prime }^\dagger]\}
=
\delta_{nn^{\prime }} ,\qquad\!\!
 \tfrac{1}{4}\mbox{Tr}
\{{\cal R}(g)
[\theta_n^\dagger,\theta_{n^\prime }^\dagger]\}
=
0 ,\qquad\!\!
\tfrac{1}{4}\mbox{Tr}
\{{\cal R}(g)[\theta_n,\theta_{n^\prime }]\}
=
0 .\!
\label{restriction}
\end{gather}
Using
(\ref{Lagrangebracketandweakcondition2}) and
(\ref{restriction}),
we get Lagrange brackets for canonical transformation of
$(\check{\Lambda },\check{\Lambda }^\star)$ to~$(\Lambda,\Lambda^\star)$.



\subsection{Validity of maximally-decoupled theory}\label{section2.3}

First
we transform the set of the fundamental equations
in the particle frame
into the one in the quasi-particle frame.
The $SO(2N)$ (TDHB) Hamiltonian of the system
is expressed as
\begin{gather}
H_{\rm HB}
=
 \tfrac{1}{2}[d,d^\dagger]{\cal F}_o
\left[
\begin{array}{c}
d^\dagger \\
d
\end{array}
\right] ,\qquad
{\cal F}_o
=
\left[
\begin{array}{cc}
-F_o^\star & -D_o^\star \\
 D_o       &  F_o
\end{array}
\right] ,\qquad
{\cal F}_o^\dagger = {\cal F}_o ,
\label{HB-matrixinqf}
\end{gather}
the relation of which to the original TDHB Hamiltonian ${\cal F}$
is given by ${\cal F}_o = g^\dagger {\cal F}g$, $g \in SO(2N)$.
The inf\/initesimal generators of collective submanifolds
and their integrability conditions expressed as
the Lie-algebra-valued equations are also rewritten into
the ones in the quasi-particle frame as follows:
\begin{gather}
H_{\rm c}
=
\tfrac{1}{2}[d, d^\dagger]
{\cal F}_{o-{\rm c}}
\left[ \!
\begin{array}{c}
d^\dagger \\
d
\end{array} \!
\right] ,\nonumber
\\
O_n^\dagger
=
\tfrac{1}{2}[d, d^\dagger]\theta_{o-n}^\dagger
\left[
\begin{array}{c}
d^\dagger \\
d
\end{array}
\right],\qquad
O_n
=
\tfrac{1}{2}[d, d^\dagger]\theta_{o-n}
\left[
\begin{array}{c}
d^\dagger \\
d
\end{array}
\right] ,\nonumber
\\
C_{t,\Lambda_n}
=
 \tfrac{1}{2}[d,d^\dagger]
{\cal C}_{o-t,\Lambda_n}
\left[
\begin{array}{c}
d^\dagger \\
d
\end{array}
\right]
= 0 , \nonumber\\
{\cal C}_{o-t,\Lambda_n}
=
i\partial_t\theta_{o-n}^\dagger
-
i\partial_{\Lambda_n}{\cal F}_{o-{\rm c}}/\hbar
-
[\theta_{o-n}^\dagger,{\cal F}_{o-{\rm c}}/\hbar] ,\nonumber\\
C_{t,\Lambda_n^\star }
=
\tfrac{1}{2}[d,d^\dagger]
{\cal C}_{o-t,\Lambda_n^\star }
\left[
\begin{array}{c}
d^\dagger \\
d
\end{array}
\right]
= 0 ,\nonumber\\
{\cal C}_{o-t,\Lambda_n^\star }
=
i\partial_t\theta_{o-n}
-
i\partial_{\Lambda_n^\star }{\cal F}_{o-{\rm c}}/\hbar
-
[\theta_{o-n},{\cal F}_{o-{\rm c}}/\hbar] ,\nonumber \\
C_{\Lambda_{n^\prime },\Lambda_n}
=
\tfrac{1}{2}[d,d^\dagger]
{\cal C}_{o-\Lambda_{n^\prime },\Lambda_n}
\left[
\begin{array}{c}
d^\dagger \\
d
\end{array}
\right]
= 0 ,\nonumber\\
{\cal C}_{o-\Lambda_{n^\prime },\Lambda_n}
=
i\partial_{\Lambda_{n^\prime }}\theta_{o-n}^\dagger
-
i\partial_{\Lambda_n}\theta_{o-n^\prime }^\dagger
-
[\theta_{o-n}^\dagger,\theta_{o-n^\prime }^\dagger] , \label{Copeonqp-frame}\\
C_{\Lambda_{n^\prime },\Lambda_n^\star }
=
\tfrac{1}{2}[d,d^\dagger]
{\cal C}_{o-\Lambda_{n^\prime },\Lambda_n^\star }
\left[
\begin{array}{c}
d^\dagger \\
d
\end{array}
\right]
= 0 , \nonumber\\
{\cal C}_{o-\Lambda_{n^\prime },\Lambda_n^\star }
=
i\partial_{\Lambda_{n^\prime }}\theta_{o-n}
-
i\partial_{\Lambda_n^\star }\theta_{o-n^\prime }^\dagger
-
[\theta_{o-n},\theta_{o-n^\prime }^\dagger], \nonumber\\
C_{\Lambda_{n^\prime }^\star,\Lambda_n^\star }
=
\tfrac{1}{2}[d,d^\dagger]
{\cal C}_{o-\Lambda_{n^\prime }^\star,\Lambda_n^\star }
\left[
\begin{array}{c}
d^\dagger \\
d
\end{array}
\right]
= 0 ,\nonumber\\
{\cal C}_{o-\Lambda_{n^\prime }^\star,\Lambda_n^\star }
=
i\partial_{\Lambda_{n^\prime }^\star } \theta_{o-n}
-
i\partial_{\Lambda_n^\star }\theta_{o-n^\prime }
-
[\theta_{o-n},\theta_{o-n^\prime }] .
\nonumber
\end{gather}
The quantities
${\cal F}_{o-{\rm c}}$, $\theta_{o-n}^\dagger$
$(= g^\dagger \theta_n^\dagger g)$
and
$\theta_{o-n}$ $(= g^\dagger \theta_n g)$
are def\/ined through partial dif\/ferential equations
on the $SO(2N)$ Lie group manifold,
\begin{gather}
-i\hbar\partial_t g^\dagger
=
{\cal F}_{o-{\rm c}} g^\dagger  \qquad \mbox{and} \qquad
-i\partial_{\Lambda_n} g^\dagger
=
\theta_{o-n}^\dagger g^\dagger ,\qquad
-i\partial_{\Lambda_n^\star }g^\dagger
=
\theta_{o-n} g^\dagger .
\label{TDFcSDthetaonqp-frame}
\end{gather}
In the above set of (\ref{Copeonqp-frame}),
all the curvatures ${\cal C}_{o-\bullet ,\bullet }$
should be made equal to zero.

The full TDHB Hamiltonian is decomposed into the collective
one and the residual one as
\begin{gather}
H_{\rm HB}
=
H_{\rm c} + H_{\rm res} ,
 \qquad {\cal F}_o
=
{\cal F}_{o-{\rm c}} + {\cal F}_{o-{\rm res}} ,
\label{HcHresinqf}
\end{gather}
at the reference point $g$ on the $SO(2N)$ group manifold.
Following the preceding section,
let us introduce other curvatures
$C_{t,\Lambda_n}^\prime$ and  $C_{t,\Lambda_n^\star }^\prime$
with the same forms as those in
(\ref{Copeonqp-frame})
except that ${\cal F}_{o-{\rm c}}$ is replaced by
${\cal F}_o$.
Then the corresponding curvatures
${\cal C}_{o-t,\Lambda_n}^\prime$ and
${\cal C}_{o-t,\Lambda_n^\star }^\prime$
are also divided into two terms,
\begin{gather*}
{\cal C}_{o-t,\Lambda_n}^\prime
=
{\cal C}_{o-t,\Lambda_n}^{\rm c} +
{\cal C}_{o-t,\Lambda_n}^{\rm res} ,\qquad
{\cal C}_{o-t,\Lambda_n^\star }^\prime
=
{\cal C}_{o-t,\Lambda_n^\star }^{\rm c} +
{\cal C}_{o-t,\Lambda_n^\star }^{\rm res} .
\end{gather*}
Here the collective curvatures
${\cal C}_{o-t,\Lambda_n}^{\rm c}$ and
${\cal C}_{o-t,\Lambda_n^\star }^{\rm c}$
arising from ${\cal F}_{o-{\rm c}}$
are def\/ined as the same forms as
the ones in~(\ref{Copeonqp-frame}).
The residual curvatures
${\cal C}_{o-t,\Lambda_n}^{\rm res}$ and
${\cal C}_{o-t,\Lambda_n^\star }^{\rm res}$
arising from
${\cal F}_{o-{\rm res}}$ are def\/ined as
\begin{gather}
{\cal C}_{o-t,\Lambda_n}^{\rm res}
=
-i\partial_{\Lambda_n}{\cal F}_{o-{\rm res}}/\hbar
-
[\theta_{o-n}^\dagger,{\cal F}_{o-{\rm res}}/\hbar] ,\nonumber\\
{\cal C}_{o-t,\Lambda_n^\star }^{\rm res}
=
-i\partial_{\Lambda_n^\star }{\cal F}_{o-{\rm res}}/\hbar
-
[\theta_{o-n},{\cal F}_{o-{\rm res}}/\hbar] .
\label{explicitCres}
\end{gather}
Using
(\ref{TDFcSDthetaonqp-frame}) and (\ref{HcHresinqf}),
the Lie-algebra-valued forms of the curvatures are calculated as
\begin{gather*}
C^{\rm res}_{t,\Lambda_n}
=
-i\partial_{\Lambda_n} H_{\rm res}/\hbar ,
\qquad C^{\rm res}_{t,\Lambda_n^\star }
=
-i\partial_{\Lambda_n^\star }H_{\rm res}/\hbar .
\end{gather*}
Supposing there exist the well-def\/ined collective submanifolds
satisfying
(\ref{TDFcSDthetaonqp-frame}),
we should demand that the following curvatures are made equal
to zero:
\begin{gather}
{\cal C}^{\rm c}_{o-t,\Lambda_n} = 0 ,\qquad
{\cal C}^{\rm c}_{o-t,\Lambda_n^\star } = 0 ,
\label{Ct-Lambda0onthecollectivsubman}
\\
{\cal C}_{o-\Lambda_{n^\prime },\Lambda_n} = 0 ,\qquad
{\cal C}_{o-\Lambda_{n^\prime },\Lambda_n^\star } = 0 ,\qquad
{\cal C}_{o-\Lambda_{n^\prime }^\star,\Lambda_n^\star } = 0 ,
\label{C0oncollective}
\end{gather}
the f\/irst equation (\ref{Ct-Lambda0onthecollectivsubman})
of which lead us to the Lie-algebra-valued relations,
\begin{gather}
C^\prime_{t,\Lambda_n}
=
C^{\rm res}_{t,\Lambda_n}
=
-i\partial_{\Lambda_n} H_{\rm res}/\hbar ,\qquad
C^\prime_{t,\Lambda_n^\star }
=
C^{\rm res}_{t,\Lambda_n^\star }
=
-i\partial_{\Lambda_n^\star }H_{\rm res}/\hbar .
\label{CCresgradHres}
\end{gather}
Then the curvature
$C^\prime_{t,\Lambda_n}$
and
$C^\prime_{t,\Lambda_n^\star }$
can be regarded as the {\em gradients} of
quantum-mechanical potentials due to the existence
of the residual Hamiltonian $H_{{\rm res}}$ on
the collective submanifolds.
The potentials become almost f\/lat on
the collective submanifolds, i.e.,
$H_{\rm HB}
=
H_{\rm c} + {\rm const}$,
if the proper subspace determined
is an almost invariant subspace of the full TDHB Hamiltonian.
This collective subspace is an almost
degenerate eigenspace of the residual Hamiltonian.
Therefore it is naturally deduced that,
provided there exists the well-def\/ined collective subspace,
the residual curvatures at a point on the subspace
are extremely small. Thus, the way of extracting
the collective submanifolds out of the full TDHB manifold
is made possible by the minimization of
the residual curvature, for which a deep insight
into~(\ref{CCresgradHres})
becomes necessary.

Finally, the restrictions to assure the Lagrange bracket
for the usual collective variables and Lagrange-like ones are
transformed into the following forms represented in the QPF:
\begin{gather*}
-  \tfrac{1}{4}\mbox{Tr}
\left\{
\left[
\begin{array}{rr}
-1_N & 0 \\
 0 & 1_N
\end{array}
\right]
[\theta_{o-n},\theta_{o-n^\prime }^\dagger]
\right\}
=
\delta_{nn^{\prime }} ,\nonumber\\
 \tfrac{1}{4}\mbox{Tr}
\left\{
\left[
\begin{array}{rr}
-1_N & 0 \\
 0 & 1_N
\end{array}
\right]
[\theta_{o-n}^\dagger,\theta_{o-n^\prime }^\dagger]
\right\}
=
0 , \qquad
\tfrac{1}{4}\mbox{Tr}
\left\{
\left[
\begin{array}{rr}
-1_N & 0 \\
 0 & 1_N
\end{array}
\right]
[\theta_{o-n},\theta_{o-n^\prime }]
\right\}
=
0 .
\end{gather*}

We discuss here how the Lagrange-like manner picture
is transformed into the usual one.
First let us regard any point on the collective submanifold
as a set of initial points (initial value) in the usual manner.
Suppose we observe the time evolution of the system
with various initial values. Then we have the following relations
which make a connection between the Lagrange-like manner
and the usual one
\begin{gather}
{\cal F}_{o-{\rm c}}/\hbar
=
\partial_t\check{\Lambda }_n\check{\theta }_{o-n}^\dagger
+
\partial_t{\check{\Lambda }_n^\star }\check{\theta }_{o-n} ,
\label{transferedFo-c}
\\
\theta_{o-n}^\dagger
=
\partial_{\Lambda_n}\check{\Lambda }_{n^\prime }
\check{\theta }_{o-n^\prime }^\dagger
+
\partial_{\Lambda_n}{\check{\Lambda }_{n^\prime }^\star }
\check{\theta }_{o-n^\prime } ,\qquad
\theta_{o-n}
=
\partial_{\Lambda_n^\star }\check{\Lambda }_{n^\prime }
\check{\theta }_{o-n^\prime }^\dagger
+
\partial_{\Lambda_n^\star }{\check{\Lambda }_{n^\prime }^\star }
\check{\theta }_{o-n^\prime } ,
\label{thetao-n}
\end{gather}
in which the transformation functions are set up by
the initial conditions,
\begin{gather*}
  \check{\Lambda }_n(t)|_{t=0}
=
\check{\Lambda }_n(\Lambda, \Lambda^\star, t)|_{t=0}
=
\Lambda_n ,\qquad
\check{\Lambda }_n^\star(t)|_{t=0}
=
\check{\Lambda }_n^\star(\Lambda, \Lambda^\star, t)|_{t=0}
=
\Lambda_n^\star ,\nonumber
\\
\partial_{\Lambda_n}\check{\Lambda }_{n^\prime }|_{t=0}
=
\delta_{nn^\prime }, \qquad
\partial_{\Lambda_n^\star }\check{\Lambda }_{n^\prime }^\star|_{t=0}
=
\delta_{nn^\prime } , \qquad
\partial_{\Lambda_n}\check{\Lambda }_{n^\prime }^\star|_{t=0}
= 0 , \qquad
\partial_{\Lambda_n^\star }\check{\Lambda }_{n^\prime }|_{t=0}
= 0 ,
\end{gather*}
in order to guarantee both pictures to coincide at time $t=0$.
On the other hand, our collective Hamiltonian
${\cal F}_{o-{\rm c}}$
can also be expressed in the form
\begin{gather}
{\cal F}_{o-{\rm c}}/\hbar
=
v_n(\Lambda,\Lambda^\star,t)\theta_{o-n}^\dagger+
v_n^\star(\Lambda,\Lambda^\star,t)\theta_{o-n},
\label{velocityexpless}
\end{gather}
where the expansion coef\/f\/icients $v_n$ and $v_n^\star$
are interpreted as velocity f\/ields in the Lagrange-like manner.
Substituting
(\ref{thetao-n}) into
(\ref{velocityexpless}) and comparing with
(\ref{transferedFo-c}),
we can get the relations
\begin{gather*}
\dot{\check \Lambda }_n
=
\partial_t{\check \Lambda }_n
=
v_{n^\prime }\partial_{\Lambda_{n^\prime }}{\check \Lambda }_n
+
v_{n^\prime }^\star\partial_{\Lambda_{n^\prime }^\star }
{\check \Lambda }_n ,\qquad
\dot{\check \Lambda }_n^\star
=
\partial_t{\check \Lambda }_n^\star
=
v_{n^\prime }\partial_{\Lambda_{n^\prime }}{\check \Lambda }_n^\star
+
v_{n^\prime }^\star\partial_{\Lambda_{n^\prime }^\star }
{\check \Lambda}_n^\star ,
\end{gather*}
from which the initial conditions of
the velocity f\/ields are given as
\begin{gather*}
\dot{\check \Lambda }_n(t)|_{t=0}
=
\partial_t{\check \Lambda }_n|_{t=0}
=
v_n(\Lambda,\Lambda^\star,t)|_{t=0} ,\qquad
\dot{\check \Lambda }_n^\star (t)|_{t=0}
=
\partial_t{\check \Lambda }_n^\star |_{t=0}
=
v_n^\star (\Lambda,\Lambda^\star,t)|_{t=0} .\!\!
\end{gather*}
Then we obtain the correspondence of the time derivatives
of the collective co-ordinates in the usual manner to the
velocity f\/ields in the Lagrange-like one.

Finally we impose the canonicity conditions in the usual manner,{\samepage
\begin{gather}
\langle\phi(\check g)|{\check O}_n^\dagger
|\phi(\check g)\rangle
=
i  \tfrac{1}{2} {\check \Lambda }_n^\star ,
 \qquad \langle\phi(\check g)|{\check O}_n
|\phi(\check g)\rangle
=
-i  \tfrac{1}{2} {\check \Lambda }_n ,
\label{canonicalconditioninqf}
\end{gather}
which leads us to the {\em weak} canonical
commutation relation  with the aid of
(\ref{C0oncollective})
and
(\ref{restriction}).}

The TDHBT for {\em maximally-decoupled}
collective motions can be formulated
parallel with TDHFT~\cite{Ma.80}.
The basic concept of the theory lies in an introduction
of the {\em invariance principle of the Schr\"{o}dinger
equation}, and the TDHBEQ is solved under
the canonicity condition and the vanishing of non-collective
dangerous terms. However, as we have no justif\/ication
on the validity of the {\em maximally-decoupled}
method, we must give a criterion how it extracts the collective
submanifold ef\/fectively out of the full TDHB manifold.
We are now in a position to derive some quantities by which
the criterion is established.
For this aim,
we express the collective Hamiltonian
${\cal F}_{o-{\rm c}}$
and the residual one
${\cal F}_{o-{\rm res}}$
in the same form as the one of the TDHB Hamiltonian
${\cal F}_o$ given in
(\ref{HB-matrixinqf}).
We also represent quantities $\theta_{o-n}^\dagger$,
${\cal C}_{o-t,\Lambda_n}^{{\rm res}}$ and
${\cal C}_{o-t,\Lambda_n^\star }^{{\rm res}}$
which consist of $N  \times  N$ block matrices
as follows:
\begin{gather}
\theta_{o-n}^\dagger
=
\left[
\begin{array}{rr}
\xi_o & \varphi_o\\
\psi_o & -\xi_o^{\rm T}
\end{array}
\right]_n,
\qquad \psi_o^{\rm T}
=
-\psi_o ,
\qquad \varphi_o^{\rm T}
=
-\varphi_o ,
\label{matirixoftheta}
\\
{\cal C}_{o-t,\Lambda_n}^{{\rm res}}
=
\left[
\begin{array}{cc}
{\cal C}_\xi^{{\rm res}}&
{\cal C}_\varphi^{{\rm res}}\\
{\cal C}_\psi^{{\rm res}}
&-{\cal C}_\xi^{{\rm res}}
{}^{\rm T}
\end{array}
\right]_n,
 \qquad {\cal C}_\psi^{{\rm res}}
{}^{\rm T}
=
-{\cal C}_\psi^{{\rm res}},
\qquad {\cal C}_\varphi^{{\rm res}}
{}^{\rm T}
=
-{\cal C}_\varphi^{{\rm res}},\nonumber\\
{\cal C}_{o-t,\Lambda_n^\star }^{{\rm res}}
=
\left[
\begin{array}{cc}
{\cal C}_{\xi^\star }^{{\rm res}}&
{\cal C}_{\varphi^\star }^{{\rm res}}\\
{\cal C}_{\psi^\star }^{{\rm res}}&
-
{\cal C}_{\xi^\star }^{{\rm res}}
{}^{\rm T}
\end{array}
\right]_n,
\qquad {\cal C}_{\xi^\star }^{{\rm res}}
=
-{\cal C}_\xi^{{\rm res}}{}^\dagger,
\qquad {\cal C}_{\psi^\star }^{\rm res}
=
{\cal C}_\psi^{{\rm res}}{}^\star,
\qquad {\cal C}_{\varphi^\star }^{{\rm res}}
=
{\cal C}_\varphi^{{\rm res}}{}^\star.
\label{matrixofCres}
\end{gather}
Substitution of the explicit form of
${\cal F}_{o-{\rm res}}$ and
equations
(\ref{matirixoftheta}) and (\ref{matrixofCres})
into
(\ref{explicitCres})
yields
\begin{gather}
{\cal C}_{\xi,n}^{{\rm res}}
=
i\partial_{\Lambda_n}F_{o-{\rm res}}^\star/\hbar
+
[\xi_{o-n},F_{o-{\rm res}}^\star/\hbar]
-
\varphi_{o,n}{\cal D}_{o-{\rm res}}/\hbar
-
D_{o-{\rm res}}^\star/\hbar\psi_{o,n} ,\nonumber\\
{\cal C}_{\psi,n}^{{\rm res}}
=
-i\partial_{\Lambda_n}D_{o-{\rm res}}/\hbar+\xi_{o,n}
^{\rm T}
D_{o-{\rm res}}/\hbar
+
D_{o-{\rm res}}/\hbar\xi_{o,n}
\!+\!
\psi_{o,n}F_{o-{\rm res}}^\star/\hbar
+
F_{o-{\rm res}}/\hbar\psi_{o,n} ,\!\!\!\label{explicitCres2}\\
{\cal C}_{\varphi,n}^{{\rm res}}
=
i\partial_{\Lambda_n}D_{o-{\rm res}}^\star/\hbar
+
\xi_{o,n}D_{o-{\rm res}}^\star/\hbar
+
D_{o-{\rm res}}^\star/\hbar\xi_{o,n}
^{\rm T}
-
\varphi_{o,n}F_{o-{\rm res}}/\hbar
-
F_{o-{\rm res}}^\star/\hbar\varphi_{o,n} .\nonumber
\end{gather}
The quantity ${\check{\cal \theta }}_{0-n}^\dagger$ can
also be expressed in the same form as the one
def\/ined in~(\ref{matirixoftheta}).
Substituting this expression into~(\ref{transferedFo-c})
and~(\ref{canonicalconditioninqf}),
we obtain the relations
\begin{gather}
F_{o-{\rm res}}/\hbar
=
F_o/\hbar+\partial_t \check{\Lambda }_n
\check{\xi }_{o,n}^{\rm T}
+
\partial_t \check{\Lambda }_n^\star\check{\xi }_{o,n}^\star ,\nonumber\\
D_{o-{\rm res}}/\hbar
=
D_o/\hbar+\partial_t \check{\Lambda }_n
\check{\psi }_{o,n}^{\rm T}
+\partial_t \check{\Lambda }_n^\star\check{\varphi }_{o,n}^\star ,
\label{Fo-resDo-res}
\end{gather}
together with their complex conjugate and
\begin{gather}
\mbox{Tr} \, \check{\xi }_{o,n}
=
i\check{\Lambda }_n^\star ,\qquad
\mbox{Tr} \, \check{\xi }_{o,n}^\dagger
=
-i\check{\Lambda }_n ,
\label{TRxiiLamba}
\end{gather}
where we have used
(\ref{HcHresinqf})
and the explicit forms of the Hamiltonian.

As was mentioned,
the way of extracting collective submanifolds
out of the full TDHB mani\-fold is made possible
by minimization of the residual curvature.
This is achieved if we require at least
expectation values of the residual curvatures
to be minimized as much as possible, i.e.,
\begin{gather}
\langle
\phi(g)|C_{t,\Lambda_n}^{{\rm res}}|\phi(g)
\rangle
=
\tfrac{1}{2}\mbox{Tr}\,
{\cal C}_{\xi,n}^{{\rm res}}
\cong 0 ,\qquad
\langle
\phi(g)|C_{t,\Lambda_n^\star }^{{\rm res}}|\phi(g)
\rangle
=
\tfrac{1}{2} \mbox{Tr}\,
{\cal C}_{\xi^\star,n}^{{\rm res}}
\cong 0 .
\label{Cres-0}
\end{gather}
We here adopt a condition similar to one of
the stationary HB method as was done in the TDHF~\cite{YK.81}:
The so-called dangerous terms in the residual Hamiltonian
${\cal F}_{o-{\rm res}}$ are made to vanish,
\begin{gather}
D_{o-{\rm res}} = 0 , \qquad
D_{o-{\rm res}}^\star = 0 .
\label{Dores0}
\end{gather}
With aid of equations
(\ref{explicitCres2}), (\ref{Fo-resDo-res})
and (\ref{TRxiiLamba}),
equations (\ref{Dores0}) and (\ref{Cres-0})
are rewritten as
\begin{gather}
D_o/\hbar
=
-\partial_t\check{\Lambda }_n\psi_{o,n}^{\rm T}
-\partial_t\check{\Lambda }_n^\star\varphi_{o,n}^\star ,\qquad
D_o^\star/\hbar
=
-\partial_t\check{\Lambda }_n\varphi_{o,n}
-\partial_t\check{\Lambda }_n^\star\psi_{o,n}^\dagger ,
\label{Do}
\\
i\partial_{\Lambda_n}\mbox{Tr}\,
F_{o-{\rm res}}/\hbar
=
i\partial_{\Lambda_n}\mbox{Tr} \,F_o/\hbar
-\partial_{\Lambda_n}
\left(
\partial_t \check{\Lambda }_{n^\prime }
\check{\Lambda }_{n^\prime }^\star
-\partial_t\check{\Lambda }_{n^\prime }^\star
\check{\Lambda }_{n^\prime }
\right)
\cong 0 ,\nonumber\\
i\partial_{\Lambda_n^\star }\mbox{Tr}\,
F_{o-{\rm res}}/\hbar
=
i\partial_{\Lambda_n^\star }\mbox{Tr} \,F_o/\hbar
-\partial_{\Lambda_n^\star }
\left(
\partial_t\check{\Lambda }_{n^\prime }
\check{\Lambda }_{n^\prime }^\star
-\partial_t\check{\Lambda }_{n^\prime }^\star
\check{\Lambda }_{n^\prime }
\right)
\cong 0 .
\label{id-LambdaTrFo-res}
\end{gather}

First, we will discuss how equation
(\ref{Do})
leads
us to the equation of path for the collective motion.
Notice that the quantities $\check{\theta }_{o-n}^\dagger$
and $\check{\theta }_{o-n}$ are subjected to
satisfy the same type of partial dif\/ferential equation
as that of
(\ref{TDFcSDthetaonqp-frame}).
Remember the explicit representation of an $SO(2N)$
matrix $g$ given in the previous section.
Then we have partial dif\/ferential equations
\begin{gather}
\check{\psi }_{0,n}
=
-i(\partial_{\check{\Lambda }_n}
\check{b}^{\rm T}\check{a}
+\partial_{\check{\Lambda }_n}
\check{a}^{\rm T}\check{b}) ,\qquad
\check{\varphi }_{0,n}
=
-i(\partial_{\check{\Lambda }_n}
\check{a}^\dagger\check{b}^\star
+\partial_{\check{\Lambda }_n}\check{b}^\dagger\check{a}^\star) .
\label{explicitpsiandvarph}
\end{gather}
together with its complex conjugate.
Putting the relation
${\cal F}_o = g^\dagger{\cal F}g$
and
(\ref{explicitpsiandvarph})
into
(\ref{Do}),
we get
\begin{gather}
\check{a}^{\rm T}
\big\{
(\check{D}\check{a}+\check{F}\check{b})/\hbar
-(i\dot{\check{\Lambda }}_n\partial_{\check{\Lambda }_n}\check{b}
+i\dot{\check{\Lambda }}_n^\star\partial_{\check{\Lambda }_n^\star }
\check{b})\big\}\nonumber\\
\qquad{} +\check{b}^{\rm T}
\big\{-(\check{F}^\star\check{a}
+\check{D}^\star\check{b})/\hbar
-(i\dot{\check{\Lambda }}_n\partial_{\check{\Lambda }_n}\check{a}
+i\dot{\check{\Lambda }}_n^\star\partial_{\check{\Lambda }_n^\star }
\check{a})
\big\}
= 0 ,
\label{Dorepongparameters}
\end{gather}
Let $H$ be an exact Hamiltonian of the system
with certain two-body interaction and let us
denote the expectation value of $H$ by
$|\phi(g)\rangle$ as $\langle H\rangle_g$.
It can be easily proved that the relations
\begin{gather}
\partial_{a^\star }\langle H\rangle_g
=
-\tfrac{1}{2}(F^\star a+D^\star b) , \qquad
\partial_{b^\star }\langle H\rangle_g
=
\tfrac{1}{2}(Fb+Da) ,
\label{relationdaH}
\end{gather}
and their complex conjugate relations do hold,
through which the well-known TDHBEQ is converted into
a matrix form as
\begin{gather}
i\dot{g}/\sqrt{2}
=
\left[
\begin{array}{rr}
\partial_{a^\star/\sqrt{2}}, &-\partial_{b/\sqrt{2}}\vspace{1mm}\\
\partial_{b^\star/\sqrt{2}}, &-\partial_{a/\sqrt{2}}
\end{array}
\right]
\langle H\rangle_{g/\sqrt{2}}/\hbar .
\label{rewrittenTDHB}
\end{gather}
The quantity $\langle H\rangle_{g/\sqrt{2}}$
means now in turn an expectation value of $H$,
being a function of $a/\sqrt{2}$,~$b/\sqrt{2}$
and their complex conjugate.
With the aid of a relation
similar to
(\ref{relationdaH}),
equation~(\ref{Dorepongparameters})
is reduced to
\begin{gather}
\check{a}^{\rm T}
\left\{
\partial_{  \check{b}^\star  /  \sqrt{2}}
\langle H\rangle_{  \check{g}/  \sqrt{2}}/   \hbar
-
\left(
i\dot{\check{\Lambda }}_n
\partial_{\check{\Lambda }_n}   \frac{\check{b}}{\sqrt{2}}
+
i\dot{\check{\Lambda }}_n^\star
\partial_{\check{\Lambda }_n^\star }   \frac{\check{b}}{\sqrt{2}}
\right)
\right\}  \nonumber\\
\qquad{} +
\check{b}^{\rm T}
\left\{
\partial_{  a^\star  /  \sqrt{2}}
\langle H\rangle_{  \check{g}/  \sqrt{2}}/   \hbar
-
\left(
i\dot{\check{\Lambda }}_n
\partial_{\check{\Lambda }_n}   \frac{\check{a}}{\sqrt{2}}
+
i\dot{\check{\Lambda }}_n^\star
\partial_{\check{\Lambda }_n^\star }   \frac{\check{a}}{\sqrt{2}}
\right)
\right\}
= 0 ,
\label{Dorepong2}
\end{gather}
As one way of satisfying
(\ref{Dorepong2}),
we may adopt the following type of
partial dif\/ferential equations:
\begin{gather}
\partial_{\check{a}^\star/\sqrt{2}}
\langle H\rangle_{\check{g}/\sqrt{2}}/\hbar
-\big(i\dot{\check{\Lambda }}_n
\partial_{\check{\Lambda }_n}\check{a}/\sqrt{2}
+i\dot{\check{\Lambda }}_n^\star
\partial_{\check{\Lambda }_n^\star }\check{a}/\sqrt{2}\big) = 0 ,\nonumber\\
\partial_{\check{b}^\star/\sqrt{2}}
\langle H\rangle_{\check{g}/\sqrt{2}}/\hbar
-\big(i\dot{\check{\Lambda }}_n
\partial_{\check{\Lambda }_n}\check{b}/\sqrt{2}
+i\dot{\check{\Lambda }}_n^\star
\partial_{{\check{\Lambda }}_n^\star }\check{b}/\sqrt{2}\big)
= 0 .
\label{Dorepong3}
\end{gather}
Here we notice
the {\em invariance principle of the Schr\"{o}dinger equation}
and the canonicity condition which leads us necessarily to
the equation of collective motion expressed in the canonical forms
\begin{gather}
i\dot{\check{\Lambda }}_n^\star
=
-\partial_{\check{\Lambda }_n}
\langle H\rangle_{\check{g}/\sqrt{2}}/\hbar ,
\qquad i\dot{\check{\Lambda }}_n
=
\partial_{\check{\Lambda }_n^\star}
\langle H\rangle_{\check{g}/\sqrt{2}}/\hbar ,
\label{collectiveequation}
\end{gather}
which can be easily derived with the use of
(\ref{rewrittenTDHB}) and (\ref{TRxiiLamba}).
Instead of solving approximately our nonlinear
time evolution equation
(\ref{Ct-Lambda0onthecollectivsubman}),
we adopt the above canonical equation.
Then, as is clear from the structure
of (\ref{Dorepong3}),
it is self-evident that equation
(\ref{Dorepong3})
becomes the equation
of path for the collective motion under
substitution of
(\ref{collectiveequation}).
In this sense,
equation
(\ref{Dorepong3})
is the natural extension of the equation of path
in the TDHF case~\cite{Ma.80,YK.81}
to the one in the TDHB case.
The set of
(\ref{Dorepong3}) and (\ref{collectiveequation})
is expected to determine the behaviour of the
{\em maximally decoupled} collective motions
in the TDHB case.
However, it means nothing else than the rewriting
of the TDHBEQ with the use of canonicity condition,
if we are able to assume only the existence
of invariant subspace in the full TDHB solution space.
The above interpretation is due to the natural consequence
of the {\em maximally decoupled} theory because
there exists, as a matter of case,
the invariant subspace, if the
{\em invariance principle of the Schr\"{o}dinger equation}
does hold true.
The {\em maximally decoupled} equation can be solved
with the additional RPA boundary condition, though its
solution is, strictly speaking, dif\/ferent
from the true motion of the system on the full $SO(2N)$
group manifold.
But how can we convince that the solution describes
the well-def\/ined {\em maximally decoupled} collective
motions from the other remaining degrees of freedom of motion?
Therefore, in order to answer such a question,
we must establish a criterion how we extract the collective
submanifolds ef\/fectively out of the full TDHB manifold.

Up to the present stage, equation
(\ref{id-LambdaTrFo-res})
remains unused yet and makes no role for approa\-ching
to our aim.
Finally with the aid of
(\ref{id-LambdaTrFo-res}),
we will derive some quantity by which the range of
the validity of the {\em maximally decoupled} theory
can be evaluated.
As was mentioned previously,
we demanded
that the expectation values of the residual curvatures
are minimized as far as possible
and adopted the canonical equation in place of
our fundamental equation
(\ref{Ct-Lambda0onthecollectivsubman}).
Then, by combining both the above propositions,
it may be expected that we can reach our f\/inal goal
of the present task.
Further substitution of the equation of motion
(\ref{collectiveequation})
(rewrite ${\langle H\rangle }_{\check{g}/\sqrt{2}}$
in the original form ${\langle H\rangle }_{\check{g}}$ again)
into
(\ref{id-LambdaTrFo-res})
yields
\begin{gather}
 \partial_{\Lambda_n}\check{\Lambda }_{n^\prime }
\mbox{Tr}
\left[
\big(
\partial_{\check{\Lambda }_{n^\prime }}
\check{\cal R}
-
\check{\Lambda }_{n^{\prime\prime }}^\star
\partial^2_{{\check{\Lambda }_{n^{\prime\prime }}^\star },
\check{\Lambda }_{n^\prime }}
\check{\cal R}
-
\check{\Lambda }_{n^{\prime\prime }}
\partial^2_{{\check{\Lambda }_{n^{\prime\prime }}},
\check{\Lambda }_{n^\prime }}
\check{\cal R}
\big)
\frac{\check{\cal F}}{\hbar }\right.\nonumber\\
\left. \qquad{} +
\big(
2\check{\cal R}
-
\check{\Lambda }_{n^{\prime\prime }}^\star
\partial_{{\check{\Lambda }_{n^{\prime\prime }}^\star }}
\check{\cal R}
-
\check{\Lambda }_{n^{\prime\prime }}
\partial_{{\check{\Lambda }_{n^{\prime\prime }}}}
\check{\cal R}
\big)
\partial_{\check{\Lambda }_{n^{\prime }}}
 \frac{\check{\cal F}}{\hbar }
\right] \nonumber\\
\qquad {}+
\partial_{\Lambda_n}  \check{\Lambda }_{n^\prime }^\star
\mbox{Tr}
\left[
\big(
\partial_{\check{\Lambda }_{n^\prime }^\star }
\check{\cal R}
-
\check{\Lambda }_{n^{\prime \prime }}^\star
\partial^2_{{\check{\Lambda }_{n^{\prime \prime }}^\star },
\check{\Lambda }_{n^\prime }^\star }
\check{\cal R}
-
\check{\Lambda }_{n^{\prime \prime }}
\partial^2_{{\check{\Lambda }_{n^{\prime \prime }}},
\check{\Lambda }_{n^\prime }^\star }
\check{\cal R}
\big)
 \frac{\check{\cal F}}{\hbar }\right.\nonumber\\
 \left. \qquad{}+
\big(
2\check{\cal R}
-
\check{\Lambda }_{n^{\prime \prime }}^\star
\partial_{{\check{\Lambda }_{n^{\prime \prime }}^\star }}
\check{\cal R}
-
\check{\Lambda }_{n^{\prime \prime }}
\partial_{{\check{\Lambda }_{n^{\prime \prime }}}}
\check{\cal R}
\big)
\partial_{\check{\Lambda }_{n^{\prime }}^\star }
{\displaystyle \frac{\check{\cal F}}{\hbar }}
\right]  \cong 0 ,
\label{id-LambdaTrFo-res2}
\end{gather}
Here we have used the transformation property
of the dif\/ferential
$
\partial_{\Lambda_n}
=
\partial_{\Lambda_n}\check{\Lambda }_{n^\prime }
\partial_{\check{\Lambda }_{n^\prime }}
+
\partial_{\Lambda_n}\check{\Lambda }_{n^\prime }^\star
\partial_{\check{\Lambda }_{n^\prime }^\star }
$
and the dif\/ferential formulae for the expectation values
of the Hamiltonians
$H$ and $H_{\rm HB}$
\begin{gather*}
\partial_{\check{\Lambda }_n}\langle H\rangle_{\check{g}}
=
- \tfrac{1}{4}
\mbox{Tr}
\left[
\partial_{\check{\Lambda }_n}\check{\cal R}
(\check{g})\check{\cal F}
\right] ,\nonumber\\
\partial_{\check{\Lambda }_n}
\langle H_{\rm HB}\rangle_{\check{g}}
=
- \tfrac{1}{2}\partial_{\check{\Lambda }_n}
\mbox{Tr}
\check{\cal F}_o
=
\partial_{\check{\Lambda }_n}\langle H\rangle_{\check{g}}
- \tfrac{1}{4}
\mbox{Tr}
\left[
\check{\cal R}(\check{g})
\partial_{\check{\Lambda }_n}\check{\cal F}
\right] .
\end{gather*}



\subsection{Nonlinear RPA theory arising from zero-curvature equation}\label{section2.4}

Our fundamental equation may work well
{\em in the large scale} beyond the RPA as the small-amplitude limit.
A linearly approximate solution of the TDHBEQ becomes the RPAEQ.
Suppose we solve the fundamental equation by expanding it
in the form of a power series of the collective variables
$\Lambda$ and $\Lambda^\star
(n = 1,\dots,m;\,  m  \ll  N(2N - 1)/2)$
def\/ined in the Lagrange-like manner.
Then we must show that the fundamental equation has
necessarily the RPA solution at the lowest power
of the collective variables which approach
in the small amplitude limit.
A paired mode amplitude
$g(\Lambda, \Lambda^\star, t)$ is separated into
stationary and f\/luctuating components as $g = g^{(o)}\tilde{g}$.
This means that the $SO(2N)$ matrix $g$ is decomposed
into a product of stationary matrix $g^{(o)}$ and
$\tilde{g}(\Lambda, \Lambda^\star,t)$ $( \simeq  \tilde{g})$.
The stationary $g^{(o)}$ satisf\/ies the usual static
$SO(2N)$(HB) eigenvalue equation.

Using the above decomposition of $g$,
an original $SO(2N)$(HB)
density matrix ${\cal R}(\Lambda, \Lambda^\star, t)$
and a HB matrix ${\cal F}(\Lambda, \Lambda^\star, t)$
are decomposed as
${\cal R}
=
g^{(o)}\widetilde{\cal R} g^{(o)}{}^\dagger$
and
${\cal F}
=
g^{(o)}\widetilde{\cal F}g^{(o)}{}^\dagger$,
respectively.
The f\/luctuating $\widetilde{\cal R}$
and the HB matrix $\widetilde{\cal F}$
in f\/luctuating QPF are
given in the following forms:
\begin{gather}
\widetilde{\cal R}(\tilde{g})
=
g^{(o)}{}^\dagger{\cal R}(g)g^{(o)},
\qquad \widetilde{\cal R}(\tilde{g})
=
\left[
\begin{array}{cc}
2\widetilde{R}(\tilde{g})-1_N &-2\widetilde{K}^\star(\tilde{g})\vspace{1mm}\\
2\widetilde{K}(\tilde{g}) & -2\widetilde{R}^\star(\tilde{g})+1_N
\end{array}
\right] ,
\label{tildeR}
\\
\widetilde{\cal F}
=
g^{(o)}{}^\dagger{\cal F}g^{(o)},
\qquad \widetilde{\cal F}
=
\left[
\begin{array}{cc}
-\epsilon^{(o)}-{\it f}^\star & -{\it d}^\star\vspace{1mm}\\
{\it d} & \epsilon^{(o)}+{\it f}
\end{array}
\right] ,
\label{tildeF}
\end{gather}
in which all the quantities are redef\/ined in
\cite{FYN.77}.
Quasi-particle energies $\varepsilon_i^{(o)}$
include a chemical potential.

Introducing f\/luctuating auxiliary quantities
$
\tilde{\theta }_n
=
g^{(o)}{}^\dagger\theta_n g^{(o)}
$
and
$
\tilde{\theta }_n^\dagger
=
g^{(o)}{}^\dagger\theta_n^\dagger g^{(o)}
$,
then under the decomposition
$g = g^{(o)}\tilde{g}$,
the zero-curvature equation
$C_{\bullet , \bullet } = 0$
in
(\ref{Cc-val})
is transformed to
\begin{gather}
i\partial_t \tilde{\theta }_n^\dagger - i\partial_{\Lambda_n}
\widetilde{\cal F}_{\rm c}/\hbar
+
[\tilde{\theta }_n^\dagger,\widetilde
{\cal F}_{\rm c}/\hbar]
= 0 ,\qquad
i\partial_t \tilde{\theta }_n-i\partial_{\Lambda_n^\star }
\widetilde{\cal F}_{\rm c}/\hbar
+[\tilde{\theta }_n,\widetilde{\cal F}_{\rm c}/\hbar]
= 0 ,
\label{fluctuatingCtLamda}
\\
i\partial_{\Lambda_{n^\prime }}\tilde{\theta }_n
-
i\partial_{\Lambda_n^\star }\tilde{\theta }_{n^\prime }^\dagger
+
[\tilde{\theta }_n,\tilde{\theta }_{n^\prime }^\dagger]
= 0 ,\nonumber\\
i\partial_{\Lambda_{n^\prime }}\tilde{\theta }_n^\dagger
-
i\partial_{\Lambda_n}\tilde{\theta }_{n^\prime }^\dagger
+
[\tilde{\theta }_n^\dagger,\tilde{\theta }_{n^\prime }^\dagger]
= 0 ,\qquad
i\partial_{\Lambda_{n^\prime }^\star }\tilde{\theta }_n
-
i\partial_{\Lambda_n^\star }\tilde{\theta }_{n^\prime }
+
[\tilde{\theta }_n,\tilde{\theta }_{n^\prime }]
= 0 ,
\label{fluctuatingCLambdaLamda}
\\
-  \tfrac{1}{4}\mbox{Tr}
\{\widetilde{\cal R}(\tilde{g})
[\tilde{\theta }_n,\tilde{\theta }_{n^\prime }^\dagger]\}
=\delta_{nn^{\prime }},\qquad\!\!
 \tfrac{1}{4}\mbox{Tr}
\{\widetilde{\cal R}(\tilde{g})
[\tilde{\theta }_n^\dagger,\tilde{\theta }_{n^\prime }^\dagger]\}
= 0 ,\qquad\!\!
\tfrac{1}{4}\mbox{Tr}
\{\widetilde{\cal R}(\tilde{g})
[\tilde{\theta }_n,\tilde{\theta }_{n^\prime }]\}
= 0 ,\!
\label{fluctuatingweakorthgonality}
\end{gather}
where the quantities
$\widetilde{\cal F}_{\rm c}$,
$\tilde{\theta }_n^\dagger$ and $\tilde{\theta }_n$
satisfy partial dif\/ferential equations,
\begin{gather}
i\hbar\partial_t\tilde{g}=\widetilde
{\cal F}_{\rm c}\tilde{g} ,\qquad
i\partial_{\Lambda_n}\tilde{g}
=
\tilde{\theta }_n^\dagger \tilde{g} \qquad \mbox{and}\qquad
i\partial_{\Lambda_n^\star }\tilde{g}
=
\tilde{\theta }_n\tilde{g} .
\label{fluctuatingSDHB}
\end{gather}
Putting
$\widetilde{\cal F}_{\rm c} = \widetilde{\cal F}$
(\ref{tildeF}) in
(\ref{fluctuatingCtLamda}),
we are able to look for a collective path ($\tilde{g}$)
and a collective Hamiltonian
($\widetilde{\cal F}_{\rm c}$)
under the minimization of the residual curvature arising
from a residual Hamiltonian
($\widetilde{\cal F}_{{\rm res}}$).
Next, for convenience of further discussion, we introduce modif\/ied
f\/luc\-tua\-ting auxiliary quantities
$
\tilde{\theta }_{o-n}^\dagger
=
\tilde{g}^\dagger\tilde{\theta }_n^\dagger \tilde{g}
$
and
$
\tilde{\theta }_{o-n}
=
\tilde{g}^\dagger\tilde{\theta }_n\tilde{g}
$.
Then we can rewrite our fundamental equations
(\ref{fluctuatingCtLamda}),
(\ref{fluctuatingCLambdaLamda}) and
(\ref{fluctuatingweakorthgonality})
in terms of the above quantities as follows:
\begin{gather}
i\partial_t\tilde{\theta }_{o-n}^\dagger
-i\tilde{g}^\dagger
\left(
\partial_{\Lambda_n}
\tilde{\cal F}_{\rm c}/\hbar
\right)
\tilde{g} = 0 ,\qquad
i\partial_t\tilde{\theta }_{o-n}-
i\tilde{g}^\dagger
\left(\partial_{\Lambda_n^\star }
\tilde{\cal F}_{\rm c}/\hbar
\right)
\tilde{g} = 0 ,
\label{CtLambdaonquasi-frame}
\\
i\partial_{\Lambda_{n^\prime }}\tilde{\theta }_{o-n}
-
i\partial_{\Lambda_n^\star}\tilde{\theta }_{o-n^\prime }^\dagger
-
[\tilde{\theta }_{o-n},\tilde{\theta }_{o-n^\prime }^\dagger]
= 0 ,\qquad
i\partial_{\Lambda_{n^\prime }}\tilde{\theta }_{o-n}^\dagger
-
i\partial_{\Lambda_n}\tilde{\theta }_{o-n^\prime }^\dagger
-
[\tilde{\theta }_{o-n}^\dagger,\tilde
{\theta }_{o-n^\prime }^\dagger]
= 0 , \nonumber\\
i\partial_{\Lambda_{n^\prime }^\star } \tilde{\theta }_{o-n}
-
i\partial_{\Lambda_n^\star }\tilde{\theta }_{o-n^\prime }
-
[\tilde{\theta }_{o-n},\tilde{\theta }_{o-n^\prime }]
= 0 ,
\label{CLambdaLamdaonquasi-frame}
\\
-  \tfrac{1}{4}\mbox{Tr}
\left\{
\left[
\begin{array}{rr}
-1_N & 0 \\
 0 & 1_N
\end{array}
\right]
[\tilde{\theta }_{o-n},\tilde{\theta }_{o-n^\prime }^\dagger]
\right\}
=
\delta_{nn^{\prime }} ,\nonumber\\
 \tfrac{1}{4}\mbox{Tr}
\left\{
\left[
\begin{array}{rr}
-1_N & 0 \\
 0 & 1_N
\end{array}
\right]
[\tilde{\theta }_{o-n}^\dagger,
\tilde{\theta }_{o-n^\prime }^\dagger]
\right\}
= 0 ,\qquad\!\!
 \tfrac{1}{4}\mbox{Tr}
\left\{
\left[
\begin{array}{rr}
-1_N & 0 \\
 0 & 1_N
\end{array}
\right]
[\tilde{\theta }_{o-n}~,\tilde{\theta }_{o-n^\prime }]
\right\}
= 0 .\!
\label{weakorthgonalityonquasi-frame}
\end{gather}
In the derivation of equations
(\ref{CtLambdaonquasi-frame})
and
(\ref{CLambdaLamdaonquasi-frame}),
we have used
(\ref{fluctuatingSDHB}).
The equation
(\ref{weakorthgonalityonquasi-frame})
is easily obtained with the aid of another expression
for the f\/luctuating density matrix
$\widetilde{\cal R}$
(\ref{tildeR}),
\begin{equation}
\widetilde{\cal R}(\tilde{g})=\tilde{g}\left[
\begin{array}{rr}
-1_N & 0 \\
 0 & 1_N
\end{array}
\right]
\tilde{g}^\dagger .
\label{fluctuatingR2}
\end{equation}

In order to investigate the set of the matrix-valued
nonlinear time evolution equation
(\ref{CtLambdaonquasi-frame}) arising
from the zero curvature equation, we give here
the $\partial_{\Lambda_n}$
and $\partial_{\Lambda_n^\star}$ dif\/ferential
forms of the TDHB density matrix
and collective hamiltonian.
First, using
(\ref{fluctuatingR2}),
we have
\begin{gather}
\partial_{\Lambda_n}\widetilde{\cal R}(\tilde{g})
=
\partial_{\Lambda_n}{\tilde{g}}
\left[
\begin{array}{rr}
-1_N & 0 \\
 0 & 1_N
\end{array}
\right]
{\tilde{g}}^\dagger
+
{\tilde{g}}
\left[
\begin{array}{rr}
-1_N & 0\\
 0 & 1_N
\end{array}
\right]
\partial_{\Lambda_n}\tilde{g}^\dagger \nonumber\\
\phantom{\partial_{\Lambda_n}\widetilde{\cal R}(\tilde{g})}{}
=
\partial_{\Lambda_n}\tilde{g}
(\tilde{g}^\dagger\tilde{g})
\left[
\begin{array}{rr}
-1_N & 0 \\
 0 & 1_N
\end{array}
\right]
\tilde{g}^\dagger
+
\tilde{g}
\left[
\begin{array}{rr}
-1_N & 0 \\
 0 & 1_N
\end{array}
\right]
\partial_{\Lambda_n}\tilde{g}^\dagger
(\tilde{g}\tilde{g}^\dagger) ,
\label{gradLambdatildeR}
\end{gather}
where we have used the relation
$\tilde{g}^\dagger\tilde{g} = \tilde{g}\tilde{g}^\dagger = 1$.
Using
(\ref{fluctuatingSDHB}),
the above equation is written as
\begin{gather*}
\partial_{\Lambda_n}\widetilde{\cal R}(\tilde{g})
=
-i\tilde{\theta}_n^\dagger\partial_{\Lambda_n}{\tilde{g}}
\left[
\begin{array}{rr}
-1_N & 0 \\
 0 & 1_N
\end{array}
\right]
{\tilde{g}}^\dagger+i{\tilde{g}}
\left[
\begin{array}{rr}
-1_N & 0\\
 0 & 1_N
\end{array}
\right]
\tilde{g}^\dagger\tilde{\theta }_n^\dagger\nonumber\\
\phantom{\partial_{\Lambda_n}\widetilde{\cal R}(\tilde{g})}{}
=
-i(\tilde{g}\tilde{g}^\dagger)
\tilde{\theta }_n^\dagger\tilde{g}
\left[
\begin{array}{rr}
-1_N & 0 \\
 0 & 1_N
\end{array}
\right]
\tilde{g}^\dagger+i\tilde{g}
\left[
\begin{array}{rr}
-1_N & 0 \\
 0 & 1_N
\end{array}
\right]
\tilde{g}^\dagger\tilde{\theta }_n^\dagger
(\tilde{g}\tilde{g}^\dagger) .
\end{gather*}
Next, by using
$
\tilde{\theta }_{o-n}^\dagger
=
\tilde{g}^\dagger\tilde{\theta }_n^\dagger \tilde{g}
$
and
$
\tilde{\theta }_{o-n}
=
\tilde{g}^\dagger\tilde{\theta }_n\tilde{g}
$,
equation
(\ref{gradLambdatildeR})
is transformed into
\begin{equation}
\partial_{\Lambda_n}\widetilde{\cal R}(\tilde{g})
=
-i\tilde{g}
\left[
\tilde{\theta }_{o-n}^\dagger,
\left[
\begin{array}{rr}
-1_N & 0\\
0 & 1_N
\end{array}
\right]
\right]
\tilde{g}^\dagger .
\label{gradLambdatildeR3}
\end{equation}
On the other hand,
from
(\ref{tildeR}),
we easily obtain
\begin{equation}
\partial_{\Lambda_n}\widetilde{\cal R}(\tilde{g})
=
\left[
\begin{array}{cc}
2\partial_{\Lambda_n}\widetilde{R}(\tilde{g})
 &-2\partial_{\Lambda_n}\widetilde{K}^\star(\tilde{g})\\
2\partial_{\Lambda_n}\widetilde{K}(\tilde{g})
& -2\partial_{\Lambda_n}\widetilde{R}^\star(\tilde{g})
\end{array}
\right] .
\label{gradLambdatildeR4}
\end{equation}
Let us substitute explicit representations for
$
\tilde{g}$
$\left(
=
\left[
\begin{array}{rr}
\tilde{a}&\tilde{b}^\star\\
\tilde{b}&\tilde{a}^\star
\end{array}
\right]
\right)
$
and
$
\tilde{\theta }_{o-n}^\dagger$
$\left(
=
\left[
\begin{array}{rr}
\xi_{o-n}&\varphi_{o-n}\\
\psi_{o-n}&-\xi_{o-n}^{\rm T}
\end{array}
\right]
\right)
$,
into the r.h.s.\ of
(\ref{gradLambdatildeR3})
and combine it with
(\ref{gradLambdatildeR4}).
Then, we obtain the f\/inal $\partial_{\Lambda_n}$
dif\/ferential form of the TDHB ($SO(2N)$) density matrix as
follows:
\begin{gather*}
\partial_{\Lambda_n}\widetilde{R}(\tilde{g})
=
 i
\big(
\tilde{b}^\star\psi_{o-n}\tilde{a}^\dagger
-
\tilde{a}\varphi_{o-n}\tilde{b}^{\rm T}
\big) ,\qquad
\partial_{\Lambda_n}\widetilde{R}^\star (\tilde{g})
=
-i
\big(
\tilde{a}^\star\psi_{o-n}\tilde{b}^\dagger
-
\tilde{b}\varphi_{o-n}\tilde{a}^{\rm T}
\big) ,\nonumber\\
\partial_{\Lambda_n}\widetilde{K}(\tilde{g})
=
 i
\big(
\tilde{a}^\star\psi_{o-n}\tilde{a}^\dagger
-
\tilde{b}\varphi_{o-n}\tilde{b}^{\rm T}
\big) , \qquad
\partial_{\Lambda_n}\widetilde{K}^\star (\tilde{g})
=
-i
\big(
\tilde{b}^\star\psi_{o-n}\tilde{b}^\dagger
-
\tilde{a}\varphi_{o-n}\tilde{a}^{\rm T}
\big) .
\end{gather*}
The $\partial_{\Lambda_n^\star }$ dif\/ferentiation of the
$SO(2N)$ density matrix is also made
analogously to the above.

As shown in
\cite{FYN.77},
the f\/luctuating components
of the HB matrix $\widetilde{\cal F}$
(\ref{tildeF}) are linear functionals of
$\widetilde{R}(\tilde{g})$
and $\widetilde{K}(\tilde{g})$.
We can easily
calculate the $\partial_{\Lambda_n}$ dif\/ferential as follows:
\begin{gather*}
\partial_{\Lambda_n}d
=
({\bf D})\partial_{\Lambda_n}\widetilde{K}(\tilde{g})
+
(\overline{\bf D})\partial_{\Lambda_n}
\widetilde{K}^\star(\tilde{g})
+
({\bf d})\partial_{\Lambda_n}\tilde{R}(\tilde{g})
=
i({\bf D}\}\psi_{o-n}+i(\overline{\bf D}\}\varphi_{o-n} ,\nonumber\\
\partial_{\Lambda_n}f^\star
=
({\bf F})\partial_{\Lambda_n}\widetilde{K}(\tilde{g})
+
(\overline{\bf F}^\star)\partial_{\Lambda_n}\widetilde{K}^\star
(\tilde{g})
+
({\bf f}^\star)\partial_{\Lambda_n}\tilde{R}(\tilde{g})
=
i({\bf F}^\star\}\psi_{o-n}
+
i(\overline{\bf F}^\star\}\varphi_{o-n} .
\end{gather*}
Here the matrices ({\bf D}) etc.\
are given in
\cite{FYN.77}.
Similarly, new matrices ({\bf D}\} etc.\ are def\/ined by
\begin{gather}
({\bf D}\}
=
\|(ij|D|kl\}\| ,\qquad
(\overline{\bf D}\}
=
\|(ij|\overline{D}|kl\}\|  \qquad \mbox{and} \nonumber\\
({\bf F}^\star\}
=
\|(ij|F^\star|kl\}\| , \qquad
(\overline{\bf F}^\star\}
=
\|(ij|\overline{F}^\star|kl\}\| ,
\label{thedifinitionsofboldDF}
\\
(ij|D|kl\}
=
(ij|D|k^\prime l^\prime)\tilde{a}_{k^\prime k}^\star
\tilde{a}_{l^\prime l}^\star
-
(ij|\overline{D}|k^\prime l^\prime)
\tilde{b}_{k^\prime k}^\star\tilde{b}_{l^\prime l}^\star
+
(ij|d|k^\prime l^\prime)
\tilde{b}_{k^\prime k}^\star\tilde{a}_{l^\prime l}^\star ,\nonumber\\
-(ij|\overline{D}|kl\}
=
(ij|D|k^\prime l^\prime)\tilde{b}_{k^\prime k}
\tilde{b}_{l^\prime l}
-
(ij|\overline{D}|k^\prime l^\prime)
\tilde{a}_{k^\prime k}\tilde{a}_{l^\prime l}
+
(ij|d|k^\prime l^\prime)\tilde{a}_{k^\prime k}
\tilde{b}_{l^\prime l} ,\nonumber\\
(ij|F^\star|kl\}
=
(ij|F^\star|k^\prime l^\prime)\tilde{a}_{k^\prime k}^\star
\tilde{a}_{l^\prime l}^\star
-
(ij|\overline{F}^\star|k^\prime l^\prime)
\tilde{b}_{k^\prime k}^\star\tilde{b}_{l^\prime l}^\star
+
(ij|f^\star|k^\prime l^\prime)
\tilde{b}_{k^\prime k}^\star\tilde{a}_{l^\prime l}^\star ,\nonumber\\
-(ij|\overline{F}^\star|kl\}
=
(ij|F^\star|k^\prime l^\prime)
\tilde{b}_{k^\prime k}\tilde{b}_{l^\prime l}
-
(ij|\overline{F}^\star|k^\prime l^\prime)
\tilde{a}_{k^\prime k}\tilde{a}_{l^\prime l}
+
(ij|f^\star|k^\prime l^\prime)\tilde{a}_{k^\prime k}
\tilde{b}_{l^\prime l} .\nonumber
\end{gather}
The above
summation is made
with indices $k^\prime$ and $l^\prime$
$(1  \sim N)$.
Putting
$\widetilde{\cal F}_{\rm c}$
=
$\widetilde{\cal F}$,
we have {\samepage
\begin{equation}
\partial_{\Lambda_n}\widetilde
{\cal F}_{\rm c}
=
i
\left[
\begin{array}{cc}
-({\bf F}^\star\}\psi_{o-n}-(\overline{\bf F}^\star \}
\varphi_{o-n} ,
&-(\overline{\bf D}\}^\star \psi_{o-n}-({\bf D}\}^\star
\varphi_{o-n} \vspace{1mm}\\
({\bf D}\}\psi_{o-n}+(\overline{\bf D}\}\varphi_{o-n} ,
&^{\rm T}({\bf F}^\star\}\psi_{o-n}
+^{\rm T}(\overline{\bf F}^\star \}
\varphi_{o-n}
\end{array}
\right],
\label{gradLambdatildeF}
\end{equation}
where $^{\rm T}({\bf F}^\star\}$ etc.\
stand for the matrices in which the indices $i$ and $j$
in
(\ref{thedifinitionsofboldDF}) are exchanged.}

We here derive a new equation formally
analogous to the $SO(2N)$ RPA equation.
To achieve this, we f\/irst further decompose the f\/luctuating
pair mode amplitude $\tilde{g}$ into a product of a~f\/luctuating
$SO(2N)$ matrix and a $2N$-dimensional diagonal matrix
with an exponential time dependence as follows:
\begin{gather}
\tilde{g}\rightarrow\tilde{g}\tilde{g}(\varepsilon,-\varepsilon) ,
 \qquad \tilde{g}(\varepsilon,-\varepsilon)
=
\left[
\begin{array}{cc}
\exp[i\varepsilon t/\hbar],& 0 \vspace{1mm}\\
0                       ,&\exp[-i\varepsilon t/\hbar]
\end{array}
\right] ,\nonumber\\
\varepsilon=(\delta_{ij}\varepsilon_i) , \qquad
\varepsilon_i
=
\varepsilon_i(\Lambda,\Lambda^\star) ,
\label{diagonalmatrixwithexponetialtimedependence}
\end{gather}
where we redenote a new f\/luctuating pair mode as
$\tilde{g}$ and $\varepsilon_i$
is the $\Lambda$ and $\Lambda^\star$
dependent quasi-particle energy including
the chemical potential.
Next,
using
(\ref{fluctuatingSDHB}),
$
\tilde{\theta }_{o-n}^\dagger
=
\tilde{g}^\dagger\tilde{\theta }_n^\dagger \tilde{g}
$
and
$
\tilde{\theta }_{o-n}
=
\tilde{g}^\dagger\tilde{\theta }_n\tilde{g}
$,
the modif\/ied f\/luctuating auxiliary quantities
$\tilde{\theta}_{o-n}^\dagger$ can be written as
\begin{equation*}
\tilde{\theta }_{o-n}^\dagger\rightarrow
\tilde{g}^\dagger(\varepsilon,-\varepsilon)
\tilde{\theta }_{o-n}^\dagger\tilde{g}
(\varepsilon,-\varepsilon)
+
\left[
\begin{array}{cc}
-\partial_{\Lambda_n}\varepsilon t/\hbar,& 0 \\
0,&\partial_{\Lambda_n}\varepsilon t/\hbar
\end{array}
\right] ,
\end{equation*}
where we again redenote the new f\/luctuating
auxiliary quantities as $\tilde{\theta }_{o-n}^\dagger$.
Accompanying the above change,
$\partial_t \tilde{\theta }_{o-n}^\dagger$
are modif\/ied to the following forms
by using the explicit expression for~$\tilde{\theta }_{o-n}^\dagger$:
\begin{gather}
\partial_t\tilde{\theta }_{o-n}^\dagger
 \rightarrow
\tilde{g}^\dagger(\varepsilon,-\varepsilon)\label{gradttrnsfferedtheta}\\
\phantom{\partial_t\tilde{\theta }_{o-n}^\dagger
 \rightarrow }{} \times
\left[  \!\!
\begin{array}{ll}
\partial_t\xi_{o-n}-\partial_{\Lambda_n}\varepsilon/\hbar
-i[\varepsilon/\hbar,\xi_{o-n}],
&\!\!\partial_t\varphi_{o-n}-i[\varepsilon/\hbar,\varphi_{o-n}]_+ \vspace{1mm}\\
\partial_t\psi_{o-n}+i[\varepsilon/\hbar,\psi_{o-n}]_+,
&\!\!\!-\partial_t\xi_{o-n}^{\rm T}
+ \partial_{\Lambda_n}\varepsilon/\hbar
 - i[\varepsilon/\hbar,\xi_{o-n}^{\rm T}]
\end{array}  \!\!
\right]
\tilde{g}(\varepsilon,-\varepsilon).
\nonumber
\end{gather}
In the above,
hereafter
we adopt $(\Lambda,~\Lambda^\star)$-independent
$\varepsilon^{(o)}$ given in
(\ref{tildeF})
as the quasi-particle energy $\varepsilon$.
If we substitute equations
(\ref{gradLambdatildeF}),
(\ref{diagonalmatrixwithexponetialtimedependence})
and
(\ref{gradttrnsfferedtheta})
into the set of the matrix-valued nonlinear
time evolution equation, i.e.,
the equation of
(\ref{CtLambdaonquasi-frame}),
we f\/inally obtain the following set
of matrix-valued equations:
\begin{gather}
\tilde{g}^\dagger\big(\varepsilon^{(o)},-\varepsilon^{(o)}\big) \!\!
\left[ \!\!\!
\begin{array}{ll}
i\hbar\partial_t\xi_{o-n}\!+\![\varepsilon^{(o)},\xi_{o-n}]
&
i\hbar\partial_t\varphi_{o-n}\!+\![\varepsilon^{(o)},\varphi_{o-n}]_+
\\
~-\{{\bf F}^\star\}\psi_{o-n}
\!-\!
\{\overline{\bf F}^\star\}\varphi_{o-n}
&~-\{\overline{\bf D}\}^\star\psi_{o-n}
\!-\!
\{{\bf D}\}^\star\varphi_{o-n}\vspace{2mm}\\
i\hbar\partial_t\psi_{o-n}-[\varepsilon^{(o)},\psi_{o-n}]_+
&-i\partial_t\xi_{o-n}^{\rm T}
\!+\!
[\varepsilon^{(o)},\xi_{o-n}^{\rm T}]\\
~\!+\!\{{\bf D}\}\psi_{o-n}
\!+\!\{\overline{\bf D}\}\varphi_{o-n}
&~\!+\!^{\rm T}
\{{\bf F}^\star\}\psi_{o-n}
\!+\!
^{\rm T}\{\overline{\bf F}^\star\}\varphi_{o-n}
\end{array} \!\!\!
\right] \nonumber\\
\phantom{\tilde{g}^\dagger\big(\varepsilon^{(o)},-\varepsilon^{(o)}\big)}{}\times
\tilde{g}
\big(\varepsilon^{(0)},-\varepsilon^{(0)}\big) = 0,
\label{matrixformofformalRPA}
\end{gather}
with the modif\/ied new matrices $\{{\bf D}\}$ etc.
def\/ined through
\begin{gather*}
\{{\bf D}\}
=
\|\{ij|D|kl\}\| ,\qquad
\{\overline{\bf D}\}
=
\|\{ij|\overline{D}|kl\}\| \qquad \mbox{and} \nonumber\\
\{{\bf F}^\star\}
=
\|\{ij|F^\star|kl\}\| ,\qquad
\{\overline{\bf F}^\star\}
=
\|\{ij|\overline{F}^\star|kl\}\| ,
\end{gather*}
whose matrix elements are given by
\begin{gather*}
\{i j|D|kl\}
=
\tilde{a}_{i^\prime i}\tilde{a}_{ j^\prime  j}
 (i^\prime  j^\prime|D|kl\}
-
\tilde{b}_{i^\prime i}\tilde{b}_{ j^\prime  j}
 (i^\prime  j^\prime|\overline{D}|kl\}^\star\nonumber\\
\phantom{\{i j|D|kl\}=}{}
-
\tilde{b}_{i^\prime i}\tilde{a}_{ j^\prime  j}
 (i^\prime  j^\prime|F^\star|kl\}
+
\tilde{a}_{i^\prime i}\tilde{b}_{ j^\prime j}
 (i^\prime  j^\prime|\overline{F}^\star|kl\}^\star,\nonumber\\
-\{i j|\overline{D}|kl\}
=
\tilde{b}_{i^\prime i}\tilde{b}_{ j^\prime  j}
 (i^\prime  j^\prime|D|kl\}^\star
-
\tilde{a}_{i^\prime i}\tilde{a}_{ j^\prime  j}
 (i^\prime  j^\prime|\overline{D}|kl\}\nonumber\\
\phantom{-\{i j|\overline{D}|kl\}=}{}
-
\tilde{a}_{i^\prime i}\tilde{b}_{ j^\prime  j}
 (i^\prime  j^\prime|F^\star|kl\}^\star
+
\tilde{b}_{i^\prime i}\tilde{a}_{ j^\prime  j}
 (i^\prime  j^\prime|\overline{F}^\star|kl\},\nonumber\\
-\{i j|F^\star|kl\}
=
\tilde{b}^\star_{i^\prime i}\tilde{a}_{ j^\prime  j}
 (i^\prime  j^\prime|D|kl\}
-
\tilde{a}^\star_{i^\prime i}\tilde{b}_{ j^\prime  j}
 (i^\prime  j^\prime|\overline{D}|kl\}^\star\nonumber\\
\phantom{-\{i j|F^\star|kl\}=}{}
-
\tilde{a}^\star_{i^\prime i}\tilde{a}_{ j^\prime  j}
 (i^\prime  j^\prime|F^\star|kl\}
+
\tilde{b}^\star_{i^\prime i}\tilde{b}_{ j^\prime  j}
 (i^\prime  j^\prime|\overline{F}^\star|kl\}^\star,\nonumber\\
\{i j|\overline{F}^\star|kl\}
=
\tilde{a}^\star_{i^\prime i}\tilde{b}_{ j^\prime  j}
 (i^\prime  j^\prime|D|kl\}^\star
-
\tilde{b}^\star_{i^\prime i}\tilde{a}_{ j^\prime  j}
 (i^\prime  j^\prime|\overline{D}|kl\}\nonumber\\
\phantom{\{i j|\overline{F}^\star|kl\}=}{}
-
\tilde{b}^\star_{i^\prime i}\tilde{b}_{ j^\prime  j}
 (i^\prime  j^\prime|F^\star|kl\}^\star
+
\tilde{a}^\star_{i^\prime i}\tilde{a}_{ j^\prime  j}
 (i^\prime  j^\prime|\overline{F}^\star|kl\} ,
\end{gather*}
in which summation is made over indices
${\it i}^\prime$ and ${\it j}^\prime$
running from 1 to $N$.
In
(\ref{matrixformofformalRPA})
by making block of\/f-diagonal matrices vanish,
we get a TD equation with respect to
$\psi_{o-n}$ and $\varphi_{o-n}$
which is formally analogous to that of the $SO(2N)$ RPA,
though our TD amplitude and matrices $\{{\bf D}\}$ etc.\
have ($\Lambda, \Lambda^\star, t$)-dependence.


\subsection{Summary and discussions}\label{section2.5}

We have studied integrability conditions of
the TDHBEQ to determine
collective submanifolds from the group theoretical viewpoint.
As we have seen above,
the basic idea lies in
the introduction of the Lagrange-like manner to
describe the collective coordinates.
It should be noted that
the variables are nothing but the parameters
to describe the symmetry of TDHBEQ.
By introducing the one-form, we gave the integrability conditions,
the vanishing of the curvatures of the connection,
expressed as the Lie-algebra-valued equations.
The full TDHB Hamiltonian~$H_{\rm HB}$
is decomposed
into the collective Hamiltonian $H_{\rm c}$ and
the residual one $H_{\rm res}$. To search for
the {\em well-defined} collective submanifold,
we have demanded that the expectation value
of the curvature is minimized so as to satisfy
$H_{{\rm res}}
\cong
{\rm const}$ or $H_{\rm c} + {\rm const}
=
H_{\rm HB}$
as far as possible.
Further
we have imposed the restriction
to assure the Lagrange bracket for the usual variables
and Lagrange-like ones.
Our fundamental equation together with
the restricted condition describes
the collective motion of the system.

We have proposed
the minimization of the residual curvature arising from
the residual part of the full TDHB Hamiltonian to
determine the collective submanifold.
With our theory it is also
possible to investigate the range of the validity of the
{\em maximally decoupled} theory of the TDHBT with use of
the condition to satisfy~(\ref{id-LambdaTrFo-res2}).
This condition makes an essential role to give the criterion
how we extract well the collective submanifold out of the full
TDHB manifold.
The reason why the condition occurs in our theory
which did not appear in the {\em maximally decoupled}
theory lies in the consideration of the $d^\dagger d$-type
in the residual Hamiltonian to calculate
the residual curvature and in the adoption of
the canonical equation.
Since the {\em maximally decoupled} theory has
no consideration of such type from the outset,
the condition is trivially fulf\/illed.
This is the essential dif\/ference between
the {\em maximally decoupled} theory and ours.

We have investigated the nonlinear
time-evolution equation arising from zero-curvature
equation on TDHB ($SO(2N)$ Lie group) manifold.
It is self-evident that the new equation has an $SO(2N)$ RPA
solution as a small-amplitude limit.
The new equation depends on the collective variables
($\Lambda, \Lambda^\star$) def\/ined in a Lagrange-like manner.
It works well {\em in the large scale} beyond
the $SO(2N)$ RPA under appropriate boundary and initial conditions.
The integrability condition is just the inf\/initesimal condition
to transfer a solution to another solution for the evolution equation
under consideration.
The usual treatment of the RPA
for small amplitude around ground state is nothing but
a method of determining an inf\/initesimal transformation of symmetry
under the assumption that
f\/luctuating f\/ields are composed of only normal-modes.
We conclude that the set of equations def\/ining the symmetry
of the SCF equation and the weak boson commutation relations
on the QPF becomes the nonlinear RPA theory.

Finally,
following Rajeev~\cite{Rajeev.94},
we also show the existence of the homogeneous symplectic 2-form $\omega$.
From
(\ref{densityRg}),
using the $\frac{SO(2N)}{U(N)}$ coset variable
$q$ $(= ba^{-1}) = - q^{\rm T}$,
the $SO(2N)$ (HB) density matrix ${\cal R}(g)$
is expressed as
\begin{gather}
{\cal R}(g)
=
g
\left[
\begin{array}{cc}
-1_N & 0 \\
0 & 1_N
\end{array}
\right]
g^\dagger
=
\left[
\begin{array}{cc}
2R(g)-1_N &-2K^\star(g)\\
2K(g) & -2R^\star(g)+1_N
\end{array}
\right] ,
\nonumber\\
R(g)
=
q^\dag q \big(1_N  +  q^\dag q \big)^{-1}, \qquad
K(g)
=
- q \big(1_N  +  q^\dag q \big)^{-1}.
\label{densityRg2}
\end{gather}
Introducing a new $N  \times  2N$ matrix ${\cal Z}(g)$ as
$
{\cal Z}(g)
=
\left(1_N + q^\dag q \right)^{-\frac{1}{2}}
\left[1_N, q^\dag \right]
$
with
$
{\cal Z}(g)^\dag {\cal Z}(g)
 =
1_{2N}
$,
we have a very simple expression for ${\cal R}(g)$ as
\begin{gather*}
{\cal R}(g)
=
1_{2N}
-
2 {\cal Z}(g) {\cal Z}(g)^\dag
=
\left[
\begin{array}{cc}
1_N &   0 \\
0 &   1_N
\end{array}
\right]
- 2
\left[
\begin{array}{cc}
\left(1_N + q^\dag q \right)^{-1} &
\left(1_N  +  q^\dag q \right)^{-1}   q^\dag\\
q   \left(1_N  +  q^\dag q \right)^{-1} &
q   \left(1_N  +  q^\dag q \right)^{-1}   q^\dag
\end{array}
\right]   ,\nonumber\\
\mbox{Tr}\,{\cal R}(g)
=
0 ,
\end{gather*}
which has quite the same form as the one given by Rajeev
\cite{Rajeev.94}.
The two-form $\omega$ is given as
\begin{gather*}
\omega
=
-
\tfrac{i}{8}
\mbox{Tr}\big\{\left(d{\cal R}(g)\right)^3\big\}
=
-
\tfrac{i}{8}
\mbox{Tr}\big\{\left(d{\cal R}(g)\right)^3{\cal R}(g)^2\big\} ,\nonumber\\
d\omega
=
- d\omega
=
0  \quad (\mbox{closed form}) .
\end{gather*}
If we introduce hermitian matrices
$
U
=
\left[
\begin{array}{cc}
0 &   u \\
u^\dag &   0
\end{array}
\right]
$
and
$
V
=
\left[
\begin{array}{cc}
0 &   v \\
v^\dag &   0
\end{array}
\right]
$,
then
we have
\begin{gather*}
\omega(U, V)
=
-
\tfrac{i}{8}
\mbox{Tr}
\left\{
\left[
\begin{array}{cc}
1_N &   0 \\
0 &   -1_N
\end{array}
\right]
[U, V]
\right\}
=
\tfrac{i}{4}
\mbox{Tr}
\left\{u^\dag v - v^\dag u \right\} ,
\end{gather*}
which is a symplectic form and
makes it possible to discuss geometric quantization on
a f\/inite/in\-f\/i\-ni\-te-dimensional Grassmannian
\cite{RajeevTurgut.98,ToprakTugurt.02}.




\section[SCF method and $\tau$-functional method on group manifolds]{SCF method and $\boldsymbol{\tau}$-functional method on group manifolds}\label{section3}


\subsection{Introduction}\label{section3.1}

Despite the dif\/ference
due to the dimension of fermions mentioned in Section~\ref{section1},
we ask the follo\-wing:
How is a {\em collective submanifold},
truncated through the SCF equation,
related to a~{\em subgroup orbit} in
the inf\/inite-dimensional Grassmannian by
the $\tau$-FM?
To get a microscopic understan\-ding of
cooperative phenomena,
the concept of collective motion
is introduced in relation to
TD variation of a SCF.
Independent-particle (IP) motion is described
in terms of particles
referring to a stationary MF.
The TD variation of the TD SCF
is attributed to couplings
between the collective and
the IP motions
and couplings among quantal f\/luctuations
of the TD SCF~\cite{YK.87}.
There is a one-to-one correspondence between
{\em MF potentials} and vacuum states of the system.
Decoupling of collective motion out of
full-parameterized TDHF dynamics corresponds to
truncation of the {\em integrable sub-dynamics}
from a full-parameterized TDHF manifold.
The collective submanifold
is a collection of collective paths developed
by the SCF equation.
The collectivity of each path ref\/lects
the {\em geometrical attribute of the Grassmannian},
which is independent of the characteristic of the SCF Hamiltonian.
Then the collective submanifold
should be understood in relation to the collectivity of
various subgroup orbits in the Grassmannian.
The collectivity arises
through interference among interacting fermions
and links with the concept of the MF potential.
The perturbative method
has been considered to be useful to describe
the {\em periodic} collective motion with large amplitude
\cite{Ma.80,YK.87}.
If we do not break the group structure of the Grassmannian
in the perturbative method,
the {\em loop group} may work under that treatment.

Thus we notice the following point in both methods:
Various subgroup orbits consisting of {\em loop} path
may {\em infinitely exist}
in the full-parameterized TDHF manifold.
They must satisfy an inf\/inite set of
Pl\"{u}cker relations to hold the Grassmannian.
As a result, the f\/inite-dimensional Grassmannian on the circle $S^1$
is identif\/ied with an inf\/inite-dimensional one.
Namely the $\tau$-FM works
as an algebraic tool to classify the subgroup orbits.
The SCF Hamiltonian is able to exist
in the inf\/inite-dimensional Grassmannian.
Then the SCFT
can be rebuilt on the inf\/inite-dimensional fermion
Fock space and also
on the $\tau$-functional space.
The inf\/inite-dimensional fermions are introduced
through Laurent expansion of
the f\/inite-dimensional fermions with respect to
the degrees of freedom of the fermions related to
the MF potential.
Inversely, the collectivity of the MF potential
is attributed to gauges of interacting
inf\/inite-dimensional fermions and
interference among fermions is
elucidated via the Laurent parameter.
These are described with the use of af\/f\/ine KM algebra
according to the Dirac theory
\cite{Dirac.58}.
Algebro-geometric structure of
{\em infinite}-dimensional fermion many-body systems
is realized in the {\em finite}-dimensional ones.


\subsection[Bilinear differential equation in SCF method]{Bilinear dif\/ferential equation in SCF method}\label{section3.2}

Owing to the anti-commutation relations
$
\{ c_\alpha,c_\beta^\dagger \}
=
\delta_{\alpha \beta }$,
$
\{ c_\alpha,c_\beta \}
=
\{ c_\alpha^\dagger,c_\beta^\dagger \}
= 0
$,
fermion pair operators
$e_{\alpha \beta }  \equiv  c_\alpha^\dagger c_\beta$
satisfy a Lie commutation relation
$
[ e_{\alpha \beta }, e_{\gamma \delta } ]
=
\delta_{\beta \gamma } e_{\alpha \delta }
-
\delta_{\alpha \delta } e_{\gamma \beta }
$
and span the $u(N)$ Lie algebra.
A canonical transformation
$
U(g)
=
e^{\gamma _{\alpha \beta }
c_\alpha^\dagger c_\beta }$
$(\gamma^\dagger
= -{\gamma }$, $
g = e^\gamma
\ni U(N))
$
generates a transformation such that
\begin{gather}
 U(g)c_\alpha^\dagger U^{-1} (g)
=
c_\beta^\dagger
g_{\beta \alpha }, \qquad
U(g) c_\alpha U^{-1} (g)
=
c_\beta
g_{\beta \alpha }^* , \nonumber\\
U^{-1}(g)
=
U(g^{-1})
=
U(g^{\dagger }), \qquad
U(gg')
=
U(g)U(g'), \qquad g^\dagger g = gg^\dagger
= 1_N .
\label{CanoTra}
\end{gather}
Let $|0\rangle$ be a free vacuum
and $|\phi_M \rangle$ be an $M$ particle S-det
\begin{gather}
 c_\alpha |0\rangle
=
0, \qquad \alpha = 1 , \dots , N, \qquad
|\phi_M \rangle
=
c_M^\dagger \cdots c_1^\dagger
|0 {\rangle },\nonumber \\
U(g) |\phi_M \rangle
=
(c^\dagger g)_M \cdots (c^\dagger g)_1
|0 \rangle
\stackrel{d}{=}
|g {\rangle }, \qquad
U(g) |0 \rangle
=
|0 \rangle ,
\label{ThoulessTra}
\end{gather}
where the $c^\dagger$ is an $N$-dimensional row vector
$c^\dagger = ( c_1^\dagger, \dots, c_N^\dagger )$.
Equation~(\ref{ThoulessTra})
shows that
the $M$ particle S-det
is an exterior product of $M$
single-particle states and that
$U(g)$ transforms~$|\phi_M \rangle$ to another S-det
(Thouless transformation)~\cite{Th.60}
under~(\ref{CanoTra}).
Such states are called ``simple'' states.
The set of all the ``simple'' states of unit modulus
together with the equivalence relation,
identifying distinct states
only in phases with the same state,
constitutes a manifold
known as Grassmannian ${\rm Gr}_M$.
The ${\rm Gr}_M$ is an orbit of the group
given through
(\ref{ThoulessTra}).
Any simple state $| \phi_M \rangle  \in  {\rm Gr}_M$
def\/ines a decomposition
of single-particle Hilbert spaces into
sub-Hilbert spaces of occupied and unoccupied states~\cite{RR.81}.
Thus, the ${\rm Gr}_M$ corresponds to a coset space
$
{\rm Gr}_M \sim \frac{U(N)}{(U(M) \times U(N  -  M))}
$.
Using a variable $p$ of the coset space,
following
\cite{Fu.77,Fu.81} and \cite{NMO.04},
we express the third equation of~(\ref{ThoulessTra})
as
\begin{gather}
U(g)|\phi_M \rangle
=
\langle \phi_M |U(g_\zeta g_w )|\phi_M \rangle
e^{p_{ia} c_i^\dagger c_a}|\phi_M {\rangle } , \qquad
 g = g_\zeta g_w  ,
\label{Slaterdet1}
\end{gather}
where we have used the relations
\begin{gather}
 1 + \sum_{\rho = 1}^{M_{\max}}
\sum_{\substack{1 \leq a_1 <\cdots <a_\rho \leq M,\\
M+1 \leq i_1 <\cdots <i_\rho \leq N}}
{\cal A}(p_{{i_1}{a_1}} \cdots
p_{{i_\rho }{a_\rho }})
c_{i_1}^\dagger c_{a_1} \cdots
c_{i_\rho }^\dagger c_{a_\rho }
=
e^{p_{ia} c_i^\dagger c_a} ,\nonumber\\
\langle \phi_M | U(g_\zeta g_w ) |\phi_M \rangle
=
[\det (1+p^\dagger p)]^{-\frac{1}{2}}
\cdot \det w
\label{Slaterdet1p}
\end{gather}
and the def\/inition
\begin{gather*}
{\cal A}(p_{{i_1}{a_1}} \cdots p_{{i_\rho }
{a_\rho }})
\stackrel{d}{=}
\det
\left[
\begin{array}{ccc}
p_{{i_1}{a_1}}& \cdots &p_{{i_1}{a_\rho }} \\
\vdots&&\vdots \\
p_{i_\rho a_1}& \cdots &
p_{{i_\rho }{a_\rho }}
\end{array}
\right] .
\end{gather*}
In
(\ref{Slaterdet1p})
a maximum value $M_{\max}$
is given by $M_{\max}
=
\min (N  -  M,M)$
and ${\cal A}(\cdots)$ is an anti-symmetrizer.
$\det w$ is a determinant of matrix $w$
and is a phase appearing in the decomposition of
any $U(N)$ matrix as $g = g_\zeta g_w$.
The indices $i$ and $a$ denote
unoccupied states ($M  +  1,\dots,N$)
and
occupied states ($1,\dots, M$),
respectively.
The matrices $p$ and $w$ are def\/ined in Appendix~\ref{appendixA}.

In the ${\rm Gr}_M$ we can introduce an expression
called the Pl\"{u}cker coordinate which has played
important roles for an algebraic construction
of a soliton theory in its early stage
\cite{Sa.81},
\begin{gather}
 U(g)|\phi_M \rangle
=
\sum_{1 \leq \alpha_1 , \dots ,\alpha_M \leq N}
v_{\alpha_1 ,\dots ,\alpha_M }^ {1,\dots ,M}(g)
c_{\alpha_M}^\dagger \cdots
c_{\alpha_1}^\dagger |0 \rangle ,\nonumber\\
v_{\alpha_1 ,\dots ,\alpha_M}^{1,\dots,M}(g)
=
\det
\left[
\begin{array}{rrl}
g_{\alpha_1 ,1} & \cdots & g_{\alpha_1 ,M} \\
\vdots & & \vdots \\
g_{\alpha_M ,1} & \cdots & g_{\alpha_M ,M}
\end{array}
\right]
\qquad \mbox{(Pl\"{u}cker~coordinate).}
\label{Plueckercoordinates}
\end{gather}
From elementary determinantal calculus,
we prove easily
the Pl\"{u}cker coordinate has a relation
\begin{gather*}
\sum_{i=1}^{M+1} (-1)^{i-1}
v _{\alpha_1 , \dots , \alpha_{M-1}, \beta_i}^
{1, \dots ,M}
\cdot
v _{\beta_1 , \dots , \beta_{i-1} ,
\beta_{i+1} , \dots , \beta_{M+1}}^
{1 ,\dots ,M} =0\qquad
\mbox{(Pl\"{u}cker relation),}
\end{gather*}
where the indices denote
the distinct sets
$1  \leq  \alpha_1 , \dots ,\alpha_{M-1}  \leq  N$
and
$1  \leq  \beta_1 ,\dots ,\beta_{M+1}  \leq  N$.
The Pl\"{u}cker relation
is equivalent to a bilinear identity equation
\begin{gather*}
\sum_{\alpha =1}^N c_\alpha^\dagger U(g) |
\phi\rangle \otimes c_\alpha U(g)
| \phi \rangle
=
\sum_{\alpha =1}^N U(g) c_\alpha^\dagger |
\phi \rangle \otimes U(g) c_\alpha
| \phi \rangle = 0 .
\end{gather*}
The bilinear equation has a more general form
\begin{gather*}
\sum_{\alpha =1}^N c_{\alpha }^\dagger
U(g) |\phi_k
\rangle \otimes c_\alpha U(g)
|\phi_l \rangle
=
\sum_{\alpha=1}^N U(g)
c_\alpha^\dagger |\phi_k \rangle
\otimes U(g) c_\alpha
|\phi_l \rangle = 0 , \qquad N  \geq  k  \geq  l  \geq  0,\!\!
\end{gather*}
where $| \phi_k \rangle$ and $|
\phi_l \rangle$
denote $k$-particle simple state and $l$-one,
respectively.
It is noted that
the ${\rm Gr}_M$ is essentially an $SU(N)$
group manifold
since the phase equivalence theorem does hold.

Now we study the relation
between the coset coordinate appeared
in
(\ref{Slaterdet1})
and
the Pl\"{u}cker coordinates
in
(\ref{Plueckercoordinates}).
Both the well-known coordinates make
a crucial role
to clarify the algebraic relation
between the SCFT, i.e.\ TDHFT,
and the soliton theory.

Using the expressions
for unoccupied and occupied states
in
(\ref{Slaterdet1}),
we can rewrite
(\ref{Plueckercoordinates})
as
\begin{gather}
  U(g)|\phi_M \rangle
= |\phi_M \rangle + \sum_{\rho = 1}^{M_{\max}}
\sum_{\substack{1 \leq a_1 < \cdots < a_\rho \leq M,\\
M+1 \leq i_1 < \cdots <i_\rho \leq N}}
v_{1, \dots, a_1 -1, a_1 +1,
\dots, a_\rho -1, a_\rho +1,
\dots, M, i_1 , \dots , i_\rho}^
{1 , \dots , M} (g_\zeta g_w)  \nonumber\\
\phantom{U(g)|\phi_M \rangle=}{}\times c_{i_\rho}^\dagger
\cdots c_{i_1}^\dagger
c_M^\dagger \cdots c_{a_\rho +1}^\dagger
c_{a_\rho -1}^\dagger
\cdots c_{a_1 +1}^\dagger
c_{a_1 -1}^\dagger \cdots
c_1^\dagger |0 \rangle \nonumber\\
\phantom{U(g)|\phi_M \rangle}{} =
|\phi_M \rangle  +
v_{1 , \dots , M}^{1 , \dots , M}
(g_\zeta g_w )
\sum_{\rho = 1}^{M_{\max}}
\sum_{\substack{1 \leq a_1 < \cdots < a_\rho \leq M,\\
M+1 \leq i_1 < \cdots < i_\rho \leq N}}
\frac{
v_{1 , \dots , i_1, \dots , i_\rho, \dots , M}^
{1 , \dots , a_1 , \dots , a_\rho , \dots , M}
(g_\zeta g_w)}
{v_{1 , \dots , M}^{1 , \dots , M} (g_\zeta g_w)}
 \nonumber\\
\phantom{U(g)|\phi_M \rangle=}{}\times  c_{i_1}^\dagger c_{a_1} \cdots
c_{i_\rho }^\dagger
c_{a_\rho }|\phi_M \rangle .
\label{Plueckercoordinates2}
\end{gather}
The last line of the above is recast again into the form
of (\ref{Plueckercoordinates})
after many time exchanges bet\-ween
$c_{a_1}\cdots c_{a_\rho }$ and
all creation operators so that
all the annihilation operators are ordered
in such a way that
they are to the right of all the creation operators
including the ones in $|\phi_M \rangle$.
Then we have the relation
\begin{gather*}
v_{1, \dots, a_1 -1, a_1 +1,
\dots, a_\rho -1, a_\rho +1,
\dots, M, i_1 , \dots , i_\rho }^
{1,  \dots , M}  (g_\zeta g_w)
=
(-1)^{\sum\limits_{j=0}^{\rho -1} (M-j-a_{\rho -j})}
v_{1 , \dots , i_1, \dots , i_\rho, \dots , M}^
{1 , \dots , a_1 , \dots , a_\rho , \dots , M}
(g_\zeta g_w) ,
\end{gather*}
and the following decompositions:
\begin{gather*}
v_{1 , \dots , i_1, \dots , i_\rho, \dots , M}^
{1 , \dots , a_1 , \dots , a_\rho , \dots , M}
(g_\zeta g_w)
=
v_{1 , \dots , i_1, \dots , i_\rho, \dots , M}^
{1 , \dots , a_1 , \dots , a_\rho , \dots , M}
(g_\zeta)
v_{1 , \dots , M}^{1 , \dots ,M} (g_w) ,\\
v_{1 , \dots , M}^
{1 , \dots , M}
(g_\zeta g_w)
= v_{1 , \dots , M}^
{1 , \dots , M} (g_\zeta)
v_{1 , \dots , M}^{1 , \dots , M}
(g_w) , \qquad
v_{1 , \dots , M}^
{1 , \dots , M}
(g_\zeta)
=
\det C(\zeta)
=
[\det (1+p^\dagger p)]^{-\frac{1}{2}} ,
\end{gather*}
where
\begin{gather}
v_{1 , \dots , i_1, \dots , i_\rho, \dots , M}^
{1 , \dots , a_1 , \dots , a_\rho , \dots , M}
(g_\zeta)
=
\det\!\!
\left[ \!\!\begin{array}{ccc}
C(\zeta)_{1,1} & \cdots &C(\zeta)_{1,M}\\
\vdots && \vdots \\
C(\zeta)_{a_1 -1,1} & \cdots &C(\zeta)_{a_1 -1,M} \\
S(\zeta)_{i_1 ,1} & \cdots &S(\zeta)_{i_1 ,M} \\
C(\zeta)_{a_1 +1,1} & \cdots &C(\zeta)_{a_1 +1,M} \\
\vdots && \vdots \\
C(\zeta)_{a_\rho -1,1} & \cdots &
C(\zeta)_{a_\rho -1,M}\\
S(\zeta)_{i_\rho ,1}& \cdots &S(\zeta)_{i_\rho,M} \\
C(\zeta)_{a_\rho +1,1} & \cdots &
C(\zeta)_{a_\rho +1,M} \\
\vdots &&\vdots \\
C(\zeta)_{M,1} & \cdots &C(\zeta)_{M,M}
\end{array}\!\!
\right]\! ,\qquad
v_{1 , \dots , M}^{1 , \dots , M} (g_w )
=
\det w .\!\!\!
\label{Slaterdet3}
\end{gather}
Here matrix elements in the $a_1$-th, $\dots $ and $a_\rho$-th rows,
$C(\zeta)_{a_1 ,1 \sim M}, \dots $ and $C(\zeta)_{a_\rho ,1 \sim M}$
are replaced with
$S(\zeta)_{i_1 ,1 \sim M},\dots $ and $S(\zeta)_{i_\rho ,1 \sim M}$
to describe $\rho~(1 <\rho < M)$ times particle-hole excitations
from hole state $a_1$ to particle state~$i_1 ,\dots$ and those of
hole state~$a_\rho$ to particle state~$i_\rho$, respectively.

Equating equations
(\ref{Slaterdet1})
and
(\ref{Plueckercoordinates})
with equations
(\ref{Plueckercoordinates2})
and
(\ref{Slaterdet3}),
respectively,
we obtain the anti-symmetrized ${\cal A}(\cdots)$ and
the coset variable
expressed in terms of Pl\"{u}cker coordinates as
\begin{gather}
{\cal A} (p_{i_1 a_1} \cdots
p_{i_\rho a_\rho})
=
\frac{v_{1 , \dots , i_1, \dots , i_\rho, \dots , M}^
{1 , \dots , a_1 , \dots , a_\rho , \dots , M}
(g_\zeta)}
{v_{1 , \dots , M}^{1 , \dots , M} (g_\zeta )}
, \qquad
p_{ia}
\!=\!
[S(\zeta) C^{-1} (\zeta)]_{ia}
=
\frac{
v_{1 , \dots , i , \dots , M}^
{1 , \dots , a , \dots , M}
(g_\zeta)}
{v_{1 , \dots , M}^{1 , \dots , M}
(g_\zeta )}
,
\label{Slaterdet4}
\end{gather}
in the second Pl\"{u}cker coordinate of which,
only one row matrix elements of its determinantal form
(\ref{Slaterdet3})
$C(\zeta)_{a ,1 \sim M}$ are replaced with
$S(\zeta)_{i ,1 \sim M}$.
Expanding the anti-symmetrized ${\cal A}(\cdots)$
in the left-hand side of
the f\/irst equation of (\ref{Slaterdet4})
with respect to, for example, the f\/irst column,
we have a decomposition rule
\begin{gather*}
\frac{v_{1 , \dots , i_1, \dots , i_\rho, \dots , M}^
{1 , \dots , a_1 , \dots , a_\rho , \dots , M}
(g_\zeta)}
{v_{1 , \dots , M}^{1 , \dots , M}
(g_\zeta)}
=
\sum_{j=1}^\rho (-1)^{j+1} p_{i_j a_1}
{\cal A}(p_{i_1 a_2} \cdots p_{i_{j-1},a_j}
p_{i_{j+1} a_{j+1}} \cdots p_{i_\rho a_\rho })\\ 
\hphantom{\frac{v_{1 , \dots , i_1, \dots , i_\rho, \dots , M}^
{1 , \dots , a_1 , \dots , a_\rho , \dots , M}
(g_\zeta)}
{v_{1 , \dots , M}^{1 , \dots , M}
(g_\zeta)}} {}=
\sum_{j=1}^\rho (-1)^{j+1}
\frac{
v_{1 , \dots , i_j , \dots , M}^{1 , \dots a_1 , \dots , M}
(g_\zeta)}
{v_{1 , \dots , M}^{1 , \dots , M}
(g_\zeta)}
\frac{
v_{1 , \dots , a_1 , \dots , i_1 , \dots ,
i_{j-1} , \dots , i_{j+1}  , \dots ,
i_\rho , \dots , M}^{1 , \dots , a_1 , \dots ,
a_2 , \dots ,
a_j , \dots , a_{j+1} , \dots , a_\rho , \dots , M}
(g_\zeta)}
{v_{1 , \dots , M}^{1 , \dots ,M} (g_\zeta)}
, \nonumber
\end{gather*}
which is rewritten to another form
(the second Pl\"{u}cker relation)
\begin{gather}
v_{1 , \dots , M}^{1 , \dots , M}
( g_\zeta )
v_{1 , \dots , i_1, \dots , i_\rho, \dots , M}^
{1 , \dots , a_1 , \dots , a_\rho , \dots , M}
( g_\zeta ) \nonumber\\
\qquad{}+
\sum_{j=1}^\rho (-1)^j
v_{1 , \dots , i_j , \dots , M}^{1 , \dots, a_1 , \dots , M}
( g_\zeta )
v_{1 , \dots , a_1 , \dots , i_1 , \dots ,
i_{j-1}, \dots,  i_{j+1}  , \dots ,
i_\rho , \dots , M}^
{1 , \dots , a_1 , \dots , a_2 , \dots ,
a_j , \dots , a_{j+1} , \dots , a_\rho , \dots , M}
( g_\zeta )
= 0 ,
\label{Plueckerrelation2}
\end{gather}
in which
hole state $a_1$ in the last Pl\"{u}cker coordinate
make no changes ($a_1 \rightarrow a_1$)
since in the second one
particle-hole excitation already occurred
from hole state $a_1$ to particle state $i_j$~\cite{WigSym99}.

It is well-known that
the Pl\"{u}cker relation
is equivalent to a bilinear identity equation
\begin{gather*}
\sum_{\alpha =1}^N c_\alpha^\dagger U(g) |
\phi_M \rangle \otimes c_\alpha U(g)
| \phi_M \rangle
=
\sum_{\alpha =1}^N U(g) c_\alpha^\dagger |
\phi_M \rangle \otimes U(g) c_\alpha
| \phi_M \rangle
= 0 ,
\end{gather*}
which have made an important role to construct
many kinds of solitons on various group manifolds
\cite{JM.83}.

Parallel to the regular representation method
by Fukutome
\cite{Fu.77,Fu.81},
we can prove that
the Lie commutation relation
is also satisf\/ied by the dif\/ferential operators for
particle-hole pairs in Appendix~\ref{appendixB}:
\begin{gather}
{\ve}^{ia}
\stackrel{d}{=}
-
\left(
p_{ja}^* p_{ib}^*
\frac{\partial }{\partial p_{jb}^*}
+ \frac{\partial }{\partial p_{ia}}
- \frac{i}{2} p_{ia}^* \frac{\partial }
{\partial \tau }
\right) ,\qquad
{\ve}_{ai}
\stackrel{d}{=}
-
\left(
p_{ja}p_{ib}
\frac{\partial }{\partial p_{jb}}
+ \frac{\partial }{\partial p_{ia}^*}
+ \frac{i}{2}p_{ia} \frac{\partial }
{\partial \tau }
\right)
, \nonumber\\
{\ve}_{ab}
\stackrel{d}{=}
p_{ia} \frac{\partial }{\partial p_{ib}}
- p_{ib}^* \frac{\partial }{\partial p_{ia}^*}
+ i \delta_{ab} \frac{\partial }{\partial \tau }
,\qquad
{\ve}_{ij}
\stackrel{d}{=}
p_{ia}^* \frac{\partial }{\partial p_{ja}^*}
- p_{ja} \frac{\partial }{\partial p_{ia}}
.
\label{p-hdifferentialoperators}
\end{gather}
From the calculations in Appendix~\ref{appendixB},
these dif\/ferential operators
are also proved to satisfy relations
\begin{gather}
{\ve}^{ia} \Phi_{M,M}(p,~p^*,\tau)
=
p_{ia}^*
\Phi_{M,M}(p,~p^*,\tau) ,\qquad
{\ve}_{ai} \Phi_{M,M}(p,p^*,\tau) = 0 , \nonumber\\
{\ve}_{ab} \Phi_{M,M}(p,p^*,\tau) = \delta_{ab}
\Phi_{M,M}(p,p^*,\tau)  ,\qquad
{\ve}_{ij} \Phi_{M,M}(p,p^*,\tau) = 0 ,
\label{p-hdifferentialoperatorsontovacuum}
\end{gather}
and a commutator
$[{\ve}^{ia},p^* _{jb}]
=
- p^* _{ib} p^* _{ja}$.
A free particle-hole vacuum function
$\Phi_{M,M}(p,p^*,\tau)$
is given as
\begin{gather}
\Phi_{M,M}(p,p^*,\tau)
= [\det (1 + p^\dagger p)]^{-\frac{1}{2}}e^{-i\tau }.
\label{freeparticle-holevacuumfunction}
\end{gather}
Further we can introduce
higher order dif\/ferential operators
obeying the relation
\begin{gather*}
D_{1 , \dots ,  i_1 ,  \dots , i_\mu ,  \dots , M}^
{1 , \dots ,  a_1 ,  \dots , a_\mu ,  \dots , M}
(p,\partial_p ,\partial_{p^*},\partial_\tau )
\stackrel{d}{=}
{\ve}^{i_1 a_1} \cdots {\ve}^{i_\mu a_\mu } , \nonumber\\
D_{1 , \dots ,  i_1 ,  \dots , i_\mu ,  \dots , M}^
{1 , \dots ,  a_1 ,  \dots , a_\mu ,  \dots , M}
(p,\partial_p ,\partial_{p^*},
\partial_\tau )\Phi_{M,M}(p,p^* ,\tau )
=
{\cal A}(p^* _{i_1 a_1}
\cdots p^* _{i_\mu a_\mu })\Phi_{M,M}(p,p^* ,\tau ) ,
\end{gather*}
which show that
by operating the dif\/ferential operator $D$
on the vacuum function $\Phi$,
we obtain the Pl\"{u}cker coordinate ${\cal A}$.
The Pl\"{u}cker relation
(\ref{Plueckerrelation2})
becomes a f\/inite set
of partial dif\/ferential equations
satisfying
\begin{gather*}
\Phi_{M,M} (p,p^* ,\tau)
D_{1 , \dots ,  i_1 ,  \dots , i_\rho ,  \dots , M}^
{1 , \dots ,  a_1 ,  \dots , a_\rho ,  \dots , M}
\Phi_{M,M} (p,p^* ,\tau)\nonumber\\
+
\sum_{j=1}^\rho   (-1)^j
D_{1 , \dots , i_1 , \dots , M}^
{1 , \dots , a_1 , \dots , M}
\Phi_{M,M} (p,p^* ,\tau)
D_{1 , \dots , a_1 , \dots , i_1 , \dots ,
i_{j-1} , \dots , i_{j+1}  ,
\dots , i_\rho , \dots , M}^
{1 , \dots , a_1 , \dots , a_2 , \dots ,
a_j , \dots , a_{j+1} , \dots , a_\rho , \dots , M}
\Phi_{M,M} (p,p^* ,\tau)
 =  0 , \nonumber\\
\left(
v_{1 , \dots ,  i_1 ,  \dots , i_\mu ,  \dots , M}^
{1 , \dots ,  a_1 ,  \dots , a_\mu ,  \dots , M}
(g_\zeta g_w)
\right)^{*}
=
\left(
v_{1 , \dots ,  i_1 ,  \dots , i_\mu ,  \dots , M}^
{1 , \dots ,  a_1 ,  \dots , a_\mu ,  \dots , M}
(g_\zeta) \det w
\right)^{*} \nonumber\\
\qquad{} =
D_{1 , \dots ,  i_1 ,  \dots , i_\mu ,  \dots , M}^
{1 , \cdots ,  a_1 ,  \dots , a_\mu ,  \dots , M}
\Phi_{M,M} (p,p^* ,\tau) .
\end{gather*}
Thus, in both the SCFT and
the soliton theory on a group,
we can f\/ind the common feature that
the Grassmannian is just identical
with the solution space of
the bilinear dif\/ferential equation.
The solution space of each dif\/ferential equation
becomes
an integral surface
\cite{KN.00,NK.02,WigSym99}.


\subsection[SCF method in $F_\infty$]{SCF method in $\boldsymbol{F_\infty}$}\label{section3.3}

We will give here a brief sketch of the SCF equation,
i.e.,
the TDHFM.
According to Rowe et al.~\cite{RR.81},
we start with a geometrical aspect of the method
in the following way:

Let us consider the time dependent
Schr\"{o}dinger equation
$i\hbar \partial_t \Psi =H\Psi$
with a Hamiltonian
\begin{gather}
H
=
h_{\beta\alpha } c_\beta^\dagger c_\alpha
+
\tfrac{1}{2}
\langle \gamma \alpha | \delta \beta \rangle
c_\gamma^\dagger c_\delta^\dagger c_\beta c_\alpha ,
\label{originalhamiltonian}
\end{gather}
where
$\langle \gamma \alpha | \delta \beta \rangle$
denotes
a matrix element of an interaction potential.
The starting point for the TDHFT
lies in an extremal condition of an action integral
\begin{gather}
 \delta
 \int_{t_1}^{t_2}
dt{\cal L}(g(t)) = 0 ,\qquad
{\cal L}(g(t))
\stackrel{d}{=}
\langle
\phi_M|U(g^\dagger (t))
(i\hbar\partial_t -H)U(g(t))|\phi_M
\rangle .
\label{actionintegral0}
\end{gather}
To get an explicit expression
for the TDHFEQ,
we calculate an expectation value
of one- and two-body operators for
the S-det
(\ref{ThoulessTra}).
Using the canonical transformation
(\ref{CanoTra}),
we have
\begin{gather}
 W_{\alpha \beta }
\stackrel{d}{=}
\langle
\phi_M|U(g^\dagger)
c_\beta^\dagger c_\alpha U(g)|\phi_M
\rangle
\!=\!
(g^\dagger)_{\beta'\beta }
(g^{\rm T})_{\alpha'\alpha }
\langle
\phi_M|c_{\beta'}^\dagger c_{\alpha'}
|\phi_M
\rangle
=
\sum_{\alpha' =1}^M g_{\alpha \alpha'}
g_{\alpha'\beta }^\dagger ,
\label{expectationonebody}
\\
\langle
\phi_M|U(g^\dagger)c_\gamma^\dagger
c_\delta^\dagger c_\beta c_\alpha U(g)
|\phi_M
\rangle
=(g^\dagger)_{\gamma'\gamma }
(g^\dagger)_{\delta'\delta }
(g^{\rm T})_{\beta'\beta }
(g^{\rm T})_{\alpha'\alpha }
\langle
\phi_M|c_{\gamma'}^\dagger
c_{\delta'}^\dagger
c_{\beta'} c_{\alpha'}|\phi_M
\rangle \nonumber\\
\phantom{\langle
\phi_M|U(g^\dagger)c_\gamma^\dagger
c_\delta^\dagger c_\beta c_\alpha U(g)
|\phi_M
\rangle}{} =W_{\alpha \gamma }W_{\beta \delta }
-W_{\alpha \delta }W_{\beta \gamma } .
\label{expectationtwobody}
\end{gather}
Introducing triangular matrices $C(\zeta )$ and $S(\zeta )$
in ${\rm Gr}_{M}$
\cite{NMO.04,Fu.81}
and  using an isometric matrix~$u^{\rm T}$
\begin{gather*}
u^{\rm T}
=
\left[
C^{\rm T}(\zeta ), S^{\rm T}(\zeta )
\right],\qquad
u^\dagger u = 1_M ,
\end{gather*}
$W$ in
(\ref{expectationtwobody})
is expressed as
$
W = uu^\dagger
$
and satisf\/ies
$
W^2 = W$ (idempotency relation).
Then, it turns out that
the above matrix $W$ is just the density matrix.
From (\ref{expectationtwobody}),
we get an energy functional, i.e.,
an expectation value of the Hamiltonian~(\ref{originalhamiltonian})
\begin{gather}
H[W]
\stackrel{d}{=}
\langle
\phi_M |U(g^\dagger )
HU(g)|\phi_M
\rangle
=
h_{\beta \alpha }W_{\alpha \beta }
+
\tfrac{1}{2}
[\gamma \alpha|\delta \beta]
W_{\alpha \gamma }
W_{\beta \delta },\nonumber\\
[\gamma \alpha | \delta \beta]
 =
\langle \gamma \alpha | \delta \beta \rangle
-
\langle \gamma \beta | \delta \alpha \rangle .
\label{energyfunctionalHW}
\end{gather}
By projecting the original hamiltonian onto the ${\rm Gr}_M$,
we obtain also a HF Hamiltonian
$H_{\rm HF}[W]$
\begin{gather}
H_{\rm HF}[W]
=
{\cal F}_{\alpha \beta }[W]
c_\alpha^\dagger c_\beta , \qquad
{\cal F}_{\alpha \beta }
=
\frac{\delta H[W]}{\delta W_{\beta \alpha }}
=
h_{\alpha \beta }
+
[\alpha \beta | \gamma \delta]W_{\delta \gamma } .
\label{HFHamiltonian}
\end{gather}
The Lagrange function ${\cal L}(g(t))$
in
(\ref{actionintegral0})
is computed as
\begin{gather}
{\cal L}(g(t))
=
\frac{i\hbar }{2}
(g_{ab}^\dagger \dot{g}_{ba}
+g_{ai}^\dagger \dot{g}_{ia}
-\dot{g}_{ab}^\dagger g_{ba}
-\dot{g}_{ai}^\dagger g_{ia})-H[W] ,
\label{Lagrangefunction}
\end{gather}
using
$
\partial_t
U(g^\dagger (t)) U(g(t))
+
U(g^\dagger (t))\cdot\partial_t U(g(t)) = 0
$.
The condition (\ref{actionintegral0})
gives the TDHFEQ
\begin{gather*}
\frac{d}{dt}
\left(
\frac{\partial {\cal L}}{\partial \dot{g}^\dagger }
\right)
-\frac{\partial {\cal L}}
{\partial g^\dagger } = 0 ,\qquad
\frac{d}{dt}
\left(
\frac{\partial {\cal L}}{\partial \dot{g}}
\right)
-\frac{\partial {\cal L}}{\partial g} = 0 ,
\end{gather*}
and then we obtain a compact form
of the TDHFEQ
$i\hbar \partial_t g(t) = {\cal F}[W\{ g(t)\} ]g(t)$.
The time evolution of the S-det
(\ref{ThoulessTra}) is given by
\begin{gather}
i\hbar\partial_t U(g(t))|\phi_M \rangle
=
H_{\rm HF}
[W(g(t))]U(g(t))|\phi_M \rangle.
\label{TDHFeq}
\end{gather}
On the other hand,
using a $\upsilon$-dependent fermion operator
given soon later,
from (\ref{actionintegral0})
we also obtain a compact form of $\upsilon$-dependent HF equation,
instead of time $t$,
as
\begin{gather}
i\hbar \partial_\upsilon g(\upsilon)
=
{\cal F}[W\{ g(\upsilon)\}] g(\upsilon),\qquad
\left(
i\hbar \partial_\upsilon u(\upsilon)
=
{\cal F}[W\{ g(\upsilon)\}] u(\upsilon)
\right) .
\label{upsilondepHFeq}
\end{gather}

Following the observation
by D'Ariano and Rasetti's
\cite{AR.85}
for the relation between
``soliton equations and coherent states'',
we may assert that
the SCFM presents the theoretical scheme for
an integrable sub-dynamics
on a certain inf\/inite-dimensional fermion Fock space,
by identifying~$|\phi_M \rangle$ with the highest weight vector
and by regarding the TDHF-manifold ${\rm Gr}_M$
as the projection onto
a subspace of the $\tau$-function.

We reconstruct a $\upsilon$-dependent SCFM
in a $F_\infty$ and
study a relation between soliton equation and
$\upsilon$-dependent HF equation~\cite{NishiProviKoma2.07}.
We start from a single-particle Schr\"{o}dinger equation
with a~$\upsilon$-dependent and a $\Upsilon$-periodic potential
$V(\vr ,\upsilon)
=
V(\vr ,\upsilon + \Upsilon)
$
\begin{gather*}
h_{\rm sp} (\vr ,\upsilon)
=
-   \frac{{\hbar }^2}{2m}  \Delta
+ V(\vr ,\upsilon) ,\qquad
h_{\rm sp} (\vr ,\upsilon)
\psi_\alpha (\vr ,\upsilon)
=
\epsilon_\alpha \psi_\alpha (\vr ,\upsilon) .
\end{gather*}
Here
we have supposed that an eigen-spectrum
$\epsilon_\alpha$ is $\upsilon$ independent,
though the potential is dependent on $\upsilon$.
It holds an iso-spectrum
under a $\upsilon$-evolution of the potential.
An eigen-function
$\psi_\alpha (\vr, \upsilon)$
constitutes an orthonormal complete set
and satisf\/ies the same periodicity,
$\psi_\alpha (\vr ,\upsilon + \Upsilon)
=
\psi_\alpha (\vr ,\upsilon)$
(Floquet's theorem).
{\em This picture is very different from that in}~\cite{WigSym99}
and~\cite{KN.00}.
According to Goddard and Olive
\cite{GO.86},
we can make Laurent expansion of
a fermion-f\/ield creation-operator
$\psi ^\dagger (\vr ,\upsilon)$
with a parameter $\upsilon$ as
\begin{gather*}
\psi ^\dagger (\vr ,\upsilon)
=
\sum_\alpha \sum_{r \in \mathbb{Z}}
 \left(\frac{1}{\Upsilon }\right)^{\frac{1}{2}}
\psi_{Nr + \alpha } z^{-r}
\psi_\alpha ^* (\vr ,\upsilon) ,
\end{gather*}
where
$
z
=
\exp
\left(
i 2 \pi  \frac{\upsilon }{\Upsilon }
\right)
$
given on a unit circle.
Thus,
the $\psi_{Nr+\alpha}$ can be regarded as
a new fermion creation-operator.
We obtain also a new fermion annihilation-operator in the same way.
The anti-commutation relations can be rewritten as
\begin{gather}
\{ c_\alpha (\upsilon), c_\beta^\dagger (\upsilon') \}
=
\delta_{\alpha \beta } \delta(\upsilon - \upsilon'), \qquad
\{ c_\alpha (\upsilon), c_\beta(\upsilon') \}
=
\{ c_\alpha^\dagger (\upsilon), c_\beta^\dagger (\upsilon') \}
=
0.
\label{tAntiCommu}
\end{gather}
Through Laurent expansion of the fermion-f\/ield operators,
inf\/inite-dimensional fermion operators
with particle spectra and Laurent spectra
can be obtained as
\begin{gather}
 c_\alpha (\upsilon)
=
\sum_{r \in \mathbb{Z}}
 \left(\frac{1}{\Upsilon }\right)^{\frac{1}{2}}
\psi_{Nr + \alpha }^* z^{r}, \qquad
c_\alpha^\dagger (\upsilon)
=
\sum_{r \in \mathbb{Z}}
  \left(\frac{1}{\Upsilon }\right)^{\frac{1}{2}}
\psi_{Nr + \alpha } z^{-r}, \nonumber \\
\delta (\upsilon - \upsilon')
=
  \frac{1}{\Upsilon }
\sum_{r \in \mathbb{Z}}
\exp
\left\{
i 2 \pi  \frac{(\upsilon - \upsilon')}{\Upsilon }  r
\right\},
\label{infFermion}
\end{gather}
where $\mathbb{Z}$ means the set of the integers.
The indices $\alpha$ and $r$
are called the label on particle spectra and that on Laurent spectra,
respectively.
Substitution of
(\ref{infFermion})
into
(\ref{tAntiCommu})
leads to the anti-commutation relations
\begin{gather}
\{ \psi_{Nr + \alpha }^*, \psi_{Ns + \beta } \}
=
\delta_{\alpha \beta } \delta_{rs}, \qquad
\{ \psi_{Nr + \alpha }^*, \psi_{Ns + \beta }^* \}
=
\{ \psi_{Nr\!+\!\alpha }, \psi_{Ns\!+\!\beta } \}
=
0.
\label{infAntiFermion}
\end{gather}
If the canonical transformation
(\ref{CanoTra})
has the $\upsilon$-dependence
and generates the $\upsilon$-evolution of the potential,
it is possible to embed a $U(N)$ group
induced from~(\ref{CanoTra})
into a group which can be induced from
a canonical transformation
of the inf\/inite-dimensional fermion operators
(\ref{infAntiFermion}).

According to Kac and Raina
\cite{Kac.83,KR.87}
in Appendix~\ref{appendixC},
we introduce a $F_\infty$
and an associative af\/f\/ine Kac--Moody algebra.
Here we restrict ourselves to the case of the Lie algebra $u(N)$.
The corresponding perfect vacuum $|\mbox{Vac} \rangle$
and
``simple'' state $|M \rangle$ are def\/ined, respectively as
\begin{gather}
 \psi_{Nr + \alpha } |\mbox{Vac} \rangle
= 0,
\qquad \langle \mbox{Vac}|\psi_{Nr + \alpha }^*
= 0
\quad (r  \leq  -1), \nonumber\\
\psi_{Nr + \alpha }^* |\mbox{Vac}\rangle
= 0,
\qquad \langle \mbox{Vac}|\psi_{Nr + \alpha }
= 0
\qquad (r  \geq  0), \nonumber\\
|M \rangle
\!=\!
\psi_M \cdots \psi_1 |\mbox{Vac} {\rangle },\qquad
\langle M|M \rangle = 1,\qquad
\langle \mbox{Vac}|\mbox{Vac} \rangle  =  1 .
\label{vacuum2}
\end{gather}
We embed the free vacuum $|0 \rangle$ and
simple state $|\phi _M \rangle$ into $F_\infty$ as
$
|0 \rangle  \mapsto  |\mbox{Vac} {\rangle }$,
$| \phi _M \rangle  \mapsto  |M {\rangle }$
$(M  =  1, \dots , N)
$.
Assume that a state with Laurent spectrum
corresponding to $|0 \rangle$
is a stable state with minimal energy.
This means a choice of gauge
under which
$|0 \rangle$ corresponds to $|\mbox{Vac} {\rangle }$.
The matrix~$\gamma$ $( \in  u(N))$ in $U(g)$
has also the periodicity $\Upsilon$.
Mapping from a unit circle~$S^1$ to~$u(N)$~\cite{GO.86},
we make Laurent expansion of $\gamma$ as
$
\gamma (z)
=
\sum\limits_{r \in \mathbb{Z}} \gamma_r z^r
$
and impose
$\gamma ^\dagger (z) = - \gamma (z)
 \mapsto
\gamma_r ^\dagger  =  - \gamma_{-r}$
and
$z^{-1}  =  z^* $ $(|z|  =  1)$.
Using the {\em correspondence}
between basic elements:
$c_\alpha^\dagger c_\beta z^r \mapsto
\uptau\{e_{\alpha \beta }(r)\}
\stackrel{d}{=}
\sum\limits_{s \in \mathbb{Z}}
\psi_{N(s-r) + \alpha } \psi_{Ns + \beta }^*$
and
normal-ordered product
$: \psi_{Nr + \alpha }\psi_{Ns + \beta }^* :
 \stackrel{d}=
\psi_{Nr + \alpha }\psi_{Ns + \beta }^*
-
\delta_{\alpha \beta }\delta_{rs}$ \mbox{$(s  < 0)$},
let us def\/ine the following
$\widehat{su}(N) ( \subset  \widehat{sl}(N))$ Lie algebra
\cite{WigSym99,KN.00}:
\begin{gather}
 X_\gamma
=
\widehat{X}_\gamma + \mathbb{C} \cdot c ,
~\mathbb{C}^*
=
- \mathbb{C} \quad (\mbox{pure~imaginary}), \nonumber\\
\widehat{X}_\gamma
=
\sum_{r \in \mathbb{Z}} \sum_{s \in \mathbb{Z}}
(\gamma_r)_{\alpha, \beta }
:\psi_{N(s-r) + \alpha } \psi_{Ns\!+\!\beta }^*: ,
\qquad \mbox{Tr}\, \gamma_r = 0, \nonumber\\
[X_\gamma,c]_{\rm KM}
= 0,\qquad
[X_\gamma, X_{\gamma '}]_{\rm KM}
=
\widehat{X}_{[\gamma, \gamma ']}
+ \alpha (\gamma ,\gamma ') \cdot c ,\qquad
c|M \rangle = 1 \cdot |M \rangle ,\nonumber\\
\alpha (\gamma ,\gamma ')
=
-\alpha^* (\gamma,\gamma ')
=
\sum_{r \in \mathbb{Z}} r \,\mbox{Tr}\,
(\gamma_r {\gamma '}_{-r}) ,
\label{SUnLieAlgebra1}
\end{gather}
where $c$ denotes a center.
As for the $\uptau$ representation (rep) and KM bracket,
see Appendix~\ref{appendixC}.

Using
equations
(\ref{infAntiFermion})
and
(\ref{SUnLieAlgebra1}),
adjoint actions of $X_\gamma$ for
$\psi$ and $\psi^*$ are computed as
\begin{gather}
[X_\gamma, \psi_{Nr + \alpha } ]
=
\sum_{s \in \mathbb{Z}} \psi_{N(r-s) + \beta }
(\gamma_s)_{\beta \alpha } ,\qquad
[X_\gamma, \psi_{Nr + \alpha }^* ]
=
\sum_{s \in \mathbb{Z}} \psi_{N(r-s) + \beta }^*
(\gamma _s^*)_{\beta \alpha } .
\label{adjointaction2}
\end{gather}
Let us introduce a canonical transformation
$U(\hat{g})  =  e^{X_{\gamma }}$
satisfying
$U^{-1}(\hat{g})
= U(\hat{g}^{-1})
= U(\hat{g}^\dagger)$ and
$U(\hat{g}\hat{g}')
=
U(\hat{g})U(\hat{g}')$.
The $\hat{g}$ $(=  e^\gamma)$
has a form analogous to $g$
but with inf\/inite dimension
and satisf\/ies
$\hat{g}^\dagger \hat{g}
=
\hat{g}\hat{g}^\dagger
=
1_\infty$.
Further, using
(\ref{adjointaction2}) and
the operator identity
$
e^{X_{\gamma }} A e^{-X_{\gamma }}
=
A + [X_\gamma,A]
+  \frac{1}{2!}
[X_\gamma,[X_\gamma,A]]
+ \cdots
$,
the inf\/inite-dimensional fermion operator
is transformed into
\begin{gather}
\psi_{Nr + \alpha } (\hat{g})
\stackrel{d}{=}
U(\hat{g})
\psi_{Nr + \alpha } U^{-1}(\hat{g})
=
\sum_{s \in \mathbb{Z}} \psi_{N(r-s) + \beta }
(g_s)_{\beta \alpha } , \qquad
\hat{g}_{Nr + \alpha ,Ns + \beta }
\equiv
(g_{s-r})_{\alpha \beta } ,\!\!\!
\label{CanoTraInfFermionandghat}
\end{gather}
where $g_s$ means the $s$-th block matrix of $\hat{g}$
on each diagonal parallel to the principal diagonal
and satisf\/ies the ortho-normalization relation.
Using the correspondence
$| \phi_M \rangle   \mapsto  |M \rangle$,
$U(g)  \mapsto  U(\hat{g})$ $(= e^{X_\gamma })$ and
\begin{gather*}
\sum_{\alpha = 1}^N c_\alpha^\dagger
\otimes c_\alpha
\mapsto
\sum_{\alpha =1}^N \sum_{r \in \mathbb{Z}}
c_\alpha^\dagger z^{-r}
\otimes c_\alpha z^r
\simeq
\sum_{\alpha =1}^N
\sum_{r \in \mathbb{Z}} \psi_{Nr\!+\!\alpha }
\otimes \psi_{Nr + \alpha }^* ,
\end{gather*}
the bilinear equation
on the f\/inite-dimensional Fock space
is embedded into the one on $F_\infty$ as
\begin{gather}
 \sum_{\alpha =1}^N \sum_{r \in \mathbb{Z}}
\psi_{Nr + \alpha } U(\hat{g}) |M \rangle
 \otimes
\psi_{Nr + \alpha }^* U(\hat{g})|M \rangle \nonumber\\
\qquad{} =
\sum_{\alpha =1}^N \sum_{r \in \mathbb{Z}}
U(\hat{g}) \psi_{Nr + \alpha }
|M \rangle
 \otimes
U(\hat{g}) \psi_{Nr + \alpha }^*
|M \rangle
 =
0 ,
\label{bilineareqc2}
\\
 \sum_{\alpha =1}^N \sum_{r \in \mathbb{Z}}
\psi_{Nr + \alpha } U(\hat{g}) |k \rangle \otimes
\psi_{Nr + \alpha }^* U(\hat{g})|l \rangle \nonumber\\
\qquad{} =
\sum_{\alpha =1}^N \sum_{r \in \mathbb{Z}}
U(\hat{g}) \psi_{Nr + \alpha } |k \rangle
\otimes U(\hat{g}) \psi_{Nr + \alpha }^*
|l \rangle = 0 ,\qquad k \geq l, \quad k,l=1, \dots, N .
\label{bilineareqdp}
\end{gather}
where $M = 1 \sim N$ and $k \geq l$
$(k,l = 1 \sim N)$.
Thus we arrive at the following picture:
An algebra of extracting sub-group orbits made of
$loop$ path from ${\rm Gr}_M$ belongs to the $sl(N)$-reduction of
$gl(\infty)$ in the soliton theory.
Relieving from restrictions of $su(N)$
and
(\ref{bilineareqc2})
and taking $\gamma \in sl(N)$
with $M$ and $k \geq l$ $(\in \mathbb{Z}$),
(\ref{bilineareqc2})
and
(\ref{bilineareqdp})
can be regarded
as the bilinear equations
of the reduced KP (Kadomtsev--Petviashvili) hierarchy and
the modif\/ied KP in the soliton theory~\cite{JM.83}.
The algebra of extracting various sub-group manifolds made of
several {\em loop-group} pathes~\cite{PS.86}
from ${\rm Gr}_M$ belongs to an $sl(N)$-reduction of~$gl(\infty)$.
This picture suggests us the possibility
to construct a $\upsilon$-dependent HF soliton equation
by using the $\upsilon$-dependent SCFM
governed by conditions~(\ref{bilineareqc2}) and~(\ref{bilineareqdp})
on the space $sl(N)$ bigger than $su(N)$.
However we must note that in the SCFM
bilinear equations~(\ref{bilineareqc2}) and (\ref{bilineareqdp})
are considered to play roles of
conditions for assuring of existence of
sub-group orbits on ${\rm Gr}_M$ dif\/ferent from
the soliton theory in which boson expressions for them
become an inf\/inite set of dynamical equations.
Notice also that the concept of quasi-particle and
vacuum in the SCFM on $S^1$ is connected to
the Pl\"{u}cker relation.

Following
\cite{WigSym99}
and
\cite{KN.00},
we embed
the original Hamiltonian
(\ref{originalhamiltonian}) into $F_\infty$.
By replacing the annihilation-creation operators of fermions as
$c_\beta^\dagger
 \mapsto
\sum\limits_{s \in \mathbb{Z}}
\psi_{Ns+\beta }$
and
$c_\alpha
\mapsto
\sum\limits_{r \in \mathbb{Z}}
\psi_{Nr+\alpha }^*$,
we get
\begin{gather}
H_{F_\infty }
=
h_{\beta \alpha }
\sum_{r,s \in \mathbb{Z}}
\psi_{Nr + \beta }
\psi_{Ns + \alpha }^* +
\tfrac{1}{2}
\langle \gamma \alpha| \delta \beta \rangle
\sum_{k,l \in \mathbb{Z}, \  r,s \in \mathbb{Z}}
\psi_{Nk + \gamma } \psi_{Nl + \delta }
\psi_{Ns + \beta }^* \psi_{Nr + \alpha }^* .
\label{psi-form-hamiltonian}
\end{gather}
Previously we had the $\upsilon$-dependent SCF Hamiltonian
${\cal F}[W\{ g(\upsilon)\}]$
through equations
(\ref{HFHamiltonian})
and
(\ref{upsilondepHFeq})
where
$\upsilon$-dependence is directly brought from~$g(\upsilon)$.
This is contrast that in the usual TDHFEQ
(\ref{TDHFeq}),
the time $t$-dependence of course arises from~$g(t)$.
To embed this SCF Hamiltonian,
we introduce a general Hamiltonian on $F_\infty$ as
\begin{gather}
H_{F_\infty }
=
\sum_{r,s \in \mathbb{Z}}
h_{Ns + \beta ,Nr + \alpha }
\psi_{Ns + \beta }
\psi_{Nr + \alpha  }^* \nonumber\\
\phantom{H_{F_\infty }=}{} +
\tfrac{1}{2}
\sum_{r,s \in \mathbb{Z}, \ k,l \in \mathbb{Z}}
\langle Nk + \gamma ,Nr + \alpha |
Nl + \delta ,Ns + \beta\rangle
\psi_{Nk + \gamma }\psi_{Nl + \delta }
\psi_{Ns + \beta }^*\psi_{Nr + \alpha }^* ,
\label{generalhamiltonian}
\end{gather}
which
is equivalent to
(\ref{psi-form-hamiltonian})
if
$h_{Ns + \beta ,Nr + \alpha }
=
h_{\beta\alpha }$ and
$
\langle
Nk + \gamma ,Nr + \alpha | Nl + \delta,Ns + \beta
\rangle
=
\langle \gamma \alpha|\delta \beta \rangle$
(equivalence conditions for~$H_{F_\infty }$) hold.
To calculate the formal expectation value
of~(\ref{generalhamiltonian})
for the vector $U(\hat{g})|M \rangle$,
f\/irst we do it for
one-body and two-body operators.
Using~(\ref{CanoTraInfFermionandghat})
we obtain
\begin{gather*}
\langle M|
\psi_{Ns+\beta }
\psi_{Nr+\alpha }^*|M
\rangle
=
\delta_{sr} \delta_{\beta \alpha }
\quad (\mbox{for}~r  =  0, \ \alpha  =  1,\dots, M
 \ \mbox{and for} \ r  <  0, \ \alpha =1,\dots, N) , \nonumber\\
 \langle
M| \psi_{Nk + \gamma }\psi_{Nl + \delta }
\psi_{Ns + \beta }^* \psi_{Nr + \alpha }^* |M
\rangle
 =
\delta_{kr}\delta_{\gamma \alpha }
\cdot
\delta_{ls}
\delta_{\delta \beta }-\delta_{ks}
\delta_{\gamma \beta }
\cdot
\delta_{lr}
\delta_{\delta \alpha } , \nonumber\\
(\mbox{for} \ r(s)  =  0, \ \alpha (\beta )
 =  1,\dots,M \ \mbox{and for} \ r(s)  <  0, \ \alpha (\beta )
 =  1,\dots,N ).\nonumber
\end{gather*}
Then,
for one-body and two-body type operators
we obtain
\begin{gather}
\widehat{W}_{Nr + \alpha ,Ns + \beta }
=
\langle
M|U(\hat{g}^\dagger)
\psi_{Ns + \beta }\psi_{Nr + \alpha }^*
U(\hat{g})|M
\rangle \nonumber\\
\phantom{\widehat{W}_{Nr + \alpha ,Ns + \beta }}{}
=
\sum_{\gamma =1}^M
(g_{-r})_{\alpha\gamma }
(g_{-s}^\dagger)_{\gamma \beta }
+
\sum_{t<0} \sum_{\gamma = 1}^N
(g_{t-r})_{\alpha \gamma }
(g_{t-s}^\dagger)_{\gamma \beta } ,
\label{one-bodyterm0}
\\
\langle
M| U   (\hat{g}^\dagger)
\psi_{Nk + \gamma }\psi_{Nl + \delta }
\psi_{Ns + \beta }^*\psi_{Nr + \alpha }^*
U   (\hat{g}) |M
\rangle\nonumber\\
\qquad{}
=
\widehat{W}_{  Nr + \alpha ,Nk + \gamma }
\widehat{W}_{  Ns + \beta ,Nl + \delta }
\-
\widehat{W}_{  Nr + \alpha ,Nl + \delta }
\widehat{W}_{  Ns + \beta ,Nk + \gamma } .\nonumber
\end{gather}
The $\widehat{W}$ is just the so-called density matrix
since it is easily proved to
satisfy the idempotency relation
$\widehat{W}^2 = \widehat{W}$
which has not been given explicitly in
\cite{WigSym99}
and
\cite{KN.00}.
It provides a strong tool to develop
our SCF scenario on the $F_\infty$.
Taking summation over inf\/inite integers,
inevitably we have an {\em anomaly}
in the above expectation value.
To avoid this {\em anomaly},
the one-body operator~(\ref{one-bodyterm0})
must be changed to a normal-ordered product as
\begin{gather}
 ({\cal{W}}_k)_{\alpha \beta }
 \stackrel{d}{=}
\langle
M|
U(\hat{g}^\dagger)
 : \uptau\{e_{\beta \alpha } (-k)\}
 \!: U(\hat{g})|M
\rangle
 =
\sum_{r\in \mathbb{Z}}
\langle
M|U(\hat{g}^\dagger)
 : \psi_{N(r+k) + \beta }\psi_{Nr + \alpha }^*
 \!: U(\hat{g})|M
\rangle\nonumber\\
\phantom{({\cal{W}}_k)_{\alpha \beta }}{}  =
\sum_{r \in \mathbb{Z}}
\widehat{W}_{Nr + \alpha,N(r+k) + \beta }
-\sum_{r<0}\delta_{k,0}\delta_{\beta \alpha }
=
\sum_{r \in \mathbb{Z}}\sum_{\gamma = 1}^M
(g_{r})_{\alpha \gamma }
(g_{r-k}^\dagger)_{\gamma \beta } ,
\label{normalproductone-bodyWK}
\end{gather}
where we have used the correspondence relation
between basic elements.
The ${\cal{W}}_k$ is identical with
a coef\/f\/icient of the Laurent expansion
of the density matrix $W$
(\ref{expectationonebody})
\begin{gather*}
{\cal{W}}_{\alpha \beta } (z)
=
\sum_{k \in \mathbb{Z}}
({\cal{W}}_k)_{\alpha \beta } z^k
=
\sum_{k \in \mathbb{Z}} \sum_{s \in \mathbb{Z}}
\sum_{\gamma =1}^M (g_s)_{\alpha \gamma }
(g_{s-k}^\dagger )_{\gamma \beta } z^k .
\end{gather*}

We compute formal expectation value of
(\ref{generalhamiltonian})
for the state $U(\hat{g})|M \rangle$.
Introducing a new integer $K$,
due to
the equivalence conditions for $H_{F_\infty }$,
from
(\ref{psi-form-hamiltonian})
we get
\begin{gather}
\langle H_{F_\infty }\rangle
[\widehat{W}]
 =
\sum_{K \in \mathbb{Z}}(h_K)_{\beta \alpha }
\sum_{s \in \mathbb{Z}}
\widehat{W}_{Ns + \alpha ,N(s-K) + \beta }\nonumber\\
\phantom{\langle H_{F_\infty }\rangle
[\widehat{W}]
 =}{} +
\tfrac{1}{2}
\sum_{K,L \in \mathbb{Z}}
[(K,\gamma ),\alpha|(L,\delta),\beta ]
\sum_{r,s \in \mathbb{Z}}
\widehat{W}_{Nr + \alpha ,N(r-K) + \gamma }
\sum_{s \in \mathbb{Z}}
\widehat{W}_{Ns + \beta,N(s-L) + \delta }, \nonumber\\
(h_k)_{\beta \alpha }\equiv h_{\beta \alpha },
\qquad [(k,\gamma),\alpha|(l,\delta),\beta]
\equiv
[\gamma \alpha|\delta \beta] \qquad \forall \; k,\, l.
\label{HamiltonianFinfinity}
\end{gather}
Changing $\widehat{W}$
in
(\ref{HamiltonianFinfinity})
into its normal-ordered product
and using
(\ref{normalproductone-bodyWK}),
we obtain
\begin{gather*}
\langle H_{F_\infty }\rangle [{\cal{W}}]
=
\sum_{k \in \mathbb{Z}}
\left\{
h_{\beta \alpha }
({\cal{W}}_{-k})_{\alpha \beta }
+
\tfrac{1}{2}
[\gamma \alpha | \delta \beta]
\sum_{l \in \mathbb{Z}}
({\cal{W}}_{-k+l})_{\alpha \gamma }
({\cal{W}}_{-l})_{\beta \delta }
\right\} .
\end{gather*}
This result coincides with
formal Laurent polynomials
of $H[W]$
(\ref{energyfunctionalHW})
in the sense of the expansion
\begin{gather}
H[{\cal{W}}(z)]
=
\sum_{l \in \mathbb{Z}}
\left\{
h_{\beta \alpha }
({\cal{W}}_l)_{\alpha \beta }
+
\tfrac{1}{2}
[\gamma \alpha | \delta \beta]
\sum_{k \in \mathbb{Z}}
({\cal{W}}_{l-k})_{\alpha \gamma }
({\cal{W}}_k)_{\beta \delta }
\right\}
z^l .
\label{energyfunctionalHWz}
\end{gather}
To get a $\upsilon$-independent Hamiltonian,
in
(\ref{energyfunctionalHWz})
it is enough for us to pick up only the term
with Laurent spectrum $l=0$.
Then, we may select a density-functional Hamiltonian
\begin{gather}
\langle H_{F_\infty } \rangle [{\cal{W}}]
=
h_{\beta \alpha }
({\cal{W}}_0)_{\alpha \beta }
+
\tfrac{1}{2}
[\gamma\alpha|\delta\beta]\sum_{k \in \mathbb{Z}}
({\cal{W}}_k)_{\alpha \gamma }
({\cal{W}}_{-k})_{\beta \delta }.
\label{energyfunctionalHFinfinityW}
\end{gather}
This is the extraction of the sub-Hamiltonian
$H_{F_\infty }^{\rm sub}$
out of the original Hamiltonian
(\ref{psi-form-hamiltonian})
as shown below
\begin{gather}
H_{F_\infty }^{\rm sub}
=
h_{\beta \alpha }\!\sum_{s \in \mathbb{Z}}
\psi_{Ns + \beta }\psi_{Ns + \alpha }^*\!
 +
\tfrac{1}{2}
[\gamma \alpha | \delta \beta]\!
\sum_{K \in \mathbb{Z}}\sum_{r,s \in \mathbb{Z}}\!
\psi_{N(r-K) + \gamma }\psi_{N(s+K) + \delta }
\psi_{Ns + \beta }^* \psi_{Nr + \alpha }^* .\!\!\!
\label{sub-Hamiltonian}
\end{gather}
We here adopt
(\ref{energyfunctionalHFinfinityW})
as an energy functional for the $u(N)$ HF system
on the $F_\infty$.
Through the variation
\begin{gather}
\delta \langle H_{F_\infty }
\rangle [{\cal{W}}]
=
\sum_{k \in \mathbb{Z}}
({\cal F}_{-k})_{\alpha \beta }
\delta ({\cal{W}}_k)_{\beta \alpha } ,\qquad
({\cal F}_k)_{\alpha \beta }
\stackrel{d}{=}
h_{\alpha \beta }\delta_{k,o}
+
[\alpha \beta| \gamma \delta]
({\cal{W}}_k)_{\delta \gamma },
\label{SCFHamiltonianFinfinity}
\end{gather}
we get a SCF Hamiltonian on the $F_\infty$
similar to formal Laurent expansion of
$H_{\rm HF}$
(\ref{HFHamiltonian})
on the ${\rm Gr}_M$ as
\begin{gather}
H_{F_\infty ;{\rm HF}}
=
\sum_{K \in \mathbb{Z}} \sum_{s \in \mathbb{Z}}
({\cal F}_K)_{\alpha \beta }
:\psi_{N(s-K) + \alpha }
\psi_{Ns + \beta }^*:.
\label{HFHamitonianonFinfinity}
\end{gather}

For the $\upsilon$-dependent HF equation on the $F_\infty$,
the state vector $U(\hat{g})|M \rangle$
is required to satisfy the variational principle
\begin{gather}
\delta S
=
\delta_{\hat{g}}
\int_{\upsilon_1}^{\upsilon_2}
d\upsilon L(\hat{g}) = 0 , \qquad
L(\hat{g})
=
\langle
M|U(\hat{g}^\dagger )
(i\partial_\upsilon -H_{F_\infty })U(\hat{g})|M
\rangle ,
\label{TDHFactionintegral}
\end{gather}
where we use $\hbar = 1$ here and hereafter.
First by using $U(\hat{g})  =  e^{X_\gamma }$
we get the following relations:
\begin{gather}
\delta_{\hat{g}}\int d\upsilon
\langle
M|U(\hat{g}^\dagger)i\partial_\upsilon
U(\hat{g})|M
\rangle
=
\delta_{\hat{g}}\int d\upsilon
\langle
M|i\partial_\upsilon |M
\rangle
 +
\delta_{\hat{g}}\int d\upsilon
\langle
M|i\partial_\upsilon X_\gamma \nonumber\\
\phantom{\delta_{\hat{g}}\int d\upsilon
\langle
M|U(\hat{g}^\dagger)i\partial_\upsilon
U(\hat{g})|M
\rangle
=}{} -
\tfrac{1}{2!}
[X_\gamma, i\partial_\upsilon X_\gamma ]
 +
\cdots
|M
\rangle ,
\label{variationtofstatevector}
\\
i\partial_\upsilon X_\gamma
 =
\sum_{r \in \mathbb{Z}}
\sum_{s \in \mathbb{Z}}
\left\{
( i\partial_\upsilon \gamma_r )_{\alpha \beta }
 : \psi_{N(s-r) + \alpha }\psi_{Ns\!+\!\beta }^*  :
 +
( \gamma_r )_{\alpha \beta }i\partial_\upsilon
 : \psi_{N(s-r) + \alpha }\psi_{Ns + \beta }^*  :
\right\}\nonumber\\
\phantom{i\partial_\upsilon X_\gamma
 = }{}
 +
i\partial_\upsilon (\mathbb{C}  \cdot  1),
\label{timeevolutionXgamma}
\end{gather}
where we have used
(\ref{CanoTraInfFermionandghat}).
From the def\/inition of $\uptau\{e_{\alpha \beta }(r)\}$
and the normal-ordered product,
we can calculate the $\upsilon$-dif\/ferentiation
of the second term in the curly bracket
of
(\ref{timeevolutionXgamma})
\begin{gather*}
i\partial_\upsilon \sum_{s \in \mathbb{Z}}
:\psi_{N(s-r) + \alpha }
\psi_{Ns + \beta }^*:
=
i\partial_\upsilon :\uptau\{e_{\alpha \beta }(r)\}:
=
ir\partial_\upsilon \ln z:\uptau\{e_{\alpha \beta }(r)\}:.
\end{gather*}
Assume the parameter of Laurent expansion
to be $z \!=\! e^{-i\omega_c \upsilon }$.
Then (\ref{timeevolutionXgamma})
is rewritten as
\begin{gather}
i\partial_\upsilon X_\gamma
=
\sum_{r \in \mathbb{Z}} \sum_{s \in \mathbb{Z}}
(D_{r;\upsilon } (\gamma_r)_{\alpha \beta }):
\psi_{N(s-r) + \alpha } \psi_{Ns + \beta }^*:
+i\partial_\upsilon (\mathbb{C} \cdot 1) ,\nonumber\\
D_{r;\upsilon }
\stackrel{d}{=}
i\partial_\upsilon + r \omega_c .
\label{timeevolutionofXgamma}
\end{gather}
It is seen that in the f\/irst term of the r.h.s.\
of
(\ref{variationtofstatevector})
a $\upsilon$-evolution of the reference vacuum
through~$z(\upsilon)$ has no inf\/luence
on variation with respect to $\hat{g}$.
Concerning the $\upsilon$-evolution of~$X_\gamma$~(\ref{timeevolutionofXgamma})
a $\upsilon$-dif\/ferential
$\partial_\upsilon$ acting on
$\gamma_r (\upsilon)$ and on $\psi$ and $\psi^*$
through $z(\upsilon)$ is
transformed to a covariant dif\/ferential
$D_{r;\upsilon }$ with a connection
$r\omega_c$ which acts only on the
$\gamma_r (\upsilon)$
from the gauge theoretical viewpoint.
We denote simply
the covariant dif\/ferential as $D$.
Therefore we can put
$\langle M|i\partial_\upsilon |M\rangle = 0$
and
$\mathbb{C} = 0$
since it has no inf\/luence on the energy functional
(\ref{energyfunctionalHFinfinityW}).
Then the $\upsilon$-dif\/ferential term in
equation~(\ref{TDHFactionintegral})
is calculated as
\begin{gather}
U(\hat{g}^\dagger )i\partial_\upsilon U(\hat{g})
=
 i\partial_\upsilon X_\gamma +  \tfrac{1}{2!}
[i\partial_\upsilon X_\gamma,
X_\gamma] + \tfrac{1}{3!}
[[i\partial_\upsilon X_\gamma, X_\gamma], X_\gamma ]
+
\cdots \nonumber \\
 \phantom{U(\hat{g}^\dagger )i\partial_\upsilon U(\hat{g})}{}
=
 \widehat{X}_{D\gamma }
+
\sum_{k\geq 2}
\tfrac{1}{k!}
[\cdots [i\partial_\upsilon X_\gamma, X_\gamma ],\cdots],
X_\gamma ]
+
\cdots .
\label{timedifferentialterm}
\end{gather}
Using
(\ref{SUnLieAlgebra1})
and the symbol $D$
for the covariant dif\/ferential,
each commutator is calculated as
\begin{gather}
[i\partial_\upsilon X_\gamma, X_\gamma ]
=
\widehat{X}_{[D\gamma, \gamma ]}
+
\sum_{r \in \mathbb{Z}} r \,\mbox{Tr}
\left\{
(D\gamma)_r \gamma_{-r}
\right\} , \nonumber\\
[[i\partial_\upsilon X_\gamma, X_\gamma ], X_\gamma ]
=
\widehat{X}_{[[D\gamma, \gamma ], \gamma ]}
+
\sum_{r \in \mathbb{Z}} r \,\mbox{Tr}
\left\{
([D\gamma, \gamma])_r \gamma_{-r}
\right\} , \nonumber \\
\cdots\cdots\cdots\cdots\cdots\cdots\cdots\cdots\cdots\cdots
\cdots\cdots\cdots\cdots\cdots\cdots\cdots\cdots\cdots\cdots\nonumber \\
[\cdots[i\partial_\upsilon X_\gamma, X_\gamma],\cdots], X_\gamma ]
=
\widehat{X}_{[\cdots[D\gamma, \gamma ],\cdots],~\gamma ]}
+
\sum_{r \in \mathbb{Z}} r \,\mbox{Tr}
\left\{
([ \cdots [D\gamma, \gamma],\cdots ])_r \gamma_{-r}
\right\} .
\label{commutators}
\end{gather}
Substituting
(\ref{commutators})
into
(\ref{timedifferentialterm})
and using $D_r
 \stackrel{d}{=}
D_{r;\upsilon }$
and $\hat{g} = e^\gamma$,
we get
\begin{gather}
 U(\hat{g}^\dagger) i\partial_\upsilon U(\hat{g})
=
\widehat{X}_{\hat{g}^\dagger D\hat{g}}
+
\mathbb{C} (\hat{g}^\dagger D\hat{g}) , \nonumber \\
\mathbb{C} (\hat{g}^\dagger D\hat{g})
=
\sum_{r \in \mathbb{Z}} r \, \mbox{Tr}
\left\{
\left(
\sum_{k \geq 2} \tfrac{1}{k!}
[\overline{\cdots [D\gamma , \gamma ] ,
\cdots }], \gamma ]
\right)_{r}
\gamma_{-r}
\right\} .
\label{timedependentUghat}
\end{gather}
The expectation value for the reference vacuum
is expressed as
\begin{gather*}
\langle
M|U(\hat{g}^\dagger )i\partial_\upsilon U(\hat{g})|M
\rangle
=
\sum_{s \in \mathbb{Z}} \sum_{\alpha =1}^M
\sum_{\gamma =1}^N
(g_s^\dagger)_{\alpha \gamma }
(D_{s;\upsilon }g_s)_{\gamma \alpha }
+
\mathbb{C} (\hat{g}^\dagger D\hat{g}) .
\end{gather*}
Using
$
\mathbb{C} (\hat{g}^\dagger D\hat{g})
-
\mathbb{C}(D\hat{g}^\dagger  \cdot  \hat{g}) = 0
$
which is proved later,
we obtain an explicit expression
for the $L(\hat{g})$ as
\begin{gather*}
L(\hat{g})
=
\tfrac{1}{2}
\sum_{s\in \mathbb{Z}}
\sum_{\alpha=1}^M \sum_{\gamma=1}^N
\left\{
(g_s^\dagger)_{\alpha \gamma }
(D_{s;\upsilon }g_s )_{\gamma \alpha }
-
(D_{-s;\upsilon }g_s^\dagger)_{\alpha \gamma }
(g_s)_{\gamma \alpha }
\right\}
-\langle H_{F_\infty } \rangle [W] .
\end{gather*}
Thus $L(\hat{g})$ is nothing less than
the coef\/f\/icient of $z^0$
in the Laurent expansion of $L(g(z))$
(\ref{Lagrangefunction}).

We give another $\upsilon$-dependent HF equation for $\hat{g}$: Laurent expansion of
\begin{gather*}
i\partial_\upsilon g( \upsilon )
=
{\cal F}[W   \{ g( \upsilon )\} ]g( \upsilon ) \qquad
\mbox{and}
\qquad
i\partial_\upsilon   U(g( \upsilon ))|\phi
\rangle
=
H_{\rm HF}
[W   (g( \upsilon ))]   U(g( \upsilon ))|\phi
\rangle
.
\end{gather*}
Demand the extremal condition of
(\ref{HFHamitonianonFinfinity})
leads to
$
D_\upsilon   \hat{g}
=
{\cal F}(\hat{g})   \hat{g}
$
where
${\cal F}(\hat{g})$
has an inf\/inite $N$-periodic sequence of block form
\{$ \dots , {\cal F}_{ -1}, {\cal F}_{0}, {\cal F}_{ 1},
 \dots $\}
like
(\ref{CorreMatrix})
in Appendix~\ref{appendixC}.
Def\/ining 
$
({\cal F}_r^c)_{\alpha \beta }(\hat{g},\omega_c)
 \stackrel{d}{=}
\omega_c \sum\limits_{s \in \mathbb{Z}}
s(g_s g_{s-r}^\dagger)_{\alpha \beta }
$,
the
$
D_\upsilon \hat{g}
=
{\cal F}(\hat{g}) \hat{g}
$
is transformed to
\begin{gather*}
 i\partial_\upsilon \hat{g}
=
{\cal F}^p (\hat{g})\hat{g} ,\qquad
{\cal F}^p (\hat{g})
\stackrel{d}{=}
{\cal F}(\hat{g})
-{\cal F}^c (\hat{g}) , \nonumber\\
({\cal F}_r^p)_{\alpha \beta }
\stackrel{d}{=}
({\cal F}_r
-{\cal F}_r ^c)_{\alpha \beta }
=
h_{\alpha \beta }\delta_{r,0}
+
[\alpha \beta | \gamma \delta](W_r)_{\delta \gamma }
-\omega_c \sum_{s\in \mathbb{Z}}
s(g_s g_{s-r}^\dagger)_{\alpha \beta } ,
\end{gather*}
introducing
$
\widehat{D}_\upsilon
 \stackrel{d}{=}
i\partial_\upsilon
+
H_{F_\infty ;{\rm HF}}^c
$,
this time which is cast into that
on the state vector $U(\hat{g})|M \rangle$~as
\begin{gather}
 \widehat{D}_\upsilon U(\hat{g})|M
\rangle
=
H_{F_\infty ;{\rm HF}}
U(\hat{g})|M
\rangle , \qquad
H_{F_\infty ;{\rm HF}}^c
 \stackrel{d}{=}
\sum_{r,s \in \mathbb{Z}}
({\cal F}_r^c)_{\alpha \beta }
:\psi_{N(s-r) + \alpha }
\psi_{Ns + \beta }^*: ,\nonumber\\
i\partial_\upsilon U(\hat{g})|M
\rangle
=
H_{F_\infty ;{\rm HF}}^p
U(\hat{g})|M
\rangle , \qquad
H_{F_\infty ;{\rm HF}}^p
 \stackrel{d}{=}
\sum_{r,s\in \mathbb{Z}}({\cal F}_r^p)_{\alpha \beta }:
\psi_{N(s-r)+\alpha }
\psi_{Ns+\beta }^*: ,
\label{bilineareqforTDHF}
\end{gather}
which suggest symmetry breaking and
arising of collective motion
due to recovery of symmet\-ry.
Suppose that
$\hat{g}$ to diagonalize ${\cal F}^p$ in
$H_{F_\infty ;{\rm HF}}^p$
and
$U(\hat{g})|M \rangle$ to do ${\cal F}^c$
in
$H_{F_\infty ;{\rm HF}}^c$
are deter\-mined spontaneously when
$\hat{g}  \simeq  \hat{g}^0   e^{-i\hat{\e} \upsilon }$ and
$\partial_\upsilon \hat{g}^0  =  0$.
Using the def\/inition of ${\cal F}^c$ we have
$\omega_c \Gamma(\hat{g}^0)
 =
{\cal F}
( \hat{g}^0 )\hat{g}^0 - \hat{g}^0   \hat{\e}$
where
$\Gamma( \hat{g}^0 )$
has an inf\/inite $N$-periodic sequence of block form
\{$\dots,-g^0 _{-1},0,g^0_{1},\dots$\}
like
(\ref{CorreMatrix})
and
$\hat{\e}  =  \mbox{diag}\{\dots,\e ,\dots\}$.
We also obtain
$g_r z^r \propto e^{-i(\e+ \omega_c I_N)\upsilon }$.
Thus the quasi-particle energy
$\e (\e_{\alpha \beta }
=
\epsilon_\alpha\delta_{\alpha \beta })$
and the boson energy $\omega_c$
are unif\/ied into gauge phase.
The static $\upsilon$-HFT on ${\rm Gr}_M$
has obviously no collective term
and
leads inevitably to $\omega_c \Gamma(\hat{g}^0 )  =  0$.
$\hat{g}^0$ should compose of
only a block-diagonal
$g_0^0  =  e^{\gamma_0}$,
$\gamma_0$
being a block-diago\-nal~$su(N)$ matrix.

Equation
(\ref{bilineareqforTDHF})
brings a $\upsilon$-evolution of
particle degrees of freedom
and a common language,
{\it infinite-dimensional} ${\rm Gr}_\infty$
{\it and affine KM algebra},
to discuss the relation between
SCFT and soliton theory.
The SCFT on $F_\infty$
is nothing else than the zero-th order
of the Laurent expansion on ${\rm Gr}_M$.
Through the construction of the SCFT
an explicit algebraic structure of
the SCFT on $F_\infty$ is made clear
since it is just the gauge theory
inherent in the SCFT.
The mean-f\/ield potential degrees of freedom
occur from
the gauge degrees of freedom of fermions
and the fermions make pairs among them absorbing
a change of gauges.
The sub-Hamiltonian
(\ref{sub-Hamiltonian})
exhibits such a phenomenon in~$u(N)$ algebra,
which allows us to interpret
absorption of gauge as
a coherent property of fermion pairs.
Thus the SCFTM is regarded as a method
to determine self-consistently
both quasi-particle energy
$\epsilon_\alpha (\hat{g})$
and boson energy $\omega_c$,
to the $\upsilon$-evolution of
the ``fermion gauge".
Then we can say that
{\it both the energies have been unif\/ied
into the gauge phase}.

Let $\epsilon$ and $\epsilon^*$
be parameters specifying
a continuous deformation of
$loop$ path on the ${\rm Gr}_M$
and independent on $z$.
Using the notation in
(\ref{SUnLieAlgebra1})
and calculating in a similar way to
(\ref{timedependentUghat}),
$e^{-X_\gamma }\partial_\epsilon e^{X_\gamma }$
is obtained as
\begin{gather}
 e^{-X_\gamma }\partial_\epsilon e^{X_\gamma }
=
\widehat{X}_{\hat{g}^{-1}
\partial_\epsilon \hat{g}}
+
\partial_\epsilon (\mathbb{C} \cdot 1)
+
\mathbb{C} (\hat{g}^{-1}\partial_\epsilon \hat{g}) , \nonumber\\
\mathbb{C} (\hat{g}^{-1} \partial_\epsilon \hat{g})
=
\partial_\epsilon
+
\partial_\epsilon \gamma
+
\sum_{k \geq 2}   \tfrac{1}{k!}
[\cdots[\partial_\epsilon
\gamma , \gamma ] ,\cdots] , \gamma ] .
\label{expression1forexponentialformula}
\end{gather}
To avoid the $anomaly$,
$\widehat{X}_\gamma$ reads
\[
\sum_{r \in \mathbb{Z}}
(\gamma_r)_{\alpha \beta }
\left\{
\sum_{s \in \mathbb{Z}}
\psi_{N(s-r) + \alpha }
\psi_{Ns + \beta }^*
 +
\delta_{r 0}
\sum_{s<0}
\delta_{\alpha \beta }
\right\} (\mbox{Tr}\,\gamma_r  =  0)
.
\]
Then equation
(\ref{expression1forexponentialformula})
is computed to be
\begin{gather}
e^{-X_\gamma }\partial_\epsilon e^{X_\gamma }
=
\sum_{r,s \in \mathbb{Z}}
(\hat{g}^{-1}_r \partial_\epsilon \hat{g}_r)_{\alpha \beta }
:\psi_{N(s-r) + \alpha } \psi_{Ns + \beta }^*:
+
\sum_{s<0} \mbox{Tr}
(\hat{g}^{-1}_0 \partial_\epsilon \hat{g}_0) .
\label{expression2forexponentialformula}
\end{gather}
From
(\ref{expression2forexponentialformula}),
we get
$
\mathbb{C} (\hat{g}^{-1}
\partial_\epsilon \hat{g})
=
\sum\limits_{s < 0}   \mbox{Tr}
(\hat{g}^{-1}_0   \partial_\epsilon \hat{g}_0)  =  0$,
$
\hat{g}^{-1}_0 \partial_\epsilon \hat{g}_0  \in  sl(N,C)$
 and $
\mathbb{C}(\hat{g}^\dagger
\partial_{\epsilon^*} \hat{g})
=
\mathbb{C}
(\partial_{\epsilon^*}\hat{g}^\dagger  \cdot  \hat{g})  =  0
$.
We obtain also
$
\mathbb{C}(\hat{g}^\dagger D
\hat{g})
 =
\mathbb{C}(D\hat{g}^\dagger  \cdot \hat{g})  =  0
$
in the same way as the above.
For subsequent discussion
it is convenient
to def\/ine inf\/initesimal generators of
the collective submanifold as follows
\cite{NK.89,NK.84.87}:
\begin{gather}
X_{\theta^\dagger }
\stackrel{d}{=}
i\partial_\epsilon
U(\hat{g})\cdot U(\hat{g})^\dagger
=
\widehat{X}_{\theta^\dagger }
+
\mathbb{C}(i\partial_\epsilon
\hat{g}\cdot\hat{g}^\dagger)
=
\widehat{X}_{\theta^\dagger } ,\qquad
\theta^\dagger
\stackrel{d}{=}
i\partial_\epsilon
\hat{g}\cdot\hat{g}^\dagger ,\nonumber\\
X_\theta
\stackrel{d}{=}
i\partial_{\epsilon^*}
U(\hat{g})\cdot U(\hat{g})^\dagger
=
\widehat{X}_\theta
+
\mathbb{C}(i\partial_{\epsilon^*}
\hat{g}\cdot\hat{g}^\dagger)
=
\widehat{X}_\theta ,\qquad
\theta
\stackrel{d}{=}
i\partial_{\epsilon^*}
\hat{g}\cdot\hat{g}^\dagger .
\label{infinitesimalgenerators}
\end{gather}


\subsection[SCF method in $\tau$-functional space]{SCF method in $\boldsymbol{\tau}$-functional space}\label{section3.4}

 Along the soliton theory
in the inf\/inite-dimensional fermion Fock space
\cite{JM.83,Kac.83,KR.87,KP.86},
we transcribe the $\upsilon$-dependent HFT in $F_\infty$
to the one in $\tau$-functional space.
We restrict ourselves mainly to the cases of $sl(N)$ and $su(N)$
and
the group orbit of the fundamental highest weight vector
$|M \rangle$.
Let us consider
inf\/inite-dimensional charged fermions
$\psi_i$  and $\psi_i^*$ $(i \in \mathbb{Z})$
satisfying
the canonical anti-commutation relation
$
\{ \psi_i^* ,\psi_j \}
=
\delta_{ij}$
 and $
\{ \psi_i^* ,\psi_j^* \}
=
\{ \psi_i ,\psi_j \}
=
0
$.
The perfect vacuum $|\mbox{Vac} \rangle $ and
the simple state $|M \rangle $ given by~(\ref{vacuum2})
are represented in terms of
another basis $\nu_i$ $(i \in \mathbb{Z})$
and the present fermions
$\psi_i$  and $\psi_i^*$
as
\begin{gather*}
 |\mbox{Vac} \rangle  \simeq  \nu_0 \wedge
\nu_{-1} \wedge \nu_{-2} \wedge \cdots
\quad  (\wedge :\mbox{exterior product}) ,\qquad
\langle \mbox{Vac} |\mbox{Vac} \rangle  =  1, \nonumber\\
\psi_i |\mbox{Vac} \rangle  =  0 \quad (i  \leq  0),
\qquad \psi_i^* |\mbox{Vac} \rangle  =  0 \quad (i  > 0), \nonumber\\
\langle \mbox{Vac} |\psi_i^*  =  0 \quad (i  \leq  0),
\qquad \langle \mbox{Vac} |\psi_i  =  0 \quad (i  >  0) ,\nonumber
\\
 |M \rangle \simeq \nu_M \wedge \nu_{M-1}
\wedge \cdots , \qquad
\langle M | M \rangle = 1, \nonumber\\
|M \rangle
=
\psi_M \cdots \psi_1
|\mbox{Vac} \rangle \quad (M > 0), \qquad
|M \rangle
=
\psi_{M+1}^* \cdots \psi_{0}^*
|\mbox{Vac} \rangle \quad (M < 0).
\end{gather*}
The basis $\{ \nu_i | i  \in  \mathbb{Z} \}$
is given by
the column vector with 1
as the $i$-th row and 0 elsewhere.
The number $M$ is called the  {\it charge number}.
The fermions
$\psi_i$  and $\psi_i^*$ $(i  \in  \mathbb{Z})$
generate
an 
algebra
$
gl(\infty)
=
\{ a_{ij}\psi_i \psi_j^*;~
\mbox{all but a f\/inite number of}~a_{ij}'s~\mbox{are 0} \}
$
satisfying
$
[\psi_i \psi_j^*, \psi_k \psi_l^* ]
=
\delta_{jk}\psi_i \psi_l^*
-
\delta_{il}\psi_k \psi_j^*
$.
We consider further a bigger Lie algebra than
$gl(\infty)$ so as to include
a Heisenberg subalgebra (bosons).
Following Appendix~\ref{appendixC},
it is def\/ined as the vector space
$
a_\infty
=
\{ \sum\limits_{i,j \in \mathbb{Z}} a_{ij}$
$:\psi_i \psi_j^*: +  \mathbb{C} \cdot c ,~
a_{ij}  =  0,~\mbox{for}~|i  -  j|  >  \mathbb{N} \}
$
and
$:\psi_i \psi_j^*:
 =
\psi_i \psi_j^*
 - \delta_{ij}  \cdot  c$ $(j  \leq  0)$.
Def\/ine a KM bracket among such elements
$X_a   =   \widehat{X}_a   +  \mathbb{C} \cdot c$
as
\begin{gather*}
[X_a, X_b ]_{\rm KM}
=
\widehat{X}_{[a, b]}
+
\alpha (a, b) \cdot c, \qquad
[X_a, c]_{\rm KM} = 0, \qquad
\widehat{X}_a
=
\sum_{i.j \in \mathbb{Z}} a_{ij}
:\psi_i \psi_j^* :.
\end{gather*}
For detail see Appendix~\ref{appendixC}.
A Heisenberg subalgebra $S$
\cite{KR.87}
is def\/ined as
\begin{gather*}
{\cal S}
=
\oplus_{k\neq 0}\Lambda_k + \mathbb{C}\cdot c ,\qquad
\Lambda_k
\stackrel{d}{=}
\sum_{i \in \mathbb{Z}} :\psi_i \psi_{i+k}^*:
\qquad ( k \in \mathbb{Z}).
\end{gather*}
From the def\/inition of the normal-ordered product,
the boson algebra is obtained as
$[\Lambda_k, \Lambda_l]
=
k\delta_{k+l, 0} \cdot  c$.
$\Lambda_k$ is called the shift operator and
$\Lambda_0$ belongs to the center.
Suppose
level-one $c = 1$.
Let $F^{(M)}$ be a linear span of semi-inf\/inite
monomials with charge number $M$.
For the representation of $\Lambda_k$ on $F^{(M)}$,
$\Lambda_k |M \rangle  =  0$ holds for $k  >  0$.
All the elements
$\Lambda_{-k_s} \cdots \Lambda_{-k_1}
|M \rangle$ $(0  <  k_1
 \leq  k_2  \leq  \cdots  \leq  k_s )$
are linear independent with each other.
Thus, we have obtained
an irreducible representation of the algebra ${\cal S}$
in the fermion space $F^{(M)}$.
This is isomorphic to the representation of
${\cal S}$ in the corresponding boson space $B^{(M)}$ below.

Let $\sigma_M$ denote this isomorphism
\begin{gather*}
\sigma_M : \ \  F^{(M)}
\mapsto
B^{(M)}
=
\mathbb{C}(x_1,x_2,\dots) \qquad
(\mbox{deg}(x_j)  =  j), \qquad
|M \rangle \mapsto 1, \nonumber\\
\phantom{\sigma_M :{}}{} \ \ \Lambda_k
\mapsto
\frac{\partial }{\partial x_k} ,\qquad
\Lambda_{-k}
\mapsto kx_k  \qquad (k>0) ,\qquad
\Lambda _0 \mapsto M .
\end{gather*}
A mapping operator is introduced as
$\sigma_M
 \stackrel{d}{=}
\langle M|e^{H(x)}$,
$H(x)
 \equiv
\sum\limits_{j \geq 1} x_j \Lambda_j$
({\em Hamiltonian} in $\tau$-FM)
\cite{JM.83}.
Then we see the correspondence of
$F^{(M)}  =  \oplus_{k \geq 0} F_k^{(M)}$ with
$B^{(M)}  =  \oplus_{k \in  \mathbb{Z}} B_k^{(M)}$
and
$
B^{(M)}
=
\tilde{z}^M \mathbb{C}(x_1 ,x_2, \dots ),
$
where we have def\/ined the direct sum of maps
$\sigma
 =
\oplus_{M \in \mathbb{Z}}\sigma_M$
and have
introduced a new variable $\tilde{z}$ to keep
a track of the index $M$ and then
$\sigma (\psi ^{(M)})  =  \tilde{z}^M$ for
$\psi ^{(M)}  \mapsto  F^{(M)}$.

The contravariant hermitian form
on the $B^{(M)}$ is given as
\begin{gather}
{\displaystyle
\langle 1|1\rangle = 1, \qquad
\left(\frac{\partial }{\partial x_k}\right)^\dagger = kx_k, \qquad
\langle P|Q \rangle = P^* \left(\frac{\partial }
{\partial x_1},\frac{1}{2}
\frac{\partial }{\partial x_2}
,\dots \right) Q(x)|_{x=0},
}
\label{contravarianthermitianform}
\end{gather}
where the $P^*$ means the complex conjugation of
all the coef\/f\/icients of the polynomial $P$
and
$x=(x_1,x_2, \dots )$.

We construct a representation in $B^{(M)}$
in reduction to $\widehat{sl}(N)$.
Let the generating series be
\begin{gather*}
\Psi (p)
=
\sum_{j \in \mathbb{Z}} p^j\psi_j ,\qquad
\Psi^* (p)
=
\sum_{j \in \mathbb{Z}} p^{-j} \psi_j^*  \qquad
(p \in \mathbb{C} \backslash 0).
\end{gather*}
and introduce Schur polynomials
$S_k (x)$ given in Appendix~\ref{appendixD}.
It should be emphasized that in~\cite{WigSym99}
and~\cite{KN.00},
we already have obtained explicit expressions
for the basic elements
\begin{gather}
\sigma_M  : \ \  \psi_i \psi_j^*
 \mapsto
z_{ij}(  x,\widetilde{\partial }_x  ),\qquad
\sigma_M : \ \  :\psi_i \psi_j^* :
 \mapsto
\widetilde{z}_{ij}(  x,\widetilde{\partial }_x  ) \qquad
\left(  = z_{ij} - \delta_{ij}, \ j  \leq  0 \right),\nonumber\\
  \widetilde{\partial }_x
 \stackrel{d}{=} \left(
 \frac{\partial }{\partial x_1} ,
\frac{1}{2}\frac{\partial }{\partial x_2},
\cdots
\right),\nonumber\\
  z_{ij}(x,\widetilde{\partial }_x)
=
\sum_{\mu,\nu \geq 0,~k \geq 0}
S_{i+k+\mu - M}(x)S_{-j-k+\nu + M}(-x)
S_\mu (-\widetilde{\partial }_x)
S_\nu(\widetilde{\partial }_x) ,
\label{expressionsforzandztilde}
\end{gather}
which makes a crucial role to construct
a $\upsilon$-dependent HFT on
$U(\hat{g}) |M \rangle$
as shown later.
For any element of
the $gl(\infty)$ and the $a_\infty$
in the $B^{(M)}$,
we have got
\begin{gather}
\sigma_M : \ \  X_a
=
\sum_{i,j \in \mathbb{Z}} a_{ij}
\psi_i \psi_j^*
\mapsto
\sum_{i,j \in \mathbb{Z}}
a_{ij} z_{ij} (x,\widetilde{\partial }_x) , \nonumber\\
\sigma_M : \ \  X_a
=
\sum_{i,j \in \mathbb{Z}}a_{ij}
:\psi_i \psi_j^* :+\mathbb{C} \cdot 1
\mapsto
\sum_{i,j}
a_{ij}\widetilde{z}_{ij}
(x, \widetilde{\partial }_x) +\mathbb{C} \cdot 1 .
\label{expressionsforanyelemnt}
\end{gather}

Using $\exp\{H(x)\}  =  \exp   \big\{\sum\limits_{j  \geq  1}
x_j \Lambda_1 ^j \big\} =  \sum\limits_{j  \geq\  0}
\Lambda_j S_j (x) $ and $H(x) |M \rangle  =  0$
due to $\Lambda_j |M \rangle  =  0$,
$x$-evolution of
the inf\/inite-dimensional fermion operator
is given in terms of the Schur polynomials~as
\begin{gather}
 e^{H(x)} \psi_i e^{-H(x)}
=
\sum_{j=0}^\infty \psi_j S_{i-j} (x) , \qquad
e^{H(x)} \psi_i^* e^{-H(x)}
=
\sum_{j=0}^\infty \psi_j^* S_{j-i} (-x) ,\nonumber \\
e^{H(x)} \psi (p)e^{-H(x)}
=
\psi (p)e^{\sum\limits_{j\geq 1}p^j x_j} ,\qquad
e^{H(x)} \psi^* (p)e^{-H(x)}
=
\psi^* (p)e^{\sum\limits_{j\geq 1}p^j x_j} .
\label{transformationofpsiandpsi}
\end{gather}
Following Appendix~\ref{appendixD},
under the action $U(g)$,
the group orbit of the highest weight vector
$|M \rangle$ is mapped to a space of $\tau$-function
$
\tau_M (x, g)
 =
\langle M |e^{H(x)}U(g)| M \rangle
$.
Let the Pl\"{u}cker coordinates
$v_{i_M ,i_{M-1},\dots,i_1}^{M,~M-1,\dots,1} (g)$
be
$
v_{i_M ,i_{M-1},\dots,i_1}^{M,M-1,\dots,1} (g)
=
\det |g_{i_M ,i_{M-1},\dots,i_1}^{M,~M-1,\dots,1}|
$,
$g_{ i_M ,i_{M-1},\dots,i_1}^{M,M-1,\dots,1}$:
Matrix located on intersection of rows
$i_M, i_{M-1},\dots,i_1$ and columns $M, M - 1,\dots,1$
of $g$ and $i_M = Nr + \alpha$ etc.
The Schur-polynomial expression
for the $\tau$-function is given in a compact form as
\begin{gather}
 \tau_M (x, g)
=
\sum_{i_M >i_{M-1} > \cdots >i_1 }
v_{i_M ,i_{M-1},\dots,i_1}^{M,M-1,\dots,1} (g)
S_{i_M -M,i_{M-1} -(M-1),\dots,i_1 -1}(x) .
\label{Schur-polynomialexpressionforTau-function}
\end{gather}
By using
(\ref{Plueckercoordinates})
and
(\ref{transformationofpsiandpsi}),
the derivation of the above is made as follows:
\begin{gather}
\tau_M (x, g)
=
\langle \mbox{Vac} |\psi_1^* \cdots \psi_M^*
e^{H(x)}
\sum_{i_M >i_{M-1} > \cdots >i_1 }
v_{i_M ,i_{M-1},\dots,i_1}^{M,M-1,\dots,1} (g)
\psi_{i_M} \cdots \psi_{i_1}
|\mbox{Vac} \rangle \nonumber\\
=
\sum_{i_M >i_{M-1} > \cdots >i_1 }
v_{i_M ,i_{M-1},\dots,i_1}^{M,M-1,\dots,1} (g) \nonumber\\
\times \!\!
\sum_{I_M = M}^1   \sum_{I_{M-1} = M}^1\!\!
\!\!\cdots\!\!
\sum_{I_1 = M}^1\!\!
S_{i_M{-}I_M}(x) S_{i_{M{-}1}{-}I_{M{-}1}}(x)
\cdots S_{i_1{-}I_1}(x)
\langle \mbox{Vac} |
\psi_1^* \!\cdots\! \psi_M^* \psi_{I_M}\!\cdots\! \psi_{I_1}
|\mbox{Vac} \rangle \nonumber\\
=
\sum_{i_M >i_{M-1} > \cdots >i_1 }
v_{i_M ,i_{M-1},\dots,i_1}^{M,M-1,\dots,1} (g) \label{Schur-polynomialexpressionforTau-function0}\\
\times \det \!
\left|
\begin{array}{@{}c@{\,}c@{\,}c@{\,}c@{\,}c@{}}
S_{i_M-M}(x) & S_{i_M-M+1}(x) & S_{i_M-M+2}(x) & \cdots & S_{i_M-M+(M-1)}(x) \\
S_{i_{M-1}-(M-1)-1}(x)  & S_{i_{M-1}-(M-1)}(x) &
S_{i_{M-1}-(M-1)+1}(x) & \cdots  & S_{i_{M-1}-(M-1)+(M-2)}(x) \\
\cdots & \cdots & \cdots & \cdots & \cdots \\
S_{i_1-1-(M-1)}(x) & S_{i_1-1-(M-1)+1}(x) & S_{i_1-1-(M-1)+2}(x) & \cdots & S_{i_1-1-(M-1)+(M-1)}x)
\end{array}
\right|\!  .
\nonumber
\end{gather}
This equation reads
(\ref{Schur-polynomialexpressionforTau-function}),
the generalization of which to inf\/inite-dimension
is given in~\cite{KR.87}.

To see that the af\/f\/ine Kac--Moody algebra
associated with the Lie algebra $\widehat{gl}_\infty$
is contained as a subalgebra,
we give a reduction of $\widehat{gl}_\infty$ to $\widehat{sl}_n$.
A subalgebra
$X_a$ $\big( =  \sum\limits_{i,j \in \mathbb{Z}} a_{ij}
:\psi_i \psi_j^* :
 +  \mathbb{C}  \cdot  1 \big)$ of $a_\infty$
is called $n$-reduced if and only if
the following two conditions are satisf\/ied:
\[
{\rm (i)} \ \ a_{i+N,j+N} = a_{ij} \quad (i,j \in \mathbb{Z}) \qquad
{\rm and}\qquad {\rm (ii)} \ \ \sum_{i=1}^N a_{i,i+Nj} = 0 \quad (j \in \mathbb{Z}).
\]
From (i) and
$\Lambda_{Nj}
=
\sum\limits_{i \in \mathbb{Z}}
:\psi_i \psi_{i+Nj}^*:$ $(j  \in  \mathbb{Z})$,
$[X_a ,\Lambda_{Nj} ]  =  0$ is proved.
This means
$\tau_M (x,\hat{g})$ $(\hat{g}  \in  \widehat{sl}(N) )$
is independent on $x_{Nj}$,
though $\Lambda_{Nj}$ does not satisfy (ii).
As a result,
the Hirota's equation includes no $x_{Nj}$.
Adopting the prescription for $\tau$-FM,
we transcribe the fundamental equation
(\ref{bilineareqc2})
for $\upsilon$-dependent HFT on
$U(\hat{g}) |M \rangle  \subset  F^{(M)}$
into the corresponding function
$ \subset  B^{(M)}$ in the following forms:
\begin{gather*}
(1)  \ \ U(\hat{g}) |M \rangle
\big(
U(\hat{g})
 =
e^{X_\gamma }; \;
X_\gamma
 \in
\widehat{sl}(N)
\big)
\mapsto \mbox{$N$-reduced KP $\tau$-function,}\nonumber\\
\phantom{(1)}{} \ \
\tau_M (x,\hat{g})_{N{\rm KP}}
=
\langle M |e^{ H(x)}U(\hat{g}) |M \rangle ,\qquad
\frac{\partial }{\partial {x}_{Nj}}
\tau_M (x,\hat{g})_{N{\rm KP}}
= 0 .
\end{gather*}
The $\widehat{gl}(\infty)$ group symmetry transformation
acting on
the $\tau$-function of the KP hierarchy was studied in~\cite{JM2.83,JM.83}.
The KP hierarchy was also studied extensively
together with the KdV hierarchy
by Gelfand and Dickey
\cite{Dickey}.
The generalized KP hierarchy was obtained out by making use of
the Gelfand--Dickey approach via
the algebra of pseudo-dif\/ferential operators~\cite{OrlovWinternitz.97}.
The $\tau$-function
$\tau_M (x,\hat{g})_{N{\rm KP}}$
has a $\upsilon$-dependence through $\hat{g}$ $(= e^\gamma)$
in which the anti-hermitian matrix $\gamma$ is given as
$
\gamma (z)
 =
\sum\limits_{r \in \mathbb{Z}} \gamma_r z^r$
 with
$z
=
\exp
\left(
i 2 \pi  \frac{\upsilon }{\Upsilon }
\right)
$.

(2) Quasi-particle and vacuum state
$\mapsto$
Hirota's bilinear equation
(see Appendix~\ref{appendixE}
\cite{JM.83,Hi.76}):
\begin{gather}
 \sum_{\alpha =1}^N \sum_{r \in \mathbb{Z}}
\psi_{Nr+\alpha }U(\hat{g})|M \rangle
 \otimes
\psi_{Nr+\alpha }^*
U(\hat{g})|M \rangle
 =  0 \nonumber\\
\qquad {}\mapsto
 \sum_{j \geq 0} S_j (2y)S_{j+N+1}
(-\tilde{D})
\exp
\left(
\sum_{s \geq 1}y_s D_s
\right)
\tau_M (x,\hat{g})_{N{\rm KP}}
\cdot
\tau_M (x,\hat{g})_{N{\rm KP}}
 =  0 .
\label{Hirotasbilinearequationonhat-g}
\end{gather}
$D  =  (D_1 ,D_2 ,\dots )$ denotes
the Hirota's bilinear dif\/ferential operator
and $\widetilde{D}  =  \big(D_1, \frac{1}{2}D_2 ,\dots\big)$.

(3) $\upsilon$-dependent HF equation on
$U(\hat{g}) |M \rangle \mapsto \upsilon$-dependent HF equation on
$\tau_M (x,\hat{g})_{N{\rm KP}}$:
\begin{gather}
 i\partial_\upsilon U\{ \hat{g}(\upsilon) \} |M \rangle
=
H_{F_\infty } \{ \hat{g}(\upsilon) \}
U\{ \hat{g}(\upsilon) \} |M \rangle \nonumber\\
\qquad{}
\mapsto
 i\partial_\upsilon \tau_M
\{ x,\hat{g} (\upsilon) \}_{N{\rm KP}}
=
H_{F_\infty ;{\rm HF}}
\{
x,\widetilde{\partial }_x,\hat{g} (\upsilon)
\}
\tau_M \{ x,\hat{g} (\upsilon) \}_{N{\rm KP}} ,
\label{TDHFequationonhat-g}
\end{gather}
in which it is seen that
an explicit and important role of the $\upsilon$-dependence of
$\tau$-function appears.
Using
(\ref{expressionsforzandztilde})
and
(\ref{expressionsforanyelemnt}),
$H_{F_\infty ;{\rm HF}} (x,\widetilde{\partial }_x ,
\hat{g})$
is given as
\begin{gather}
 H_{F_\infty ;{\rm HF}} (x,\widetilde{\partial }_x ,
\hat{g})
 =
\sum_{r,s \in \mathbb{Z}}
\{ {\cal F}_r (\hat{g}) \}_{\alpha \beta }
\widetilde{z}_{N(s-r)+\alpha,Ns+\beta }
(x,\widetilde{\partial }_x) , \nonumber\\
\{ {\cal F}_r (\hat{g}) \}_{\alpha \beta }
 =
h_{\alpha \beta }\delta_{r,0}
+
[\alpha \beta | \gamma \delta]
({\cal W}_r)_{\delta \gamma } ,\qquad
({\cal W}_r)_{\alpha \beta }
 =
\sum_{\gamma =1}^M
\sum_{s \in \mathbb{Z}} (g_s)_{\alpha \gamma }
(g_{s-r}^\dagger)_{\gamma \beta } .
\label{infinite-dimensionalSCF-hamiltonian}
\end{gather}


\subsection[Laurent coef\/f\/icients of soliton solutions
for $\widehat{sl}(N)$ and for $\widehat{su}(N)$]{Laurent coef\/f\/icients of soliton solutions
for $\boldsymbol{\widehat{sl}(N)}$ and for $\boldsymbol{\widehat{su}(N)}$}\label{section3.5}

We here show typical $\tau$-functions
called $n$-soliton solutions.

On $\widehat{gl}(\infty)$
\cite{JM2.83}:
We get a $\tau$-function for $\widehat{gl}(\infty)$ as
\begin{gather}
\tau_{M;n;a,p,q}(x)
=
\langle M|e^{{H(x)}}
e^{\sum\limits_{\mu \geq 1}^n
a_\mu\psi(p_\mu)\psi^*(q_\mu)}|M \rangle ,
\label{tau-glinfty}
\end{gather}
which is a famous solution of the KP hierarchy
obtained from~(\ref{Hirotasbilinearequationonhat-g})
\cite{KP.86,Kac.83,KR.87}.
As was shown in
\cite{JM2.83},
from the second line of
(\ref{transformationofpsiandpsi})
and the Wick's theorem,
we get a determinantal formula for $\tau$-function as
\begin{gather*}
\tau_{M;n;a,p,q}(x)
=
\det
\left\{
\delta_{\mu \nu }
+
a_\mu
\frac{p_\mu }{p_\mu - q_\nu }
\left(
\frac{p_\mu }{q_\nu }
\right)^{\!M}
e^{\xi (x,p_\mu)-\xi(x,q_\nu)}
\right\} .
\end{gather*}
If we use
$\Gamma(p,q)^2 \tau=0$ for
a {\em good} formal power series of $\tau$,
we have an explicit form of
$\tau_{M;n;a,p,q}(x)$
\begin{gather*}
 \tau_{M;n;a,p,q}(x)
=
e^{\sum\limits_{\mu=1}^n
\left\{
a_\mu
\frac{p_\mu }{p_\mu - q_\nu }
\left(
\frac{p_\mu }{q_\nu }
\right)^{M}
\Gamma(p_\mu ,q_\mu)
\right\}
}
\cdot 1\nonumber\\
\qquad{}=
1 + \sum_{\mu=1}^n
\left(\frac{p_\mu }{q_\mu }
\right)^M e^{ \eta_\mu }
+
\sum_{1\leq \mu<\nu \leq n}
\left(\frac{p_\mu }{q_\mu }
\right)^{M}
\left(\frac{p_\nu }{q_\nu }
\right)^{M}
\frac{(p_\mu - p_\nu)(q_\mu - q_\nu)}
{(p_\mu - q_\nu)(q_\mu - p_\nu)}
e^{\eta_\mu + \eta_\nu } + \cdots ,
\end{gather*}
where
$\xi(x,p)
=
\sum\limits_{j \geq 1}x_j p^j$ and
$\eta_\mu = \xi(x,p_\mu )-\xi(x,q_\mu)
+ \ln
\left(
a_\mu  \frac{p_\mu }{p_\mu - q_\mu }
\right)$.

On $\widehat{u}_\infty$:
We get a $\tau$-function for $\widehat{u}_\infty$ as
\begin{gather*}
\tau_{M;n;a,p,q}(x)
=
\langle M |e^{H(x)}
e^{\sum\limits_{\mu =1}^n
\left\{
a_\mu \psi(p_\mu)\psi^* (q_\mu)-a_\mu^*
\psi
\left(\frac{1}{q_\mu^*}\right)
\psi^*
\left(\frac{1}{p_\mu^*}\right)
\right\}
}|M \rangle .
\end{gather*}

On $\widehat{sl}(N)$
\cite{KN.00,Kac.83}:
Making a special choice of parameters $p_\mu$ and $q_\mu$ in
(\ref{tau-glinfty})
as
$q_\mu  =  \epsilon^{s_\mu } p_\mu$,
with
$\epsilon  =  e^{2\pi i /N}$ and
$s_\mu  =  1,  \dots, N  -  1$,
and using
$[X_a, \Lambda_{Nj}]  =  0$,
we get a $\tau$-function for~$\widehat{sl}(N)$~as
\begin{gather}
\tau_{M;n,a,p,\epsilon^s p}(x)
=
\langle M |e^{H(x)}
e^{\sum\limits_{\mu \geq 1}^n
a_\mu \psi(p_\mu)\psi^*
(\epsilon^{s_\mu } p_\mu)}|M \rangle .
\label{tau-sln}
\end{gather}

On $\widehat{su}(N)$:
Taking the exponent in
(\ref{tau-sln})
anti-Hermitian,
a $\tau$-function for $\widehat{su}(N)$
is given as
\begin{gather*}
\tau_{M;n;a,p,\epsilon^s p}(x)
=
\langle M |
e^{H(x)}
e^{\sum\limits_{\mu =1}^n
\left\{
a_\mu \psi(p_\mu)\psi^*
(\epsilon^{s_\mu } p_\mu)
-
a_\mu^* \psi
\left(
\frac{1}{(\epsilon^{s_\mu } p_\mu)^*}
\right)
\psi^*
\left(
\frac{1}{p^*}
\right)
\right\}
}
|M \rangle .
\end{gather*}
We have a one-soliton solution
on $\widehat{su}(N)$ and
on the simplest $\widehat{su}(2)$,
respectively as
\begin{gather*}
\widehat{su}(N): \
\tau_{M;1;a,p,\epsilon^s p}(x)
=
\langle M |
e^{H(x)}
e^{
\left\{
a\psi (p)\psi^*(\epsilon^s p)
-a^* \psi
\left(\frac{\epsilon^s}{p^*}\right)
\psi^*
\left(\frac{1}{p^*}\right)
\right\}
}
|M \rangle \nonumber\\
\qquad{}=
1
+\epsilon^{-Ms}
e^{
\eta
\left(x,p,\epsilon^s p
\right)
}
+\epsilon^{Ms}
e^{
\eta
\left(x,\epsilon^s/p^*,
1/p^*
\right)
}\nonumber\\
\qquad{}+
\frac{(pp^*-\epsilon^s )(pp^*
-\epsilon^{-s})}{(pp^* -1)(pp^* -1)}
e^{
\sum\limits_j
x_j
(p^j-{p^*}^{-j})
(1-\epsilon^{sj})
+
{\rm Log} \frac{ -|a|^2}
{(1-\epsilon^s)(1-\epsilon^{-s})}
} ,\nonumber\\
\eta(x,p,\epsilon^s p)
 =
\sum_{j\geq 1}x_j p^j
(1-\epsilon^{sj})+
{\rm Log}
\left\{
a/(1-\epsilon^s)
\right\} ,\nonumber\\
\eta
\left(x,\epsilon^s/p^*,1/p^*
\right)
 =
\sum_{j\geq 1} -x_j
{p^*}^{-j}(1-\epsilon^{sj})
+
{\rm Log}
\left\{
- a^*/(1-\epsilon^{-s})
\right\},\nonumber
\\
\widehat{su}(2): \ \
\tau_{M;1;a,p,-p}(x)
=
\langle M |e^{H(x)}
e^{
\left\{
a\psi(p)\psi^* (-p)-a^*
\psi
\left(-\frac{1}{p^*}\right)
\psi^*
\left(\frac{1}{p^*}\right)
\right\}
}
|M \rangle \nonumber\\
\qquad{}=
1
+
e^{
2\sum\limits_{j\in{\rm odd}}
x_j p^j
+ \mbox{Log} \frac{a}{2}
}\!\!
+
e^{
-
2\sum\limits _{j\in{\rm odd}}
x_j \frac{1}{p^{*j}}
+ \mbox{Log} \frac{ - a^*}{2}
} \!\! +\frac{(|p|^2 + 1)^2}{(|p|^2 - 1)^2}
e^{
2\sum\limits_{j\in{\rm odd}}
x_j
\frac{|p|^j - 1}{p^{*j}}
+ \mbox{Log} \frac{- |a|^2}{4}
}\!.
\end{gather*}

Finally, we give a reduction of soliton solution in $a_\infty$
to the simplest case of
$\widehat{sl}(2)$ (KdV).
A~subalgebra
$X_a$ $( = \sum\limits_{r,s \in \mathbb{Z}} a_{rs}
:\psi_ri \psi_s^* : + \mathbb{C} \cdot 1 )$ of $a_\infty$
and the Chevalley bases for $sl(2)$
\cite{Kac.83,JM2.83}
are expressed as
\begin{gather*}
X_a
 =
\sum_{r,s \in \mathbb{Z}}
(a_r)_{\alpha \beta } :\psi_{2(s-r)+\alpha }
\psi_{2s+\beta }^* :+ \mathbb{C} \cdot 1 \qquad
(\alpha , \beta=1 , 2 ;\mbox{Tr}\,a_r =0,~a_r \in \widehat{sl}(2) ), \nonumber\\
\widehat{e} =
\left[
\begin{array}{cc}
0 & 1 \\
0 & 0
\end{array}
\right],\!\quad
\widehat{f}  =
\left[
\begin{array}{cc}
0 & 0 \\
1 & 0
\end{array}
\right],\!\quad
\widehat{h}  =  \widehat{h}_+ + \widehat{h}_-
 =
\left[
\begin{array}{cc}
1 & 0 \\
0 &-1
\end{array}
\right] \!\quad \mbox{(Chevalley bases for}~ sl(2)).
\end{gather*}
Using
$\psi(p)  =  \sum\limits_{r  \in  \mathbb{Z}} \psi_r p^r$
and
$\psi(p)^*  =  \sum\limits_{r  \in  \mathbb{Z}}\psi_r^* p^{-r}$,
the algebra $X_a^n$ for
$n$-soliton in
(\ref{tau-sln})
is compu\-ted~as
\begin{gather}
X_a^N
=
\sum_{\mu =1}^n a_\mu
\psi (p_\mu ) \psi^* ( -p_\mu)
=
\sum_{r,s \in \mathbb{Z}}
\sum_{\mu =1}^n a_\mu \nonumber\\
{}\times
\left\{
\widehat{h}_+ \cdot
\psi_{2(s-r)+1 } p_\mu ^{2(s-r)+1 }\psi_{2s+1}^*(- p_\mu )^{-(2s+1)}
+
\widehat{e} \cdot
\psi_{2(s-r)+1 } p_\mu ^{2(s-r)+1 }\psi_{2s+2}^*(- p_\mu )^{-(2s+2)}
\right. \nonumber\\
\left.
{}+
\widehat{f} \cdot
\psi_{2(s-r)+2 } p_\mu ^{2(s-r)+2 }\psi_{2s+1}^*(- p_\mu )^{-(2s+1)}
+
\widehat{h}_- \cdot
\psi_{2(s-r)+2 } p_\mu ^{2(s-r)+2 }\psi_{2s+2}^*(- p_\mu )^{-(2s+2) }
\right\}\nonumber\\
{}=
\sum_{r,s \in \mathbb{Z}}
\sum_{\mu =1}^N
a_\mu
\left[
\begin{array}{cc}
-p_\mu^{-2r} ,&p_\mu^{-(2r+1)} \vspace{1mm}\\
-p_\mu^{-(2r-1)},&p_\mu^{-2r}
\end{array}
\right]_{\alpha \beta }
\psi_{2(s-r)+\alpha } \psi_{2s+\beta }^* .
\label{N-solitonsolution}
\end{gather}
Thus, we obtain an $n$-soliton solution
$a^n (z)
 =
\sum\limits_{r \in \mathbb{Z}} a _{r}^n z^r  +  \mathbb{C} \cdot 1$
$\big(\mathbb{C}
 =
\sum\limits_{\mu =1}^n
a_\mu /2\big)
$
for $\widehat{sl}(2)$ in a~mat\-rix as
\begin{gather*}
a_r^n
=
\sum_{\mu =1}^n a_r (p_\mu ,a_\mu ), \qquad
a_r (p,a) = a \cdot
\left[
\begin{array}{cc}
-p^{-2r} ,&p^{-(2r+1)} \vspace{1mm}\\
-p^{-(2r-1)} ,& p^{-2r}
\end{array}
\right] \qquad
(a_\mu = {\rm const}).
\end{gather*}
Restricting a solution to the case of $\widehat{su}(2)$,
along the similar way as the above,
from
(\ref{N-solitonsolution})
we also get
\begin{gather*}
X_\gamma^n
=
X_a^n  - X_a^{n \dagger }
=
\sum_{\mu =1}^n
\left\{
a_\mu \psi (p_\mu ) \psi^* (-p_\mu )
-
a_\mu^* \psi
\left(
-
\frac{1}{p_\mu ^*}
\right)
\psi^*
\left(
\frac{1}{p_\mu ^*}
\right)
\right\} \\
\phantom{X_\gamma^n}{}
=
\sum_{r,s \in \mathbb{Z}} (\gamma_r^n )_{\alpha \beta }
: \psi_{2(s-r)+ \alpha } \psi_{2s+\beta }:
+
\sum_{\mu =1}^n
\left(a_\mu - a_\mu^* \right)   /2,
\end{gather*}
which reads
\begin{gather*}
\gamma_r^n
 =
\sum_{\mu =1}^n \gamma_r (p_\mu , a_\mu ),\qquad
\gamma_r (p,a)
 =
\left[  \!\!
\begin{array}{cc}
-(ap^{-2r} - a^* p^{* 2r}) ,
&ap^{-(2r+1)}+ a^* p^{* (2r+1)} \vspace{1mm}\\
-(ap^{-(2r-1)}+ a^* p^{*(2r-1)})
,&ap^{-2r} - a^* p^{* 2r}
\end{array} \!\!
\right]\!.
\end{gather*}
We can generalize the above $n$-soliton solutions to the cases of
$\widehat{sl}(N)$ and $\widehat{su}(N)$.
Using the Chevalley bases for $sl(N)$ and for $su(N)$
\cite{Kac.83,JM2.83},
the Laurent coef\/f\/icients
can be derived for each case
as
\begin{gather*}
\widehat{sl}(N): \ \
a_{r}^n\big|_{{\rm soliton}}
 =
\sum_{\mu =1}^n a_{r}
\left(
p_\mu ,
\epsilon^{s_\mu } p_\mu,b_\mu
\right)
 +
\sum_{\mu =1}^n
{\displaystyle \frac{b_\mu }{1 - \epsilon }},\\
a_r
\{
 =
a_{r}
\left(
p,\epsilon^s p,b
\right)_{\alpha \beta } (\alpha, \beta   =  1, \dots , N)
\}
 =
(bp^{-Nr + \alpha - \beta }\epsilon^{-s \beta })\\
{} \simeq  bp^{-Nr}
\left [
\begin{array}{ccccc}
1 &p^{-1} & \cdots  & \cdots  &p^{-(N-1)}  \\
p & 1 &\ddots &  &\vdots  \\
\vdots &\ddots &\ddots &\ddots &\vdots  \\
\vdots & &\ddots &\ddots &p^{-1}  \\
p^{N-1} &\cdots &\cdots &p &1\\
\end{array}
\right]
\left [
\begin{array}{ccccc}
\epsilon^{-s}& & & & \\
&\epsilon^{-2s}& & & \\
& & \ddots  & & \\
& & & \ddots  & \\
& & & &\epsilon^{-Ns}
\end{array}
\right]   ,
\\
\widehat{su}(N): \ \
\gamma_{r}^n|_{{\rm soliton}}
 =
\sum_{\mu =1}^n \gamma_{r}
\left(
p_\mu ,
\epsilon^{s_\mu } p_\mu,b_\mu
\right)
 +
\sum_{\mu =1}^n
\left(
  \frac{b_\mu }{1 - \epsilon }
 -
\frac{b_\mu^*}{1 - \epsilon^*}
\right),\\
\gamma_{r}
\{
 =
\gamma_{r}
\left(p,\epsilon^s p,b\right)_{\alpha \beta }
(\alpha,\beta   =  1, \dots , N)
\}
 =
a_{r}  -  a_{-r}^{* {\rm T}}
 =
(bp^{-Nr + \alpha - \beta }\epsilon^{-s \beta }
\!-\!
b^* \epsilon^{-s \alpha }{p^*}^{Nr + \beta - \alpha })\\
\simeq  bp^{-Nr}
\left [
\begin{array}{ccccc}
1 &p^{-1} & \cdots  & \cdots  &p^{-(N-1)}  \\
p & 1 &\ddots &  &\vdots  \\
\vdots &\ddots &\ddots &\ddots &\vdots  \\
\vdots & &\ddots &\ddots &p^{-1}  \\
p^{N-1} &\cdots &\cdots &p &1
\end{array}
\right]
\left [
\begin{array}{ccccc}
\epsilon^{-s}& & & & \\
&\epsilon^{-2s}& & & \\
& & \ddots  & & \\
& & & \ddots  & \\
& & & &\epsilon^{-Ns}
\end{array}
\right]\\
-b^* {p^*}^{Nr}
\left [
\begin{array}{ccccc}
\epsilon^{-s}& & & & \\
&\epsilon^{-2s}& & & \\
& & \ddots  & & \\
& & & \ddots  & \\
& & & &\epsilon^{-Ns}
\end{array}
\right]
\left [
\begin{array}{ccccc}
1 &p^* & \cdots  & \cdots  &{p^*}^{(N-1)}  \\
{p^*}^{-1} & 1 &\ddots &  &\vdots  \\
\vdots &\ddots &\ddots &\ddots &\vdots  \\
\vdots & &\ddots &\ddots &p^*  \\
{p^*}^{-(N-1)} &\cdots &\cdots &{p^*}^{-1} &1
\end{array}
\right]   .
\end{gather*}

Suppose the external parameter $\upsilon$ to be a time $t$
and solve the TDHFEQ
by restricting a~solution space to the above spaces.
We get a soliton solution, i.e.,
a solitary wave
propagating on a surface rather than a colliding soliton~\cite{NW.78}.
This is in contrast with a two-dimensional soliton
\cite{BLMP.88}, i.e.,
{\em dromion}
\cite{FS.88},
of the Davey--Stewartson equation (DSE)~\cite{DS.74}
which provides a~two-dimensional generalization of
the celebrated nonlinear Schr\"{o}dinger equation (NLSE)
\cite{FS.88}.
The {\em dromion} for $gl(2\infty)$
was derived from the standpoint
of a Clif\/ford algebra with generators of inf\/inite free fermions,
in terms of a reduction of the two-component KP hierarchy
\cite{HH.90,JMMA.90,KacLeur.03}.
As Kac and van der Leur pointed out
\cite{KacLeur.03},
the {\em dromion} solution of the DSE was f\/irst studied
from the point of view of the spinor formalism by Heredero et al.
\cite{HMM.91}.


\subsection{Summary and discussions}\label{section3.6}

We have transcribed a bilinear equation
for $\upsilon$-HFT
into the corresponding $\tau$-function
using regular representations for groups
\cite{Fu.Int.J.Quantum.Chem.81}
and Schur polynomials.
The concept of quasi-particle and vacuum
in SCFT is connected with
bilinear dif\/ferential equations.
So far SCFM has focused mainly on
construction of various types of boson expansions
for quantum f\/luctuations of mean-f\/ield(MF)
rather than
taking the bilinear dif\/ferential equations(Pl\"{u}cker relations)
into account.
These methods turn out to be
essentially equivalent with each other.
Various subgroup-manifolds consisting of
several {\em loop-group} paths
\cite{PS.86}
exist innumerably in ${\rm Gr}_{\!M}$
relating to collective motions.
To go beyond the perturbative method
in terms of the collective variables,
we have aimed to construct $\upsilon$-HFT
on 
af\/f\/ine Kac--Moody algebras
along soliton theory,
using inf\/inite-dimensional fermions.
These fermions
have been introduced through
Laurent expansion of f\/inite-dimensional fermions
with respect to the degrees of freedom of
fermions related to MF.
Consequently
$\upsilon$-SCFT on $F_{ \infty }$
leads to dynamics on
inf\/inite-dimensional Grassmannian ${\rm Gr}_{ \infty }$.
${\rm Gr}_M$ is identif\/ied with ${\rm Gr}_{ \infty }$
af\/f\/iliated with a manifold obtained
by reduction of $gl_{ \infty }$ to $sl(N)$ and $su(N)$(reduction of KP hierarchy to DS, NLS and KdV hierarchies).
We have given explicit expressions for
Laurent coef\/f\/icients of
soliton solutions for $\widehat{sl}(N)$ and $\widehat{su}(N)$
using Chevalley bases for $sl(N)$ and $su(N)$.
In this sence
the algebraic treatment of
extracting subgroup-orbits with $z(|z|  =  1)$
from ${\rm Gr}_{ M}$ exactly forms
the dif\/ferential equation(Hirota's bilinear equation).
The $\upsilon$-SCFT on $F_{ \infty }$ results in
{\em gauge theory of fermions} and
{\em collective motion}
due to quantal f\/luctuations of
$\upsilon$-dependent SCMF potential
is attributed to {\em motion of the gauge of fermions}
in which {\em common gauge factor} causes interference
among fermions.
{\em The concept of particle and collective motions} is
regarded as the compatible condition for particle
and collective modes.
The collective variables may have close relation with
a spectral parameter in soliton theory.
These show that $\upsilon$-SCFT on $F_{ \infty }$
presents us {\em new algebraic method on $S^1$} for
microscopic understanding of fermion many-body systems.

We have studied the relation between
$\upsilon$-SCFT and soliton theory
on group manifold and shown
that both the theories describe
dynamics on each Grassmannian ${\rm Gr}$ which is
the group orbit of highest weight vector.
The former stands on
the f\/inite-dimensional fermion operators
but the latter does on
the inf\/inite-dimensional ones.
Each ${\rm Gr}$ is just identical with
the solution space for respective f\/inite and inf\/inite
set of bilinear dif\/ferential equations on
the boson space mapped from those on the fermion space.
We have investigated the dynamics
on $\upsilon$-dependent HF manifold
using regular representations for groups~\cite{Fu.81}.
A picture of quasi particle and vacuum
in $\upsilon$-SCFT is connected with
the bilinear dif\/ferential equation.
$\upsilon$-HFT on f\/inite-dimensional Fock space
is embedded into $\upsilon$-HFT on inf\/inite one.
The wave function
in an SC $\Upsilon$-periodic MF potential
becomes dependent on
the Laurent parameter $z$ on a unit circle $S^1$.
This owes to the introduction
of
af\/f\/ine Kac--Moody algebra
by inf\/inite-dimensional fermion operators,
Laurent expansion of
f\/inite-dimensional fermions with respect to~$z$.
The Pl\"{u}cker relation
on {\em coset va\-riab\-les} becomes
analogous to Hirota's bilinear form.
The $\upsilon$-SCFM has been mainly devoted to
the construction of boson-coordinate systems
rather than
that of soliton solution by $\tau$-FM.
It turns out that
both the methods are equivalent with each other
due to the Pl\"{u}cker relation
def\/ining~${\rm Gr}$.
From {\it loop} group viewpoint
and with clearer physical picture
we have proposed
description of particle and collective motions
in $\upsilon$-SCFT on $F_{ \infty }$ in relation to
iso-spectral equation in soliton theory.
Then the $\upsilon$-SCFT on $F_{ \infty }$ may be regarded
as soliton theory in the sense that
it bases on ${\rm Gr}_{ \infty }$
and may describe dynamics on
inf\/inite set of real fermion-harmonic oscillators
though the soliton theory describes dynamics on complex ones.
The soliton equation is nothing but
the bilinear equation and
the boson coordinate $x_k$ with highest degree
plays a role of an evolutional variable on
$\tau$-functional space (FS)
on which
in $\upsilon$-HFT,
the bilinear equation provides algebraic means
to extract subgroup orbits parametrized with $z$
from ${\rm Gr}_M$.
The inf\/inite set of $x_k$ becomes
coordinates on $\tau$-FS
and their $\upsilon$-evolution
yield trajectories of the SCF Hamiltonian
$H_{F_{ \infty } }^p$.

Though we have started with
a periodic potential
to introduce inf\/inite-dimensional fermions,
it is easy to see that
$\upsilon$-dependence with periodicity $\Upsilon$ is by no means
a necessary condition.
The fact that
Schr\"{o}dinger function
is dependent on an unit circle $S^1$, however,
makes a crucial role for construction
of inf\/inite-dimensional fermions.
As pointed out in
\cite{KN.00},
it turns out that
the fully parametrized $\upsilon$-dependent SCF Hamiltonian
is made up of
only the $\upsilon$-dependent Hamiltonian~$H_{F_\infty;{\rm HF}}$.
Then, we have a very important question
why inf\/inite-dimensional Lie algebras
work well in fermion systems.
As concerns this problem,
Pan and Draayer (PD)
\cite{PD.98.99}
have developed
an inf\/inite-dimensional algebraic approach
using af\/f\/ine Lie algebras~$\widehat{su}(2)$ and~$\widehat{su}(1,1)$.
They have introduced
fermion pair operators with two parameters
for the general pairing Hamiltonian
and
boson operators through
Jordan--Schwinger fermion-boson mapping
for an exactly solvable
$su(2)$ Lipkin--Meshkov--Glick (LMG) model
\cite{LMG.65}.
They have obtained analytical expressions
for exact eigenvalues and eigenfunctions of this Hamiltonian
based on the Bethe anzatz~(BA), from which
BA equation
\cite{BA.31}
or
Richardson equation
\cite{Ri.65}
is derived.

\looseness=1
It is interesting to study
a relationship between
various subgroup-manifolds
of ${\rm Gr}_\infty$
and
collective sub-manifolds of $\upsilon$-SCF Hamiltonian
by using a simple and exactly solvable LMG model.
Notwithstanding, it is possible to provide
a theoretical frame of formal RPA
\cite{NK.02,NPK.05}
as a~tool of truncating
a collective motion with only one normal mode, i.e.,
a collective submanifold
out of ${\rm Gr}_\infty$.
As mentioned in~\cite{NK.02,NPK.05},
the collective submanifold may be interpreted
as a rotator on curved surface
in ${\rm Gr}_\infty$.
It is stressed that
the $\upsilon$-HFT on $F_\infty$
describes a dynamics
on real fermion-harmonic oscillators
while soliton theory does
the same but on complex oscillators.
This remark gives us an attractive task
to extend the $\upsilon$-HFT on
real space $\widehat{su}(N)$
to the theory on complex space $\widehat{sl}(N,C)$
removing the restriction $|z|=1$.
We have discussed a close connection between
$\upsilon$-SCFM and $\tau$-FM on an abstract
fermion Fock space and denoted them independently on $S^1$.
It means that algebro-geometric structures of
{\em infinite}-dimensional fermion many-body systems
is also realisable in {\em finite}-dimensional ones.
The $\upsilon$-dependent HF equation on $\tau_M (x,\hat{g})$,
however, should lead to multi-circles,
relating closely
to a problem of construction of
multi-dimensional soliton theory
\cite{DJKM.81,KBK.97,KacLeur.03}.
It is also a very exciting problem to investigate such new motions
on the multi-circles ($\mathbf{d}$: Number of circles)
in f\/inite fermion many-body systems.
As suggested by the referee,
the motions, on the other hand,
may be related to the coupled $(\mathbf{d}  +  1)$D systems
\cite{KacLeur.03}
or the linear f\/lows
on the Birkhof\/f strata of the universal Sato Grassmannian~\cite{AdlerMoerbeke.94}.


\section[RPA equation embedded into inf\/inite-dimensional Fock space]{RPA equation embedded into inf\/inite-dimensional Fock space}\label{section4}


\subsection{Introduction}\label{section4.1}

The purpose of this section is to give a geometrical aspect of
RPA equation (RPAEQ)
\cite{NK.89,Komatsu.00}
and an explicit expression for the RPAEQ with a normal mode
on $F_\infty$.
We also argue about the relation between a {\it loop} collective path
and a formal RPAEQ (FRPAEQ).
Consequently, it can be proved that the usual perturbative method
with respect to periodic collective variables $\eta$ and $\eta ^*$
in TDHFT
\cite{Ma.80},
is involved in the present method which
aims for constructing TDHFT
on the 
af\/f\/ine KM algebra.
It turns out that the collective submanifold
is exactly a rotator on a curved surface
in the ${\rm Gr}_{ \infty }$.
If we could arrive successfully at our f\/inal goal
of clarifying relation between the SCFT
and the soliton theory on a group,
the present work may give us important clues for description of
large-amplitude collective motions in nuclei and molecules
and for construction of multi-dimensional soliton equations
\cite{KacLeur.03,DJKM.81}
since the collective motions usually
occur in multi-dimensional {\it loop} space.


\subsection[Construction of formal RPA equation on $F_\infty$]{Construction of formal RPA equation on $\boldsymbol{F_\infty}$}\label{section4.2}

We construct the FRPAEQ on $F_\infty$.
We put the following canonicity conditions
which guarantee the variables $(\epsilon ,\epsilon^*)$
to be an orthogonally canonical coordinate system
\cite{YK.87,Ma.80,KN.00}:
\begin{gather}
\langle \hat{g}|\partial_\epsilon|\hat{g}\rangle
\stackrel{d}{=}
\langle M
|U(\hat{g}^\dagger)
\partial_\epsilon U(\hat{g})|
M \rangle
=
\tfrac{1}{2} \epsilon^* ,\qquad
\langle \hat{g}|\partial_{\epsilon^*}|\hat{g}\rangle
\stackrel{d}{=}
\langle M
|U(\hat{g}^\dagger)
\partial_{\epsilon^*} U(\hat{g})|
M \rangle
=
-
 \tfrac{1}{2} \epsilon .
\label{canonicitycondition}
\end{gather}
Previously
we def\/ine the inf\/initesimal generators of the collective submanifold
$X_\theta$  and $X_{\theta^\dagger }$
(\ref{infinitesimalgenerators})
in which the term
$
\mathbb{C} (\hat{g}^{-1}\partial_\epsilon \hat{g})
$
is proved to vanish.
From these inf\/initesimal generators and
$\partial_{\epsilon^*} \langle \hat{g}|\partial_\epsilon
|\hat{g}\rangle
 -
\partial_\epsilon
\langle\hat{g}|\partial_{\epsilon^*}|\hat{g}\rangle$,
we obtain the {\it weak} orthogonality condition
\begin{gather}
1
=
\langle \hat{g}|
[X_\theta, X_{\theta^\dagger }]
|\hat{g}\rangle
=
\sum_{\alpha=1}^M \sum_{\gamma =1}^N
\sum_{r \in \mathbb{Z}}
([\theta, \theta^\dagger]_r)_{\alpha \gamma }
({\cal W}_{-r})_{\gamma \alpha }
+
\sum_{r \in \mathbb{Z}} r \,\mbox{Tr}
({\theta }_r {\theta }_{-r} ^\dagger) ,
\label{weakorthogonalitycondition}
\end{gather}
where we have used
(\ref{SUnLieAlgebra1})
and~(\ref{normalproductone-bodyWK}).

As shown in
\cite{KN.00},
using Lax's ideas
\cite{Lax.68}
we recast
(\ref{timedependentUghat})
and
$
D_t \hat{g}
 =
{\cal F}(\hat{g}) \hat{g}
$,
and
(\ref{infinitesimalgenerators})
into
\begin{gather}
 D_t \hat{g}
=
{\cal F}(\hat{g})\hat{g},
\qquad \partial_t \hat{g}^0 = 0,\qquad
{\cal F}(\hat{g}) = {\cal F}(\hat{g}^0) , \nonumber\\
i\partial_\epsilon \hat{g}
=
\theta^\dagger
(\hat{g})\hat{g},
\qquad \theta^\dagger (\hat{g})
=
\theta^\dagger (\hat{g}^0)+\hat{g}^0
(\partial_\epsilon \hat{\e})
\hat{g}^{0\dagger }\cdot t, \nonumber\\
i\partial_{\epsilon^*} \hat{g}
=
\theta (\hat{g})\hat{g},
\qquad \theta (\hat{g})
=
\theta (\hat{g}^0)+\hat{g}^0
(\partial_{\epsilon^*} \hat{\e})
\hat{g}^{0\dagger }\cdot t.
\label{evolutioneq}
\end{gather}
Upon introduction of
$E
=
\sum\limits_{\alpha =1}^M \epsilon_\alpha (\epsilon ,\epsilon^*)$,
the canonicity condition~(\ref{canonicitycondition})
transforms into
\begin{gather}
 \langle \hat{g}|\partial_\epsilon
|\hat{g}\rangle
=
\langle \hat{g}^0 |\partial_\epsilon
|\hat{g}^0\rangle
-
i\partial_\epsilon E\cdot t
=
\tfrac{1}{2}
\epsilon^* -i\partial_\epsilon E \cdot t , \nonumber\\
\langle \hat{g}|\partial_{\epsilon^*}
|\hat{g}\rangle
=
\langle \hat{g}^0 |\partial_{\epsilon^*}
|\hat{g}^0\rangle
-
i\partial_{\epsilon^*} E \cdot t
=
-
\tfrac{1}{2}
\epsilon -i\partial_{\epsilon^*} E \cdot t .
\label{canonicitycondition2}
\end{gather}
From
(\ref{canonicitycondition2}),
the {\it weak} orthogonality condition
(\ref{weakorthogonalitycondition})
is expressed as
\begin{gather*}
1
=
\partial_{\epsilon^*} \langle \hat{g}
|\partial_\epsilon |\hat{g}\rangle
-
\partial_\epsilon \langle \hat{g}
|\partial_{\epsilon^*}|\hat{g} \rangle
=
\partial_{\epsilon^*} \langle \hat{g}^0
|\partial_\epsilon |\hat{g}^0 \rangle
-
\partial_\epsilon \langle \hat{g}^0
|\partial_{\epsilon^*}|\hat{g}^0 \rangle
=
\langle
\hat{g}^0 |
[X_{\theta (\hat{g}^0)}, X_{\theta^\dagger (\hat{g}^0)}]
|\hat{g}^0
\rangle .
\end{gather*}
To satisfy integrability conditions for
$\epsilon$, $\epsilon^*$ and $t$,
curvatures obtained from~(\ref{evolutioneq})
should vanish;
\begin{gather}
 {\cal C}_{t,\epsilon }
\stackrel{d}{=}
D_t \theta^\dagger
(\hat{g}) - i\partial_\epsilon
{\cal F}(\hat{g})
+
[\theta^\dagger (\hat{g}),
{\cal F}(\hat{g})] = 0 , \nonumber\\
 {\cal C}_{t,{\epsilon^*}}
\stackrel{d}{=}
D_t \theta
(\hat{g}) - i\partial_{\epsilon^*}
{\cal F}(\hat{g})
+[\theta (\hat{g}),
{\cal F}(\hat{g})] = 0 , \nonumber\\
{\cal C}_{\epsilon ,{\epsilon^*}}
\stackrel{d}{=}
i\partial_\epsilon
\theta (\hat{g})
-
i\partial_{\epsilon^*}
\theta^\dagger (\hat{g})
+
[\theta (\hat{g}),\theta^\dagger (\hat{g})] = 0 ,
\label{zerocurvatures}
\end{gather}
and
$\partial_t \hat{g}^0 = 0$.
Here
$D_t \theta$ and $D_t \theta^\dagger$
are def\/ined as
\begin{gather*}
 (D_t \theta )_r
=
D_{r;t}
\theta_r
=
(i\partial_t + r\omega_c )\theta_r,
\qquad (D_t \theta^\dagger )_r
=
D_{r;t}
\theta_{-r}^\dagger
=
(i\partial_t + r\omega_c )\theta_{-r}^\dagger.
\end{gather*}

The expressions for the curvatures on
the quasi-particle frame (QPF)
have the same form as those of RPAEQs in the f\/inite Fock space~\cite{NK.89}.
As mentioned before, the TDHFEQ on the~$F_\infty$
leads to the RPAEQ
if we take into account only a small f\/luctuation
around a stationary ground-state solution.
The form of RPAEQ on the QPF
has a following simple geometrical interpretation:
Relative vector f\/ields made of the SCF Hamiltonian
around each point on $loop$ pathes
also take the form of RPAEQ around
the same point which is in turn a f\/ixed point in the QPF.
Thus, the curvature equation in the QPF is regarded
as the FRPAEQ on the ${\rm Gr}_\infty$.
Using~(\ref{CanoTraInfFermionandghat}),
the canonical transformation for $\hat{g}$ is given by
\begin{gather*}
 \psi_{Nr+\alpha } (\hat{g})
=
\sum_{s \in \mathbb{Z}}\sum_{\beta =1}^N
\psi_{N(r-s)+\beta }(g_s^0)_{\beta \alpha }
e^{-i\epsilon_\alpha t} ,
\end{gather*}
together with its hermitian conjugate.
Owing to
\cite{NK.89},
(\ref{evolutioneq})
is rewritten on the above QPF as
\begin{gather}
 -D_t \hat{g}^\dagger
=
{\cal F}
(\hat{g}^\dagger)|_{\rm qpf}
\hat{g}^\dagger ,\qquad
{\cal F} (\hat{g}^\dagger)
|_{\rm qpf}
 \stackrel{d}{=}
\hat{g}^\dagger
{\cal F}({\hat{g}}) \hat{g} ,\nonumber \\
 -i\partial_\epsilon \hat{g}^\dagger
 =
\theta^\dagger
(\hat{g}^\dagger )|_{\rm qpf}
\hat{g}^\dagger ,\qquad
\theta^\dagger (\hat{g}^\dagger)
|_{\rm qpf}
 \stackrel{d}{=}
\hat{g}^\dagger
\theta^\dagger (\hat{g})\hat{g} , \nonumber\\
-i\partial_{\epsilon^*} \hat{g}^\dagger
 =
\theta
(\hat{g}^\dagger )|_{\rm qpf}
\hat{g}^\dagger , \qquad
\theta (\hat{g}^\dagger)
|_{\rm qpf}
 \stackrel{d}{=}
\hat{g}^\dagger
\theta (\hat{g}) \hat{g} ,
\label{evolutioneqonqpf}
\end{gather}
The subscript ``qpf'' means the quasi-particle frame (QPF).
For~({\ref{zerocurvatures}})
we obtain also another expression on this QPF as
\begin{gather}
 (D_t \theta^\dagger -i\partial_\epsilon
{\cal F}-[\theta^\dagger ,{\cal F}])
|_{\rm qpf} = 0 , \qquad
(D_t \theta -i\partial_{\epsilon^*}
{\cal F}-[\theta, {\cal F}])
|_{\rm qpf} = 0 , \nonumber\\
(i\partial_\epsilon
\theta -i\partial_{\epsilon^*}
\theta^\dagger -[\theta,\theta^\dagger])
|_{\rm qpf} = 0 .
\label{curvatureeq2}
\end{gather}
Further, using (\ref{evolutioneqonqpf}) and the relation
$
i\partial_\epsilon {\cal F}
|_{\rm qpf}
=
i\partial _\epsilon
(\hat{g}^\dagger {\cal F}(\hat{g})\hat{g})
=
-
[\theta^\dagger, {\cal F}]|_{\rm qpf}
+
\hat{g}^\dagger i\partial_\epsilon
{\cal F}\hat{g}
$,
one can rewrite equations
in the f\/irst line of
(\ref{curvatureeq2})
as
\begin{gather}
D_t \theta^\dagger |_{\rm qpf}
- \hat{g}^\dagger i\partial_\epsilon
{\cal F}(\hat{g})\hat{g}
= 0 ,
\qquad D_t \theta |_{\rm qpf}
- \hat{g}^\dagger i\partial_{\epsilon^*}
{\cal F}(\hat{g})\hat{g}
= 0 .
\label{curvatureeq3}
\end{gather}
From
(\ref{evolutioneqonqpf})
and
(\ref{evolutioneq}),
the inf\/initesimal operators are expressed as
\begin{gather}
\theta^\dagger
(\hat{g}^\dagger)|_{\rm qpf}
=
-i\partial_\epsilon
\hat{g}^\dagger \cdot \hat{g}
=
e^{i\hat{\se}t}
\big\{
\partial_\epsilon \hat{\e}\cdot t
+
\theta^\dagger (\hat{g}^0{}^\dagger )
|_{\rm qpf}
\big\}
e^{-i\hat{\se}t} ,
\label{generatoronqpf}
\end{gather}
together with the same relation for
$\theta(\hat{g}^\dagger)|_{\rm qpf}$.
We have also
$
\theta^\dagger (\hat{g}^0{}^\dagger )|_{\rm qpf}
 =
-i\partial_\epsilon \hat{g}^0{}^\dagger  \cdot  \hat{g}^0
$
and
$
\theta (\hat{g}^0{}^\dagger )|_{\rm qpf}
 =
-i\partial_{\epsilon^*} \hat{g}^0{}^\dagger  \cdot  \hat{g}^0
$.
Then,
from
(\ref{curvatureeq3})
we can derive the FRPAEQ on the ${\rm Gr}_\infty$ in the form
\begin{gather}
\omega_c \Gamma
\left\{
\theta^\dagger
(\hat{g}^0{}^\dagger ) |_{\rm qpf}
\right\}
+
i\partial_\epsilon \hat{\e}
-
\left[
\hat{\e}, \theta^\dagger
(\hat{g}^0{}^\dagger) |_{\rm qpf}
\right]
-
i\hat{g}^0{}^\dagger \partial_\epsilon
{\cal F} (\hat{g}^0) \hat{g}^0
= 0 ,
\label{formalRPAequationontheinfinite-dimensionalGrassmannian}
\end{gather}
To obtain an explicit expression for the last term
of the l.h.s.\ of~(\ref{formalRPAequationontheinfinite-dimensionalGrassmannian}),
we introduce an auxi\-lia\-ry density matrix
$
\widehat{R}
 =
\hat{g}^0
\mbox{diag}
\left[
\cdots I_{M \otimes (N-M)} \cdots
\right]
\hat{g}^0{}^\dagger
$,
where
$
I_{M \otimes (N-M)}
\stackrel{d}{=}
\left[
\begin{array}{cc}
-I_M & \\ & I_{N-M}
\end{array}
\right]
$.
The~$\widehat{R}$ is related to
density matrix $\widehat{W}$ as
$\widehat{R} = \widehat{I}-2\widehat{W}$
($\widehat{I}$: inf\/inite-dimensional unit matrix).
Then, we obtain
\begin{gather}
i\partial_\epsilon \widehat{W}
=
- \tfrac{1}{2}  \hat{g}^0
\big \{
- i\partial_\epsilon \hat{g}^0{}^\dagger
\cdot \hat{g}^0 \widehat{I}_{M \otimes (N-M)}
-\widehat{I}_{M \otimes (N-M)}
(-i\partial_\epsilon \hat{g}^0{}^\dagger
\cdot \hat{g}^0 )
\big\}
\hat{g}^0{}^\dagger\nonumber\\
\phantom{i\partial_\epsilon \widehat{W}}{} =
-
\tfrac{1}{2}  \hat{g}^0
\big[
\theta^\dagger
(\hat{g}^0{}^\dagger) |_{\rm qpf} ,
\widehat{I}_{M \otimes (N-M)}
\big]
\hat{g}^0{}^\dagger ,
\label{densitygrad}
\end{gather}
and we have used
(\ref{generatoronqpf}).
Further we introduce the following quantities:
\begin{gather*}
\theta_r^{0\dagger }
|_{\rm qpf}
\stackrel{d}{=}
\left[
\begin{array}{cc}
\xi_r ^0 & \phi_r ^0\vspace{1mm}\\
\psi_r ^0 & \bar{\xi }_r ^0
\end{array}
\right]
,\qquad
B_r^\dagger |_{\rm qpf}
\stackrel{d}{=}
-
\tfrac{1}{2}
\big[
\theta_r^{0\dagger } |_{\rm qpf},~
I_{M \otimes (N-M)}
\big]
=
\left[
\begin{array}{cc}
0 &-\phi_r ^0 \\
\psi_r ^0 & 0
\end{array}
\right] ,
\end{gather*}
Using these, we rewrite
(\ref{densitygrad}) as
\begin{gather}
 i\partial_\epsilon \widehat{W} = \hat{g}^0
\widehat{B}^\dagger |_{\rm qpf}
\hat{g}^0{}^\dagger
=\sum_{r \in \mathbb{Z}}(i\partial_\epsilon W_r)z^r, \nonumber\\
i\partial_\epsilon W_r
=
\sum_{k,l \in \mathbb{Z}}
g_k^0 B_{k-l-r}^\dagger|_{\rm qpf}
g_l^0{}^\dagger
=
\sum_{k,l \in \mathbb{Z}} g_k^0
\left[
\begin{array}{cc}
0&-\phi_{k-l-r}^0\\
\psi_{k-l-r}^0&0
\end{array}
\right]
g_l^0{}^\dagger .
\label{differentialofW}
\end{gather}

Let $a$ $(\bar{a})$ and $i$ $(\bar{i})$ be
$1, \ldots , m$ hole-states and $m+1, \ldots , N$ particle-states
of the QPF, respectively.
Substituting the second equation of
(\ref{differentialofW})
into (\ref{SCFHamiltonianFinfinity}),
for $r \ne 0$ we get
\begin{gather*}
i\partial_\epsilon ({\cal F}_r)_{\alpha \beta }
 =
[\alpha \beta | \gamma \delta]
\sum_{k,l \in \mathbb{Z}}
\big\{
\left(g_k^0\right)_{\delta i}
\big(g_l^0{}^\dagger\big)_{a \gamma }
\left(\psi_{k-l-r}^0\right)_{ia}
 -
\left(g_k^0\right)_{\delta a}
\big(g_l^0{}^\dagger\big)_{i \gamma }
\left(\phi_{k-l-r}^0\right)_{ai}
\big\} .
\end{gather*}
Thus, we can reach the desired form of the equation,
part of the FRPAEQ on the ${\rm Gr}_\infty$
(\ref{formalRPAequationontheinfinite-dimensionalGrassmannian}),
\begin{gather*}
 i
\left(
\hat{g}^0{}^\dagger \cdot \partial_\epsilon
{\cal F}\cdot \hat{g}^0
\right)_r
=
\sum_{k,l \in \mathbb{Z}}
g_k^0{}^\dagger \cdot i\partial_\epsilon
{\cal F}_{k-l+r} \cdot g_l^0\\
=
\sum_{k,l \in \mathbb{Z},~\bar{k},\bar{l} \in \mathbb{Z}}
\left[
\begin{array}{l}
\left[
\begin{array}{c} kl\\ab
\end{array}
\right. |{\F}|
\left.
\begin{array}{c}
\bar{k}\bar{l}\\
\bar{i}\bar{a}
\end{array}
\right]
\left(\psi^0_{(\bar{k}-\bar{l})-(k-l)-r}\right)_{\bar{i}\bar{a}}
-
\left[
\begin{array}{c} kl\\ab
\end{array}
\right. |\overline{{\F}}|
\left.
\begin{array}{c}
\bar{k}\bar{l}\\ \bar{a}\bar{i}
\end{array}
\right]
\left(\phi^0_{(\bar{k}-\bar{l})-(k-l)-r}\right)_{\bar{a}\bar{i}} , \\
\left[
\begin{array}{c} kl\\ia
\end{array}
\right. |{\D}|
\left.
\begin{array}{c} \bar{k}\bar{l}\\
\bar{i}\bar{a}
\end{array}
\right]
\left(\psi^0_{(\bar{k}-\bar{l})-(k-l)-r}\right)_{\bar{i}\bar{a}}
-
\left[
\begin{array}{c} kl\\ia
\end{array}
\right. |\overline{{\D}}|
\left.
\begin{array}{c}
\bar{k}\bar{l}\\ \bar{a}\bar{i}
\end{array}
\right]
\left(\phi^0_{(\bar{k}-\bar{l})-(k-l)-r}\right)_{\bar{a}\bar{i}},
\end{array}
\right.
\\
\qquad \qquad \qquad \left.
\begin{array}{l}
\left[
\begin{array}{c} kl\\ai
\end{array}
\right. |{\D}|
\left.
\begin{array}{c} \bar{k}\bar{l}\\
\bar{i}\bar{a}
\end{array}
\right]
\left(\psi^0_{(\bar{k}-\bar{l})-(k-l)-r}\right)_{\bar{i}\bar{a}}
-
\left[
\begin{array}{c} kl\\ai
\end{array}
\right. |\overline{{\D}}|
\left.
\begin{array}{c}
\bar{k}\bar{l}\\ \bar{a}\bar{i}
\end{array}
\right]
\left(\phi^0_{(\bar{k}-\bar{l})-(k-l)-r}\right)_{\bar{a}\bar{i}} \vspace{1mm}\\
\left[
\begin{array}{c} kl\\ij
\end{array}
\right. |{\F}|
\left.
\begin{array}{c} \bar{k}\bar{l}\\
\bar{i}\bar{a}
\end{array}
\right]
\left(\psi^0_{(\bar{k}-\bar{l})-(k-l)-r}\right)_{\bar{i}\bar{a}}
-
\left[
\begin{array}{c} kl\\ ij
\end{array}
\right. |\overline{{\F}}|
\left.
\begin{array}{c}
\bar{k}\bar{l}\\ \bar{a}\bar{i}
\end{array}
\right]
\left(\phi^0_{(\bar{k}-\bar{l})-(k-l)-r}\right)_{\bar{a}\bar{i}}
\end{array}
\right] .
\end{gather*}
Substituting the above result into
(\ref{formalRPAequationontheinfinite-dimensionalGrassmannian}),
we can derive the FRPAEQ on $F_\infty$.

Finally we show the following equations to determine
the collective submanifold and motion:

The canonicity condition
(\ref{canonicitycondition}):
\begin{gather*}
\langle \hat{g}^0 |\partial_{
\mbox{\scriptsize
$\left[ \!\!\!
\begin{array}{c}
\epsilon\\
\epsilon^* \!\!\!
\end{array}
\right] $}}
|\hat{g}^0 \rangle
=
\sum_{\alpha =1}^M \sum_{s \in \mathbb{Z}}
\left(
g_s^0{}^\dagger \partial_{
\mbox{\scriptsize
$\left[ \!\!\!
\begin{array}{c}
\epsilon\\
\epsilon^* \!\!\!
\end{array}
\right] $}}
g_s^0
\right)_{\alpha \alpha }
=
\tfrac{1}{2}
\left[ \!\!\!
\begin{array}{c}
\epsilon^*\\
-\epsilon \!\!\!
\end{array}
\right].
\end{gather*}

The FRPAEQ
(\ref{formalRPAequationontheinfinite-dimensionalGrassmannian}):
\begin{gather*}
\omega_c \Gamma
\big\{
\theta^\dagger
(\hat{g}^0{}^\dagger )
|_{\rm qpf}
\big\}
 +
i\partial_\epsilon \hat{\e}
 -
\big[
\hat{\e}, \theta^\dagger (\hat{g}^0{}^\dagger )
|_{\rm qpf}
\big]
 -
i\hat{g}^0{}^\dagger
\partial_\epsilon{\cal F}(\hat{g}^0)\hat{g}^0  =  0 ,\qquad
\hat{g}
 =
\hat{g}^0 (\epsilon , \epsilon^*)
e^{-i\hat{\se}(\epsilon , \epsilon^*)t} .
\end{gather*}
Through constructions of the TDHFT
and the FRPAEQ on $F_{ \infty }$,
the following become apparent:
The ordinary perturbative method for
collective variables
$\eta$ and $\eta ^*$
\cite{Ma.80}
is involved in
the way of construction of
the TDHFT on the af\/f\/ine KM algebra
if we restrict ourselves to $\widehat{su}(N)$.
When the $\eta$ and $\eta ^*$ are represented as
$\eta  =  \sqrt{\Omega }e^{i\varphi }$,
we can always express
\[
\gamma (\eta, \eta ^* )
 =
\sum\limits_{r,s \in \mathbb{Z}}
\bar{\gamma }_{r,s}\eta ^{*r}\eta ^s
 =
\sum_r \gamma_r z^r
\]
on the Lie algebra if we put
$z  =  e^{i\varphi }$.
This means that
the inf\/inite-dimensional Lie algebra in the SCFT
is introduced in a natural way and is useful
to study various motions of fermion many-body systems.


\subsection{Summary and discussions}\label{section4.3}

FRPAEQ has been provided as a tool for truncating
a collective submanifold with only one normal mode out of
an ${\rm Gr}_{\!\infty }$.
We have given a simple geometrical interpretation
for FRPAEQ.
The collective submanifold is interpreted as
a rotator on a curved surface
in the ${\rm Gr}_{ \infty }$.
In $F_{ \infty }$,
to study motions of f\/inite fermion systems,
it is manifestly natural and useful to introduce
an inf\/inite-dimensional Lie algebra arising from
anti-commutation relations among fermions.
In order to discuss the relation
between TDHFT and soliton theory,
we have given expressions for TDHFT on $\tau$-FS
along soliton theory.
From the $loop$ group viewpoint and with a clearer physical picture,
we have proposed a way of describing particle and collective motions
in SCFT on $F_{ \infty }$ in relation to
an iso-spectral equation in soliton theory.
Then, SCFT on $F_{\ \infty }$ may be regarded as soliton theory
in the sense that it is based on the ${\rm Gr}_{ \infty }$
and may describe dynamics on an inf\/inite set
of {\it real fermion-harmonic oscillators}.
On the other hand,
soliton theory describes dynamics on
{\it complex fermion-harmonic oscillators}.
It is one of the most challenging problem
to extend real space $\widehat{su}(N)$
to complex space $\widehat{sl}(N)$
in TDHFT on $F_{ \infty }$
together with removal of the restriction $|z|  =  1$.
Concerning the construction of
soliton theory on multi-dimensional space~\cite{KacLeur.03,DJKM.81},
we have an interesting future problem that is
to extend the Pl\"{u}cker relation (Hirota's form)
with only one circle to the case of multi-circles such that
SCFM on $F_{ \infty }$ can describe
dynamics of fermion systems in terms of multi-RPA bosons.




\section[Inf\/inite-dimensional KM algebraic approach to LMG model]{Inf\/inite-dimensional KM algebraic approach to LMG model}\label{section5}


\subsection{Introduction}\label{section5.1}

To go beyond the maximaly-decoupled method,
we have aimed to construct an SCF theory, i.e., $\upsilon$-HFT.
In constructing the $\upsilon$-HF theory we must observe, however,
the following two dif\/ferent points
between the maximaly-decoupled method and
the $\upsilon$-HF SCFM
(i) The former is built on the f\/inite-dimensional Lie algebra
but the latter on the inf\/inite-dimensional one.
(ii) The former has an SCF Hamiltonian
consisting of a fermion one-body operator,
which is derived from
a functional derivative of an expectaion value of
a fermion Hamiltonian by a ground-state wave function.
The latter has a fermion Hamiltonian with a
one-body type operator brought artif\/icially
as an operator which maps states
on a fermion Fock space into corresponding ones on a~$\tau$-FS.
Toward such an ultimate goal,
the $\upsilon$-HFT has been reconstructed on
an 
af\/f\/ine KM algebra along the soliton theory,
using inf\/inite-dimensional fermion.
An inf\/inite-dimensional fermion operator is introduced through
a Laurent expansion of f\/inite-dimensional fermion operators
with respect to degrees of freedom of the fermions
related to a $\upsilon$-dependent potential with a~$\Upsilon$-periodicity.
A bilinear equation for the $\upsilon$-HFT
has been transcribed
onto the corresponding $\tau$-function
using the regular representation for the group and the Schur-polynomials.
The $\upsilon$-HF SCFM on an inf\/inite-dimensional Fock space $F_\infty$
leads to a dynamics on an inf\/inite-dimensional Grassmannian ${\rm Gr}_\infty$
and may describe more precisely such a dynamics on the group manifold.
A f\/inite-dimensional Grassmannian
is identif\/ied with a ${\rm Gr}_\infty$
which is af\/f\/iliated with the group manifold obtained
by reducting $gl(\infty)$ to $sl(N)$ and $su(N)$.
We have given explicit expressions for
Laurent coef\/f\/icients of
soliton solutions for $\widehat{sl}(N)$ and $\widehat{su}(N)$
on the ${\rm Gr}_\infty$
using Chevalley bases for $sl(N)$ and $su(N)$.
As an illustration
we make the $\upsilon$-HFT approach to
an inf\/inite-dimensional matrix model
extended from the f\/inite-dimensional $su(2)$ LMG model
and represent an inf\/inite-dimensional matrix LMG model
in terms of the Schur polynomials.


\subsection[Application to Lipkin--Meshkov--Glick model]{Application to Lipkin--Meshkov--Glick model}\label{section5.2}

To show the usefulness of the inf\/inite KM algebra and
to avoid an unnecessary complication,
we apply it to a simple model, the LMG model
consisting of $N$ $( =  M)$ particles.
Let us introduce the LMG Hamiltonian
which has two $N$-fold degenerate levels with  energies
$\frac{1}{2}\varepsilon $
and~$-\frac{1}{2}\varepsilon $, respectively
\begin{gather*}
H = \varepsilon {\widehat{K}}_{0}
-
\tfrac{1}{2}
V({\widehat{K}}_{+}^{2} + {\widehat{K}}_{-}^{2}) .
\end{gather*}
The operators
${\widehat{K}}_{0}$, ${\widehat{K}}_{+}$ and ${\widehat{K}}_{-}$
are def\/ined by
\begin{gather}
{\widehat{K}}_{0}
\equiv
\tfrac{1}{2}
\left(
\sum_{i=1}^{N}{c}_{i}^{\dagger }{c}_{i}
-
\sum_{a=1}^{N}{c}_{a}^{\dag }{c}_{a}
\right) ,\qquad
{\widehat{K}}_{+}
\equiv
\sum_{i=a=1}^{N} {c}_{i}^{\dagger }{c}_{a}
= {\widehat{K}}_{-}^{\dagger },
\label{Koperators}
\end{gather}
where the indices $i$ and $a$ stand for
particle-state and hole-state, respectively
and satisfy the $SU(2)$ quasi-spin algebra
\begin{gather*}
[{\widehat{K}}_{0} , {\widehat{K}}_{\pm }]
= \pm {\widehat{K}}_{\pm },\qquad
[{\widehat{K}}_{+} , {\widehat{K}}_{-}]
= 2{\widehat{K}}_{0}.
\end{gather*}
Then the S-det, $|{S}^{N}\rangle$,
in which the $N$ particles f\/ill the lower level, satisf\/ies
\begin{gather*}
c_i|S^N\rangle = 0 , \qquad
c_a^{\dag }|S^N\rangle = 0  \qquad
(i,a = 1,2,\dots,N) ,\qquad
\hat{K}_- |S^N\rangle = 0 .
\end{gather*}
We here use a notation
$[c^\dag]$
denoting a $2N$-dimensional row vector
$[c_a^\dag,c_i ^\dag]$ $(i  =  1,2,\dots,N$; $a  =  1,2,\dots,N)$.
We introduce the following $SU(2N)$ Thouless transformation:
\begin{gather}
U(g) [c^\dag] {U}^{-1}(g)
= [c^\dag] g ,\qquad
U(g) = {e}^{i\psi {\widehat{K}}_{0}}
{e}^{\frac{\theta }{2} ({\widehat{K}}_{+} - {\widehat{K}}_{-})}
{e}^{i\varphi {\widehat{K}}_{0}},\nonumber
\\
g
 =
\left[
\begin{array}{cc}
\cos \frac{\theta }{2}
\cdot {e}_{\ }^{-i\frac{1}{2}(\psi  +\varphi )}
\cdot {1}_{N}  &- \sin \frac{\theta }{2}
\cdot {e}_{\ }^{-i\frac{1}{2}(\psi  -\varphi )}
\cdot {1}_{N}\vspace{1mm}\\
\sin \frac{\theta }{2} \cdot{e}_{\ }^{i\frac{1}{2}(\psi  -\varphi )}
\cdot {1}_{N}\  & \cos \frac{\theta }{2}
\cdot {e}_{\ }^{i\frac{1}{2}(\psi  +\varphi )}
\cdot {1}_{N}
\end{array}
\right] ,\nonumber\\
{g}^{\dagger } g
 =  g{g}^{\dagger }
 =  {1}_{2N},
\qquad \det g  =  1 ,
\label{Thoulesstransformation2}
\\
U(g)U(g' ) = U(gg' ) ,
\qquad U({g}^{-1}) = {U}^{-1}(g) = {U}^{\dagger }(g) ,
\qquad U(1) = 1.\nonumber
\end{gather}
The above $SU(2N)$ matrix is essentially
the direct sum of the $SU(2)$ matrix.
Any $N$-particle S-det is constructed by
the Thouless transformation of a reference S-det, $|{S}^{N}\rangle$
(the Thouless theorem) as
\begin{gather}
|g \rangle = U(g)|{S}^{N}\rangle
=
\langle{S}^{N}|U(g)|{S}^{N}\rangle
\exp [p
\exp (i\psi) \widehat{K}_{+}] |{S}^{N}\rangle ,\nonumber\\
 \langle{S}^{N}|U(g)|{S}^{N}\rangle
=
\left( \cos   \tfrac{\theta }{2} \right)^{\!N} ,\qquad
p
=
(p_{ia})
=
\tan
\left( \tfrac{\theta }{2} \right)
e^{i\psi } \cdot 1_N ,
\label{Thoulessthorem0}
\end{gather}
which is the CS rep of fermion state vector on the $SU(2N)$ group~\cite{Pere.72}.

The HF density matrix is given as
\begin{gather*}
W  =
\left[ \begin{array}{cc}
 \cos^2 \tfrac{\theta }{2}
\cdot {1}_{N}
& \tfrac{1}{2}  \sin \theta e^{-i \psi }
\cdot {1}_{N}\vspace{1mm}\\
\tfrac{1}{2}  \sin \theta e^{i \psi }
\cdot {1}_{N}
&  \sin^2 \tfrac{\theta }{2}
\cdot {1}_{N}
\end{array}
\right]\nonumber\\
\phantom{W}{}
 =
\tfrac{1}{2}
\hat{I}_{2N}
 +
\cos \theta \tfrac{1}{2}
\hat{h}_{2N}
 +
\tfrac{1}{2}  \sin \theta e^{-i \psi }
\hat{e}_{2N}
 +
\tfrac{1}{2}  \sin \theta e^{i \psi }
\hat{f}_{2N} ,
\end{gather*}
where
$\hat{I}_{2N}$ is a $2N$-dimensional unit matrix and
$\hat{h}_{2N}$, $\hat{e}_{2N}$ and $\hat{f}_{2N}$
are def\/ined in the next subsection.
The usual HF energy
$\langle g|H|g \rangle$ $(= H[W])$
is obtained as
\begin{gather*}
H[W]
=
 \tfrac{\varepsilon N }{2}
\left[
 {\sin }^2 \tfrac{\theta }{2}
-{\cos }^2  \tfrac{\theta }{2}
-
\chi
\big\{
\tfrac{1}{2} \sin \theta e^{-i \psi }
\big\}^2
-
\chi
\big\{
\tfrac{1}{2} \sin \theta e^{i \psi }
\big\}^2
\right ] ,\qquad
\chi
\equiv
\tfrac{(N-1)V}{\varepsilon }
.
\end{gather*}
The Fock operator
$F[W]$
$(=\delta H[W]/\delta W^{{\rm T}})$,
in the HF approximation is represented as
\begin{gather}
F[W]
=
\left[
\begin{array}{cc}
-  \varepsilon   \tfrac{1}{2}
 \cdot  {1}_{N}
&
- \varepsilon \chi \tfrac{1}{2} \sin \theta e^{i \psi }
 \cdot  {1}_{N}\vspace{1mm}\\
- \varepsilon \chi \tfrac{1}{2}  \sin \theta e^{-i \psi }
 \cdot  {1}_{N}
&   \varepsilon \tfrac{1}{2}
 \cdot  {1}_{N}
\end{array}
\right] \nonumber\\
\phantom{F[W]}{}
=
-
\tfrac{\varepsilon }{2}
\hat{h}_{2N}
 -    \tfrac{\varepsilon \chi }{2}  \sin \theta e^{i \psi }
\hat{e}_{\!2N}
 -    \tfrac{\varepsilon \chi }{2} \sin \theta e^{-i \psi }
\hat{f}_{ 2N} .
\label{HFockoperator0}
\end{gather}


\subsection[Inf\/inite-dimensional Lipkin--Meshkov--Glick model]{Inf\/inite-dimensional Lipkin--Meshkov--Glick model}\label{section5.3}

Using
(\ref{vacuum2}),
we introduce the following
inf\/inite-dimensional ``particle''- and ``hole''-annihilation operators
$\psi_{Nr + i}^*$ for particle-state
and
$\psi_{Nr + a}$ for hole-state,
respectively,
and
a vacuum sta\-te~$|M \rangle$:
\begin{gather*}
\psi_{Nr + i}^*|M \rangle
= 0 \qquad (i = 1,\dots,N  ,\  r\geq 0) ,\qquad
\psi_{Nr + a}|M \rangle
= 0 \qquad (a = 1,\dots,N ,\  r\leq 0) ,\\
|M \rangle
=
\psi_M \cdots \psi_1 |\mbox{Vac} \rangle \qquad (M = N).
\end{gather*}
Then the
${\widehat{K}}_{0}$ and ${\widehat{K}}_{\pm }$
(\ref{Koperators})
are expressed in terms of the above inf\/inite-dimensional operators
as follows:
\begin{gather*}
\widehat{K}_0
=
\sum_{r,s \in \mathbb{Z}}
\left(\tfrac{1}{2} \hat h_{2N}\right)_{\a \b }
:\psi_{N(s-r) + \a}\psi_{Ns + \b}^* : \\
\phantom{\widehat{K}_0}{}
=
\tfrac{1}{2}
\sum_{r,s \in \mathbb{Z}}
\left(
\sum_{i=1}^N
:\psi_{N(s-r) + i}\psi_{Ns + i}^* :
-
\sum_{a=1}^N
:\psi_{N(s-r) + a}\psi_{Ns + a}^* :
\right) , \\
\widehat{K}_+
=
\sum_{r,s \in \mathbb{Z}}
(\hat e_{2N})_{\a \b } :\psi_{N(s-r) + \a}\psi_{Ns + \b }^* :
=
\sum_{r,s \in \mathbb{Z}}
\sum_{i=a=1}^N : \psi_{N(s-r) + i}\psi_{Ns + a}^* :, \\
\widehat{K}_-
=
\sum_{r,s \in \mathbb{Z}}
(\hat f_{2N})_{\a \b } :\psi_{N(s-r) + \a}\psi_{Ns + \b }^* :
=
\sum_{r,s \in \mathbb{Z}}
\sum_{a=i=1}^N : \psi_{N(s-r) + a}\psi_{Ns + i}^* : ,
\end{gather*}
and
\[
\widehat{K}_-|M \rangle
=
\sum_{r,s \in \mathbb{Z}}
\sum_{a=i=1}^N : \psi_{N(s-r) + a}\psi_{Ns + i}^* :|M \rangle
= 0  \qquad (s \geq 0)
.
\]

Let us introduce the following $2N$-dimensional dual elements of
the direct sum of the algebra $sl(2, C)$ multiplied by $z^r$:
\begin{gather*}
\widehat{K}_0(r)
=
\tfrac{1}{2}
\hat{h}_{2N}z^r
=
\tfrac{1}{2}
\left[\begin{array}{cc} 1_N & 0 \\ 0 & -1_N \end{array} \right] z^r
=
\tfrac{1}{2}
\hat{h}(r) ,\nonumber\\
\widehat{K}_+(r)
=
\hat{e}_{2N}z^r
=
\left[\begin{array}{cc} 0 & 1_N \\ 0 & 0 \end{array} \right] z^r
=
\hat{e}(r) ,\nonumber\\ 
\widehat{K}_-(r)
=
\hat{f}_{2N}z^r
=
\left[\begin{array}{cc} 0 & 0 \\ 1_N & 0 \end{array} \right] z^r
=
\hat{f}(r) ;
\nonumber\\
K_0(r)
=
\big\{\widehat{K}_0(r)\big\}_{\a \b }e_{\a \b }
=
\left(
\tfrac{1}{2} \hat{h}_{2N}z^r
\right)_{\a \b }e_{\a \b }
=
\left(
\tfrac{1}{2} \hat{h}_{2N}
\right)_{\a \b }e_{\a \b }(r) , \nonumber\\
K_\pm(r)
=
\big\{\widehat{K}_\pm(r)\big\}_{\a \b }e_{\a \b }
=
 \binom{\hat{e}_{2N}z^r}{\hat{f}_{2N}z^r}
_{\a \b }e_{\a \b }
=
 \binom{\hat{e}_{2N}}{\hat{f}_{2N}}
_{\a \b }e_{\a \b }(r) .
\end{gather*}
Using the formulas in Appendix~\ref{appendixC},
the $\uptau$ reps of the operators
${\widehat{K}}_{0}$ and ${\widehat{K}}_{\pm }$
are given, respectively, in the following forms:
\begin{gather*}
\uptau(K_0)
 =
\uptau
\left\{
\sum_{r \in \mathbb{Z}} K_0(r)
\right\}
 =
\sum_{r \in \mathbb{Z}}
\left( \tfrac{1}{2} \hat{h}_{2N} \right)_{\a \b }
\uptau
\{e_{\a \b }(r)\} , \nonumber\\
\phantom{\uptau(K_0)}{}
=
\sum_{r,s \in \mathbb{Z}}
\left( \tfrac{1}{2} \hat{h}_{2N} \right)_{\a \b }
E_{N(s-r) + \a , Ns + \b }
\simeq
\sum_{r,s \in \mathbb{Z}}
\left( \tfrac{1}{2} \hat{h}_{2N} \right)_{\a \b }
\psi_{N(s-r) + \a}\psi^*_{Ns + \b } , \nonumber\\
\uptau(K_\pm)
 =
\uptau
\left\{
\sum_{r \in \mathbb{Z}}
K_\pm(r)
\right\}
 =
\sum_{r \in \mathbb{Z}}
 \binom{\hat{e}_{2N}}{\hat{f}_{2N}}_{\a \b }
\uptau
\{e_{\a \b }(r)\} \nonumber\\
\phantom{\uptau(K_\pm)}{} =
\sum_{r,s \in \mathbb{Z}}
{\displaystyle \binom{\hat{e}_{2N}}{\hat{f}_{2N}}_{\a \b }}
E_{N(s-r) + \a, Ns + \b }
\simeq
\sum_{r,s \in \mathbb{Z}}
\binom{\hat{e}_{2N}}{\hat{f}_{2N}}_{\a \b }
\psi_{N(s-r) + \a}\psi^*_{Ns + \b } ,
\end{gather*}
from which
we have the KM brackets among the operators
$K_0(r)$ and $K_\pm(r)$.
For detailed calculations see Appendix~\ref{appendixF}.
\begin{gather*}
\left[K_0(r),K_0(s)\right]_{\rm KM}
=
\tfrac{1}{2}
Nr \delta_{r+s,0} \cdot c , \qquad
\left[K_\pm(r), K_\pm(s)\right]_{\rm KM}
=
0 , \nonumber\\
\left[K_0(r), K_\pm(s)\right]_{\rm KM}
=
\pm
K_\pm(r+s) , \qquad
\left[K_+(r),K_-(s)\right]_{\rm KM}
=
2K_0(r+s) + Nr \delta_{r+s,0} \cdot c .
\end{gather*}

Following Lepowsky and Wilson
\cite{LW.78},
we introduce the elements of
$\widehat{sl}(2,C)$
in the linear-combination forms of
$\hat{e}_{2N}$ and $\hat{f}_{2N}$:
\begin{gather*}
\binom{X_k}{Y_k}
\equiv
(\hat{e}_{2N})_{\a \b } e_{\a \b }(r)
\pm
(\hat{f}_{2N})_{\a \b } e_{\a \b }(r + 1)
=
  \binom{\widehat{X}_k}{\widehat{Y}_k}
_{\a \b }e_{\a \b } ,\\
\widehat{X}_k
\equiv
\left[ \begin{array}{cc} 0 & 1_N \\ z 1_N & 0 \end{array} \right] z^r ,\qquad
\widehat{Y}_k
\equiv
\left[ \begin{array}{cc} 0 & 1_N \\ -z 1_N & 0 \end{array} \right] z^r  \qquad
(k=2r+1),
\end{gather*}
the 2-{\em cocycle} $\alpha$'s on a pair of the above elements
read
\begin{gather*}
\a
\left(
\frac{1}{\sqrt{N}}X_k, \frac{1}{\sqrt{N}}X_l
\right)
=
k\delta_{k+l,0} ,\qquad
\a
\left(
\frac{1}{\sqrt{N}}Y_k, \frac{1}{\sqrt{N}}Y_l
\right)
=
-k\delta_{k+l,0} , \\
\a
\left(
\frac{1}{\sqrt{N}}X_k, \frac{1}{\sqrt{N}}Y_l
\right)
=
\delta_{k+l,0} ,\qquad
\a
\left(
\frac{1}{\sqrt{N}}Y_k, \frac{1}{\sqrt{N}}X_l
\right)
=
-\delta_{k+l,0} \nonumber\\
(k  =  2r  +  1, \, l  =  2s  +  1) .
\end{gather*}
For details see Appendix~\ref{appendixF}.
Then we have the following KM brackets and the map $\sigma_K$:
\begin{gather*}
 \left[\frac{1}{\sqrt{N}}X_k, \frac{1}{\sqrt{N}}X_l \right]_{\rm KM}
=
k
\delta_{k+l,0} \cdot c ,\qquad
\left[\frac{1}{\sqrt{N}}Y_k, \frac{1}{\sqrt{N}}Y_l \right]_{\rm KM}
=
-k
\delta_{k+l,0} \cdot c , \nonumber\\
\left[\frac{1}{\sqrt{N}}X_k, \frac{1}{\sqrt{N}}Y_l \right]_{\rm KM}
=
\frac{2}{\sqrt{N}}\frac{1}{\sqrt{N}}Y_{k+l}
+
\delta_{k+l,0} \cdot c ,
\\
\sigma_K: \ \ \frac{1}{\sqrt{N}}X_k \to \frac{\partial }{\partial x_k} , \qquad
\sigma_K: \ \ \frac{1}{\sqrt{N}}Y_k \to k y_k ,\nonumber\\
\sigma_K: \ \ \frac{1}{\sqrt{N}}X_{-k} \to k x_k ,
\qquad
\sigma_K: \ \ \frac{1}{\sqrt{N}}Y_{-k} \to \frac{\partial }{\partial y_k} .\nonumber
\end{gather*}
Then
$\frac{\sqrt{2}}{\sqrt{N}} K_0(r)$, $\frac{1}{\sqrt{N}} X_k$ $(k  =  2r  +  1)$
 and $\frac{1}{\sqrt{N}} Y_k$ $(k  =  2r  +  1)$
are clearly an inf\/inite-dimensional Heisenberg subalgebra of
the KM algebra
$\widehat{sl}(2, C)$.
We also introduce the element in the form of $\hat{h}_{2N}$ as
\begin{gather*}
Y_{2r}
\equiv
\big(
(-\hat{h}_{2N})_{\a \b }e_{\a \b }(r)
\big)
=
(\widehat{Y}_{2r})_{\a \b }e_{\a \b } ,\qquad
\widehat{Y}_{2r}
\equiv
\left[\begin{array}{cc} -1_N & 0 \\ 0 & 1_N \end{array} \right] z^r .
\end{gather*}


\subsection[Representation of inf\/inite-dimensional LMG model in terms of Schur polynomials]{Representation of inf\/inite-dimensional LMG model\\ in terms of Schur polynomials}\label{section5.4}

The expressions for the operators
${\widehat{K}}_{0}$ and ${\widehat{K}}_{\!\pm }$
(\ref{Koperators})
in terms of the operators
$
\frac{1}{\sqrt{N}} X_{ 2r  +  1}$, $\frac{1}{\sqrt{N}} Y_{ 2r  +  1}
$
and
$
\frac{1}{\sqrt{N}} Y_{ 2r}
$
and the expressions for the map $\sigma_K$ for the operators
$
\frac{1}{\sqrt{N}} X_{ 2r  +  1}
$
and
$
\frac{1}{\sqrt{N}} X_{ -(2r  +  1)}
$
are given as follows:
\begin{gather}
\widehat{K}_0
=
\sum_{r \in \mathbb{Z}}K_0(r)
=
-
 \frac{\sqrt{N}}{2}
\sum_{r \in \mathbb{Z}}
  \frac{1}{\sqrt{N}} Y_{2r} , \nonumber\\
 \binom{\widehat{K}_+}{\widehat{K}_-}
=
\sum_{r \in \mathbb{Z}}
  \binom{K_+(r)}{K_-(r+1)}
=
 \frac{\sqrt{N}}{2}
\sum_{r \in \mathbb{Z}}
\left(
\frac{1}{\sqrt{N}}X_{2r+1} \pm \frac{1}{\sqrt{N}}Y_{2r+1}
\right) , \nonumber\\
\sigma_K: \ \
 \frac{1}{\sqrt{N}} X_{2r+1}
\to
  \frac{\partial }{\partial x_{2r+1}}  ,\qquad
\sigma_K:  \ \
\frac{1}{\sqrt{N}}X_{-(2r+1)}
\to
(2r+1)x_{2r+1}  \qquad (r>0) .
\label{operatorKofXandYmapK}
\end{gather}
Consequently
we can obtain important sum-rules for the operators
$Y_{2r  +  1}$  and $Y_{2r}$
as
\begin{gather}
2   \sum_{r  \in  \mathbb{Z}}
\binom{Y_{2r}}{Y_{2r+1}}
=
\sum_{r  \geq  0}
 \binom{S_{2r}(2x)}{S_{2r+1}(2x)}
\sum_{s  \geq  0}   S_{2s}(-2\widetilde{\partial })
 +
\sum_{r  \geq  0}
 \binom{S_{2r+1}(2x)}{S_{2r}(2x)}
\sum_{s  \geq  0}   S_{2s+1}(-2\widetilde{\partial }) \nonumber\\
=
\tfrac{1}{2}
\left\{
e^{2\sum\limits_{m=0}^{\infty }x_{2m+1}
}
e^{-2\sum\limits_{n=0}^{\infty }\frac{1}{2n+1}
\frac{\partial }{\partial x_{2n+1}}
}
\pm
e^{-2\sum\limits_{m=0}^{\infty }x_{2m+1}
}
e^{2\sum\limits_{n=0}^{\infty }\frac{1}{2n+1}
\frac{\partial }{\partial x_{2n+1}}
}
\right\} ,
\label{sumruleforY}
\\
\left(
2  \! \sum_{r \in \mathbb{Z}}Y_{2r+1}
\right)^{ 2}\!\!
 =
\tfrac{1}{4} \!
\left[ \!
e^{4\!\!   \sum\limits_{m=0}^{\infty }x_{2m+1}}
e^{ -  4 \!\!  \sum\limits_{n=0}^{\infty }\frac{1}{2n+1}
\frac{\partial }{\partial x_{2n+1}}}
 -
e^{ -  4 \!\!  \sum\limits_{m=0}^{\infty }x_{2m+1}}
e^{4 \!\!  \sum\limits_{n=0}^{\infty }\frac{1}{2n+1}
\frac{\partial }{\partial x_{2n+1}}}
 -
2  \!
\right]\!   ,
\label{sumruleforY2}
\end{gather}
where we have used the relation
\begin{gather*}
\sum_{n \geq 0}  \binom{S_{2n}(x)}{S_{2n+1}(x)}
 =
\tfrac{1}{2}
\left\{
\exp \left(2\sum_{n=0}^\infty x_{2n+1}\right)
 \pm
\exp \left(-2\sum_{n=0}^\infty x_{2n+1}\right)
\right\}   \qquad
(x_{2n}  =  0, \ n  \geq  1) ,
\end{gather*}
which is derived from
the def\/inition of the Schur polynomial $S_k(x)$ in Appendix~\ref{appendixD}.
Further we have an expression for a quadratic operator
$\widehat{K}_+^2  +  \widehat{K}_-^2$
as
\begin{gather}
\widehat{K}_+^2  +  \widehat{K}_-^2
 =
  \frac{N}{4}
\left\{
\sum_{r \in \mathbb{Z}}
\left(
\frac{1}{\sqrt{N}}X_{2r+1} \! + \! \frac{1}{\sqrt{N}}Y_{2r+1}
\right)
\right\}^2\!\!
 +
\frac{N}{4}
\left\{
\sum_{r \in \mathbb{Z}}
\left(
\frac{1}{\sqrt{N}}X_{2r+1} \! -  \!\frac{1}{\sqrt{N}}Y_{2r+1}
\right)
\right\}^2 \! \nonumber\\
 \phantom{\widehat{K}_+^2  +  \widehat{K}_-^2}{} =
  \frac{N}{2}
\left\{
\left(
\sum_{r \in \mathbb{Z}}
 \frac{1}{\sqrt{N}}X_{2r+1}
\right)^2
 +
\left(
\sum_{r \in \mathbb{Z}}
  \frac{1}{\sqrt{N}}Y_{2r+1}
\right)^2
\right\} .
\label{quadraticoperatorK2}
\end{gather}
Finally
from
(\ref{operatorKofXandYmapK}), (\ref{sumruleforY}), (\ref{sumruleforY2})
and
(\ref{quadraticoperatorK2})
we get an expression for the LMG Hamiltonian
as
\begin{gather}
H
 =
\epsilon \widehat{K}_0
-
  \frac{V}{2}
\left(\widehat{K}_+^2 + \widehat{K}_-^2 \right) \nonumber\\
= -
\epsilon  \frac{\sqrt{N}}{2}
\sum_{r \in \mathbb{Z}}
\frac{1}{\sqrt{N}}Y_{2r} - \frac{V}{2}\frac{N}{2}
\left\{
\left(
\sum_{r \in \mathbb{Z}}
 \frac{1}{\sqrt{N}}X_{2r  +  1}
\right)^{  2}
+
\left(
\sum_{r \in \mathbb{Z}}
 \frac{1}{\sqrt{N}}Y_{2r  +  1}
\right)^{  2}
\right\} \nonumber\\
=
-
  \frac{1}{4} \epsilon
\exp \left(2   \sum_{m=0}^{\infty }x_{2m  +  1}\right)
\exp
\left(
-2   \sum_{n=0}^{\infty }
\frac{1}{2n  +  1}
\frac{\partial }{\partial x_{2n  +  1}}
\right)
 -
 \frac{V}{2}\frac{N}{2}
\left[
\sum_{m=0}^{\infty }
  \frac{\partial }{\partial x_{2m  +  1}}
\sum_{n=0}^{\infty }
 \frac{\partial }{\partial x_{2n  +  1}}
\right. \nonumber\\
+
\sum_{m=0}^{\infty }
\left[
 \frac{\partial }{\partial x_{2m  +  1}},
\sum_{n=0}^{\infty }(2n  +  1)x_{2n  +  1}
\right]
+
\sum_{m=0}^{\infty }(2n  +  1)x_{2m  +  1}
\sum_{n=0}^{\infty }
 \frac{\partial }{\partial x_{2n  +  1}} \nonumber \\
\left.
+
\sum_{m=0}^{\infty }(2m  +  1)x_{2m  +  1}
\sum_{n=0}^{\infty }
 \frac{\partial }{\partial x_{2n  +  1}}
+
\sum_{m=0}^{\infty }(2m  +  1)x_{2m  +  1}
\sum_{n=0}^{\infty }(2n  +  1)x_{2n  +  1}
\right]
+
 \frac{V}{32} \nonumber \\
-
 \frac{V}{64}
\left\{
\exp (4   \sum_{m=0}^{\infty }   x_{2m  +  1}  )
\exp
\left(
-4   \sum_{n=0}^{\infty }
\frac{1}{2n  +  1}
\frac{\partial }{\partial x_{2n  +  1}}
\right)
\right. \nonumber\\
\left.
 -
\exp \left(-4   \sum_{m=0}^{\infty }   x_{2m  +  1}  \right)
\exp
\left(
4   \sum_{n=0}^{\infty }
\frac{1}{2n  +  1}
\frac{\partial }{\partial x_{2n \!+\! 1}}
\right)
\right\}   .
\label{HamiltonianHofXandYmapK}
\end{gather}


\subsection[Inf\/inite-dimensional representation of $SU(2N)_{\infty}$ transformation]{Inf\/inite-dimensional representation of $\boldsymbol{SU(2N)_{\infty}}$ transformation}\label{section5.5}

We prepare operators
$\widehat{K}_0(\varphi )$, $\widehat{K}_0(\psi )$ and $\widehat{K}_\pm(\theta )$
to generate an inf\/inite-dimensional representation of
an $SU(2N)_{\infty}$ transformation-matrix.
First we give
$\hat{K}_0(\varphi )$ and $\hat{K}_0(\psi )$
as
\begin{gather*}
\widehat{K}_0
 \binom{\varphi }{\psi }
=
\sum_{r,s \in \mathbb{Z}}
\left\{
\binom{-\varphi_r}{-\psi_r}
 \tfrac{1}{2}
\hat{h}_{2N}
\right\}_{\a \b }
: \psi_{N(s-r) + \a}\psi_{Ns + \b }^*: \nonumber\\
\phantom{\widehat{K}_0
 \binom{\varphi }{\psi }}{}=
\sum_{r,s \in \mathbb{Z}}
\binom{-\varphi_r}{-\psi_r}
{\rm Tr}
\left[
\BA{cc}
\frac{1}{2}1_N & 0 \\
0 & -\frac{1}{2}1_N
\EA
\right]
\left[
\BA{cc}
: \psi_{N(s-r) + i}\psi_{Ns + j}^*: &
: \psi_{N(s-r) + i}\psi_{Ns + b}^*: \\
: \psi_{N(s-r) + a}\psi_{Ns + j}^*: &
: \psi_{N(s-r) + a}\psi_{Ns + b}^*:
\EA
\right] \nonumber\\
\phantom{\widehat{K}_0
 \binom{\varphi }{\psi }}{}=
\tfrac{1}{2}
\sum_{r,s \in \mathbb{Z}}
\binom{-\varphi_r}{-\psi_r}
\left\{
\sum_{i=1}^N
:\psi_{N(s-r)+i}\psi_{Ns+i}^*:
-
\sum_{a=1}^N
:\psi_{N(s-r)+a}\psi_{Ns+a}^*:
\right\} ,
\end{gather*}
adjoint actions of which for
$\psi_{Nr + \a}$  and $\psi_{Nr + \a}^*$
are computed as
\begin{gather*}
\left[
\widehat{K}_0 (\varphi (\psi)),
\binom{\psi_{Nr\!+\!\a}}{\psi_{Nr\!+\!\a}^*}
\right]
=
\sum_{s \in \mathbb{Z}}\psi_{N(r-s)\!+\!\b }
\left\{
\binom{-\varphi_s (\psi_s)}{-\varphi_s (\psi_s)}
\tfrac{1}{2}\hat{h}_{2N}
\right\}_{\b \a} .
\end{gather*}
Then the inf\/inite-dimensional fermion operator
$\psi_{Nr + \a}(\hat{g}_{\binom{\varphi }{\psi }})$
is transformed into
\begin{gather}
\binom{\psi_{Nr+\a}(\hat{g}_{\binom{\varphi }{\psi }})}
{\psi_{Nr+\a}^*(\hat{g}_{\binom{\varphi }{\psi }})}
\stackrel{d}{=}
U(\hat{g}_{\binom{\varphi }{\psi }})
\binom{\psi_{Nr+\a}(\hat{g}_{\binom{\varphi }{\psi }})}
{\psi_{Nr+\a}^*(\hat{g}_{\binom{\varphi }{\psi }})}
U^{-1}(\hat{g}_{\binom{\varphi }{\psi }})
=
e^{i\widehat{K}_0{\binom{\varphi }{\psi }}}
\binom{\psi_{Nr + \a}(\hat{g}_{\binom{\varphi }{\psi }})}
{\psi_{Nr + \a}^*(\hat{g}_{\binom{\varphi }{\psi }})}
e^{-i\widehat{K}_0{\binom{\varphi }{\psi }}} \nonumber\\
\qquad{}=
\sum_{s \in \mathbb{Z}}
\binom{\psi_{N(r-s) + \b}(g_{\binom{\varphi_s}{\psi_s}})}
{\psi_{N(r-s) + \b }^*(g_{\binom{\varphi_s}{\psi_s}})}_{\!\!\b \a} , \qquad
\binom{
g_{\binom{\varphi_s}{\psi_s}}
=
e^{-i {\binom{\varphi_s}{\psi_s}} \frac{1}{2}\hat{h}_{2N}}
}
{
g_{\binom{\varphi_s}{\psi_s}}^*
=
e^{i {\binom{\varphi_s}{\psi_s}} \frac{1}{2}\hat{h}_{2N}}
}
.
\label{transhatgphaipsi}
\end{gather}
In a similar way as the above
we also give
$\widehat{K}_\pm(\theta)$
as
\begin{gather*}
\widehat{K}_+(\theta)
=
\widehat{K}^\dagger _-(\theta)
=
\sum_{r,s \in \mathbb{Z}}
\left(- \theta_r \hat{e}_{2N} \right)_{\a \b }
: \psi_{N(s-r) + \a}\psi_{Ns + \b }^*: \\
\phantom{\widehat{K}_+(\theta)}{}
=
\sum_{r,s \in \mathbb{Z}}
\left(-\theta_r \right) {\rm Tr}
\left[
\BA{cc}
0 & 0 \\
1_N & 0
\EA
\right]
\left[
\BA{cc}
: \psi_{N(s-r) + i}\psi_{Ns + j}^*: &
: \psi_{N(s-r) + i}\psi_{Ns + b}^*: \\
: \psi_{N(s-r) + a}\psi_{Ns + j}^*: &
: \psi_{N(s-r)  + a}\psi_{Ns + b}^*:
\EA
\right] \\
\phantom{\widehat{K}_+(\theta)}{}
=
\sum_{r,s \in \mathbb{Z}}
\left(-\theta_r \right)
\sum_{i=a=1}^N
:\psi_{N(s-r) + i}\psi_{Ns + a}^*: ,
\end{gather*}
and from
(\ref{adjointactionKpm})
adjoint actions of which for
$\psi_{Nr + \a}$
are given as
\begin{gather*}
\left[
\binom{\widehat{K}_+(\theta)}{\widehat{K}_-(\theta)}
,
\psi_{Nr\!+\!\a}
\right]
=
\sum_{s \in \mathbb{Z}}\psi_{N(s-r) + \b }
\binom{- \theta_r\hat{e}_{2N}}{- \theta_{-r}\hat{f}_{2N}}
_{\b \a} ,\qquad
(\theta_r = \theta_{-r}) .
\end{gather*}
Then the transformed inf\/inite-dimensional fermion operator
$\psi_{Nr + \a}(\hat{g}_{\theta })$
is transformed into
\begin{gather}
\psi_{Nr + \a}(\hat{g}_{\theta })
=
U(\hat{g}_{\theta })\psi_{Nr + \a} U^{-1}(\hat{g}_{\theta })
=
e^{\frac{1}{2}
\left\{
\widehat{K}_+(\theta) - \widehat{K}_-(\theta)
\right\}}
\psi_{Nr + \a}e^{-\frac{1}{2}
\left\{
\widehat{K}_+(\theta) - \widehat{K}_-(\theta)
\right\}} \nonumber\\
\phantom{\psi_{Nr + \a}(\hat{g}_{\theta })}{}
=
\sum_{s \in \mathbb{Z}}\psi_{N(r-s) + \b } (g_{\theta_s})_{\b \a} , \qquad
g_{\theta_s}
=
e^{\frac{1}{2}(\hat{e}_{2N} - \hat{f}_{2N})(- \theta_s)} ,
\label{transformedfermionoperatr}
\end{gather}
The whole transformation of the inf\/inite-dimensional fermion operator
$\psi_{Nr + \a}(\hat{g})$
is given as
\begin{gather}
\psi_{Nr+\a}(\hat{g})
=
U(\hat{g}_{\psi } )U(\hat{g}_{\theta } )U(\hat{g}_{\varphi } )
\psi_{Nr + \a}
U^{-1}(\hat{g}_{\varphi } )U^{-1}
(\hat{g}_{\theta })
U^{-1}(\hat{g}_{\psi } ) \nonumber\\
\phantom{\psi_{Nr+\a}(\hat{g})}{} =
U(\hat{g}_{\psi } )
\sum_{s \in \mathbb{Z}}
U(\hat{g}_{\theta } ) \psi_{N(r-s) + \b}
U^{-1}(\hat{g}_{\theta } )U^{-1}
(\hat{g}_{\psi } )(g_{\varphi_s} )_{ \b \a} \nonumber\\
\phantom{\psi_{Nr+\a}(\hat{g})}{}=
\sum_{s,t \in \mathbb{Z}}
U   (\hat{g}_{\psi } )
\psi_{N(r-s-t) + \g}
U^{ -1}   (\hat{g}_{\psi } )(g_{\theta_t})_{\g\b }
(g_{\varphi_s} )_{ \b \a}\nonumber\\
\phantom{\psi_{Nr+\a}(\hat{g})}{}=
\sum_{s,t,u \in \mathbb{Z}}
\psi_{N(r-s-t-u) + \delta }
(g_{\psi_u} )_{\delta\g}(g_{\theta_t} )_{\g\b }
(g_{\varphi_s} )_{ \b \a}\nonumber \\
\phantom{\psi_{Nr+\a}(\hat{g})}{}=
\sum_{s,t,u \in \mathbb{Z}}
\psi_{N(r-s-t-u) + \b }(g_{u,t,s} )_{ \b \a} .
\label{transhatgphaipsitheta}
\end{gather}
The block matrix
$g_{u,t,s}$
of an $SU(2N)_{\infty }$ transformation-matrix is given by~(\ref{blockmatrixguts}).
For simplicity,
in the sum over $u$ and $t$ of
(\ref{transhatgphaipsitheta}),
we pick up only the term $u = s$ and $t = -s$,
then we have a~simple transformation
of the inf\/inite-dimensional fermion operator
$\psi_{Nr + \a}(\hat{g})$
as
\begin{gather*}
\psi_{Nr + \a}(\hat{g})
 =
\sum_{s \in \mathbb{Z}}\psi_{N(r-s) + \b } (g_s)_{\b \a} , \nonumber\\
g_{s}
 =
\left[
\BA{cc}
\cos
\left( \frac{\theta_s}{2} \right)
e^{-\frac{i}{2}(\psi_s + \varphi_s)}  \cdot  1_{N} &
-
\sin
\left( \frac{\theta_s}{2} \right)
e^{-\frac{i}{2}(\psi_s - \varphi_s)}  \cdot  1_{N}\vspace{1mm} \\
\sin
\left( \frac{\theta_s}{2} \right)
e^{\frac{i}{2}(\psi_s - \varphi_s)}  \cdot  1_{N} &
\cos
\left( \frac{\theta_s}{2} \right)
e^{\frac{i}{2}(\psi_s+ \varphi_s)}  \cdot  1_{N}
\EA
\right] .
\end{gather*}


\subsection[Representation of inf\/inite-dimensional HF Hamiltonian in terms of Schur polynomials]{Representation of inf\/inite-dimensional HF Hamiltonian\\ in terms of Schur polynomials}\label{section5.6}

Using the $\uptau$ rep of
(\ref{HFockoperator0})
and
(\ref{operatorKofXandYmapK}),
an inf\/inite-dimensional HF Hamiltonian for the LMG model is represented
in terms of the Schur polynomials as
\begin{gather*}
\widehat{F}[\widehat{W}]
=
-
\sum_{r \in \mathbb{Z}}
\left[
\varepsilon \tfrac{1}{2}
(\hat{h}_{2N})_{\alpha \beta }
 +
\varepsilon \chi \tfrac{1}{2}  \sin \theta e^{i \psi }
(\hat{e}_{2N})_{\alpha \beta }
 +
\varepsilon \chi \tfrac{1}{2}  \sin \theta e^{-i \psi }
(\hat{f}_{2N})_{\alpha \beta }
\right]
\uptau\{e_{\alpha \beta }(r)\} \nonumber\\
=
-
\sum_{r \in \mathbb{Z}}
\left[
\varepsilon K_0 (r)
 +
\varepsilon \chi \tfrac{1}{2}  \sin \theta e^{i \psi }
K_+ (r)
 +
\varepsilon \chi \tfrac{1}{2} \sin \theta e^{-i \psi }
K_- (r)
\right] \nonumber\\
=
\tfrac{\sqrt{N}}{2}
\sum_{r \in \mathbb{Z}}
\left[
\varepsilon  \tfrac{1}{\sqrt{N}}Y_{2r}
-
\varepsilon \chi  \sin \theta
\tfrac{1}{2}
\big(e^{i \psi }  +  e^{-i \psi }\big)
\tfrac{1}{\sqrt{N}}X_{2r+1}
-
\varepsilon \chi  \sin \theta
\tfrac{1}{2}
\big(e^{i \psi }  -  e^{-i \psi }\big)
 \tfrac{1}{\sqrt{N}}Y_{2r+1}
\right]  \nonumber\\
=
\tfrac{1}{2}
\varepsilon
\tfrac{1}{4}
\left\{
e^{2\sum\limits_{m=0}^{\infty }x_{2m+1}
}
e^{-2\sum\limits_{n=0}^{\infty }\frac{1}{2n+1}
\frac{\partial }{\partial x_{2n+1}}
}
 +
e^{-2\sum\limits_{m=0}^{\infty }x_{2m+1}
}
e^{2\sum\limits_{n=0}^{\infty }\frac{1}{2n+1}
\frac{\partial }{\partial x_{2n+1}}
}
\right\}\nonumber \\
-
\tfrac{1}{2}
\varepsilon \chi \sin \theta \cdot \cos \psi
\sum_{r \in \mathbb{Z}}
\frac{\partial }{\partial x_{2r+1}}\\ 
-
\tfrac{1}{2}
\varepsilon \chi \sin \theta \cdot i \sin \psi
\tfrac{1}{4}
\left\{
e^{2\sum\limits_{m=0}^{\infty }x_{2m+1}
}
e^{-2\sum\limits_{n=0}^{\infty }\frac{1}{2n+1}
\frac{\partial }{\partial x_{2n+1}}
}
 -
e^{-2\sum\limits_{m=0}^{\infty }x_{2m+1}
}
e^{2\sum\limits_{n=0}^{\infty }\frac{1}{2n+1}
\frac{\partial }{\partial x_{2n+1}}
}
\right\}   ,\nonumber
\end{gather*}
the second line of which has a form very similar to the Hamiltonian given
by Mansf\/ield~\cite{Mansfield.85}.
The HF density matrix for the LMG model is also represented in terms of
$K_0 (r)$ and $K_\pm (r)$ as
\begin{gather*}
\widehat{W}
-
\tfrac{1}{2}
\hat{I}_{2N}
=
\sum_{r \in \mathbb{Z}}
\left[
\cos \theta \tfrac{1}{2}
(\hat{h}_{2N})_{\alpha \beta }
 +
\tfrac{1}{2}  \sin \theta e^{-i \psi }
(\hat{e}_{2N})_{\alpha \beta }
 +
\tfrac{1}{2}  \sin \theta e^{i \psi }
(\hat{f}_{2N})_{\alpha \beta }
\right]
\uptau\{e_{\alpha \beta }(r)\} \nonumber\\
 \phantom{\widehat{W}
-
\tfrac{1}{2}
\hat{I}_{2N}}{} =
\sum_{r \in \mathbb{Z}}
\left[
\cos \theta K_0 (r)
 +
\tfrac{1}{2}  \sin \theta e^{-i \psi } K_+ (r)
 +
\tfrac{1}{2}  \sin \theta e^{i \psi } K_- (r)
\right] ,
\end{gather*}
The inf\/inite-dimensional HF operator for the LMG model
$H_{F_\infty;{\rm HF}}
(x,\widetilde{\partial }_x ,\hat{g})$,
corresponding to (\ref{infinite-dimensionalSCF-hamiltonian}),
is expressed in terms of
$
\widetilde{z}_{2N(s-r)+\alpha,2Ns+\beta }
(x,\widetilde{\partial }_x  )$
$(\alpha, \beta  =  1,2,  \dots , N, N  +  1,  \dots , 2N)
$ as
\begin{gather}
H_{F_\infty;{\rm HF}}
(x,\widetilde{\partial }_x ,\hat{g})
 =
\sum_{r,s \in \mathbb{Z}}
\left\{ {\widehat{F}[\widehat{W}]}(r) \right\}_{\alpha \beta }
\widetilde{z}_{2N(s-r)+\alpha,2Ns+\beta }
(x,\widetilde{\partial }_x  ) \nonumber\\
 =
\sum_{r,s \in \mathbb{Z}}
\left\{
-\varepsilon \tfrac{1}{2}
\hat{h}(r)
 -
\varepsilon \chi \tfrac{1}{2}  \sin \theta e^{i \psi }
\hat{e}(r)
 -
\varepsilon \chi \tfrac{1}{2}  \sin \theta e^{-i \psi }
\hat{f}(r)
\right\}_{\alpha \beta }
\widetilde{z}_{2N(s-r)+\alpha,2Ns+\beta }
(x,\widetilde{\partial }_x  ) \nonumber\\
 =
\sum_{r,s \in \mathbb{Z}}
\left[
-  \tfrac{1}{2}  \varepsilon
\sum_{\alpha = 1}^N
\widetilde{z}_{2N(s-r)+\alpha,2Ns+\alpha }
(x,\widetilde{\partial }_x  )
 +
\tfrac{1}{2} \varepsilon
\sum_{\alpha = N +1}^{2N}
\widetilde{z}_{2N(s-r)+\alpha,2Ns+\alpha }
(x,\widetilde{\partial }_x  )
\right.\nonumber \\
- \tfrac{1}{2} \varepsilon \chi \sin \theta e^{i\psi }
 \sum_{\alpha = 1}^{N}
\widetilde{z}_{2N(s-r)+\alpha,2Ns+\alpha \!+\! N}
(x,\widetilde{\partial }_x  ) \nonumber\\
\left.
-
\tfrac{1}{2} \varepsilon \chi \sin \theta e^{-i\psi }
 \sum_{\alpha = 1}^{N}
\widetilde{z}_{2N(s-r)+\alpha  +   N,2Ns+\alpha }
(x,\widetilde{\partial }_x  )
\right]   z^r   ,
\label{HFockoperatorKM2}
\end{gather}
in which the explicit form of
$
\widetilde{z}_{2N(s-r)+\alpha,2Ns+\beta } (x,\widetilde{\partial }_x)
$
is given by
(\ref{expressionsforzandztilde}).
For example,
the f\/irst term in the last line of
(\ref{HFockoperatorKM2}),
picking up only the f\/irst order in
$S_k(-\widetilde{\partial }_x  )$ and $S_k(\widetilde{\partial }_x  )$,
is given as
\begin{gather}
\widetilde{z}_{2N(s-r)+\alpha,2Ns+\alpha +N}
(x,\widetilde{\partial }_x  )
 =
S_{2N(s-r)+\alpha+k+1-N}( x )S_{-2Ns-\alpha -k}(- x )
S_1(-\widetilde{\partial }_x  )\nonumber\\
 \qquad{}+
S_{2N(s-r)+\alpha+k-N}( x )S_{-2Ns-\alpha -k+1}(- x )
S_1(\widetilde{\partial }_x  )  +
\cdots \nonumber\\
\qquad{}+
S_{2N(s-r)+\alpha+k}( x )S_{-2Ns-\alpha -k}(- x )
S_N(-\widetilde{\partial }_x  )\nonumber\\
 \qquad{}+
S_{2N(s-r)+\alpha+k-N}( x )S_{-2Ns-\alpha -k+N}(- x )
S_N(\widetilde{\partial }_x  ) +
\cdots .
\label{HFockzoperator}
\end{gather}

On the other hand,
from
(\ref{Schur-polynomialexpressionforTau-function0}),
we have
f\/irst leading terms of the $\tau$-function, $\tau_{M\!=\!N} (x, g)$,
for the LMG model with indices of the Pl\"{u}cker coordinates,
$i_M = N,N + 1,\dots$, $i_{M-1} = N - 1,\dots,i_1=1$ and etc.,
in the following form:
\begin{gather}
\tau_N (x, g)
=
v_{N,N-1,\dots,1}^{N,N-1,\dots,1} (g)
\det
\left|
\begin{array}{ccccc}
S_0(x) & S_1(x) & S_2(x) & \cdots & S_{N-1}(x) \\
0 & S_0(x) & S_1(x) & \cdots & S_{N-2}(x) \\
\cdots & \cdots & \cdots & \cdots & \cdots \\
0      & 0      & 0      & \cdots & S_0(x)
\end{array}
\right|\nonumber\\
\phantom{\tau_N (x, g)=} {}+
v_{N +1,N-1,\dots,1}^{N,N-1,\dots,1} (g)
\det
\left|
\begin{array}{ccccc}
S_1(x) & S_2(x) & S_3(x) & \cdots & S_N(x) \\
0      & S_0(x) & S_1(x) & \cdots & S_{N-2}(x) \\
\cdots & \cdots & \cdots & \cdots & \cdots \\
0      & 0      & 0      & \cdots & S_0(x)
\end{array}
\right|  +
\cdots \nonumber\\
\phantom{\tau_N (x, g)} {}=
v_{N,N-1,\dots,1}^{N,N-1,\dots,1} (g)
+
v_{N+1,N-1,\dots,1}^{N,N-1,\dots,1} (g)
\left(
S_1(x)
+
\cdots
+
S_N(x)
\right)
+
\cdots \nonumber\\
\phantom{\tau_N (x, g)} {}=
\left(\cos \tfrac{\theta }{2} \right)^N
e^{-\frac{i}{2}N(\psi + \varphi)}
+
\left(\cos \tfrac{\theta }{2}\right)^{N-1}
e^{-\frac{i}{2}(N-1)(\psi + \varphi)}
\left(\sin \tfrac{\theta }{2}\right)
e^{\frac{i}{2}(\psi - \varphi)}\nonumber\\
\phantom{\tau_N (x, g)=} {}\times
\left(
S_1(x)
+
\cdots
+
S_N(x)
\right)
+
\cdots .
\label{Schur-polynomialexpressionforTau-function1}
\end{gather}
The $\upsilon$-dependent HF equation
(\ref{TDHFequationonhat-g}),
$
i\partial_\upsilon \tau_N
\{  x, \hat{g} (\upsilon)  \}
=
H_{F_\infty ;{\rm HF}}
\{
x,\widetilde{\partial }_x, \hat{g} (\upsilon)
\}
\tau_N \{  x, \hat{g} (\upsilon)  \}
$,
on the
$\tau_N (x, \hat{g} (\upsilon))$
may be expected to give a new $\upsilon$-dependent HF solution,
where
the $\upsilon$-dependent HF operator
is given by
(\ref{HFockoperatorKM2}) and (\ref{HFockzoperator})
and the
$\tau$-function
is provided by~(\ref{Schur-polynomialexpressionforTau-function1}).



\subsection{Summary and discussions}\label{section5.7}

In the preceding section,
for the LMG model
we have an approximate HF operator and a $\tau$-function
up to the f\/irst order in
$S_k(-\widetilde{\partial }_x )$, $S_k(\widetilde{\partial }_x ),~S_k(x)$
and $S_k(-x)$
which should satisfy the $\upsilon$-dependent HF equation.
Then, we meet inevitably with an interesting and exciting problem of
solving the $\upsilon$-dependent HF equation on the
$\tau_N (x, g (\upsilon))$.
After determining HF parame\-ters~$\theta$ and $\psi$ self-consistently,
a further study should be made to obtain
a soliton solution derived from a~$\upsilon$-dependent Hirota's bilinear equation
regarding $\upsilon$ as time $t$
and relationship between a~collective motion and the soliton solution.
Such attractive problems have not been treated yet and just begin to open.

Here, we will recur to the representation of the inf\/inite-dimensional LMG model.
In the previous section,
we already have obtained the expression for the LMG Hamiltonian
(\ref{HamiltonianHofXandYmapK})
in which, however,
the commutator term in the f\/irst term in the forth line of the equation
brings an {\it anomaly} (an inf\/initely divergent result) for us,
as shown below,
\begin{gather*}
\sum_{m=0}^{\infty }
\left[
\frac{\partial }{\partial x_{2m + 1}},
\sum_{n=0}^{\infty }(2n + 1)x_{2n + 1}
\right]
=
\sum_{n=0}^{\infty }(2n + 1) \to \infty   \qquad
(\textit{anomaly}).
\end{gather*}
For the present,
we ought to discard this anomalous term to construct
the {\it anomaly-free} inf\/inite-dimensional LMG model.

Finally, we will point out the possibility of the extension of
the present algebra $sl(2,C)$ to
the af\/f\/ine Lie algebra $A_1^{(1)}$.
Corresponding to an extension of the adopted simple LMG model
to the so-called coupled LMG model
\cite{HayaIwa.80,KuriYamaIida.84},
we have a very interesting problem of constructing
an~$A_1^{(1)}$ LMG model for which the idea given in the paper~\cite{JM2.83}
is considered to be very suggestive and useful.
According to
\cite{JM2.83},
we can choose the Chevalley basis for $A_1^{(1)}$ as follows:
\begin{alignat*}{3}
& h_1 =
[e_1,f_1], \qquad &&  h_2 =[e_2,f_2], & \nonumber\\
& e_1 =
\sum_{\nu \in \mathbb{Z}}
\psi_{-1  +  2\nu }\psi_{2\nu }^* , \qquad && e_2
= \sum_{\nu \in \mathbb{Z}}
\psi_{2\nu }\psi_{1  +  2\nu }^* , & \nonumber\\
&f_1
=
\sum_{\nu \in \mathbb{Z}}
\psi_{2\nu }\psi_{-1  +  2\nu }^* ,
\qquad &&
f_2
=
\sum_{\nu \in \mathbb{Z}}
\psi_{1  +  2\nu }\psi_{2\nu }^* , &
\end{alignat*}
in which the total Hamiltonian of the coupled-LMG system is composed of
two LMG model Hamitonians
$H_1  =  H_1\big( \widehat{K}_{1;0},\widehat{K}_{1;\pm } \big)$
and
$H_2  =  H_2\big( \widehat{K}_{2;0},\widehat{K}_{2;\pm } \big)$
and an interaction term
$
\widehat{K}_{1;+ }\widehat{K}_{2;- }$
$ +
\widehat{K}_{2;+ }\widehat{K}_{1;- }
$.
We will discuss the above $A_1^{(1)}$ LMG model elsewhere
using the present inf\/inite-dimensional SCF method in $\tau$-functional space
on~$F_\infty$.




\section{Summary and future problems}\label{section6}

In Section~\ref{section1},
from algebro-geometric viewpoint,
we have given a brief history of microscopic understanding
of theoretical nuclear physics.
It is summarized that we seek for
an optimal coordinate-system describing dynamics
on a group manifold based on
a Lie algebra of fermion pairs.
The TDHF/TDHB are nonlinear dynamics owing to their SCF characters.
Seeking for collective coordinates in
a fully parametrized dynamical system
is exactly f\/inding a symmetry of an evolution equation
in nonlinear dynamics.
In dif\/ferential geometrical approaches for nonlinear problems,
the integrability conditions are stated as
the zero curvature of connection on the corresponding
Lie groups of systems.
Nonlinear evolution equations, e.g., the famous KdV
and sine/sinh-Gordan equations and etc.,
come from the well-known Lax equation
\cite{Lax.68}
which
arises as the zero curvature
\cite{Satt.82}.
These soliton equations describe motions of
the tangent space of local gauge f\/ields
on a time $t$ and a space $x$,
which are Lie group/algebra-valued-equations
arising from the integrability condition of gauge f\/ield
with respect to $t$ and $x$.
In the TDHFT/TDHBT,
the corresponding Lie groups
are unitary transformation groups of their ortho-normal bases
dependent on $t$ but not on $x$.

In Section~\ref{section2},
along the Lax form for integrable systems,
we have studied essential {\em curvature equations}
to extract collective submanifolds
out of the full TDHF/TDHB manifold and shown the following:

(i) Expectation values of the zero curvatures
for a state function become a set of equations of motion,
imposing  weak orthogonal conditions among inf\/initesimal generators,
i.e., equations for {\em tangent vector fields}
on the group submanifold.
Those of non-zero curvatures
become gradients of a potential
arising from a residual Hamiltonian
along collective variables.
These quantities are expected to give a criterion
how the collective submanifold is truncated well.

(ii) The zero-curvature equation in QPF
is nothing but the FRPAEQ imposed by
the weak orthogonal conditions
and has a simple geometrical interpretation:
Relative vector f\/ields made of the SCF Hamiltonian
around each point on an integral curve
constitute solutions for the FRPA around
the same point which is in turn a f\/ixed point in QPF.
It means the FRPA
is a~natural extension of the usual RPA
for small-amplitude 
f\/luctuations
around a ground state to RPA at any point
on the collective submanifold.
The enveloping curve,
made of a solution of the FRPA at each point
on an integral curve,
becomes another integral curve.
The integrability condition is the inf\/initesimal condition
to transfer a solution to another solution
for the evolution equation.
Then the usual RPAEQ becomes nothing but
a method of determining an inf\/initesimal transformation of symmetry
if f\/luctuating f\/ields are composed of only normal modes.

In Section~\ref{section3},
to go beyond a perturbative method
with respect to collective variables
to extract large-amplitude collective motions,
we have studied an algebro-geometric relation
between SCFM and $\tau$-FM,
method of constructing integrable equations (Hirota's equations)
in soliton theory.
At the beginning,
descriptions of dynamical fermion systems
in both the methods had looked
very dif\/ferent manners at f\/irst glance.
In abstract fermion Fock spaces,
each solution space of dynamics in both the methods
is the corresponding Grassmannian.
There is, however, a dif\/ference between
f\/inite-dimensional and inf\/inite-dimensional fermion systems.
In spite of such a dif\/ference,
we have aimed at closely connecting
the concept of mean-f\/ield potential with gauge of fermions
and at making a role of {\em loop group} clear
and consequently we have shown the relation between both the methods:

(i) The Pl\"{u}cker relation on the coset variable
becomes analogous to the Hirota's bilinear form.
The SCFM has been mainly devoted to
the construction of boson-coordinate systems rather than
to the construction of soliton solution by the $\tau$-FM.
It turns out that
both the methods are equivalent with each other from the viewpoint of
the Pl\"{u}cker relation or the bilinear identity equation
def\/ining Grassmannian.

(ii) The inf\/inite-dimensional fermion operators are introduced
through Laurent expansion of the f\/inite-dimensional fermions
with respect to degrees of freedom of fermions
related to a~$\upsilon$ dependent mean-f\/ield potential.
Inversely,
the mean-f\/ield potential
is attributed to gauges of cooperating inf\/inite-dimensional fermions.
The construction of fermion operators can be contained
in that of a Clif\/ford algebra.
This fact permits us to introduce an af\/f\/ine KM algebra.
It means that the usual perturbative method with respect to
collective variables with time periodicity
has implicitly stood on a ${\rm Gr}_\infty$.
Then we rebuilt the $\upsilon$-HFT with the use of
the af\/f\/ine KM algebra and
map it to the corresponding $\tau$-functional space.
As a result,
the $\upsilon$-HFT becomes {\em a gauge theory of fermions} and
the collective motion appears as the motion of fermion gauges
with a common factor.
The physical concept of {\em quasi-particle and vacuum}
in the SCFM on $S^1$ is connected with
{\em the Pl\"{u}cker relations}.
Extracting sub-group orbits consisting of loop paths
out of the ${\rm Gr}_M$ is just the formation of
the Hirota's bilinear equation for the reduced KP hierarchy
to $su(N)$ $( \subset  sl(N))$.
The present theory gives the manifest structure of
gauge theory of fermions inherent in SCFM and
provides a {\em new algebraic tool for microscopic understanding}
of the fermion many-body system.

(iii) Through the investigation of physical meanings for the
inf\/inite-dimensional shift operators and the conditions of reduction
to $sl(N)$ in $\tau$-FM from the $loop$ group viewpoint,
it is induced that there is
the close connection between {\em collective variables}
and {\em spectral parameter} in soliton theory and that
the algebraic mechanism bringing
the physical concept of particle and collective motions
arises from the reduction from $u(N)$ to $su(N)$
for the $\upsilon$-dependent HF Hamiltonian.

(iv) It must be stressed that
though the $\upsilon$-HFT describes a dynamics
on {\em real fermion-harmonic oscillators},
the soliton theory does
on {\em complex fermion-harmonic oscillators}.
This suggests us an important task
to extend the $\upsilon$-HFT on
real space $\widehat{su}(N)$
to that on complex space $\widehat{sl}(N)$.
It gives us a deeper understanding
of the concept of quasi-particle energies and
boson ones, in other words,
independent particles and mean-f\/ield potential,
in a microscopic treatment.
Recently Wiegmann et al.\ have developed an approach
in which the theory of classical integrable systems is applied to
studies of 1D-fermion systems and
the so-called orthogonality catastrophe in a Fermi gas.
They have introduced a {\em boundary condition changing operator}
\cite{Wiegmann.07}
but have made no map
$
\sigma_M : \  :\psi_i \psi_j^* :
 \mapsto
\widetilde{z}_{ij}(x,\widetilde{\partial }_x)$
(\ref{expressionsforzandztilde})
contrary to the present $\upsilon$-HFT.

In Section~\ref{section4},
we have given a geometrical aspect of RPAEQ
\cite{NK.89,Komatsu.00}
and an explicit expression for the RPAEQ with a normal mode
on $F_{ \infty }$.
We also have argued about the relation between a~{\it loop} collective path
and a FRPAEQ.
Consequently,
the usual perturbative method
is shown to be involved in the present method which
aims for constructing TDHFT on the 
af\/f\/ine KM algebra.
It turns out that the collective submanifold
is interpreted as a rotator on a curved surface in the~${\rm Gr}_{ \infty }$.
The present theory may lead to multi-circles occurring
multiple parameterized collective motions.
If we could arrive successfully at such a f\/inal goal,
the present work may give us important clues for description of
large-amplitude collective motions in nuclei and molecules
and for construction of multi-dimensional soliton equations
\cite{KacLeur.03,DJKM.81}
since the collective motions usually
occur in multi-dimensional {\it loop} space.

In Section~\ref{section5},
as an illustration
we have attempted to make a $\upsilon$-HFT approach to
an inf\/inite-dimensional matrix model
extended from the f\/inite-dimensional $su(2)$ LMG model
\cite{LMG.65}.
For this aim, we have given an af\/f\/ine KM algebra~$\widehat{sl}(2,C)$
(complexif\/ication of $\widehat{su}(2))$
to which the LMG generators subject
and their $\uptau$ representations and the $\sigma_K$ mappings for them.
Further we have introduced inf\/inite-dimensional
``particle'' and ``hole'' operators
and operators
${\widehat{K}}_{0}$ and ${\widehat{K}}_{\pm }$
def\/ined by the inf\/inite-dimensional ``particle-hole'' pair operators.
Using these operators,
we have constructed the inf\/inite-dimensional Heisenberg subalgebra of
the af\/f\/ine KM algebra $\widehat{sl}(2,C)$.
Thus the LMG Hamiltonian and its HF Hamiltonian have been represented in terms of
the Heisenberg basic-elements whose representations
are isomorphic to those in the corresponding boson space.
They have been expressed in terms of inf\/inite numbers of
the variables $x_k$ and the derivatives $\partial_{x_k}$ through
the Schur polynomials $S_k (x)$.
Further we have obtained an approximate HF operator and a $\tau$-function
up to the f\/irst order in
$S_k(-\widetilde{\partial }_x  )$, $S_k(\widetilde{\partial }_x  )$, $S_k(x)$
and $S_k(-x)$.

In Appendices,
we have given the inf\/inite-dimensional representation of
$SU(2N)_{\infty }$ transformation of the ``particle'' and ``hole'' operators.
The expression for $\uptau$ rep of
$(
{\rm Y} _{-(2i + 1)}
 +
{\rm Y}_{(2i + 1)}
)$,
i.e.,
$
g_{{\rm Y}_{-(2i + 1)}
+
{\rm Y}_{(2i + 1)}}(z)
$
has been f\/irst given in terms of the Bessel functions.
We have also shown an explicit expression for Pl\"{u}ker coordinate
and calculated a quantity,
$\det (1_N  +  p^\dag p)$,
in terms of the Schur polynomials.

Finally
intimate relation of SCFT to soliton theory
has been shown to
come from ways of constructing a closed system of solution spaces.
The ordinary SCFM has been almost devoted to approach
cooperative phenomena in f\/inite fermion systems.
We must contrive construction of the optimal coordinate-system
on the group manifold.
For this purpose the relation between
the boson expansion method for f\/inite fermion systems
and the $\tau$-FM for inf\/inite ones
should be intensively investigated to
clarify algebro-geometric structures of integrable systems.
Such algebro-geometric approach will make a bridge between
f\/inite fermion systems and inf\/inite ones.
Various physical concepts and mathematical methods
will work well also in the inf\/inite ones.
The SCFM based on global symmetry
should be much improved
noticing local symmetry of the inf\/inite ones
and then may open a new area
in vigorous pursuit of wider f\/ields of physics.

We have many future problems in connection with the above discussions,
which are itemized as follows:

(i) To study the relation between the quantity of
non-zero curvature and the collectivity:
It is interesting to study the relation
using the simple LMG model,
which leads to an investigation of the ef\/fective condition
for the collective submanifold extracted
by the zero-curvature equation.
Temporarily digressing from the integrability condition,
adopting the Bethe anzatz (BA)
we have obtained exact solutions
for the LMG model solving the BA equation
\cite{MOPN.06}.
Contrary to Pan and Draayer's work
\cite{PD.98.99} and our previous works
\cite{Nishi.94},
we do not use any bosonization nor inf\/inite-dimensional techniques
and hence have no restrictions on interaction-strengths of LMG Hamiltonian.
Considering the advantage of the integrability condition,
the famous Gaudin model plays an important role
to solve ef\/fectively the BA equation
\cite{Gau76}.
From the {\em loop group} viewpoint,
as shown by Sklyanin,
with the use of
the exactly-solvable Gaudin model obeying the Gaudin algebra,
an exponential generating function of correlators
is obtained from the Gauss decomposition for $sl(2,C)$
{\em loop algebra},
which gives correlators including
the Richardson--Gaudin determinant formula
for the Bethe eigen-function
\cite{Skl99}.
A generalization of the Gaudin algebra is given by Ortiz et al.~\cite{OSDR.05}.
These works may have an intimate relation with our recent work~\cite{NPK.05}
and the present work.

(ii) To clarify the explicit relation between
spectral parameter and collective variable
and the physical concept of the geometrical connection:
The spectral parameter of the iso-spectral equation
in soliton theory and the collective variable in SCFM,
though showing dif\/ferent aspects at a glance,
work as scaling parameters on $S^1$.
The former relates to a scaling
by analytical continuation of $S^1$, i.e.,~$z$.
The latter makes roles of deformation parameters
of {\em loop paths} in~${\rm Gr}_M$.

(iii) To study the relation between weak boson operators
and boson mapping operators,
i.e., the shift operators in $\tau$-FM:
The generators for collective variables
in $F_\infty$ can never
take exact boson commutation relations because of
the f\/inite-dimensional matrices.

\looseness=1
(iv) To study a relation hidden behind
gauge of state functions and construction of fermion pairs:
In the usual algebraic treatment of fermion many-body systems,
we assign an abstract number
to each of the set of quantal numbers
and let their fermions make the Lie algebras
($u(N),~so(2N),~so(2N+1)$ and etc.).
For the pair-constructions we have an interpretation
as classif\/ications of Laurent spectra
in the inf\/inite-dimensional fermions,
although we did not manifestly state.
On the other hand, as well known, electron spin can be
described as a geometrical phase of gauge
with the help of M\"{o}bius band.
Then we inversely start from fermions
with the abstract numbers and through any way we could return
to the original fermions with the physical quantal numbers.
We think it not so wrong to attempt
to understand quantal numbers
as the geometrical attribute of the Grassmannian
made of the abstract fermions.

(v) To establish mathematical tools
to obtain subgroup-orbits of loop paths in ${\rm Gr}_M$,
basing on the Pl\"{u}cker relations on $S^1$,
i.e., soliton equations.
This problem is most fundamental to solve the new theory:
For this aim we must know
sub-group orbits or corresponding sub-Lie algebras
and establish
mathematical tools to extract them out of ${\rm Gr}_M$.
In this concern,
we are intensively interested in the algebraic mechanism
for spontaneous decision of a f\/ixed point
and a collective submanifold around the point.

(vi) To study a relation between
nonlinear superposition principle in soliton theory
and generator coordinate method (GCM) in SCFM:
The GCM may provide a superposition principle
on a nonlinear space
\cite{Boyd90,Fu.81}.
Standing on the viewpoint of local symmetry of
inf\/inite fermion systems behind global symmetry of f\/inite ones,
we might reconstructed the GCM and nonlinear superposition methods
using the inf\/inite-dimensional shift operators.
What relation does exist between
the construction of exact solutions based on the idea of
the imbricate series in soliton equations
\cite{TW.97}
and resonating mean-f\/ield theories
\cite{FukuNishi.88.91}?

(vii) To study why soliton solutions for
classical wave equations show fermion-like behaviours in
quantum dynamics and about what symmetries are hidden in soliton
equations.
To both the questions suggested by Tajiri et al.,
we cannot give a satisfactory answer yet within the present framework.
Because both the methods are {\em a priori}
based on the fermion system from the outset.
That is to say:
SCFM describes a quasi classical dynamics
on Grassmannian (S-det orbit) which
is induced owing to the anti-commutative property of fermions.
On the other hand,
$\tau$-FM also uses the fermions
to explain Grassmannian of solution space which
ref\/lects the fermion-like behaviours of soliton solutions.
Therefore we should study further why extraction of
soliton equations out of classical wave equations
brings out Grassmannian.
In the reductive perturbation method
\cite{RPM-Farf},
soliton equations
appear as the symmetry space in a classical wave equation
with respect to a transformation and a scaling transformation
with a common parameter
for independent and dependent variables.
We can see a relation between
the power exponents of the parameter and
the {\em degree} of shift operators.
At any point on extracting way of soliton equations
from classical wave equations did we introduce
the ``anti-commutative character'',
in other words, did we introduce the structure of Grassmannian?




\appendix

\section{Coset variables}\label{appendixA}

Following \cite{Fu.81},
let us introduce the triangular matrix functions
$S(\zeta ) $, $C(\zeta )$ and $\tilde{C}(\zeta )$
def\/ined as
\begin{gather*}
 S(\zeta )
=
(S_{ia}(\zeta ))
 =
\sum^\infty_{k=0}(-1)^k
  \frac{1}{(2k+1)!}
\zeta (\zeta ^{\dagger }\zeta)^k, \nonumber\\
C(\zeta )
=
(C_{ab}(\zeta ))
 =
1_M + \sum^\infty_{k=1}(-1)^k
  \frac{1}{(2k)!}
(\zeta ^{\dagger }\zeta)^k
 =
C^{\dagger }(\zeta ),\nonumber \\
\tilde C(\zeta )
=
(\tilde C_{ij}(\zeta ))
 =
1_{N-M} + \sum^\infty_{k=1}(-1)^k
  \frac{1}{(2k)!}
 (\zeta \zeta ^{\dagger })^k
 =
\tilde C^{\dagger }(\zeta ),
\end{gather*}
which have the properties analogous to
the usual triangular functions
\begin{gather*}
C^2(\zeta )+S^{\dagger }(\zeta )S(\zeta ) = 1_M,
\qquad \tilde C^2(\zeta )+
S(\zeta)S^{\dagger }(\zeta ) = 1_{N-M},
\qquad S(\zeta )C(\zeta )=\tilde{C}(\zeta )S(\zeta ).
\end{gather*}
Then the matrix $p$ in (\ref{Slaterdet1})
is def\/ined as
$p=(p_{ia})
=
S(\zeta )C^{-1}(\zeta )$. The matrix $g$
in (\ref{Slaterdet1})
is decomposed as
$g  =  g_{\zeta } g_{w}$ using
the matrices given below
\begin{gather*}
 g_{\zeta }
=
e^{\gamma '}
=
\left[
\BA{cc} C(\zeta ) &
-S(\zeta )^\dagger \\ S(\zeta )  &
\bar {C}(\zeta )
\EA
\right], \qquad
\gamma '
=
\left[
\BA{cc} 0&
-\zeta ^\dagger \\ \zeta &0
\EA
\right] , \qquad
 g_{w}
=
e^{\gamma ''}
=
\left[
\BA{cc} w & 0 \\ 0  &
\bar {w}
\EA
\right],\nonumber\\
w
=
e^{\eta },\qquad
\bar {w}
=
e^{\bar{\eta }},\qquad
\gamma ''
=
\left[
\BA{cc} \eta &0 \\ 0 &
\bar {\eta }
\EA
\right], \qquad
\eta^{\dagger }
=
-\eta ,\qquad
\bar {\eta }^{\dagger }
=
-\bar {\eta },
\end{gather*}
where $\zeta$ is a $(N-M) \times M$
matrix $(\zeta_{ia})$
and $\eta$ and $\bar {\eta }$ are
$M \times M$ and $(N-M) \times (N-M)$
anti-hermitian matrices $(\eta_{ab})$ and
$(\bar {\eta }_{ij})$, respectively.
The indices $i$ and $a$ denote
unoccupied ($M+1 \sim N$) states and
occupied (1$ \sim M$) states, respectively.

\section[Properties of the dif\/ferential operator $e_{ia}$ acting on $\Phi_{M,M}$]{Properties of the dif\/ferential operator $\boldsymbol{e_{ia}}$ acting on $\boldsymbol{\Phi_{M,M}}$}\label{appendixB}

Let us introduce a dif\/ferential operator
and a function
corresponding to
the so-called particle-hole operator
and
the free particle vacuum
in the physical fermion space,
respectively as
\begin{gather}
 {\ve}_{ia}^*
\stackrel{d}{=}
-
\left(
p_{ib} p_{ja}
\frac{\partial }{\partial p_{jb}}
+
\frac{\partial }{\partial p_{ia}^* }
+
\frac{i}{2}p_{ia}
\frac{\partial }{\partial \tau }
\right) , \nonumber\\
\Phi_{M,M} (p, p^* ,\tau )
=
[\det (1+p^\dagger p)
]^{-\frac{1}{2}} e^{i \tau } .
\label{differentialeia}
\end{gather}
By using the formula for
the dif\/ferential of a determinant,
we can easily calculate as
\begin{gather}
\frac{\partial }{\partial p_{jb}}
[\det (1+p^\dagger p)]^{-\frac{1}{2}}
=
-\tfrac{1}{2}
p_{jc}^* [(1+p^\dagger p)^{-1}
]^{\rm T}_{cb}
[\det (1+p^\dagger p)]^{-\frac{1}{2}}
 , \nonumber\\
\frac{\partial }{\partial p_{ia}^*}
[\det (1+p^\dagger p)]^{-\frac{1}{2}}
=
-\tfrac{1}{2}
p_{id} [(1+p^\dagger p)^{-1}]_{da}
[\det (1+p^\dagger p)]^{-\frac{1}{2}}
 ,
\label{differentialdet}
\end{gather}
where $a^{-1}$ denotes an inverse matrix of a.
Then from equations~(\ref{differentialeia}) and
(\ref{differentialdet})
we get
\begin{gather*}
{\ve}_{ia}^* \Phi_{M,M}
 =
\left\{ \tfrac{1}{2}
p_{ib} p_{ja} \cdot p_{jc}^*
[(1+p^\dagger p ) ^{-1}
]^{\rm T}_{cb}
+\tfrac{1}{2} p_{id}
[(1+p^\dagger p )^{-1}]_{da}
+\tfrac{1}{2} p_{ia} \right\}
\Phi_{M,M}\nonumber\\
\phantom{{\ve}_{ia}^* \Phi_{M,M}}{} =
\tfrac{1}{2} \left[\{ p(1+p^\dagger p )^{-1}
(1+p^\dagger p)\}_{ia} + \tfrac{1}{2} p_{ia}\right]
\Phi_{M,M} = p_{ia} \Phi_{M,M} ,
\\
{\ve}_{ia}^* p_{jb} = -p_{ib} p_{ja} +p_{jb} e_{ia}^* .
\end{gather*}
Thus we can prove
\begin{gather*}
{\ve}_{ia}^* \Phi_{M,M} = p_{ia} \Phi_{M,M} ,\qquad
[{\ve}_{ia}^*, p_{jb} ] = -p_{ib}p_{ja} .
\end{gather*}
These are just the relation
given in the f\/irst line of
(\ref{p-hdifferentialoperatorsontovacuum}).
The other relations in
(\ref{p-hdifferentialoperatorsontovacuum})
can be proved in a similar manner
\cite{NMO.04}.

\section[Af\/f\/ine Kac--Moody algebra]{Af\/f\/ine Kac--Moody algebra}\label{appendixC}

According to Kac and Raina
\cite{KR.87},
let $gl(N)$ be
the Lie algebra of all $N  \times  N$
matrices with complex
entries acting in ${\mathbb{C}}^N$ and
let ${\mathbb{C}}[z,z^{-1}]$ be
the ring of Laurent polynomials
in indeterminate~$z$ and~$z^{-1}$.
The $loop$ algebra
\cite{PS.86}
$\widetilde{gl}(N)
({\supset }\widetilde{u}(N))$ is def\/ined as
$gl(N) ({\mathbb{C}} [z,z^{-1}])$
$({\supset } u(N) ({\mathbb{C}}[z,z^{-1}]))$, i.e.,
as the complex Lie algebra
of $N \times N$ matrices with
Laurent polynomials as entries.
An element of $\widetilde{gl}(N)$
is given in the form
\begin{gather}
a(z) = \sum_{r \in \mathbb{Z}} z^r a_r \qquad (a_r \in gl(N)).
\label{loop1}
\end{gather}
As was pointed out by Kac and Raina,
we regard the fermion pair- and
single-operators as
\begin{gather*}
 c_\alpha^\dagger c_\beta \mapsto e_{\alpha \beta }
\quad (1 \leq \alpha,~\beta \leq N), \nonumber\\
c_\alpha^\dagger \mapsto u_\alpha
\quad  (1 \leq \alpha \leq N),\qquad
c_\alpha \mapsto u_\alpha^{\rm T}
\quad (1 \leq \alpha \leq N),
\end{gather*}
where the matrix elements $e_{\alpha \beta }$
are equal to 1
in the ($\alpha ,\beta$) entry and 0 elsewhere.
The matrix elements $e_{\alpha \beta }$ form
a basis of $gl(N)$.
The components of $N \times 1$ column vectors
$u_\alpha$ are equal to 1
in the $\alpha$-th row and 0 elsewhere.
They span the vector space
${\mathbb{C}}^N$ in which the $gl(N)$ acts.
The symbol \mbox{\scriptsize T} means the transpose of
a vector or a matrix.
The matrices
$e_{\alpha \beta } (r)
\stackrel{d}{=}
z^r e_{\alpha \beta }$
constitute a basis of $\widetilde{gl}(N)$.
The {\it loop} algebra $\widetilde{gl}(N)$ acts
in the vector space
${\mathbb{C}}[z,z^{-1}]^N$ consisting of
$N \times 1$ column vectors with
the Laurent polynomials
in $z$ and $z^{-1}$ as entries.
The Lie bracket on $\widetilde{gl}(N)$
is the commutator
\begin{gather}
\left[
e_{\alpha \beta } (r), e_{\gamma \delta } (s)
\right]
=
\delta_{\b \g }e_{\a \delta }(r+s)
-
\delta_{\a \delta }e_{\gamma \beta }(r+s) .
\label{Liebracketgln}
\end{gather}
The vectors def\/ined as
\begin{gather*}
\nu_{Nr + \alpha } = z^{-r}u_{\alpha } ,\qquad
\nu_{Nr + \alpha }^*
=
u_{\alpha }^{\rm T} z^{r} ,
\end{gather*}
also form a basis of the vector space
${\mathbb{C}}[z,z^{-1}]^N$ indexed by $\mathbb{Z}$
and its dual space.
The $\{ \nu_i \,|\, i = Nr + \alpha \in \mathbb{Z} \}$
is given by
the column vector with 1
as the $i$-th row and 0 elsewhere.
Thus it is possible to identify
${\mathbb{C}} [z,z^{-1}]^N$ with ${\mathbb{C}}^\infty$.
The relation
$e_{\alpha \beta }(r) \nu_{Ns\!+\!\beta }
=
\nu_{N(s-r) + \alpha }$
is easily derived.

For $a(z)\in \widetilde{gl}(N)$
we denote
the corresponding matrix
in $\overline{a}_\infty$ by $\uptau\{a(z)\}$.
Then we deduce
a~matrix representation
for $\uptau\{e_{\alpha \beta }(r)\}$ in
$\overline{a}_\infty$ as
\begin{gather}
\uptau\{e_{\alpha \beta } (r)\}
=
\sum_{s\in \mathbb{Z}}
E_{N(s-r) + \alpha,Ns + \beta } .
\label{tau}
\end{gather}
$E_{ij}$ $(i,j \in \mathbb{Z})$ have 1
as the $(i ,j)$ entry and 0 elsewhere
and form a $gl_\infty$.
Suppose a bigger algebra~$\overline{a}_\infty$
\begin{gather}
\overline{a}_\infty = \{ (a_{ij})\,| \, i,j\in \mathbb{Z},~a_{ij}
= 0~ \mbox{for}~|i-j| \gg  \mathbb{N} \} .
\label{biggerLiealgebra}
\end{gather}
There exists such an $\mathbb{N}$
satisfying the above condition.
The corresponding matrix of the $a(z)$ in~(\ref{loop1})
in $\overline{a}_\infty$
has an inf\/inite $N$ periodic sequence of block form
\begin{gather}
\uptau\{a(z)\}
=
\left[
\begin{array}{ccccccc}
 \ddots& \ddots& \ddots&       & \ddots&       &       \\
       & a_{-1}& a_0   &  a_1  &       & \ddots&       \\
 \ddots&       & a_{-1}&  a_0  &  a_1  &       & \ddots\\
       & \ddots&       & a_{-1}&  a_0  &  a_1  &       \\
       &       & \ddots&       & \ddots& \ddots&\ddots
\end{array}
\right] .
\label{CorreMatrix}
\end{gather}
We regard
(\ref{CorreMatrix})
as a representation of
the matrix $\overline{a}_\infty$ in which
elements on each diagonal
parallel to the principal diagonal
form a periodic sequence with period $N$.
Let $X(k)  =  z^k X $ $(X  \in  gl(N))$ be
an element of $\widetilde{gl}(N)$.
Def\/ine an antilinear anti-involution $\omega$ on
$\widetilde{gl}(N)$ by
$\omega[X(k)]  =  z^{-k} X^\dagger$.
Then in the $\overline{a}_\infty$
we get
$\uptau\{\omega[X(k)]\}  =  \uptau ^\dagger \{X(k)\}$.

Using the fundamental idea of the Dirac theory
\cite{Dirac.58},
we def\/ine the vacuum state
in which the state
labeled by the ``Laurent spectrum''
with positive energy is empty
but all the negative energy states
labeled by the ``Laurent spectra'' are occupied.
Denoting an exterior product of vectors
as $\wedge$,
a perfect vacuum $\Psi_0$ and
a reference vacuum $\Psi_M$ are expressed as
\begin{gather*}
\Psi_0 = \nu_0 \wedge \nu_{-1}
\wedge \nu_{-2} \wedge\cdots ,\qquad
\Psi_M = \nu_M \wedge \nu_{M-1}
\wedge \nu_{M-2}\wedge  \cdots   .
\end{gather*}

Let the space
$V  =  \oplus_{i  \in  \mathbb{Z}}{\mathbb{C}}\nu_i$
and
its dual
$V^*  =  \oplus_{j\in \mathbb{Z}}{\mathbb{C}}\nu_j^*$
be an  inf\/inite-dimensional complex space
with a basis $\{ \nu_i,\nu_j^* \,| \, i,j  \in \mathbb{Z} \}$
giving a linear functional $\nu_j^*$ on $V$ by
$\nu_j^*(\nu_i )  =  \delta_{ij}$ $(i,j\in \mathbb{Z})$.
Each $\nu \in V$ and $\nu^* \in V^*$ def\/ines
a {\it wedging operator} $\hat{\nu }$ and
a {\it contracting operator} $\check{\nu^*}$ on
the $F_\infty$ as
\begin{gather}
\hat{\nu }(\nu_{i_1} \wedge \nu_{i_2}
\wedge \cdots )
=\nu \wedge \nu_{i_1} \wedge \nu_{i_2}\wedge
\cdots ,\nonumber\\
\check{\nu^*}(\nu_{i_1} \wedge \nu_{i_2}
\wedge \cdots ) =
\nu^* (\nu_{i_1}) \nu_{i_2} \wedge \nu_{i_3}
\wedge \nu_{i_4} \wedge \cdots
-\nu^* (\nu_{i_2}) \nu_{i_1}
\wedge \nu_{i_3} \wedge \nu_{i_4}
\wedge \cdots \nonumber\\
\phantom{\check{\nu^*}(\nu_{i_1} \wedge \nu_{i_2}
\wedge \cdots ) =}{}
+\nu^* (\nu_{i_3}) \nu_{i_1}
\wedge \nu_{i_2} \wedge \nu_{i_4}
\wedge \cdots .
\label{Clifford}
\end{gather}
Then the operators
$\{ \hat{\nu_i} ,\ \check{\nu^* _j}\,|\, i,j  \in  \mathbb{Z} \}$
generate a Clif\/ford algebra
\begin{gather}
\{ \check{\nu_i^*} ,  \hat{\nu_j} \}
=
\delta_{ij} , \qquad \{ \check{\nu_i^*} ,  \check{\nu_j^*} \}
=
\{ \hat{\nu_i} , \hat{\nu_j} \} = 0 .
\label{Cliffordalgebra}
\end{gather}
Thus the anti-commutation relations lead us to
the identif\/ication of the new fermion
annhilation-creation operators~(\ref{infFermion})
with the present operators
at a pointwise $z$ with $|z|=1$ as
\begin{gather}
\psi_{Nr + \alpha }^* \mapsto
\check{\nu }_{Nr + \alpha }^* ,\qquad
\psi_{Nr + \alpha } \mapsto \hat{\nu }_{Nr + \alpha } .
\label{identification}
\end{gather}
Using the above identif\/ication,
the corresponding perfect vacuum and
reference vacuum, it can be shown that
\begin{gather*}
\Psi_0 \mapsto |\mbox{Vac} {\rangle },\qquad
\Psi_M \mapsto |M \rangle
=
\psi_M \cdots \psi_1 |\mbox{Vac} {\rangle } , \nonumber\\
\psi_{Nr+\alpha } |\mbox{Vac} \rangle = 0 ,
\qquad \langle \mbox{Vac}|\psi_{Nr + \alpha }^*  =  0
\qquad (r  \leq  -1),\nonumber\\
\psi_{Nr + \alpha }^* |\mbox{Vac} \rangle  =  0 ,\qquad
 \langle \mbox{Vac}|\psi_{Nr + \alpha }  =  0\qquad
 (r  \geq  0).
\end{gather*}

From (\ref{tau}), (\ref{Clifford}),
(\ref{identification}) and the Clif\/ford algebra
(\ref{Cliffordalgebra})
we obtain representations
\begin{gather}
\uptau\{e_{\alpha \beta }(r)\}
=
\sum_{s \in \mathbb{Z}} E_{N(s-r)\!+\!\alpha ,Ns\!+\!\beta }
\simeq
\sum_{s \in \mathbb{Z}}
\hat{\nu }_{N(s-r) + \alpha }
\check{\nu }_{Ns + \beta }^*
\simeq
\sum_{s \in \mathbb{Z}}
\psi_{N(s-r) + \alpha } \psi_{Ns + \beta }^* .
\label{tau-expression}
\end{gather}
Construction of the SCFM
on the inf\/inite-dimensional Fock space
is an interesting illustrative problem
in order to clarify an algebraic structure
among the original fermionic f\/ield,
the vacuum f\/ield def\/ined in
the SCF potential and
the bosonic f\/ield associated with Laurent spectra.

It is well known that the operator of
(\ref{tau-expression})
acting in the $F_\infty$ has
in general an ``anomaly''.
To avoid the ``anomaly'',
we had better use either of
the normal-ordered products given below
\begin{gather}
 :E_{Nr + \alpha ,Ns + \beta } :
\stackrel{d}{=}
E_{Nr + \alpha ,Ns + \beta }
-\delta_{\alpha \beta } \delta_{rs}
\qquad (s<0),\nonumber \\
:\hat{\nu }_{Nr + \alpha }
\check{\nu }_{Ns + \beta }^* :
\stackrel{d}{=}
\hat{\nu }_{Nr + \alpha }
\check{\nu }_{Ns + \beta }^*
-\delta_{\alpha \beta } \delta_{rs} \qquad (s<0), \nonumber\\
:\psi_{Nr + \alpha } \psi_{Ns + \beta }^* :
\stackrel{d}{=}
\psi_{Nr + \alpha } \psi_{Ns + \beta }^*
-\delta_{\alpha \beta } \delta_{rs} \qquad (s<0).
\label{normalorderedop}
\end{gather}
We def\/ine important shift operators
$\Lambda_{j }$ and $\Lambda_{-j }$ $(j \in \mathbb{Z}_+)$
in the soliton theory on a group as
\begin{gather}
 \Lambda_{j }
 =
\uptau
\left(
\left[
\begin{array}{rrrrl}
0 & 1  &\cdots &\cdots &0  \\
\vdots &\ddots & 1  &  &\vdots  \\
\vdots & &\ddots &\ddots &\vdots  \\
\vdots & & &\ddots & 1   \\
z &\cdots &\cdots &\cdots & 0  \\
\end{array}
\right]^j
\right),\nonumber\\
\Lambda_{-j}
 =
\uptau
\left(
\left [
\begin{array}{rrrrl}
0 &\cdots &\cdots &\cdots &z^{-1} \\
1 &\ddots &  &  & \vdots  \\
\vdots & 1  &\ddots &  & \vdots  \\
\vdots & &\ddots &\ddots & \vdots  \\
0 &\cdots &\cdots & 1  & 0 \\
\end{array}
\right]^j
\right)
 =
\Lambda_j^\dagger ,
\label{lambdamatrix}
\end{gather}
which are images under $\uptau$ as
\[
\Lambda_j
 =
\uptau\left\{\left[ \sum_{\alpha = 1}^{N-1}
e_{\alpha ,\alpha +1}  +  ze_{N,1}\right]^j \right\}
 \simeq
\uptau\left\{\left[\sum_{\alpha = 1}^{N-1}
c_\alpha^\dagger c_{\alpha + 1}  +
c_n^\dagger c_1 z \right]^j\right\}.
\]
Through an easy calculation with the use of
(\ref{tau-expression}) and
(\ref{normalorderedop}),
the 2-{\it cocycle}~$\alpha$
on $\bar{a}_\infty$
(\ref{biggerLiealgebra}) induces
the following 2-{\it cocycle} on a pair of
basis elements of $\widetilde{gl}_n$:
\begin{gather}
\alpha (\uptau\{e_{\alpha \beta }(k)\},
\uptau\{e_{\gamma \delta }(l)\})
=
\delta_{\alpha \delta }
\delta_{\beta \gamma }  \delta_{k+l,0} \cdot k ,
\label{cocycle}
\end{gather}
where the Kac--Peterson 2-{\it cocycle}~$\alpha$
\cite{Mickel.89,Kac.83,KP.86}
is def\/ined as
\begin{gather*}
\alpha (E_{ij},~E_{kl})
 =
\delta_{jk} \delta_{il}~(i \leq 0)
 -
\delta_{li} \delta_{kj}~(j \leq 0)
 =
\left \{
\begin{array}{rl}
1 , & \mbox{for}~j=k,~j \geq 1,~i=l,~i \leq 0, \\
 -1 , &
 \mbox{for}~i=l,~i \geq 1,~j=k,~j \leq 0, \\
0, & \mbox{otherwise}.
\end{array}
\right.
\end{gather*}
Then we have for the shift operators
(\ref{lambdamatrix})
\begin{gather*}
\alpha (\Lambda _{k}, \Lambda _{l})
=\delta_{k+l,0} \cdot k .
\end{gather*}
For any elements $a(z)$ and $b(z)$
in the
$\widetilde{gl}(N)$
the formula
(\ref{cocycle})
is written as
\begin{gather*}
\alpha (\uptau\{a(z)\}, \uptau\{b(z)\})
=
\mbox{Res}_{0}\mbox{Tr}\,a'(z)b(z) ,
\end{gather*}
where $a'(z)$ is the derivative
of $a$ with respect to
$z$ and $\mbox{Res}_0$ is the residue at $z  =  0$,
i.e., the coef\/f\/icient of $z^{-1}$.
Investigation of
the highest weight representation of
$\widetilde{gl}(N)$
leads to its central extension
$
\widehat{gl}(N)
 =
\widetilde{gl}_N  +  {\mathbb{C}}  \cdot  c
$
in which
general elements
$a(z)$ and $b(z)$ and
center $\mathbb{C}  \cdot  c$
satisfy the KM brackets
\begin{gather}
 [a(z), c]_{\rm KM} = 0 , \nonumber\\
[a(z), b(z)]_{\rm KM}
=
[a(z), b(z)]
+
\left\{
\mbox{Res}_{0}\mbox{Tr}\,a'(z)b(z)
\right\}
\cdot c .
\label{Kac-Moody}
\end{gather}
The Lie algebra $\widehat{gl}(N)$
is called the af\/f\/ine KM algebra
associated with the Lie algebra $gl(N)$.
For simplicity consider
the level one case,
$
c|M \rangle  =  1  \cdot  |M \rangle
$.
Using the one-level formula
it is possible to rewrite (\ref{Kac-Moody}) as
\begin{gather*}
 X_a
=
\widehat{X}_a + {\mathbb{C}} \cdot c , \nonumber\\
\widehat{X}_a
\stackrel{d}{=}
\sum_{r=-\mathbb{N}}^{\mathbb{N}} \sum_{s \in \mathbb{Z}}
(a_r )_{\alpha \beta }
: \psi_{N(s-r) + \alpha } \psi_{Ns + \beta }^* : , \qquad
(a_r )_{\alpha \beta }
\equiv
(a_r )_{N(s-r) + \alpha, Ns + \beta } , \nonumber\\
[X_a, c]_{\rm KM} = 0 ,\qquad
[X_a, X_b ]_{\rm KM}
=
\widehat{X}_{[a, b]} + \alpha (a,b) \cdot c .
\end{gather*}
The matrices $a$ and $b$ are any elements of
(\ref{CorreMatrix}).
The $[a,b]$ denotes the Lie bracket of their matrices.
The $\uptau$-rep of $a(z)$ is given as
\begin{gather*}
\uptau\{a(z)\}
=
\sum_{r \in \mathbb{Z}} \sum_{s \in \mathbb{Z}}
(a_r )_{N(s-r) + \alpha, Ns + \beta }
\psi_{N(s-r) + \alpha } \psi_{Ns + \beta }^* ,
\end{gather*}
from which the commutator and the 2-{\it cocycle} $\alpha $ between
$\uptau\{a(z)\}$ and $\uptau\{b(z)\}$
are calculated as
\begin{gather*}
[\uptau\{a(z)\}, \uptau\{b(z)\}]
=
\sum_{r,s \in \mathbb{Z}}\sum_{t \in \mathbb{Z}}
([a_r, a_s])_{N(t-r-s) + \alpha, Ns + \beta }
\psi_{N(s-r) + \alpha } \psi_{Nt + \beta }^* ,\nonumber\\
\alpha (\uptau\{a(z)\}, \uptau\{b(z)\})
=
\sum_{r \in \mathbb{Z}} r\,\mbox{Tr}(a_rb_{-r}) .
\end{gather*}

\section[Schur polynomials and $\tau$-function]{Schur polynomials and $\boldsymbol{\tau}$-function}\label{appendixD}

 The Schur polynomials
$S_k (x)$ belonging to
${\mathbb{C}}(x_1 ,x_2 ,\dots )$  are def\/ined by
the generating function
\begin{gather}
\exp \sum_{k\geq 1}^\infty  x_k p^k
=
\sum_{k \geq 0}S_k (x) p^k .
\label{Schurpolynomials}
\end{gather}
The Schur polynomials and
their associated recursion formulas
play important roles to evaluate matrix elements
between two number-projected
Hartree--Bogoliubov states~\cite{Nishi.99}.
The Schur polynomial is related to
a symmetric function $h_k$
$\sum\limits_{k \geq 0} h_k p^k
=
\Pi_i^N (1- \epsilon_i p) ^{-1}$.
Then the Schur polynomial $S_k (x)$
is written as
\begin{gather*}
S_k (x)
=
h_k (\epsilon_1 , \epsilon_2 , \dots , \epsilon_N ) ,\qquad
x_j
=
\tfrac{1}{j}
\left(
\epsilon_1 ^j + \epsilon_2 ^j + \cdots + \epsilon_N ^j
\right)
\end{gather*}
and is given explicitly as
\begin{gather*}
S_0(x)=1 ,\qquad
S_1(x)=x_1, \qquad
S_2(x)=x_2 + \tfrac{1}{2} x_1^2 , \qquad
S_3(x)=x_3 + x_1x_2 + \tfrac{1}{6} x_1^3, \\
S_4(x)=x_4 + x_1x_3 + \tfrac{1}{2} x_2^2 + \tfrac{1}{2} x_1^2x_2
+ \tfrac{1}{24} x_1^4, \qquad \dots.
\end{gather*}

We here construct
an explicit form of the polynomials
in the bosonic Fock space which
under the $\sigma_M$ corresponds to
the f\/inite monomials of the~$F^{(M)}$.
It is given by
\begin{gather}
\sigma_M ; \nu_{i_M}
\wedge \nu_{i_{M-1}} \wedge \nu_{i_{M-2}}
\wedge \cdots \wedge \nu_{i_1}
\mapsto
S_{i_M-M,i_{M-1}-(M-1),i_{M-2}-(M-2),\dots,i_1-1}(x) ,
\label{semi-infinitemonomials}
\end{gather}
where $i_M > i_{M-1} >  \cdots  >i_1$.
The Schur polynomial
$
S_\lambda (x)
 =
S_{\lambda_1, \lambda_2,
\lambda_3, \dots, \lambda_k}(x)
$
is given as
\begin{gather*}
S_{\lambda_1, \lambda_2,
\lambda_3, \dots, \lambda_k}(x)
=
\det
\left|
\begin{array}{ccccc}
S_{\lambda_1}(x) &  S_{\lambda_1 +1}(x) &
S_{\lambda_1 +2}(x) &  \cdots &  S_{\lambda_1 +k-1}(x) \\
S_{\lambda_2 -1}(x) &  S_{\lambda_2}(x) &
S_{\lambda_2 +1}(x) &  \cdots &  S_{\lambda_2 +k-2}(x) \\
S_{\lambda_3 -2}(x) &  S_{\lambda_3 -1}(x) &
S_{\lambda_3}(x) &  \cdots & S_{\lambda_3 +k-3}(x) \\
\cdots &  \cdots &  \cdots &  \cdots &  \cdots \\
S_{\lambda_k +1-k}(x) &  S_{\lambda_k +2-k}(x) &  S_{\lambda_k +3-k}(x) &  \cdots &  S_{\lambda_k}(x)
\end{array}
\right| ,
\end{gather*}
where $\lambda$ denotes a partition
Par$\lambda  =  \{ \lambda_1 \geq
\lambda_2 \geq \cdots  \geq \lambda_k >0 \}$
\cite{Kac.83,KR.87}.
As for the contravariant hermitian form
(\ref{contravarianthermitianform}) in $B^{(M)}$
they form an orthogonal basis
$\langle S_\lambda |S_\mu \rangle
 =
\delta_{\lambda \mu }$.
For the special partition
${\rm Par}\,
\lambda
 =
1^M
 \equiv
\{
\lambda_1  =  1, \lambda_2  =  1 , \lambda_3  =  1, \lambda_4  =  1 , \dots , \lambda_M  =  1
\}
$,
i.e., the completely anti-symmetric Young diagram,
\begin{gather*}
S_{1^M} (x)
=
\det
\left|
\begin{array}{cccccc}
S_1 (x)   & S_2 (x)   & S_3 (x)   & S_4 (x)   & \cdots & S_M (x) \\
1         & S_1 (x)   & S_2 (x)   & S_3 (x)   & \cdots & S_{M - 1} (x) \\
0         & 1         & S_1 (x)   & S_2 (x)   & \cdots & S_{M - 2} (x) \\
\cdots    & \cdots    & \cdots    & \cdots    & \cdots & \cdots \\
0         & 0         & 0         & 0         & \cdots & S_1 (x)
\end{array}
\right|
=
(-1)^M S_M (- x) .
\end{gather*}
A group orbit of the highest weight vector
$|M \rangle$ under the action
$U(g)$ is mapped to a space of $\tau$-function
$
\tau_M (x, g)
=
\langle M|e^{H(x)}
U(g)|M \rangle
$.
Using
(\ref{semi-infinitemonomials}),
the Schur-polynomial expression
for $\tau$-function is given by
(\ref{Schur-polynomialexpressionforTau-function}).
Noting the relation
$\langle S_\lambda |S_\mu \rangle
 =
S_\lambda (\widetilde{\partial }_x  ) S_\mu (x)|_{x = 0}
 =
\delta_{\lambda \mu }$,
$v_{i_M ,i_{M-1}, \dots ,i_1 }^{M, M-1, \cdots ,1} (g)$
is obtained from
$\tau_M (x, g)$ as
\begin{gather*}
 v_{i_M,i_{M-1},\dots,i_1}^{M, M-1,\ dots,1}(g)
 =
v_{i_M ,i_{M-1},\dots,i_1}^{M, M-1,\dots,1} \{g(x)\}|_{x=0}
 =
S_{i_M-M, i_{M-1}-(M-1),\dots,i_1}
(\widetilde{\partial }_x)\tau_M (x, g)|_{x=0} ,\\
v\{g(x)\}
 \equiv
e^{H(x)}v(g) e^{-H(x)},\qquad
g_{ij}(x)
 =
\sum_{k \ge i}^\infty S_{k-i}(x)g_{kj}(x).
\end{gather*}
The coef\/f\/icient
$v_{i_M,i_{M-1},\dots,i_1}^{M,M-1,\dots,1}(g)$
is called the Pl\"{u}cker coordinate~\cite{Howe.97}.
In the soliton theory
on a~group the Pl\"{u}cker relations among
$v_{i_M,i_{M-1},\dots,i_1}^{M,M-1,\dots,1} \{g(x)\}$
correspond to the Hirota's forms
in the KP hierarchy.

\section{Hirota's bilinear equation}\label{appendixE}

According to Kac
\cite{Kac.83},
a vector $|\tau_M \rangle \in F^{(M)}$
belongs to the group orbit of
the highest weight vector (the vacuum)
if and only if it satisf\/ies
the bilinear identity equation
\begin{gather*}
\sum_{i \in \mathbb{Z}} \psi_i |\tau_M \rangle
\otimes \psi_i^* | \tau_M \rangle = 0   \quad
\Longleftrightarrow \quad
| \tau_M \rangle = U(\hat{g}) |M \rangle , \qquad
\hat{g} \in GL(\infty ) .
\end{gather*}
This identity is cast
into an inf\/inite set of nonlinear dif\/ferential
equations for
$
\tau_M (x)
{\stackrel{d}{=}}
\langle M|e^{H(x)} \!|\tau_M \!\rangle\!\!$
\begin{gather}
0
=
\frac{1}{2\pi i}  \oint   \frac{dp}{p}
\langle M+1|e^{H(x')}\Psi (p)|\tau_M \rangle
\otimes
\langle M-1|e^{H(x'')}\Psi^* (p)| \tau_M \rangle \label{Psi-expression} \\
\phantom{0} {}=
  \frac{1}{2\pi i}   \oint   dp
\exp \left\{  \sum_{j \geq 1} p^j (x'_j -x''_j)  \right\}
\exp
\left\{
-\sum_{j \geq 1}
  \frac{p^{-j}}{j}
\left(
\frac{\partial }{\partial x'_j}
 -
\frac{\partial }{\partial x''_j}
\right)
\right\}
\tau_M (x') \cdot \tau_M (x'') .
\nonumber
\end{gather}
It is described by
the Hirota's bilinear dif\/ferential operator as
\begin{gather*}
P(D)f \cdot g
\stackrel{d}{=}
P
\left(
\frac{\partial }{\partial y_1},
\frac{\partial }{\partial y_2},\dots
\right)
(f(x+y) \cdot g(x-y))|_{y=0} ,
\end{gather*}
where~$P(D)$  is a polynomial in
$D  =  (D_1 ,D_2 ,\dots )$.
Note that $Pf \cdot f \equiv 0$
if and only if~$P(D)  =  -P(-D)$.
Def\/ining new variables $x$ and $y$
by the relations $x'  =  x-y$ and $x''  =  x+y$,
with the help of
(\ref{Schurpolynomials})
and the notation
$\widetilde{D}  =  (D_1, \tfrac{1}{2}D_2 ,\dots)$,
(\ref{Psi-expression})
is brought to the form
\begin{gather}
\sum_{j \geq 0} S_j (-2y) S_{j+1} (\widetilde{D})
\exp \left(\sum_{s \geq 1} y_s D_s \right)
\tau_M (x) \cdot \tau_M (x) = 0 .
\label{Hirotasform}
\end{gather}
If we expand
(\ref{Hirotasform})
into a multiple Taylor series of variables
$y_1 ,y_2 ,\dots$ and
make each coef\/f\/icient of this series vanishing,
we get an inf\/inite set of nonlinear partial
dif\/ferential equation for the KP hierarchy.

\section[Calculation of commutators and 2-cocycles among operators $K$]{Calculation of commutators and 2-cocycles among\\ operators $\boldsymbol{K}$}\label{appendixF}

Commutators and 2-{\it cocycles} among operators
$K_{ 0}(r)$ and $K_{ \pm }(r)$
are calculated
using~(\ref{Liebracketgln})
and~(\ref{cocycle})
as follows:
\begin{gather*}
\left[K_0(r), K_\pm(s)\right]
=
\left[
\left(\tfrac{1}{2}\hat{h}_{2N}\right)_{\a \b }e_{\a \b }(r) ,
\left(
\BA{c}
\hat{e}_{2N} \\ \hat{f}_{2N}
\EA
\right)_{\g \delta }
e_{\g \delta }(s)
\right]  \nonumber \\
\phantom{\left[K_0(r), K_\pm(s)\right]}{} =
\left[
\left(\tfrac{1}{2}\hat{h}_{2N}\right)_{\a \b } \times
\left(
\BA{c}
\hat{e}_{2N} \\ \hat{f}_{2N}
\EA
\right)_{\g \delta }
\right]
\left\{
\delta_{\b \g }e_{\a \delta }(r + s)-\delta_{\a \delta }
e_{\g \b }(r + s)
\right\}  \nonumber \\
\phantom{\left[K_0(r), K_\pm(s)\right]}{} =
\left(
\tfrac{1}{2}\hat{h}_{2N}
\BA{c}
\hat{e}_{2N} \\ \hat{f}_{2N}
\EA \right)_{\a \delta }e_{\a \delta }(r + s)
-
\left(
\BA{c}
\hat{e}_{2N} \\ \hat{f}_{2N}
\EA \tfrac{1}{2}\hat{h}_{2N}
\right)_{\g \b }e_{\g \b }(r+s)  \nonumber \\
\phantom{\left[K_0(r), K_\pm(s)\right]}{} =
\left(
\left[
\tfrac{1}{2}\hat{h}_{2N},
\BA{c}
\hat{e}_{2N} \\ \hat{f}_{2N}
\EA
\right]
\right)_{\a \b }\!\! e_{\a\b }(r + s)
 =
\left(\!
\BA{c}
\hat{e}_{2N} \\ -\hat{f}_{2N}
\EA\!
\right)_{\a \b }\!e_{\a\b }(r + s)
 =
\pm K_\pm(r + s) ,
\\
\a\left\{K_0(r),K_\pm(s)\right\}
=
\a\left\{
\uptau\{K_0(r)\}, \uptau\{K_\pm(s)\}
\right\}  \nonumber \\
\phantom{\a\left\{K_0(r),K_\pm(s)\right\}}{} =
\a
\left\{
\left( \tfrac{1}{2}\hat{h}_{2N} \right)_{\a \b }\uptau\{e_{\a \b }(r)\},
\left(
\BA{c}
\hat{e}_{2N} \\ \hat{f}_{2N}
\EA
\right)_{\g \delta }
\uptau\{e_{\g \delta }(s)\}
\right\}  \nonumber \\
\phantom{\a\left\{K_0(r),K_\pm(s)\right\}}{} =
\left( \tfrac{1}{2}\hat{h}_{2N}\!\right)_{\a \b }
\left(
\BA{c}
\hat{e}_{2N} \\ \hat{f}_{2N}
\EA
\right)_{\g \delta }
\delta_{\a \delta }\delta_{\b \g}
\delta_{r+s,0}  \cdot  r\\
\phantom{\a\left\{K_0(r),K_\pm(s)\right\}}{}
 =
\tfrac{1}{2}{\rm Tr}
\left( \hat{h}_{2N}
\BA{c}
\hat{e}_{2N} \\ \hat{f}_{2N}
\EA
\right)   \delta_{r+s,0}  \cdot  r
 =  0 ,
\\
\left[K_+(r), K_-(s)\right]
 =
\left[
(\hat{e}_{2N})_{\a\b }e_{\a \b }(r),
(\hat{f}_{2N})_{\g \delta }e_{\g \delta }(s)
\right]  \nonumber \\
\phantom{\left[K_+(r), K_-(s)\right]}{} =
\left[
(\hat{e}_{2N})_{\a \b } \times (\hat{f}_{2N})_{\g \delta }
\right]
\left\{
\delta_{\b \g }e_{\a\delta }(r\!+\!s)-\delta_{\a \delta }
e_{\g \b }(r + s)
\right\}  \nonumber \\
\phantom{\left[K_+(r), K_-(s)\right]}{} =
(\hat{e}_{2N}\hat{f}_{2N})_{\a \delta }e_{\a \delta }(r + s)
-
(\hat{f}_{2N}\hat{e}_{2N})_{\g \b }e_{\g \b }(r + s)  \nonumber \\
\phantom{\left[K_+(r), K_-(s)\right]}{}=
\left(
[\hat{e}_{2N}, \hat{f}_{2N}]
\right)_{\a \b }e_{\a \b }(r + s)
 =
2\left(\tfrac{1}{2}\hat{h}_{2N}\right)_{\a \b }e_{\a \b }(r + s)
 =
2K_0(r + s) ,
\\
\a\left\{K_+(r), K_-(s)\right\}
 =
\a
\left\{
\uptau\{K_+(r)\}, \uptau\{K_-(s)\}
\right\}  \nonumber \\
\phantom{\a\left\{K_+(r), K_-(s)\right\}}{} =
\a
\left\{
(\hat{e}_{2N})_{\a \b }\uptau\{e_{\a \b }(r)\},
(\hat{f}_{2N})_{\g \delta }\uptau\{e_{\g \delta }(s)\}
\right\}  \nonumber \\
\phantom{\a\left\{K_+(r), K_-(s)\right\}}{} =
(\hat{e}_{2N})_{\a \b}(\hat{f}_{2N})_{\g \delta }\delta_{\a \delta }
\delta_{\b \g}\delta_{r+s,0}  \cdot  r
 =
{\rm Tr}
\big(
\hat{e}_{2N} \hat{f}_{2N}
\big)\delta_{r+s,0}  \cdot  r
 =
Nr \delta_{r+s,0} .
\end{gather*}
Commutators and 2-{\it cocycles} among operators
$X_k$ and $Y_k$ $(k=2r+1)$
are calculated
as follows:
\begin{gather*}
\left[X_k, X_l \right]
=
\left[Y_k, Y_l \right]
=
0,
\vspace{1mm}\\
\a
\left\{
 \binom{X_k}{Y_k} ,
  \binom{X_l}{Y_l}
\right\}
=
\a
\left\{
\uptau  \binom{X_k}{Y_k} ,
\uptau  \binom{X_l}{Y_l}
\right\}
(k=2r+1,\ l=2s+1) \vspace{1mm}\\
\qquad =
\a
\left\{
(\hat{e}_{2N})_{\a \b }\uptau
\left\{
e_{\a\b }(r)
\right\}
 \pm
(\hat{f}_{2N})_{\a \b }\uptau
\left\{
e_{\a \b }(r+1)
\right\} ,\right.\vspace{1mm}\\
\left.\qquad\quad{}
(\hat{e}_{2N})_{\g \delta }\uptau
\left\{
e_{\g \delta }(s)
\right\}
 \pm
(\hat{f}_{2N})_{\g \delta }\uptau
\left\{
e_{\g \delta }(s+1)
\right\}
\right\} \vspace{1mm}\\
\qquad {}=
(\hat{e}_{2N})_{\a \b }(\hat{e}_{2N})_{\g \delta }\delta_{\a \delta }
\delta_{\b \g}\delta_{r+s,0}  \cdot  r \\
\qquad\quad{}\pm (\hat{e}_{2N})_{\a\b }(\hat{f}_{2N})_{\g\delta }\delta_{\a \delta }
\delta_{\b \g}\delta_{r+s+1,0}  \cdot  r
\pm (\hat{f}_{2N})_{\a\b }(\hat{e}_{2N})_{\g\delta }\delta_{\a \delta }
\delta_{\b \g}\delta_{r+1+s,0}  \cdot  (r+1) \\
\qquad\quad{} + (\hat{f}_{2N})_{\a \b }(\hat{f}_{2N})_{\g \delta }\delta_{\a \delta }
\delta_{\b \g }\delta_{r+1+s+1,0}  \cdot  (r+1) \vspace{1mm}\\
\qquad{}=
\pm N(2r+1)  \cdot  \delta_{r+s+1,0}
=
\left\{
\BA{l}
N  \cdot  k\delta_{k+l,0} , \\
-N  \cdot  k\delta_{k+l,0} ,
\EA
\right.
\\
\left[X_k, Y_l \right]
 =
\big[
(\hat{e}_{2N})_{\a \b } e_{\a \b }(r)
 +
(\hat{f}_{2N})_{\a \b } e_{\a \b }(r  +  1) ,
(\hat{e}_{2N})_{\g \delta } e_{\g \delta }(s)
 -
(\hat{f}_{2N})_{\g \delta } e_{\g \delta }(s \!+\! 1)
\big] \nonumber \vspace{1mm}\\
\phantom{\left[X_k, Y_l \right]}{} =
\big[
(\hat{e}_{2N})_{\a \b } \times (\hat{e}_{2N})_{\g \delta }
\big]
\left\{
\delta_{\b \g } e_{\a \delta }(r+s)-\delta_{\a \delta }
e_{\g \b }(r+s)
\right\}  \nonumber \vspace{1mm}\\
\phantom{\left[X_k, Y_l \right]=}{}
-
\big[
(\hat{e}_{2N})_{\a \b } \times (\hat{f}_{2N})_{\g \delta }
\big]
\left\{
\delta_{\b \g } e_{\a \delta }(r+s+1)-\delta_{\a \delta }
e_{\g \b }(r+s+1)
\right\}  \nonumber \vspace{1mm}\\
\phantom{\left[X_k, Y_l \right]=}{}
+
\big[
(\hat{f}_{2N})_{\a \b } \times (\hat{e}_{2N})_{\g \delta }
\big]
\left\{
\delta_{\b \g } e_{\a \delta }(r+1+s)-\delta_{\a \delta }
e_{\g \b }(r+1+s)
\right\}   \nonumber \vspace{1mm}\\
\phantom{\left[X_k, Y_l \right]=}{}
-
\big[
(\hat{f}_{2N})_{\a \b } \times (\hat{f}_{2N})_{\g \delta }
\big]
\left\{
\delta_{\b \g } e_{\a \delta }(r  +  1  +  s  +  1)
 -
\delta_{\a \delta }
e_{\g \b }(r  +  1  +  s  +  1)
\right\}  \nonumber \vspace{1mm}\\
\phantom{\left[X_k, Y_l \right]}{} =
(\hat{e}_{2N}\hat{e}_{2N})_{\a \delta }e_{\a \delta }(r+s)
-(\hat{e}_{2N}\hat{e}_{2N})_{\g \b }e_{\g \b }(r+s)  \nonumber \vspace{1mm}\\
\phantom{\left[X_k, Y_l \right]=}{}
-
(\hat{e}_{2N}\hat{f}_{2N})_{\a \delta }e_{\a \delta }(r+s+1)
+
(\hat{f}_{2N}\hat{e}_{2N})_{\g \b }e_{\g \b }(r+s+1)  \nonumber \vspace{1mm}\\
\phantom{\left[X_k, Y_l \right]=}{}
+
(\hat{f}_{2N}\hat{e}_{2N})_{\a \delta }e_{\a \delta }(r+1+s)
-
(\hat{e}_{2N}\hat{f}_{2N})_{\g \b }e_{\g \b }(r+1+s)  \nonumber \vspace{1mm}\\
\phantom{\left[X_k, Y_l \right]=}{}
-
(\hat{f}_{2N}\hat{f}_{2N})_{\a\delta }e_{\a\delta }(r+1+s+1)
+
(\hat{f}_{2N}\hat{f}_{2N})_{\g \b }e_{\g \b }(r+1+s+1)  \nonumber \vspace{1mm}\\
\phantom{\left[X_k, Y_l \right]}{} =
-
\big(
\big[
\hat{e}_{2N}, \hat{f}_{2N}
\big]
\big)_{\a \b }e_{\a \b }(r  +  s  +  1)
+
\big(
\big[\hat{f}_{2N}, \hat{e}_{2N}
\big]
\big)_{\a \b }e_{\a \b }(r  +  s  +  1) \nonumber \vspace{1mm}\\
\phantom{\left[X_k, Y_l \right]}{} =
2(-\hat{h}_{2N})_{\a \b }e_{\a \b }(r+s+1)
= 2Y_{2r+1+2s+1}
= 2Y_{k+l} ,
\\
\left[X_k, Y_{2s} \right]
 =
\big[
(\hat{e}_{2N})_{\a\b } e_{\a \b }(r)
 +
(\hat{f}_{2N})_{\a \b } e_{\a \b }(r + 1),
(-\hat{h}_{2N})_{\g \delta } e_{\g \delta }(s)
\big] \
(Y_{2r}
 \equiv
(-\hat{h}_{2N})_{\a \b }e_{\a \b }(r)) \nonumber \vspace{1mm}\\
\phantom{\left[X_k, Y_{2s} \right]}{} =
\big[
(\hat{e}_{2N})_{\a \b },
(-\hat{h}_{2N})_{\g \delta }
\big]
\left\{\delta_{\b \g } e_{\a \delta }(r+s)-\delta_{\a \delta }
e_{\g \b }(r+s)
\right\}  \nonumber \vspace{1mm}\\
\phantom{\left[X_k, Y_{2s} \right]=}{}
+
\big[
(\hat{f}_{2N})_{\a \b },
(-\hat{h}_{2N})_{\g \delta }
\big]
\left\{
\delta_{\b \g } e_{\a \delta }(r+1+s)-\delta_{\a \delta }
e_{\g \b }(r+1+s)
\right\}   \nonumber \vspace{1mm}\\
\phantom{\left[X_k, Y_{2s} \right]}{} =
-(\hat{e}_{2N}\hat{h}_{2N})_{\a \delta }e_{\a \delta }(r+s)
+(\hat{h}_{2N}\hat{e}_{2N})_{\g \b }e_{\g \b }(r+s)  \nonumber \vspace{1mm}\\
\phantom{\left[X_k, Y_{2s} \right]=}{}
-
(\hat{f}_{2N}\hat{h}_{2N})_{\a \delta }e_{\a \delta }(r+s+1)
+
(\hat{h}_{2N}\hat{f}_{2N})_{\g \b }e_{\g \b }(r+s+1)  \nonumber \vspace{1mm}\\
\phantom{\left[X_k, Y_{2s} \right]}{}=
\big(
\big[
\hat{h}_{2N}, \hat{e}
\big]
\big)_{\a\b }e_{\a\b }(r+s)
+
\big(
\big[
\hat{h}, \hat{f}_{2N}
\big]
\big)_{\a \b }e_{\a \b }(r+s+1) \nonumber \vspace{1mm}\\
\phantom{\left[X_k, Y_{2s} \right]}{}=
2
\big\{
(\hat{e}_{2N})_{\a \b }e_{\a \b }(r+s)-(\hat{f}_{2N})_{\a \b }
e_{\a\b }(r+s+1)
\big\}
=
2 Y_{2(r+s)+1}
=
2Y_{k+2s} ,
\\
 \a(X_{k},Y_{2s})
=
\a
\left\{
\uptau(X_{k}), \uptau(Y_{2s})
\right\}  \nonumber \vspace{1mm}\\
\phantom{\a(X_{k},Y_{2s})}{} =
\a
\big\{
(\hat{e}_{2N})_{\a \b } \uptau
\left\{
e_{\a \b }(r)
\right\}
 +
(\hat{f}_{2N})_{\a \b } \uptau
\left\{
e_{\a \b }(r  +  1)
\right\},
(-\hat{h}_{2N})_{\g \delta }\uptau
\left\{
e_{\g \delta }(s)
\right\}
\big\}  \nonumber \vspace{1mm}\\
\phantom{\a(X_{k},Y_{2s})}{}=
(\hat{e}_{2N})_{\a \b }(-\hat{h}_{2N})_{\g \delta } \delta_{\a \delta }
\delta_{\b\g}\delta_{r+s,0}  \cdot  r\vspace{1mm}\\
\phantom{\a(X_{k},Y_{2s})=}{}
 +
(\hat{f}_{2N})_{\a \b }(-\hat{h}_{2N})_{\g \delta }\delta_{\a \delta }
\delta_{\b \g }\delta_{r+1+s,0}  \cdot  (r  + 1)  =  0 ,
\\
\left[Y_k,Y_l \right]
 =
\big[
(\hat{e}_{2N})_{\a \b } e_{\a \b }(r)
 -
(\hat{f}_{2N})_{\a \b } e_{\a \b }(r  +  1),
(-\hat{h}_{2N})_{\g \delta } e_{\g \delta }(s)
\big]   \
(k  =  2r  +  1, \ l  =  2s) \nonumber \vspace{1mm}\\
\phantom{\left[Y_k,Y_l \right]}{} =
\big[
(\hat{e}_{2N})_{\a \b } \times (-\hat{h}_{2N})_{\g \delta }
\big]
\left\{
\delta_{\b \g } e_{\a \delta }(r+s)-\delta_{\a \delta }
e_{\g \b }(r+s)\right\}  \nonumber \vspace{1mm}\\
\phantom{\left[Y_k,Y_l \right]=}{}
-
\big[
(\hat{f}_{2N})_{\a \b } \times (-\hat{h}_{2N})_{\g \delta }
\big]
\left\{
\delta_{\b \g } e_{\a \delta }(r+1+s)-\delta_{\a \delta }
e_{\g \b }(r+1+s)
\right\}  \nonumber \vspace{1mm}\\
\phantom{\left[Y_k,Y_l \right]}{}=
\big(
\big[
\hat{h}_{2N}, \hat{e}_{2N}
\big]
\big)_{\a \b }e_{\a \b }(r+s)
-
\big(
\big[
\hat{h}_{2N}, \hat{f}_{2N}
\big]
\big)_{\a \b }e_{\a \b }(r+s+1) \nonumber \\
\phantom{\left[Y_k,Y_l \right]}{} =
2
\big\{
(\hat{e}_{2N})_{\a \b }e_{\a \b }(r+s) + (\hat{f}_{2N})_{\a \b }
e_{\a \b }(r+s+1)
\big\}
=
2X_{k+l} ,
\vspace{1mm}\\
\a
\left(
\frac{1}{\sqrt{N}}Y_{2r},
\frac{1}{\sqrt{N}}Y_{2s}
\right)
=
2r\delta_{r+s,0}  , \qquad
\left[
\frac{1}{\sqrt{N}}Y_{2r},
\frac{1}{\sqrt{N}}Y_{2s}
\right]_{\rm KM}
=
2r  \cdot  \delta_{r+s,0}  \cdot  c ,
\vspace{1mm}\\
\sigma_K: \ \
 \frac{1}{\sqrt{N}}Y_{2r}
\to
\frac{\partial }{\partial y_{2r}} , \qquad
\sigma_K: \ \
\frac{1}{\sqrt{N}}
Y_{-2r}
 \to
2r y_{2r} ,
\vspace{1mm}\\
\a(X_{k},Y_{2s})
=
0 , \qquad
\left[
\frac{1}{\sqrt{N}}X_{k},
\frac{1}{\sqrt{N}}Y_{2s}
\right]_{\rm KM}\!\!
=
\frac{2}{\sqrt{N}}\frac{1}{\sqrt{N}}Y_{k+2s}
 \cdot  c ,
\vspace{1mm}\\
\a\left(
\frac{1}{\sqrt{N}}Y_k,
\frac{1}{\sqrt{N}}Y_l
\right)\!
 =
-k\delta_{k+l,0} = 0  \  (k  \neq  -l) , \!\qquad
\left[
 \frac{1}{\sqrt{N}}Y_{k},
 \frac{1}{\sqrt{N}}Y_{l}
\right]_{\rm KM}\!\!
=
\frac{2}{\sqrt{N}}\frac{1}{\sqrt{N}}X_{k+l}
\cdot c .
\end{gather*}

\section[Sum-rules for $2(Y_i + Y_{-i})$]{Sum-rules for $\boldsymbol{2(Y_i + Y_{-i})}$}\label{appendixG}


First we give the vertex operator for $\Gamma(p)$
in terms of the Schur polynomials
\begin{gather*}
\Gamma(p)
=
\exp
\left\{
\sum_{m_{\rm odd}  \geq  1}   p^m (2x_m)
\right\}
\exp
\left\{
\sum_{n_{\rm odd}  \geq  1}
\left(\frac{1}{p}\right)^n (-2\widetilde{\partial }_n)
\right\}\nonumber\\
\phantom{\Gamma(p)}{}
 =
\sum_{l \in \mathbb{Z}}   \sum_{n=0}^\infty
S_{n-l}(2x)   S_n(-2\widetilde{\partial })p^{ -l} ,
\end{gather*}
According to Kac
\cite{Kac.83},
we have
\begin{gather}
\sum_{l \in \mathbb{Z}}p^{-l} Y_l
 \to
\tfrac{1}{2}
(\Gamma(p)  -  1),\nonumber\\
2 Y_l  +  \delta_{l0}
 =
\sum_{n=0}^\infty
S_{n-l}(2x) S_n(-2\widetilde{\partial }) \qquad
(x_n  =  \widetilde{\partial }_n  =  0 ~\textrm{for even}~n) .
\label{YSchurdelSchur}
\end{gather}
Then using
(\ref{YSchurdelSchur}),
the sum-rules for
$2(Y_i + Y_{-i})$
are derived as follows:
\begin{gather*}
2 Y_0
=
S_1(2x) S_1(-2\widetilde{\partial })
+
S_2(2x) S_2(-2\widetilde{\partial })
+
S_3(2x) S_3(-2\widetilde{\partial }) \\
\phantom{2 Y_0}{}
+
S_4(2x) S_4(-2\widetilde{\partial })
+
S_5(2x) S_5(-2\widetilde{\partial })
+
S_6(2x) S_6(-2\widetilde{\partial })
+
\cdots ,
\\
2 (Y_{-2} + Y_{2})
=
\sum_{n=0}^{\infty }
\left(
S_{n+2}(2x)
+
S_{n-2}(2x)
\right)
S_n(-2\widetilde{\partial }) \\
\phantom{2 (Y_{-2} + Y_{2})}{} =
S_2(2x)  + S_2(-2\widetilde{\partial })
+ S_{3}(2x) S_1(-2\widetilde{\partial })
+ S_{1}(2x) S_3(-2\widetilde{\partial }) + S_{4}(2x) S_2(-2\widetilde{\partial })  \\
\phantom{2 (Y_{-2} + Y_{2})=}{}
+ S_{2}(2x) S_4(-2\widetilde{\partial })
 + S_{5}(2x) S_3(-2\widetilde{\partial })
 + S_{6}(2x) S_4(-2\widetilde{\partial })
 + \cdots ,
\\
2 (Y_{-4} + Y_{4})
=
\sum_{n=0}^{\infty }
\left( S_{n+4}(2x)
+
S_{n-4}(2x)
\right)
S_n(-2\widetilde{\partial }) \\
\phantom{2 (Y_{-4} + Y_{4})}{} =
S_4(2x)  + S_4(-2\widetilde{\partial })
+ S_{5}(2x) S_1(-2\widetilde{\partial })
 + S_{6}(2x) S_2(-2\widetilde{\partial })   \\
\phantom{2 (Y_{-4} + Y_{4})=}{} + S_{7}(2x) S_3(-2\widetilde{\partial })
 + S_{8}(2x) S_4(-2\widetilde{\partial })
  + \cdots ,
\\
2 (Y_{-1} + Y_{1})
=
\sum_{n=0}^{\infty }
\left( S_{n+1}(2x)
+
S_{n-1}(2x)
\right)
S_n(-2\widetilde{\partial }) \\
\phantom{2 (Y_{-1} + Y_{1})}{} =
S_1(2x)   + S_1(-2\widetilde{\partial })
  + S_{2}(2x) S_1(-2\widetilde{\partial })
+ S_{1}(2x) S_2(-2\widetilde{\partial })      \\
\phantom{2 (Y_{-1} + Y_{1})=}{}+ S_{3}(2x) S_2(-2\widetilde{\partial })
+ S_{2}(2x) S_3(-2\widetilde{\partial })
+ S_{4}(2x) S_3(-2\widetilde{\partial }) + \cdots ,
\\
2 (Y_{-3} + Y_{3})
=
\sum_{n=0}^{\infty }
\left(
S_{n+3}(2x) + S_{n-3}(2x)
\right)
S_n(-2\widetilde{\partial }) \\
\phantom{2 (Y_{-3} + Y_{3})}{} =
S_3(2x)   + S_3(-2\widetilde{\partial })
+ S_{4}(2x) S_1(-2\widetilde{\partial })
+ S_{1}(2x) S_4(-2\widetilde{\partial })  \\
\phantom{2 (Y_{-3} + Y_{3})=}{}  + S_{5}(2x) S_2(-2\widetilde{\partial })
 + S_{6}(2x) S_3(-2\widetilde{\partial })
  + S_{7}(2x) S_4(-2\widetilde{\partial })
  + \cdots ,
\\
2 (Y_{-5} + Y_{5})
=
\sum_{n=0}^{\infty }
\left(
S_{n+5}(2x) + S_{n-5}(2x)
\right)
S_n(-2\widetilde{\partial }) \\
\phantom{2 (Y_{-5} + Y_{5})}{} =
S_5(2x)
+ S_{6}(2x) S_1(-2\widetilde{\partial })
 + S_{7}(2x) S_2(-2\widetilde{\partial })     \\
\phantom{2 (Y_{-5} + Y_{5})=}{}+ S_{8}(2x) S_3(-2\widetilde{\partial })
   + S_{9}(2x) S_4(-2\widetilde{\partial })
  + \cdots ,
\\
\cdots\cdots\cdots\cdots\cdots\cdots\cdots\cdots\cdots\cdots\cdots\cdots\cdots\cdots\cdots\cdots\cdots\cdots\cdots\cdots
\cdots
\end{gather*}

\section[Expression for $g_{{\rm Y}_{-(2i+1)}+ {\rm  Y}_{(2i+1)}}(z)$ in terms of Bessel functions]{Expression for $\boldsymbol{g_{{\rm Y}_{-(2i+1)}+ {\rm  Y}_{(2i+1)}}(z)}$ in terms of Bessel functions}\label{appendixH}

Using the formula for
$(Y_{-(2i+1)}  +  Y_{(2i+1)})$,
\begin{gather}
Y_{-(2i+1)}  +  Y_{(2i+1)}
 =
( \hat{e}_{2N} )_{\a \b }e_{\a \b }( - i  -  1)
 -
( \hat{f}_{2N} )_{\a \b }e_{\a \b }(-i)\nonumber\\
\phantom{Y_{-(2i+1)}  +  Y_{(2i+1)}=}{}
 +
( \hat{e}_{2N} )_{\a \b }e_{\a \b }(i)
 -
( \hat{f}_{2N} )_{\a \b }e_{\a \b }(i  +  1) , \nonumber\\
\hat{e}_{2N} z^{-i - 1} - \hat{f}_{2N}z^{-i}
+
\hat{e}_{2N} z^{i} - \hat{f}_{2N} z^{i+1}
=
\hat{e}_{2N} (z^{-i - 1} + z^{i})
-
\hat{f}_{2N} (z^{-i} + z^{i+1}) \nonumber\\
\phantom{\hat{e}_{2N} z^{-i - 1} - \hat{f}_{2N}z^{-i}
+
\hat{e}_{2N} z^{i} - \hat{f}_{2N} z^{i+1}}{} =
z^{-i}(z^{-1}\hat{e}_{2N} - \hat{f}_{2N})(1+z^{2i + 1}) ,
\label{Y2iplus1}
\end{gather}
we express straightforwardly the
$
g_{{\rm Y}_{-(2i+1)}
+{\rm Y}_{(2i+1)}}(z)
$,
$\uptau$ rep of
(\ref{Y2iplus1}),
in terms of the Bessel functions as,
\begin{gather*}
g_{{\rm Y}_{-(2i+1)}
+{\rm Y}_{(2i+1)}}(z)
 =
\uptau
\left[
\left\{
e^{ \frac{\theta }{4}z^{-i}(z^{-1}\hat{e}_{2N}
-
\hat{f}_{2N})(1+z^{2i+1})}
\right\}_{\a \b } e_{\a \b }
\right] \\
 =
\uptau
\left[
\left\{
\left( \tfrac{\theta }{4} \right)^0
\left[ \!\!
\BA{cc}
1_N & 0\\ 0 & 1_N
\EA  \!\!
\right]
+
\tfrac{1}{2!}
\left( \tfrac{\theta }{4} \right)^2
\left[  \!\!
\BA{cc}
-z^{-(2i+1)}(1+z^{2i+1})^2  \cdot 1_N & 0 \\
0 & -z^{-(2i+1)}(1+z^{2i+1})^2  \cdot 1_N
\EA  \!\!
\right]\!
\right.
\right. \\
+
\tfrac{1}{4!}
\left( \tfrac{\theta }{4} \right)^4
\left[
\BA{cc}
(z^{-(2i+1)})^2(1+z^{2i+1})^4  \cdot 1_N & 0 \\
0 & (z^{-(2i+1)})^2(1+z^{2i+1})^4  \cdot 1_N
\EA
\right]
+
\cdots \\
+
\left( \tfrac{\theta }{4} \right)^1
z^{-i}(z^{-1}\hat{e}_{2N} - \hat{f}_{2N})(1+z^{2i+1})
+
\tfrac{1}{3!}
\left( \tfrac{\theta }{4} \right)^3
(-z^{-3i-1})(z^{-1}\hat{e}_{2N} - \hat{f}_{2N})(1+z^{2i+1})^3 \\
\left.
\left.
+
\tfrac{1}{5!}
\left( \tfrac{\theta }{4} \right)^5
z^{-5i-2}(z^{-1}\hat{e}_{2N} - \hat{f}_{2N})(1+z^{2i+1})^5
+
\cdots
\right\}_{\a \b } e_{\a \b }
\right]  \\
=
\uptau
\left[
\left[
\left\{
\left( \tfrac{\theta }{4} \right)^0 1
 +
\tfrac{1}{2!}
\left( \tfrac{\theta }{4} \right)^2
(-z^{-(2i+1)})(1+z^{2i+1})^2\!
 +
\tfrac{1}{4!}
\left( \tfrac{\theta }{4} \right)^4
(z^{-(2i+1)})^2(1+z^{2i+1})^4
 +
\cdots \!
\right\}
\hat{I}_{2N}\!
\right.
\right. \\
+
z^i
\left\{
\left( \tfrac{\theta }{4} \right)^1
z^{-(2i+1)}(1+z^{2i+1})
+
\tfrac{1}{3!}
\left( \tfrac{\theta }{4} \right)^3
(-z^{-(2i+1)})^2(1+z^{2i+1})^3
\right. \\
\left.
+
\tfrac{1}{5!}
\left( \tfrac{\theta }{4} \right)^5
(z^{-(2i+1)})^3(1+z^{2i+1})^5
+
\cdots
\right\}
\hat{e}_{2N} \\
-
z^{i+1}
\left\{
\left( \tfrac{\theta }{4} \right)^1
z^{-(2i+1)}(1+z^{2i+1})
+
\tfrac{1}{3!}
\left( \tfrac{\theta }{4} \right)^3
(-z^{-(2i+1)})^2(1+z^{2i+1})^3
\right. \\
\left.
\left.
\left.
+
\tfrac{1}{5!}
\left( \tfrac{\theta }{4} \right)^5
(z^{-(2i+1)})^3(1+z^{2i+1})^5
+
\cdots
\right\}
\hat{f}_{2N}
\right]_{\a \b }e_{\a \b }
\right] \\
=
\sum_{r \geq 0}(-1)^r J_{2r}
\left( \tfrac{\theta }{2} \right)
(\hat{{\rm I}}_{2N})_{\a \b }\uptau
\left\{e_{\a \b }((2i\!+\!1)r)  +  e_{\a \b }(-(2i\!+\!1)r)
\right\}
 -
J_0
\left( \tfrac{\theta }{2} \right)
(\hat{{\rm I}}_{2N})_{\a \b }\uptau
\left\{
e_{\a \b }(0)
\right\} \\
+
\sum_{r \geq 0}(-1)^r J_{2r+1}
\left(\tfrac{\theta }{2} \right)
(\hat{e}_{2N})_{\a \b }\uptau
\left\{
e_{\a \b }((2i+1)r+i) + e_{\a \b }(-(2i+1)r-(i+1))
\right\} \\
-
\sum_{r \geq 0}(-1)^r J_{2r+1}
\left( \tfrac{\theta }{2} \right)
(\hat{f}_{2N})_{\a \b }\uptau
\left\{
e_{\a \b }((2i+1)r+i+1) + e_{\a \b }(-(2i+1)r-i)
\right\} ,
\end{gather*}
where we have used the relations
\begin{gather*}
\sum_{r \geq 0} (-1)^r J_{2r}
\left( \tfrac{\theta }{2} \right)
\big\{
z^{(2i+1)r}
 +
z^{-(2i+1)r}
\big\}
=
\sum_{r \in \mathbb{Z}} (-1)^{r} J_{2r}
\left( \tfrac{\theta }{2} \right)
z^{(2i+1)r} , \\
\sum_{r \geq 0} (-1)^r J_{2r+1}
\left( \tfrac{\theta }{2} \right)
\big\{
z^{(2i+1)r+i}
 +
z^{-(2i+1)r-(i+1)}
\big\}
=
\sum_{r \in \mathbb{Z}} (-1)^r J_{2r+1}
\left( \tfrac{\theta }{2} \right)
z^{(2i+1)r+i} , \\
\sum_{r \geq 0} (-1)^r J_{2r+1}
\left( \tfrac{\theta }{2} \right)
\big\{
z^{(2i+1)r+i+1}
 +
z^{-(2i+1)r-i}
\big\}
=
\sum_{r \in \mathbb{Z}} (-1)^r J_{2r+1}
\left( \tfrac{\theta }{2} \right)
z^{-(2i+1)r-i}.
\end{gather*}
Finally we can express the
$
{\mbox{\it \large g}}_{\mbox{{\scriptsize Y}}_{-(2i+1)}
+\mbox{{\scriptsize Y}}_{(2i+1)}}(z)
$
as
\begin{gather*}
g_{{\rm Y}_{-(2i+1)}
+{\rm Y}_{(2i+1)}}(z)
=
\uptau
\left[
\left[
\sum_{r \in \mathbb{Z}}(-1)^r J_{2r}
\left( \tfrac{\theta }{2} \right)
z^{(2i+1)r}\hat{I}_{2N}
\right.
\right. \\
\left.
\left.
+
\sum_{r \in \mathbb{Z}}(-1)^r J_{2r+1}
\left( \tfrac{\theta }{2} \right)
z^{(2i+1)r+i}\hat{e}_{2N}
-
\sum_{r \in \mathbb{Z}}(-1)^r J_{2r+1}
\left( \tfrac{\theta }{2} \right)
z^{-(2i+1)r-i}\hat{f}_{2N}
\right]_{\a \b }e_{\a \b }
\right] \\
=
\sum_{r \in \mathbb{Z}}(-1)^r J_{2r}
\left( \tfrac{\theta }{2} \right)
(\hat{I}_{2N})_{\a \b }\uptau
\left\{
e_{\a \b }((2i+1)r)
\right\} \\
+
\sum_{r \in \mathbb{Z}}(-1)^r J_{2r+1}
\left( \tfrac{\theta }{2} \right)
\left[
(\hat{e}_{2N})_{\a \b }\uptau
\left\{
e_{\a \b }((2i+1)r+i)
\right\}
 -
(\hat{f}_{2N})_{\a \b }\uptau
\left\{e_{\a \b }(-(2i+1)r-i)
\right\}
\right] .
\end{gather*}
This is the f\/irst expression which we give
in terms of the Bessel functions.

\section[Properties of $SU(2N)_\infty$ transformation matrix]{Properties of $\boldsymbol{SU(2N)_\infty}$ transformation matrix}\label{appendixI}

In
(\ref{transhatgphaipsi})
the $2N$-dimensional matrix
$g_{\varphi_s}$
is expressed as
\begin{gather}
g_{\varphi_s}
=
\hat{1}_{2N}
 +
i
\left(-\frac{\varphi_s}{2}\right)\hat{h}_{2N}
 +
\frac{i^2}{2!}
\left(-\frac{\varphi_s}{2}\right)^2\hat{h}_{2N}^2
 +
\frac{i^3}{3!}
\left(-\frac{\varphi_s}{2}\right)^3\hat{h}_{2N}^3
 +
\frac{i^4}{4!}
\left(-\frac{\varphi_s}{2}\right)^4\hat{h}_{2N}^4
+
\cdots \nonumber\\
\phantom{g_{\varphi_s}}{} =
\left\{
1
-
\frac{1}{2!}
\left(-\frac{\varphi_s}{2}\right)^2
+
\frac{1}{4!}
\left(-\frac{\varphi_s}{2}\right)^4
+
\cdots
\right\}
\hat{1}_{2N}\nonumber\\
\phantom{g_{\varphi_s}=}{}
+
i
\left\{
\left(-\frac{\varphi_s}{2}\right)
-
\frac{1}{3!}
\left(-\frac{\varphi_s}{2}\right)^3
+
\frac{1}{5!}
\left(-\frac{\varphi_s}{2}\right)^5
+
\cdots
\right\}
\hat{h}_{2N} \nonumber\\
\phantom{g_{\varphi_s}}{} =
\cos
\left( -  \frac{\varphi_s}{2}  \right)
\hat{1}_{2N}
 +
i\sin
\left( -  \frac{\varphi_s}{2}  \right)
\hat{h}_{2N}
=
\left[
\BA{cc}
e^{-i\frac{\varphi_s}{2}}
 \cdot  1_N &  0  \\
0 &  e^{i\frac{\varphi_s}{2}}
 \cdot  1_N
\EA
\right] .
\label{gphi}
\end{gather}
Adjoint actions
$\widehat{K}_\pm(\theta)~\mbox{for}~\psi_{Nr\!+\!\a}$
are computed as
\begin{gather}
\big[ \widehat{K}_+(\theta),\psi_{Nr + \a} \big]
 =
\sum_{s \in \mathbb{Z}}\psi_{N(s-r) + \b }
\left(  - \theta_r \hat{e}_{2N}   \right)_{\b \a} ,\nonumber\\
\big[ \widehat{K}_-(\theta),\psi_{Nr + \a} \big]
 =
\sum_{s \in \mathbb{Z}}\psi_{N(s-r) + \b }
\big(  - \theta_{-r} \hat{f}_{2N}   \big)_{\b \a}   ,
\label{adjointactionKpm}
\end{gather}
from which
we obtain
\begin{gather*}
\big[
\widehat{K}_+(\theta) - \widehat{K}_-(\theta) , \psi_{Nr + \a}
\big]
=
\sum_{s \in \mathbb{Z}}\psi_{N(s-r) + \b }
\left(- \theta_r \right)
\big(\hat{e}_{2N} - \hat{f}_{2N}\big)_{ \b \a}   \qquad
(\theta_{-r} = \theta_{r}) .
\end{gather*}
We also have the relations
\begin{gather}
(\hat{e}_{2N} - \hat{f}_{2N})^2
=
- \hat{I}_{2N} , \qquad
(\hat{e}_{2N} - \hat{f}_{2N})^3
= - (\hat{e}_{2N} - \hat{f}_{2N}) , \nonumber\\
(\hat{e}_{2N} - \hat{f}_{2N})^4
=
\hat{I}_{2N} ,\qquad
\dots .
\label{efrelations}
\end{gather}
Using
(\ref{adjointactionKpm}) and (\ref{efrelations}),
the transformed
$\psi_{Nr + \a}(\hat{g}_{\theta })$
is shown to be
(\ref{transformedfermionoperatr}).
The $2N$-dimensional matrix
$g_{\theta_s}$
is expressed as
\begin{gather}
g_{\theta_s}
 =
\hat{1}_{2N}
 +
\left( -\frac{\theta_s}{2} \right)
(\hat{e}_{2N}  -  \hat{f}_{2N})
 +
\frac{1}{2!}\left( -\frac{\theta_s}{2} \right)^2
(\hat{e}_{2N}  -  \hat{f}_{2N})^2
 +
\frac{1}{3!}\left( -\frac{\theta_s}{2} \right)^3
(\hat{e}_{2N}  -  \hat{f}_{2N})^3 \nonumber \\
\phantom{g_{\theta_s}=}{}
+
\frac{1}{4!}\left(-\frac{\theta_s}{2}\right)^4
(\hat{e}_{2N} - \hat{f}_{2N})^4
+
\frac{1}{5!}\left(-\frac{\theta_s}{2}\right)^5
(\hat{e}_{2N} - \hat{f}_{2N})^5
+
\cdots  \nonumber \\
\phantom{g_{\theta_s}}{}
=
\left\{
1  -  \frac{1}{2!}( -\frac{\theta_s}{2} )^2
 +
\frac{1}{4!}( -\frac{\theta_s}{2} )^4
 +
 \cdots
\right\}
\hat{1}_{2N}\label{gtheta}\\
\phantom{g_{\varphi_s}=}{}
 +
\left\{
\left( -\frac{\theta_s}{2}\right)  -  \frac{1}{3!}\left( -\frac{\theta_s}{2} \right)^3
 +
\frac{1}{5!}\left( -\frac{\theta_s}{2}\right)^5
 +
 \cdots
\right\}
(\hat{e}_{2N}  -  \hat{f}_{2N}) \nonumber \\
\phantom{g_{\varphi_s}}{} =
\cos
\left( -\frac{\theta_s}{2} \right)
 \cdot
\hat{1}_{2N}
 +
\sin
\left( -\frac{\theta_s}{2} \right)
\cdot(\hat{e}_{2N}  -  \hat{f}_{2N})
 =
\left[
\BA{cc}
\cos
\left( \frac{\theta_s}{2} \right)
 \cdot
\hat{1}_{N} &
-
\sin
\left( \frac{\theta_s}{2} \right)
 \cdot
\hat{1}_{N}  \vspace{1mm}\\
\sin
\left( \frac{\theta_s}{2} \right)
 \cdot
\hat{1}_{N} &
\cos
\left( \frac{\theta_s}{2} \right)
 \cdot
\hat{1}_{N}
\EA
\right] .
\nonumber
\end{gather}
Using
(\ref{gphi})
and
(\ref{gtheta}),
we obtain an explicit expression for the block matrix
$g_{u,t,s}$
of an $SU(2N)_{\infty }$ transformation-matrix
appeared in
(\ref{transhatgphaipsitheta})
as
\begin{gather}
g_{u,t,s}
 =
\left[
\BA{cc}
\cos
\left( \frac{\theta_t}{2} \right)
e^{-\frac{i}{2}(\psi_u + \varphi_s)}  \cdot  1_{N} &
-
\sin
\left( \frac{\theta_t}{2} \right)
e^{-\frac{i}{2}(\psi_u - \varphi_s)}  \cdot  1_{N} \vspace{1mm}\\
\sin
\left( \frac{\theta_t}{2} \right)
e^{\frac{i}{2}(\psi_u - \varphi_s)}  \cdot  1_{N} &
\cos
\left( \frac{\theta_t}{2} \right)
e^{\frac{i}{2}(\psi_u + \varphi_s)}  \cdot  1_{N}
\EA
\right] \qquad
(\theta_{-t} = \theta_t) .
\label{blockmatrixguts}
\end{gather}
The transformation
(\ref{transhatgphaipsitheta})
is rewritten under the change of index of the block matrix
as follows:
\begin{gather*}
\psi_{Nr + \a}(\hat{g})
=
\sum_{s \in \mathbb{Z}}
\left(
\sum_{t,u \in \mathbb{Z}}\psi_{N(r-s-t-u) + \b }
\right)
(g_{u,t,s})_{\b \a}   \quad
(g_{u,-t,s} = g_{u,t,s}) \nonumber\\
\phantom{\psi_{Nr + \a}(\hat{g})}{} \stackrel{s \to s-t-u}{=}
\sum_{s \in \mathbb{Z}}
\psi_{N(r-s) + \b}
\left(
\sum_{t,u \in \mathbb{Z}}g_{u,t,s-t-u}
\right)_{ \b \a} \nonumber\\
\phantom{\psi_{Nr + \a}(\hat{g})}{}
=
\sum_{s \in \mathbb{Z}}
\psi_{N(r-s) + \b }
\left(g_s \right)_{\b \a}  \quad
\left(
g_s
 \equiv
\sum_{t,u \in \mathbb{Z}}g_{u,t,s-t-u}
\right) .
\end{gather*}
Noting the def\/inition
$
(\hat{g})_{Nr + \a,Ns + \b }
\equiv
(g_{s-r})_{\a \b }
$
in the second equation of
(\ref{CanoTraInfFermionandghat})
and again chan\-ging the index of the block matrix,
the above transformation is further rewritten as
\begin{gather*}
\psi_{Nr + \a}(\hat{g})
=
\sum_{s \in \mathbb{Z}}
\psi_{Ns+\b }(\hat{g})_{Ns + \b, Nr + \a}
=
\sum_{s \in \mathbb{Z}}
\psi_{Ns + \b }(g_{r-s})_{\b \a}
\stackrel{s \to r-s}{=}
\sum_{s \in \mathbb{Z}}
\psi_{N(r-s) + \b }(g_{s})_{ \b \a} \nonumber\\
\phantom{\psi_{Nr + \a}(\hat{g})}{}
=
\sum_{s \in \mathbb{Z}}
\psi_{N(r-s) + \b}
\left(
\sum_{t,u \in \mathbb{Z}}g_{u,t,s-u-t}
\right)_{ \b \a}
\stackrel{t \to t-u}{=}
\sum_{s \in \mathbb{Z}}
\psi_{N(r-s) + \b }
\left(
\sum_{t,u \in \mathbb{Z}}g_{u,t-u,s-t}
\right)_{ \b \a} .
\end{gather*}
Then the matrix element of $\hat{g}$
is also represented as
\begin{gather*}
(\hat{g})_{Nr + \a,Ns + \b }
=
\left(
\sum_{t,u \in \mathbb{Z}}g_{u,t,s-t-u-r}
\right)_{ \a \b }
=
\left(
\sum_{t,u \in \mathbb{Z}}g_{u,t-u,s-t-r}
\right)_{ \a \b } .
\end{gather*}

\section[Explicit expression for Pl\"{u}cker~coordinate and calculation of $\det(1_N + p^\dag p)$\\ in terms of Schur polynomials for LMG model]{Explicit expression for Pl\"{u}cker~coordinate and calculation of $\boldsymbol{\det(1_N + p^\dag p)}$ in terms of Schur polynomials for LMG model}\label{appendixJ}


From
(\ref{Slaterdet4}),
the coset variable $p$
is expressed in terms of Pl\"{u}cker coordinates as
\begin{gather*}
p
=
(p_{ia})
=
\left([S(\zeta) C^{-1} (\zeta)]_{ia}\right)
=
\left(
\frac{
v_{1 , \dots , i , \dots , M}^{1 , \dots , a , \dots , M} (g_\zeta)}
{v_{1 , \dots , M}^{1 , \dots , M} (g_\zeta )}
\right) .
\end{gather*}
Using explicit expression for $SU(2N)$ matrix of the Thouless transformation
(\ref{Thoulesstransformation2}),
the matrix representations of the Pl\"{u}cker coordinates
$v_{1 , \dots , i , \dots , M}^
{1 , \dots , a , \dots , M} (g_\zeta)$
and
$v_{1 , \dots , M}^{1 , \dots , M} (g_\zeta )$
are given simply as
\begin{gather*}
v_{1 , \dots , i , \dots , M}^
{1 , \dots , a , \dots , M} (g_\zeta)
=
\sin \tfrac{\theta }{2} \cdot 1_N ,\qquad
v_{1 , \dots , M}^{1 , \dots , M} (g_\zeta )
=
\cos \tfrac{\theta }{2} \cdot 1_N \qquad (M = N).
\end{gather*}
Then we have
$
p
=
\tan \frac{\theta }{2}  \cdot  1_N
$,
which is identical with the expression given in
(\ref{Thoulessthorem0})
except the phase~$e^{i\psi }$.
Following~\cite{NMO.04},
using the famous formula
\[
\det (1  +  X)
 =
\exp \{ \mbox{Tr}\, \ln (1  +  X) \}
 =
\exp
\left\{
\sum\limits_{l = 1}^{\infty } (-1)^{l - 1} \mbox{Tr}\, (X^l) /l
\right\}
\]
and the Schur polynomials given in Appendix~\ref{appendixD},
we have an expression for
$\det (1 + p^{\dagger } p)$ in
(\ref{freeparticle-holevacuumfunction})
as
\begin{gather}
\det (1  +  p^{\dagger } p)
=
\sum_{l = 0}^{\infty } S_l (\chi) , \qquad
\chi _l
 \equiv
(-1)^{l - 1}\,
\mbox{Tr} [(p^{\dagger } p)^l] /l, \nonumber\\
\chi _l  =  S_l (\chi)
 =
0  \qquad (l \ge  M  +  1).
\label{Schurexpressionfordet}
\end{gather}
Using the formula
\[
[\det (1  +  X)]^{-\frac{1}{2}}
 =
\exp \{ \mbox{Tr}\, \ln (1  +  X)^{-\frac{1}{2}} \}
=
\exp
\left\{
\sum_{l = 1}^{\infty } (-1)^l \, \mbox{Tr}\, (X^l) /(2l)
\right\},
\]
the vacuum function
$\Phi_{M,M}(p,p^*  ,\tau)$
(\ref{freeparticle-holevacuumfunction})
is also expressed in terms of the Schur polynomials~as
\begin{gather}
\Phi_{M,M}(p,p^*  ,\tau)
=
\sum_{l = 0}^{\infty } S_l (\xi)  \cdot
e^{-iN\tau } , \qquad
\xi _l
 \equiv
(-1)^l\,
\mbox{Tr}\, [ (p^{\dagger } p) ]^l /(2l), \nonumber\\
\xi _l  =  S_l (\xi)
 =
0 \qquad (l  \ge  M  +  1).
\label{freeparticle-holevacuumfunction2}
\end{gather}
Rowe et al.\ have showed
the number-projected $SO(2N)$ wave function satisf\/ies some recursion relations.
They express it with the aid of the relations in a form of determinant~\cite{Rowe.91}
which is well known as the completely anti-symmetric Schur function
in the theory of group characters~\cite{Littlewood.58}.
In the present $U(N)$ case,
equation~(\ref{Schurexpressionfordet})
is also given by a determinant form
\begin{gather*}
{\varphi }_{l}(z)
=
\displaystyle \frac{1}{l!}
\det
\left|
\begin{array}{cccccc}
{\chi }_{1}&1&0&0&\cdots&0\\
2{\chi }_{2}&{\chi }_{1}&2&0&\cdots&0\\
3{\chi }_{3}&2{\chi }_{2}&{\chi }_{1}&3&\cdots &0\\
\vdots &\vdots&\vdots&\vdots&\cdots&\vdots\\
\vdots&\vdots&\vdots&\cdots&\cdots&l -1\\
l{\chi }_{l}&(l -1){\chi }_{l -1}&(l -2){\chi }_{l-2}&
(l -3){\chi }_{l -3}&\cdots&{\chi }_{1}
\end{array}
\right|
 =
(-1)^lS_l (- \chi),
\end{gather*}
which is exactly the same form as that given in~\cite{Nishi.99}.
The Schur function ${\varphi }_{l}(\chi)$ satisf\/ies the recursion relation and
the dif\/ferential formula
\begin{gather}
{\varphi }_{l}(\chi)
=
 \frac{1}{l}
\left\{
{{\chi }_{1}
-\sum _{l' =1}^{l -1} (l' + 1){\chi }_{l' + 1}
\frac{\partial }{\partial {\chi }_{l'}}
}
\right\}
{\varphi }_{l -1}(\chi) , \qquad
\frac{\partial }{\partial {\chi }_{l'}}
{\varphi }_{l}(\chi)
=
(-1)^{l' + 1}{\varphi }_{l - l'} (\chi) .
\label{recursionrelation}
\end{gather}
By using the second equation of
(\ref{recursionrelation}),
we can rewrite the above recursion relation as
\begin{gather*}
{\varphi }_{l}(\chi)
=
\frac{1}{l}
\sum_{l' = 1}^{l}
(-1)^{l' + 1} l' {\chi }_{l'} {\varphi }_{l - l'} (\chi)  \qquad ({\varphi }_0 =  1) .
\end{gather*}
With the aid of the second equation in
(\ref{Schurexpressionfordet})
with
$
p
=
\tan \frac{\theta }{2} \!\cdot\! 1_N
$
and the explicit form of the Schur polynomials given in Appendix~\ref{appendixD},
the $\chi_l$ and $S_l (\chi)$ take simple forms, respectively, as
\beq
\chi _l
=
(-1)^{l - 1}
\frac{1}{l} \left(\tan^2 \tfrac{\theta }{2}\right)^l \cdot N ,\qquad
S_l (\chi)
=
 \frac{N!}{l! (N  -  l)!}
1^{N-l}\left( \tan^2 \tfrac{\theta }{2} \right)^l  .
\label{chiandSchurpolynomial}
\eeq
Then using the f\/irst equation in
(\ref{Schurexpressionfordet})
and
(\ref{chiandSchurpolynomial}),
we have
\begin{gather*}
\det (1  +  p^{\dagger } p)
 =
\sum_{l = 0}^{\infty } S_l (\chi)
 =
\sum_{l = 0}^N
 \frac{N!}{l! (N - l)!}
1^{N-l}\left( \tan^2 \tfrac{\theta }{2} \right)^l
=
\left( 1 + \tan^2  \tfrac{\theta }{2} \right)^N
\!=\!
\left( \cos  \tfrac{\theta }{2} \right)^{-2N} ,
\end{gather*}
which coincides with the result by the direct calculation of
$\det (1  +  p^{\dagger } p)$.
The vacuum function~(\ref{freeparticle-holevacuumfunction2})
is obtained as
\begin{gather*}
\Phi_{M,M}(p, p^*,\tau)
=
\left( \cos  \tfrac{\theta }{2} \right)^{N}
e^{-iN\tau } ,\qquad
{\ve}_{ai} \Phi_{M,M}(p, p^*,\tau) = 0 \qquad
((\ref{p-hdifferentialoperators})~\mbox{and}~(\ref{p-hdifferentialoperatorsontovacuum})) .
\end{gather*}

\subsection*{Acknowledgements}

One of the authors (S.N.) would like to
express his sincere thanks to
Professor Alex H.~Blin for
kind and warm hospitality extended to him at
the Centro de F\'\i sica Te\'orica,
Universidade de Coimbra.
This work was supported by
the Portuguese Project POCTI/FIS/451/94.
The authors thank YITP,
where discussion during the YITP workshop(YITP-W-06-13) on
{\it Fundamental Problems and Applications of Quantum Field Theory
``Topological Aspects of Quantum Field Theory'' -- 2006}
was useful to complete this work.
S.N.~also would like to acknowledge partial
support from Projects PTDC/FIS/64707/2006 and CERN/FP/83505/2008.

\LastPageEnding

\end{document}